\documentclass[aps,rmp,groupedaddress,twocolumn]{revtex4}
\usepackage{graphicx,graphics}
\usepackage{amsmath,amssymb,amsfonts}
\usepackage{color}
\usepackage{subfig}
\usepackage{dcolumn,bm}
\usepackage{mathbbol}
\usepackage{comment}
\def\gsim{\buildrel {\textstyle >}\over {_\sim}}
\def\lsim{\buildrel {\textstyle <}\over {_\sim}}

\begin{document}

\title{Statistical Physics of Fracture, Friction and Earthquake}
\author{Hikaru Kawamura}
\email{kawamura@ess.sci.osaka-u.ac.jp}
\affiliation{Department of Earth \& Space Science, Osaka University, Osaka, Japan}
\author{Takahiro Hatano}
\email{hatano@eri.u-tokyo.ac.jp}
\author{Naoyuki Kato}
\email{nkato@eri.u-tokyo.ac.jp}
\affiliation{Earthquake Research Institute, University of Tokyo, Japan} 
\author{Soumyajyoti Biswas}
\email{soumyajyoti.biswas@saha.ac.in}
\author{Bikas K. Chakrabarti}
\email{bikask.chakrabarti@saha.ac.in}
\affiliation{Theoretical Condensed Matter Physics Division and Centre for 
Applied Mathematics and Computational Science, Saha Institute of Nuclear 
Physics, 1/AF Bidhan Nagar, Kolkata 700064, India.}
\begin{abstract}
We review the present status of our research and understanding regarding the dynamics and the statistical properties of earthquakes, mainly from a statistical physical viewpoint. Emphasis is put both on the physics of friction and fracture, which provides a ``microscopic'' basis of our understanding of an earthquake instability, and on the statistical physical modelling of earthquakes, which provides ``macroscopic'' aspects of such phenomena. Recent numerical results on several representative models are reviewed, with attention to both their ``critical'' and ``characteristic'' properties. We highlight some of relevant notions and related issues, including the origin of power-laws often observed in statistical properties of earthquakes,  apparently contrasting features of characteristic earthquakes or asperities, the nature of precursory phenomena and nucleation processes, the origin of slow earthquakes, {\it etc\/}.
\end{abstract}
\maketitle
\tableofcontents
\section{Introduction}
\noindent Earthquakes are large scale mechanical failure phenomena, which have still defied our complete understanding. In this century, we already experienced two gigantic earthquakes: 2004 Sumatra-Andaman earthquake (M9.1) and 2011 East Japan earthquake (M9.0). Given the disastrous nature of the phenomena, the understanding and forecasting of earthquakes have remained to be the most important issue in physics and geoscience (Carlson, Langer and Shaw, 1996; Rundle, Turcotte and Klein, 2000; Scholz, 2002; Rundle et al, 2003; Bhattacharyya and Chakrabarti, 2006; Ben-Zion, 2008; Burridge, 2006; De Rubies et al, 2006; Kanamori, 2009; Daub and Carlson, 2010). Although there is some recent progress in our understanding of the basic physics of fracture and friction, it is still at a primitive stage (Marone, 1998; Scholz, 1998, 2002; Dieterich, 2009; Tullis, 2009; Daub and Carlson, 2010). Furthermore, our lack of a proper understanding of the dynamics of earthquakes poses an outstanding challenge to both physicists and seismologists.

 While earthquakes are obviously complex phenomena, certain empirical laws have been known concerning their statistical properties, {\it e.g.\/}, the Gutenberg-Richter (GR) law for the magnitude distribution of earthquakes, and the Omori law for the time evolution of the frequency of aftershocks (Scholz, 2002; Rundle, 2003; Turcotte, 2009). The GR law states that the frequency of earthquakes of its energy (seismic moment) $E$ decays with $E$ obeying a power-law, {\it i.e.\/}, $\propto E^{-(1+B)}=E^{-(1+\frac{2}{3}b)}$ where $B$ and $b=\frac{3}{2}B$ are appropriate exponents, whereas the Omori law states that the frequency of aftershocks decays with the time elapsed after the mainshock obeying a power-law. These laws, both of which are power-laws possessing a scale-invariance, are basically of statistical nature, becoming evident only after examining large number of events. Although it is extremely difficult to give a definitive prediction for each individual earthquake event, clear regularity often shows up when one measures its statistical aspect for an ensemble of many earthquake events. This observation motivates statistical physical study of earthquakes due to the following two reasons: First,  a law appearing after averaging over many events is exactly the subject of statistical physics. Second, a power law or a scale invariance has been a central subject of statistical physics over years in the context of critical phenomena. Indeed, Bak and collaborators proposed the concept of ``self-organized criticality (SOC)'' (Bak, Tang and Wiesenfeld, 1987). According to this view, the Earth's crust is always in the critical state which is self-generated dynamically (Turcotte 1997; Hergarten, 2002; Turcotte, 2009; Pradhan, Hansen and Chakrabarti, 2010). One expects that such an SOC idea might possibly give an explanation of the scale-invariant power-law behaviors frequently observed in earthquakes, including the GR law and the Omori law. However, one should also bear in mind that real earthquakes often exhibit apparently opposite features, {\it i.e.\/}, the features represented by  ``characteristic earthquakes'' where an earthquake is regarded to possess its characteristic energy or time scale (Scholz, 2002; Turcotte, 2009). 
 
 Earthquakes also possess strong relevance to material science. It is now established that  earthquakes could be regarded as a stick-slip frictional instability of a pre-existing fault, and statistical properties of earthquakes are governed  by the physical law of rock friction (Marone, 1998; Scholz, 1998; 2002, Dieterich, 2009; Tullis, 2009). The physical law describing rock friction or fracture is  often called ``constitutive law''. As most of the major earthquakes are caused by rubbing of faults, such friction laws give the ``microscopic'' basis in analyzing the dynamics of earthquakes.  One might naturally ask: How statistical properties of earthquakes depend on the material properties characterizing earthquake faults, {\it e.g.\/}, the elastic properties of the crust or the frictional properties of the fault, {\it etc\/}. Answering such questions would give us valuable information in understanding the nature of earthquakes.

 In spite of some recent progress, we still do not have precise knowledge of the constitutive law characterizing the stick-slip dynamics of earthquake faults. In fact, law of rock friction is often quite complicated,  depending not just on the velocity or the displacement, but on the previous history and the ``state'' of contact surface, {\it etc\/}.  The rate-and-state friction (RSF) law currently occupies the standard position among friction laws in the field of tectonophysics. Although the RSF law is formulated empirically three decades ago  to account for certain aspects of rock friction experiments (Dieterich, 1979; Ruina 1983), the underlying physics was not known until very recently. While the RSF law shows qualitatively good agreement with numerous experiments, it is only good at aseismic slip velocities (slower than mm/sec).

 Among some progress made recently in the study of friction process, the most fascinating findings might be the rich variety of mechano-chemical phenomena, which comes into play at seismic slip velocities. Another important progress might be the understanding of the friction law of granular matter. This is also a very important point in understanding the friction law of faults as they consist of fine rock powder that are ground up by the fault motion of the past. The investigations on friction phenomenon at seismic slip velocities is now a frontier in tectonophysics. The RSF law no longer applies to this regime, where many mechano-chemical phenomena have been observed in experiments. The most illustrating examples are melting due to frictional heat, thermal decomposition of calcite, silica-gel lubrication and so on. There have not been any friction laws that can describe such varied class of phenomena, which significantly affect the nature of sliding friction.  In this review article, we wish to review the recent development concerning the basic physics of friction and fracture.

 Statistical physical study of earthquakes is usually based on models of various levels of simplification. There are several advantages in employing simplified models in the study of earthquakes. First, it is straightforward in the model study to control various material parameters as input parameters. A systematic field study of the material-parameter dependence of real earthquakes meets serious difficulties, because it is difficult to get precise knowledge of, or even to control, various material parameters characterizing real earthquake faults.
Second, since an earthquake is a large-scale natural phenomenon, it is intrinsically not ``reproducible''. Furthermore, large earthquakes are rare, say, once in hundreds of years for a given fault. If some observations are to be made for a given large event, it is often extremely difficult to see how universal it is and to put reliable error bars to the obtained data. In the model, on the other hand, it is often quite possible to put reliable error bars to the data under well controlled conditions, say,  by performing extensive computer simulations.
An obvious disadvantage of the model study is that the model is not the reality in itself, and one has to be careful in elucidating what aspect of reality is taken into account or  discarded in the model under study.  

 While numerous earthquake models of various levels of simplifications have been studied in the past, one may classify them roughly into two categories: The first one is of the type possessing an equation of motion describing its dynamics where the constitutive relation can be incorporated as a form of ``force''. The so-called spring-block or the Burridge-Knopoff (BK) model, which is a discretized model consisting of an assembly of blocks coupled via elastic springs, belong to this category (Burridge and Knopoff, 1967). Continuum models also belong to this category (Tse and Rice, 1986; Rice, 1993). The second category encompasses further simplified statistical physical models, coupled-lattice models,  most of which were originally introduced as a model of SOC. This category includes the so-called Olami-Feder-Christensen (OFC) model (Olami, Feder and Christensen, 1992), the fiber bundle model (Pradhan, Hansen and Chakrabarti, 2010), and the two fractal overlap model \cite{bkc1,bkc2,bkc26}. These models possesses extremely simplified evolution rule, instead of realistic dynamics and constitutive relation. Yet, one expects that its simplicity enables one to perform exact or precise analysis which might be useful in extracting essential qualitative features of the phenomena. 

 It often happens in practice that, even when the adopted model looks simple in its appearance, it is still highly nontrivial to reveal its statistical properties. Then, the strategy in examining the model properties is often to perform numerical computer simulations on the model, together with the analytical treatment. In this review article, we wish to review the recent developments concerning the properties of these models mainly studied in statistical physics.

 Earthquake forecast is an ultimate goal of any earthquake study. A crucially important ingredient playing a central role there might be various kinds of precursory phenomena. We wish to touch upon the following two types of precursory phenomena in this review article: The first type is a possible change in statistical properties of earthquakes which might occur prior to mainshocks. The form of certain spatiotemporal correlations of earthquakes might change due to the proximity effect of the upcoming mainshock. For example, it has been pointed out that the power-law exponent describing the GR law might change before the mainshock, or a doughnut-like quiescence phenomenon might occur around the hypocenter of the upcoming mainshock, etc. The second type of precursory phenomena is a possible nucleation process which might occur preceding mainshocks (Dieterich, 2009).  Namely, prior to seismic rupture of a mainshock, the fault might exhibit a slow rupture process localized to a compact ``seed'' area, with its rupture velocity orders of magnitude slower than the seismic wave velocity. The fault spends a very long time in this nucleation process, and then at some stage, exhibits a rapid acceleration process accompanied by a rapid expansion of the rupture zone, finally getting into a final seismic rupture of a mainshock. These possible precursory phenomena preceding mainshocks are of paramount importance in their own right as well as in  possible connection to earthquake forecast. We note that similar nucleation process is ubiquitously observed in various types of failure processes in material science and in engineering. 

 The purpose of the present review article is to help researchers link different branches of earthquake studies. First, we wish to link basic physics of friction and fracture underlying earthquake phenomenon to macroscopic properties of earthquakes as a large-scale dynamical instability. These two features should be inter-related as ``input versus output'' or as ``microscopic versus macroscopic'' relation, but its true connection is highly nontrivial and still remains largely unexplored. To understand an appropriate constitutive law describing an earthquake instability and to make a link between such constitutive relations and the macroscopic properties of earthquakes is crucially important in our understanding of earthquakes. Second, we wish to promote an interaction between statistical physicists and seismologists. We believe that the cooperation of scientists in these two areas would be very effective, and in some sense, indispensable in our proper understanding of earthquakes. 

 Recently, there has been some progress made by statistical physicists in characterizing the statistical aspects of the earthquake phenomena. These efforts are of course based on established literature in seismology, physics of fracture and friction. Also, there has been considerable fusion and migration of the scientists and the established knowledge bases between physics and seismology. In this article, we intend to review  the present state of our understanding regarding the dynamics of earthquakes and the statistical physical modelling of such phenomena, starting with the same for fracture and friction.

 The article is organized as follows. In section II, we deal with the basic physics of fracture and friction. After reviewing the classic Griffith theory of fracture in II.A, we review a theory of fracture as a dynamical phase transition in II.B. Rate and state dependent friction (RSF) law is reviewed in II.C, while the recent development beyond the RSF law is discussed in II.D. Section II.E is devoted to some microscopic statistical mechanical theories of friction. In section III, we deal with statistical properties of the model of our first type which includes the spring-block Burridge-Knopoff model (III.A) and the continuum model (III.B). In III.A, we examine statistical properties of earthquakes including precursory phenomena with emphasis on both their critical and characteristic properties, while, in III.B, we mainly examine characteristic properties of earthquakes and various slip behaviors including slow earthquakes. Implications of RSF laws to earthquake physics are also discussed in this section (III.B). In section IV, we deal with statistical properties of our second type of models which include the OFC model (IV.A), the fiber bundle model (IV.B) and the two fractal overlap model (IV.C). 
We also provide a a Glossary of some interdisciplinary terms as
an Appendix.

\section{Fracture and friction}
\subsection{Griffith energy balance and brittle fracture strength of solids}
\noindent In a solid, stress ($\sigma $) and strain ($S$) bear a linear relation in the Hookean region (small stress).
Non-linearity appears for further increase of stress, which finally ends in fracture or failure of the solid. In brittle
solids, failure occurs immediately after the linear region. Hence linear elastic theory can be applied to study this essentially non-linear
and irreversible phenomena. 

The failure process has strong dependence on, among other things, the disorder properties of the material \cite{bkc66}. Often, stress gets concentrated
 around the disorder~\cite{bkc5, bkc6,bkc65} where microcracks are formed. The stress values at the notches and corners of the microcracks can be
several times higher than the applied stress. Therefore the scaling properties of disorder plays an important role in breakdown properties
of solids. Although the disorder properties tell us about the location of instabilities, it does not tell us about when a microcrack 
propagates. For that detailed energy balance study is needed. 

Griffith in 1920, equating the released elastic energy (in an elastic continuum) 
with the energy of the surface newly created (as the crack grows),
arrived at a quantitative criterion for the equilibrium extension of the 
microcrack already present within the stressed material
\cite{bkc27}. The following analysis is valid effectively for two-dimensional stressed solids
with a single pre-existing crack, as for example the case of a large plate with
a small thickness. Extension to three-dimensional solids is straightforward.

\begin{figure}
 \includegraphics[width=3.0cm]{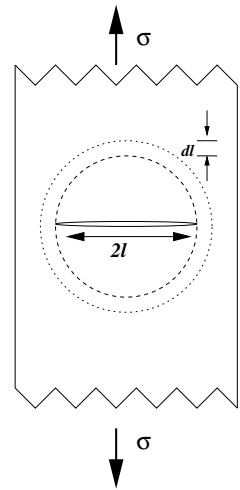}
\caption{A portion of a plate (of thickness $w$) 
under tensile stress $\sigma$ (Model I loading)
containing a linear crack of length $2l$. For a further growth of the
crack length by $2{\rm d}l$, the elastic energy released from the
annular region must be sufficient to provide the surface energy 
$4 \Gamma w {\rm d} l$ (extra elastic energy must be released 
for finite velocity of crack propagation).}
\label{bkc-slitsigma}
\end{figure}

Let us assume a thin linear crack of length $2l$ in an infinite elastic
continuum subjected to uniform tensile stress $\sigma$ perpendicular to the length
of the crack (see Fig. \ref{bkc-slitsigma}). Stress parallel to the crack does not affect the
stability of the crack and has not, therefore, been considered. Because of
the crack (which can not support any stress field, at least on its surfaces),
the strain energy density of the stress field ($\sigma^2 /2Y$; where $Y$ represents the elasticity modulus) 
is perturbed in a region
around the crack, having dimension of the length of the crack. We assume
here this perturbed or stress-released region to have a circular cross-section
with the crack length as the diameter.
The exact geometry of this perturbed region is not important here, and it
determines only an (unimportant) numerical factor in the 
Griffith formula
(see e.g. \citet{bkc5}).
Assuming for the purpose of illustration that half of the stress energy of the annular or cylindrical
volume, having the internal radius $l$ and outer radius $l + {\rm d}l$ 
and length $w$ (perpendicular to the plane of the stress; here the width 
$w$ of the plate is very small compared to the other dimensions), 
to be released as the crack propagates by a length ${\rm d}l$, 
one requires this released strain energy to be sufficient for providing 
the surface energy of the four new surfaces produced. This suggests
\[
\frac{1}{2} (\sigma^2 /2Y) (2 \pi w l {\rm d}l) \ge \Gamma (4 w {\rm d}l).
\]
Here  $\Gamma$ represents the
surface energy density of the solid, measured by the extra energy required
to create unit surface area within the bulk of the solid.

We have assumed here, on average, half of the strain energy of the
cylindrical region having a circular cross-section with diameter $2l$ to be
released. If this fraction is different or the cross-section is different, 
it will
change only some of the numerical factors, in which we are not very much
interested here. Also, we assume here linear elasticity up to the breaking
point, as in the case of brittle materials. The equality holds when energy
dissipation, as in the case of plastic deformation or for the propagation
dynamics of the crack, does not occur. One then gets
\begin{equation}
\label{bkc-sigma_f}
\sigma_f = \frac{\Lambda}{\sqrt{2l}}; \ \ \Lambda = \left( \frac{4}{\sqrt \pi} \right) \sqrt{Y \Gamma}
\end{equation}
for the critical stress at and above which the crack of length $2l$ starts
propagating and a macroscopic fracture occurs. Here $\Lambda$ is called the critical
stress-intensity factor or the fracture toughness. 

In a three-dimensional solid containing a single elliptic disk-shaped planar crack 
parallel to the applied tensile stress direction, a straightforward extension of the above analysis suggests that the maximum stress
concentration would occur at the two tips (at the two ends of the major
axis) of the ellipse. The Griffith stress 
for the brittle fracture
 of the solid
would therefore be determined by the same formula (\ref{bkc-sigma_f}), 
with the crack length $2l$ replaced by the length of the major axis 
of the elliptic planar crack.
Generally, for any dimension therefore, if a crack of length $l$ already
exists in an infinite elastic continuum, subject to uniform tensile stress
$\sigma$ perpendicular to the length of the crack, then for the onset 
of brittle fracture , Griffith equates (the differentials of) the elastic 
energy $E_l$ with the surface energy $E_s$:
\begin{equation}
\label{bkc-E_l}
E_l \simeq \left( \frac{\sigma^2}{2Y}\right) l^d = E_s \simeq \Gamma l^{d-1},
\end{equation}
where $Y$ represents the elastic modulus appropriate for the strain, $\Gamma$
the surface energy density and $d$ the dimension. Equality holds when no energy
dissipation (due to plasticity or crack propagation) occurs and one gets
\begin{equation}
\label{bkc-sigma_f1}
\sigma_f \sim \frac{\Lambda}{\sqrt l}; \ \Lambda \sim \sqrt{Y\Gamma}
\end{equation}
for the breakdown stress at (and above) which the existing crack of length $l$
starts propagating and a macroscopic fracture occurs. It may also be noted
that the above formula is valid in all dimensions ($d \ge 2$).

This quasistatic picture can be extended \cite{bkc31} to fatigue behavior of crack
propagation for $\sigma<\sigma_f$. At any stress $\sigma$ less than $\sigma_f$,
the cracks (of length $l_0$) can still nucleate for a further extension at any finite 
temperature $k_BT$ with a probability $\sim \exp[-E/k_BT]$ and consequently
the sample fails within a failure time $\tau$ given by
\begin{equation}
\tau^{-1}\sim \exp[-E(l_0)/k_BT],
\end{equation}
where
\begin{equation}
E(l_0)=E_s+E_l\sim \Gamma l_0^2-\frac{\sigma^2}{Y}l_0^3
\end{equation}
is the crack (of length $l_0$) nucleation energy. One can therefore express
$\tau$ as
\begin{equation}
\tau \sim \exp[A(1-\frac{\sigma^2}{\sigma_f^2})],
\end{equation}
where (the dimensionless parameter) $A\sim l_0^3\sigma_f^2/(Yk_BT)$ and
$\sigma_f$ is given by Eq.~(\ref{bkc-sigma_f1}). This immediately
suggests that the failure time $\tau$ grows exponentially for $\sigma<\sigma_f$
and approaches infinity if the stress $\sigma$ is much smaller than $\sigma_f$
when the temperature $k_BT$ is small, whereas $\tau$ becomes vanishingly
small   as the stress $\sigma$ exceeds $\sigma_f$; see, e.g., \citet{bkc32} and 
\citet{bkc33}.

For disordered solids, let us model the solid by a percolating system.
For the occupied bond/site concentration $p > p_c$, the percolation threshold, the typical pre-existing cracks in the solid
will have the dimension ($l$) of correlation length $\xi \sim \Delta p^{-\nu}$
and the elastic strength $Y \sim \Delta p^{T_e}$ \cite{bkc7}.
Assuming that the surface energy density $\Gamma$ scales as $\xi^{d_B}$, with
the backbone (fractal) dimension $d_B$ \cite{bkc7}, equating $E_l$
and $E_s$ as in (\ref{bkc-E_l}), one gets 
$\left( \frac{\sigma_f^2}{2Y}\right) \xi^d \sim \xi^{d_B}$.
This gives 
\[
\sigma_f \sim (\Delta p)^{T_f}
\]
with 
\begin{equation}
\label{bkc-T_f}
T_f = \frac{1}{2} [T_e + (d-d_B)\nu]
\end{equation}
for the `average' fracture strength of a disordered solid (of fixed value)
as one approaches the percolation threshold. Careful extensions of
such scaling relations (\ref{bkc-T_f}) and rigorous bounds for $T_f$ has been
obtained and compared extensively in \citet{bkc6,bkc28, bkc29}.
\subsubsection*{Extreme statistics of the fracture stress}
\label{extrm}
\noindent The fracture strength $\sigma_f$ of a disordered solid does not have
self-averaging statistics; most probable and the average $\sigma_f$
may not match because of the extreme
 nature of the statistics. This is
because, the `weakest point' of a solid determines the strength of the
entire solid, not the average weak points!
As we have modelled here, the statistics of clusters of defects are
governed by the random percolation processes.

 We have also
discussed, how the linear responses, like the elastic moduli
of such random networks, can be obtained from the averages
over the statistics of such clusters. This was possible because of the 
self-averaging property of such linear responses.
This is because the elasticity of a random network is determined by
all the `parallel' connected material portions or paths, 
contributing their share in the net
elasticity of the sample.

 The fracture or breakdown property of a 
disordered solid, however, is determined by only the weakest (often the longest) defect
cluster or crack in the entire solid. Except for some indirect effects, most
of the weaker or smaller defects or cracks in the solid do not determine the
breakdown strength of the sample. The fracture 
or breakdown statistics of
a solid sample is therefore determined essentially by the 
extreme statistics
of the most dangerous or weakest (largest) defect cluster or crack within
the sample volume. 

We discuss now more formally the origin of this 
extreme statistics.
Let us consider a solid of linear size $L$, containing $n$ cracks within
its volume. We assume that each of these cracks have a failure probability
$f_i(\sigma), i=1,2,\ldots,n$ to fail or break (independently) under an
applied stress $\sigma$ on the solid, and that the perturbed or
stress-released regions of each of these cracks are separate and do not 
overlap. If we denote the cumulative failure probability of the entire
sample, under stress $\sigma$, by $F(\sigma)$ then \cite{bkc60, bkc6}
\begin{eqnarray}
\label{bkc-1-Fsigma}
1-F(\sigma)&&=\prod_{i=1}^{n} (1-f_i(\sigma)) \simeq 
\exp \left[ -\sum_i f_i(\sigma) \right] \\ \nonumber
&&= \exp \left[ -L^d \tilde{g}(\sigma) \right]
\end{eqnarray}
where $\tilde{g}(\sigma)$ denotes the density of cracks within the sample
volume $L^d$ (coming from the sum $\sum_i$ over the entire volume),
which starts propagating at and above the stress level $\sigma$. The above
equation comes from the fact that the sample survives if each of the cracks
within the volume survives. This is the essential origin of the above
extreme statistical
 nature of the failure probability $F(\sigma)$ of the
sample.

\begin{figure}
\centering\resizebox*{8cm}{!}{\includegraphics{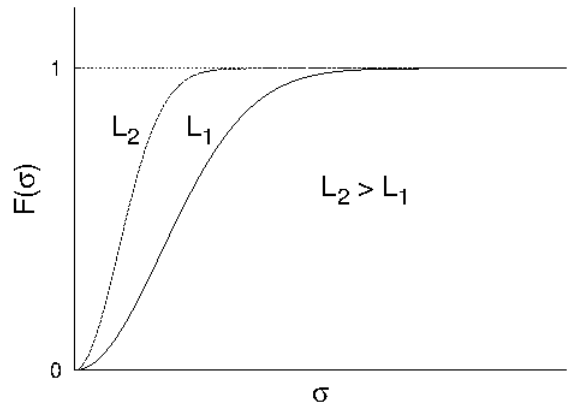}}
\caption{Schematic variation of failure probability $F(\sigma)$ with stress 
$\sigma$ for a disordered solid with volume $L_1^d$ or $L_2^d$ ($L_2 > L_1$).}
\label{bkc-weibullcurve}
\end{figure}

Noting that the pair correlation $g(l)$ of two occupied sites at distance $l$
on a percolation cluster decays as $\exp\left(-l/\xi(p)\right)$, and
connecting the stress $\sigma$ with the length $l$ by using 
Griffith's law
(Eq. (\ref{bkc-sigma_f})) that $\sigma \sim \frac{\Lambda}{l^a}$, one gets
$\tilde{g}(\sigma) \sim \exp \left( -\frac{\Lambda^{1/a}}{\xi \sigma^{1/a}} \right)$ for $p \to p_c$.
On substituting this, Eq. (\ref{bkc-1-Fsigma}) gives the Gumbel distribution  \cite{bkc6}. 
If, on the other hand, one assumes a power law decay of $g(l)$: $g(l) \sim l^{-b}$, then
using the Griffith's law (\ref{bkc-sigma_f}), one gets 
$\tilde{g}(\sigma) \sim \left( \frac{\sigma}{\Lambda} \right)^m$, giving
the Weibull distribution, 
from eqn. (\ref{bkc-1-Fsigma}),
where $m=b/a$ gives the Weibull modulus 
\cite{bkc6}.
The variation of $F(\sigma)$ with $\sigma$ in both the cases
have the generic form shown in Fig.~\ref{bkc-weibullcurve}. $F(\sigma)$ is non-zero for any 
stress $\sigma > 0$ and its value (at any $\sigma$) is higher for larger 
volume ($L^d$).
This is because, the possibility of a larger defect (due to fluctuation)
is higher in a larger volume and consequently, its failure probability
is higher. Assuming $F(\sigma_f)$ is finite for failure, the most probable
failure stress $\sigma_f$ becomes a decreasing function of volume if
extreme statistics is at work.

The precise ranges of the validity of the Weibull or 
Gumbel 
distributions for the breakdown strength of disordered solids are not 
well established yet. However, analysis of the results of detailed 
experimental and numerical studies of breakdown in disordered solids seem 
to suggest that the fluctuations of the extreme statistics
 dominate for 
small disorder \cite{bkc28, bkc29}. Very near to the percolation point,
the percolation statistics takes over and the 
statistics
become self-averaging. One can argue \cite{bkc27}, that arbitrarily 
close to the percolation threshold, the 
fluctuations
of the extreme statistics
 will probably get suppressed and the percolation 
statistics should take over and the most probable breaking stress 
becomes independent of the sample volume (its variation with 
disorder being determined, as in Eqn.(\ref{bkc-T_f}), by an 
appropriate breakdown exponent). This is because the appropriate 
competing length scales for the two kinds of statistics are the
Lifshitz scale $\ln L$ (coming from the finiteness of the volume integral of the
defect probability: $L^d(1 - p)^l$ finite, giving the typical defect size 
$l \sim \ln L$) and the percolation 
correlation length $\xi$. 
When $\xi < \ln L$, the above scenario of extreme statistics should be 
observed. For $\xi > \ln L$, the percolation statistics is expected 
to dominate. 
\subsection{Fracture as dynamical phase transition}
\begin{figure}
\centering \includegraphics[width=8.0cm]{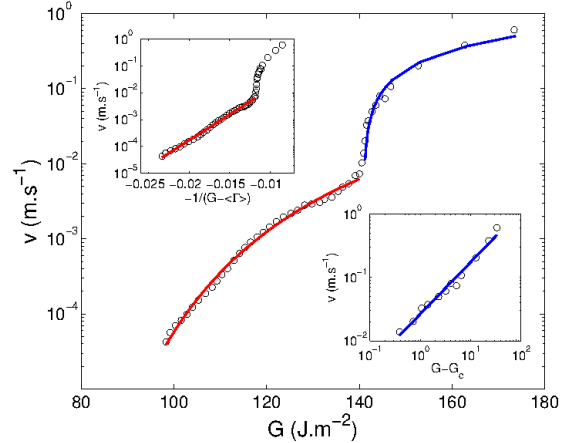}
   \caption{Variation of average crack-front velocity against the mechanical energy release rate is shown for
Botucatu sandstone \cite{bkc15}. The sub-critical creep region and supercritical power-law variations are shown in top-left
and bottom right insets respectively. For the sub-critical regime, the data is fitted with a function $v\sim e^{-C/(G-\langle \Gamma\rangle)^{\mu}}$ for
$\mu=0.60$ and $\langle \Gamma\rangle=65 Jm^{-2}$. For the power-law variation ($v\sim (G-G_c)^{\theta}$) in the super-critical region, $G_c=140 Jm^{-2}$ and $\theta=0.80$. From \cite{bkc15}.
}
\label{expt}
\end{figure}

\noindent When a material is stressed, according to the linear elastic theory discussed above, it develops a proportional amount of strain. 
Beyond a threshold, cracks appear and on further application of stress, the material is fractured as it breaks into pieces. 
In a disordered solid, however, the advancing cracks may be stopped or {\it pinned} by the defect centers present within the material.
So a competition develops between the pinning force due to disorder and the external force. Upto a critical value of the external
force, the average velocity of the crack-front will disappear in the long time limit, i.e., the crack will be pinned. However, if
the external force crosses this critical value, the crack front moves with a finite velocity. This depinning transition can be
viewed as a {\it dynamical critical phenomena} in the sense that near the criticallity universal scaling is observed which
are independent of the microscopic details of the materials concerned \cite{bkc25}. 
The order parameter for this transition is the average velocity
$\overline{v}$ of the crack front. When the external force $f^{ext}$ approaches the critical value $f^{ext}_c$ from a higher value, the order parameter vanishes as
\begin{equation}
\overline{v} \sim (f^{ext}-f^{ext}_c)^{\theta},
\end{equation}  
where $\theta$ denotes the velocity exponent. 
\begin{figure}
\centering \includegraphics[width=9.0cm]{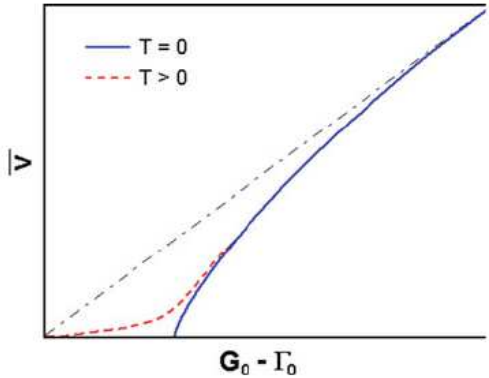}
   \caption{The average velocity of the crack front is plotted against external force ($f=G-\Gamma$, where $G$ is the mechanical
energy release rate and $\Gamma$ is the fracture energy). For $T=0$ the depinning transition is seen. For finite temperature
sub-critical creep is shown \cite{bkc64}. From \cite{bkc64}.}
\label{depinning}
\end{figure}
It is to be mentioned here that the pinning of a crack-front by disorder potential can occur at zero temperature.  At
 finite temperature, there can be healing of cracks due to diffusion or there can be
 sub-critical crack propagation (in the so called creep regime) \cite{bkc25}.
 In the later case, 
the velocity is expected to scale as
\begin{equation}
\overline{v}\sim\exp\left(-C\left(\frac{f^{ext}_c}{f}\right)^{\phi}\right).
\end{equation}
This sub-critical scaling agrees well with experiments \cite{bkc15, bkc16}. 
In Fig.~\ref{expt}, the experimental result of the 
crack propagation in the Botucatu sandstone \cite{bkc15} is shown. The average velocity of the crack is plotted against the mechanical
energy release rate $G$ ($f=G-\Gamma$, where $\Gamma$ is the fracture energy). The subcritical creep regime and the supercritical
power-law variations are clearly seen (insets), which gives the velocity exponent close to $\theta \approx 0.81$ . 

Theoretical predictions of this exponent using Functional Renormalisations Group methods
 have placed its value around $\theta=0.59$ \cite{bkc17}, where the experimental findings differ 
significantly ($\theta\approx0.80\pm0.15$). Here
we mention the numerical study of a model of the elastic crack-front propagation in a disordered solid.
The basic idea is to consider the propagation of the crack front as an elastic string driven through a random medium.
The crack front is characterised by  an array of integral height (measured in the direction of the crack propagation) 
$\{h_1,h_2,\dots,h_L\}$ with periodic boundary conditions,
where the unique values for the height profile suggest that any overhangs in the height profile is neglected. The forces acting on a site can
be written as
\begin{equation}
f_i(t)=f^{el}_i+f^{ext}+g\eta_i(h_i)
\end{equation}
where $f^{el}$ is the elastic force due to stretching, $f^{ext}$ is the applied external force and $\eta$ is due to disorder. 
 The dynamics of the driven elastic chain is then given by the simple rule
\begin{eqnarray}
h_i(t+1)-h_i(t)=v_i(t)&=& 1 \qquad \mbox{if}\quad f_i(t)>0 \nonumber \\
&=& 0 \qquad \mbox{otherwise}.
\end{eqnarray} 
The elastic force may have different forms in various contexts. When this force is short ranged (nearest neighbours) the well studied
models are Edwards -Wilkinson (EW) \cite{bkc39} (see also \citet{bkc53,bkc54}) and KPZ models \cite{bkc40,bkc55,bkc56} (see \cite{bkc41} for extensive analysis). The long range versions includes the ones where the force
decays as inverse square (see e.g., \cite{bkc18}). The velocity exponent $\theta$, as is defined before, turns out to be $0.625\pm0.005$ \cite{bkc18}.
Also, mean field models (infinite range) are studied in this context \cite{bkc42,bkc43,bkc44} (with $\theta=1/2$; exactly).
 An infinite range model, where the elastic force only
depends upon the total stretching of the string, has also been studied recently \cite{bkc45}, where the observed velocity exponent value ($\theta=0.83\pm0.01$) 
is rather close to that found in some experiments \cite{bkc15}.


%
%
%
%
%
%
%
%
%
%
%
%
%
%
%
%
%
%
%
%
%
%

\subsection{Rate- and state-dependent friction law}
\subsubsection{General remarks}
\noindent In a simplified view, an earthquake may be regarded as the rubbing of a fault.
From this standpoint, friction laws of faults play a vital role in understanding and predicting the earthquake dynamics.
In addition, it should be noted that the celebrated Coulomb-Mohr criterion for 
brittle fracture involves the (internal) friction coefficient and thus 
the role of a friction law in earthquake physics is considerable.
In this section, the phenomenology of friction and its underlying physical processes 
are briefly reviewed focusing on the recent developments.
Some recent remarkable progress in experiments shall be also introduced, 
whereas, unfortunately, theoretical understanding of such experiments is rather poor.
Thus, we try to propose the problems to be solved by physicists.

Before explaining the knowledge obtained in the 20th and 21st centuries,
it is instructive to see the ancient  (16-17th centuries) phenomenology, 
which has been referred to as the Coulomb-Amonton's law:
(i) Frictional force is independent of the apparent area of contact.
(ii) Frictional force is proportional to the normal load.
(iii) Kinetic friction does not depend on the sliding velocity and is smaller than 
static friction.
Among these three, the first two laws do not need any modification to this date, 
whereas the third law is to be modified and to be replaced by the celebrated 
rate- and state-dependent friction law, which shall be introduced in the following sections.
The Coulomb-Amonton law in its original form is just a phenomenology involving only 
the macroscopic quantities such as apparent contact area and normal load.
It is generally instructive to consider the sub-level (or microscopic) ingredients 
that underlie such a macroscopic phenomenology.

The essential microscopic ingredient in friction is {\it asperity}, which is a junction of 
protrusions of the surfaces \cite{Bowden2001, Rabinowicz1965}.
In other words, the two macroscopic surfaces in contact are indeed detached 
almost everywhere except for asperities.
The total area of asperities defines the {\it true contact area}, which is generally much 
smaller than the apparent contact area. Thus,  the macroscopic frictional behavior is 
mainly determined by the rheological properties of asperity.
We write the area of asperity $i$ as $A_i$. Then the total area of true contact reads
\begin{equation}
A_{\rm true}=\sum_{i\in{\cal S}} A_i,
\end{equation}
where ${\cal S}$ denotes the set of the asperities. This set depends on the state of 
the surfaces such as topography, and is essentially time dependent because 
the state of the surface is dynamic due to sliding and frictional healing.

Due to the stress concentration at asperities, molecules or atoms are directly pushed 
into contact so that an asperity may be viewed as a grain boundary possibly with some 
inclusions and impurities \cite{Bowden2001, Rabinowicz1965}.
Suppose that each asperity has its own shear strength $\sigma_i$, above which 
the asperity undergoes sliding. It may depend on the degree of grain-boundary misorientation 
and on the amount of impurities at asperity.
For simplicity, however, here we assume $\sigma_i =\sigma_Y$; 
i.e., the yield stress or shear strength of each asperity is the same.
Then the frictional force needed to slide the surface reads
\begin{equation}
\label{truecontact}
F=\sum_{i\in{\cal S}} A_i \sigma_i \simeq \sigma_Y\sum_{i\in{\cal S}} A_i
= \sigma_Y A_{\rm true}.
\end{equation}
The frictional force is thus proportional to the area of true contact.
Dividing Eq. (\ref{truecontact}) by the normal force $N$, one obtains the friction coefficient 
$\mu\equiv F/N$. Using $N=A_{\rm a} P$, where $A_{\rm a}$ is the apparent area of contact 
and $P$ is the normal pressure, one gets
\begin{equation}
\label{frictioncoefficient}
\mu=\sum_{i\in{\cal S}} \frac{A_i}{A_{\rm a}}\frac{\sigma_i}{P}
\simeq \frac{A_{\rm true}\sigma_Y}{A_{\rm a} P}.
\end{equation}
Alternatively, one can have $A_{\rm true} / A_{\rm a} = \mu P / \sigma_Y$.
This means that the fraction of true contact area is proportional to the pressure 
normalized by the yield stress, where the friction coefficient is the proportionality coefficient.
Assuming that the yield stress of asperity is the same as that of the bulk, 
we may set $\sigma_Y\sim 0.01G$, where $G$ is the shear modulus.
Inserting this and $\mu\simeq0.6$ into Eq. (\ref{frictioncoefficient}), one has
 $A_{\rm true} / A_{\rm a} \sim 60 P/G$. This rough estimate can be confirmed 
in experiment and numerical simulation \cite{Dieterich1996,Hyun2004}, 
where the proportionality coefficient is on the order of $10$.
For example, at the normal pressure on the order of kPa, the fraction of true contact is 
as small as $10^{-5}$.

In view of Eq. (\ref{truecontact}), the first two laws of Coulomb-Amonton can be recast
in the form that {\it frictional force is proportional to the true contact area, 
which is independent of the apparent contact area but proportional to the normal load.}
This constitutes the starting point of a theory on friction, which shall be discussed 
in the following subsections.
The third law of  Coulomb-Amonton is just a crude approximation of what we know of today.
This should be replaced by the modern law, which is now referred to as the rate- and state- 
dependent friction (RSF) law.
In the next subsection, we discuss the RSF law based on the first two laws of Coulomb-Amonton.

%
%

\subsubsection{Formulation}
\noindent Extensive experiments on rock friction have been conducted in 1970s and 80s 
in the context of earthquake physics. An excellent review on these experimental works 
is done by Marone (1998).  Importantly, these experiments reveal that kinetic friction 
is indeed not independent of sliding velocity. Thus, the third law of Coulomb-Amonton 
must be modified. Dieterich devises an empirical law that describes the behavior of 
friction coefficient (for both steady state and transient state) based on his experiments 
on rock friction \cite{Dieterich1979}. Later, the formulation is to some extent modified 
by Ruina (1983) by introducing additional variable(s) other than the sliding velocity.
A new set of variables describes the state of the frictional surfaces so that they are 
referred to as the {\it state variables}. Although in general state variable(s) may be 
a set of scalars, in most cases a single variable is enough for the purpose.
Hereafter the state variable is denoted by $\theta'$.
Using the state variable, the friction law reads
\begin{equation}
\label{rsf}
\mu = c' + a' \log \frac{V}{V_*} + b' \log\frac{V_*\theta'}{{\cal L}},
\end{equation}
where $a'$ and $b'$ are positive nondimensional constants, $c'$ is a reference friction coefficient 
at a reference sliding velocity $V_*$, 
and ${\cal L}$ is a characteristic length scale interpreted to be comparable to a typical asperity length.
In typical experiments, $a'$ and $b'$ are on the order of $0.01$, 
and ${\cal L}$ is of the order of micrometers.
Note that the state variable $\theta'$ has a dimension of time.

The state variable $\theta'$ is in general time-dependent so that one must have 
a time evolution law for $\theta'$ together with Eq. (\ref{rsf}).
Many empirical laws have been proposed so far in order to describe time-dependent properties of friction coefficient.
One of the commonly-used equations is the following \cite{Ruina1983}.
\begin{equation}
\label{dieterich}
\dot{\theta'} = 1 - \frac{V}{{\cal L}}\theta', 
\end{equation}
which is now referred to as the Dieterich's law or the aging law.
This describes the time-dependent increase of the state variable even at $V=0$.
Meanwhile, other forms of evolution law may also be possible due to the empirical nature of Eq. (\ref{rsf}).
For example, the following one is also known to be consistent with experiments \cite{Ruina1983}.
\begin{equation}
\label{ruina}
\dot{\theta'} = -\frac{V\theta'}{{\cal L}} \log\frac{V\theta'}{{\cal L}},
\end{equation}
which is referred to as the Ruina's law or the slip law.
In a similar manner, a number of other evolution laws have been proposed so far, such as the composite of the slowness law and the slip law (Kato and Tullis, 2001).

Although there have been many attempts to clarify which evolution law is the most suitable, no decisive conclusions have been made.
As most of them give the identical result if linearized around steady-sliding state,
the difference between each evolution law becomes apparent only at far from steady-sliding state.
One may immediately notice that in Eq. (\ref{ruina}) the state variable is time-independent at $V=0$
so that it is not very quantitative in describing friction processes to which the healing is relevant.
On the other hand, Eq. (\ref{ruina}) can describe a relaxation process after the instantaneous velocity switch
($V=V_1$ to $V_2$) better than Eq. (\ref{dieterich}), while Eq. (\ref{dieterich}) predicts different responses for $V=V_1$ to $V_2$ and for $V=V_2$ to $V_1$, respectively.
(Experimental data suggests that they are symmetric.)
Also, it is known that Eq. (\ref{ruina}) can describe a nucleation process better than Eq. (\ref{dieterich}) 
\cite{Rubin}.
However, we would not go further into the details of the experimental validation of evolution laws 
and leave it to the review by Marone (1998).

Irrespective of the choice of evolution law, a steady state is characterized by $\theta'_{\rm ss}={\cal L}/V$ 
so that the steady-state friction coefficient at sliding velocity $V$ reads 
\begin{equation}
\label{steadystate}
\mu_{\rm ss} = c' + (a'-b') \log\frac{V}{V_*}.
\end{equation}
Note that, as the nondimensional constants $a'$ and $b'$ are typically on the order of $0.01$, 
the velocity dependence of steady-state friction is very small; 
change in sliding velocity by one order of magnitude results in $\sim0.01$ (or even less) 
change in friction coefficient. It is thus natural that people in the 17th century overlooked 
this rather minor velocity dependence.
However, this velocity dependence is indeed not minor at all but very important to the sliding instability problem: e.g., earthquakes.

We also remark that Eq. (\ref{rsf}) together with an evolution law such as Eqs. (\ref{dieterich}) or (\ref{ruina}), well describe the behavior of friction coefficient not only for rock surfaces but also metal surfaces \cite{Popov2010}, two sheets of paper \cite{Heslot1994}, etc.
In this sense, the framework of Eq. (\ref{rsf}) is rather universal.
This universality is partially because the deformation of asperities involves atomistic processes (i.e., creep).
One can assume for creep of asperities $\sigma_Y =  k_BT/\Omega \log(V/V_0)$, where $\Omega$ is an activation volume and $V_0$ is a characteristic velocity involving the activation energy.
Then Eq. (\ref{frictioncoefficient}) leads to
\begin{equation}
\label{creep_RSF}
\mu \simeq \frac{k_BT}{\Omega P_{\rm true}}\log(V/V_0) 
\end{equation}
where $P_{\rm true} = P A_{\rm a}/ A_{\rm true}$ is the actual pressure acting on the asperities.
Comparing Eqs. (\ref{creep_RSF}) and (\ref{rsf}) with $b'=0$ (no healing), one can infer that 
$a'=\frac{k_BT}{\Omega P_{\rm true}}$, as previously derived by some authors 
\cite{Heslot1994,Nakatani2001,Rice2001}.
However, we are unaware of a microscopic expression for $b'$ to this date.
We are also unaware of any microscopic derivations of evolution laws, such as Eqs. (\ref{dieterich}), and (\ref{ruina}), whereas an interesting effort to understand the physical meaning of evolution laws can be found in \cite{Yoshioka1997}.

\subsubsection{Stability of a steady state within the framework of the RSF}
\noindent As we discuss the earthquake dynamics based on the RSF, it is essential to discuss the frictional instability within the framework of the RSF.
For simplicity, we consider a body on the frictional surface.
The body is pulled by a spring at a constant velocity $V$.
\begin{equation}
\label{onebody}
M\ddot{X} = -k (X - Vt) - \mu N,
\end{equation}
where $X$ is the position, $M$ is the mass, $k$ is the spring constant, and $N$ is the normal load. This may be regarded as the simplest model of frictional instability driven by tectonic loading. Suppose that the friction coefficient $\mu$ is given by the RSF law Eq. (\ref{rsf}) together with an evolution law.
The choice of the evolution law, i.e., Dieterich's or Ruina's, does not affect the following 
discussions as they are identical if linearized around steady state.
The motion of the block is uniform in time if the surface is steady-state velocity strengthening 
($a' > b'$) or if the spring constant is sufficiently large.
For a steady-state velocity weakening surface the steady-sliding state undergoes Hopf bifurcation below a critical spring constant.
A linear stability analysis \cite{Ruina1983,Heslot1994} shows that the steady-sliding state 
is unstable if
\begin{equation}
\label{kc}
k < k_{\rm crt} \equiv \frac{N}{{\cal L}} (b'-a').
\end{equation}
This relation plays a central role in various earthquake models, in which a constitutive law is given by the RSF law. This shall be discussed in section III.
An important consequence of Eq. (\ref{kc}) is that the tectonic motion is essentially stable if $a'-b'>0$.
Namely, steady sliding is realized in the region where $a'-b' > 0$, whereas the motion may be unstable if $a'-b'<0$.
In addition, smaller ${\cal L}$ widens the parameter range of unstable motion.

Although the above analyses involve a one-body system, the stability condition Eq. (\ref{kc})
appears to be essentially the same in many-body systems and continuum systems.
Thus, provided that Eq. (\ref{onebody}) applies, it is widely recognized in seismology that seismogenic zone has negative a′ − b′ and smaller L, whereas aseismic zone has the opposite tendency.
\subsection{Beyond the RSF law}
\noindent It should be remarked that the RSF law has certain limit of its application.
Many experiments reveal that the RSF no longer holds at high sliding velocities.
This may be due to various mechano-chemical reactions that are induced by the frictional heat, which typically lubricate surfaces to a considerable degree; the friction coefficient becomes as low as $0.2$ or even less than $0.1$ \cite{Tsutsumi1997, Goldsby2002, DiToro2004, Hirose2005, Mizoguchi2006}.
If such lubrication occurs in a fault, the fault motion is accelerated to a considerable degree and thus such effects have been paid much attention to during the last decade.
Feedback of frictional heat may be indeed very important to faults, because the normal pressure in a seismogenic zone is of the order of $100$ MPa.
(Note that, however, the presence of high-pressure pore fluid may reduce the effective pressure.)
As this area of study is relatively new, the current status of our understanding on such mechano-chemical effects is rather incomplete.
Taking the rapid development of this area into account, here we wish to mention some of the important experiments briefly.

\subsubsection{Flash heating}
\noindent Friction under such high pressure may lead to melting of rock.
There have been some reports of molten rock observed in fault zones, which implies that 
the temperature is elevated up to $2000$ K during earthquakes.

A series of pioneering works on frictional melting in the context of earthquake has been 
conducted by Shimamoto and his coworkers. They devised a facility for rock friction 
at high speed under high pressure to find a behavior very different from that of the RSF law.
Steady-state friction coefficient typically shows remarkable negative dependence 
on sliding velocity and the relaxation to steady state is twofold \cite{Tsutsumi1997,Hirose2005}.
At higher sliding velocity (e.g., $1$ m/sec), friction coefficient decreases as low as 
$0.2$ (or  even less), whereas the typical value at quasistatic regime is around $0.7$.
We wish to stress that such a drastic decrease of friction coefficient cannot be explained 
in terms of the RSF law, where the change of steady-state friction coefficient is 
of the order of $0.01$ even if the sliding velocity changes by a few orders of magnitude.
(recall Eq. (\ref{steadystate}), where $a'$ and $b'$ are both on the order of $0.01$.)
Thus, the mechanism of weakening must be qualitatively different from that of the RSF law.
Indeed, in such experiments, molten rock is produced on surfaces due to the frictional heat. 
It is considered that so produced melt lubricates the surfaces to result in unusually low 
friction coefficient.

In view of Eq. (\ref{truecontact}), frictional melting must take place at asperities,
where the frictional heat is produced. Thus, before the entire surface melts, 
asperities experience very high temperature, which may change the constitutive law.
Such asperity heating has been also known in tribology and is referred to as {\it flash heating}.
Rice applied this idea to fault friction in order to estimate the feasibility of flash heating 
in earthquake dynamics. His argument is as follows \cite{Rice2006}:
The power input to asperity $i$ is $\sigma_Y A_i V$, which is to be stored in the proximity of asperity.
As discussed later, it is essential to assume here that heat conduction is one-dimensional; 
i.e., temperature gradient is normal to the surface, whereas uniform along 
the transverse directions.
The produced heat invades toward the bulk over the distance $\sqrt{D_{\rm th} t}$, where 
$D_{\rm th}$ is thermal diffusivity. Thus, frictional heat is stored in the small volume of 
$A_i\sqrt{\alpha t}$.
Writing the average temperature of this hot volume as $T(t)$, the deposited thermal energy 
reads $c_P \rho A_i\sqrt{D_{\rm th} t}(T(t)- T_0)$, where $c_P$ is the isobaric specific heat, 
$\rho$ is the mass density, and $T_0$ is the ambient temperature.
Then the energy balance leads to 
\begin{equation}
T(t) - T_0 \simeq \frac{\sigma_Y V}{\rho c_P}\sqrt{\frac{t}{D_{\rm th}}}.
\end{equation}
This indicates that the surface temperature increases with time as $\sqrt{t}$.
Writing $T_w$ as the critical temperature above which an asperity looses its shear strength, 
then the duration $t_w$ for the temperature to be elevated up to the critical temperature reads
\begin{equation}
\label{tw}
t_w = D_{\rm th} \left[\frac{\rho c_P(T_w - T_0)}{\sigma_Y V}\right]^2.
\end{equation}
This heating process is limited to the duration or lifetime of an asperity contact.
If we write the longitudinal dimension of each asperity as ${\cal L}_i$, the lifetime of an asperity is estimated as ${\cal L}_i/V$.
Thus, weakening of an asperity occurs if and only if $t_w \le {\cal L}_i/V$.
Taking Eq. (\ref{tw}) into account, this condition may be written as
\begin{equation}
\label{fh_condition}
V \ge \frac{D_{\rm th}}{{\cal L}_i} \left[\frac{\rho c_P(T_w - T_0)}{\sigma_Y}\right]^2.
\end{equation}
Neglecting the statistics of ${\cal L}_i$, one gets the characteristic sliding velocity $V_w$ above which weakening occurs.
\begin{equation}
\label{Vw}
V_w = \frac{D_{\rm th}}{{\cal L}} \left[\frac{\rho c_P(T_w - T_0)}{\sigma_Y}\right]^2.
\end{equation}

Alternatively, from Eq. (\ref{fh_condition}), the maximum size of asperity that does not melt 
at the sliding velocity $V$ is given by
\begin{equation}
\label{Lmax}
{\cal L}_{\rm max} = \frac{D_{\rm th}}{V} \left[\frac{\rho c_P(T_w - T_0)}{\sigma_Y}\right]^2.
\end{equation}
The proportion of non-melting asperity may be approximated by ${\cal L}_{\rm max}/ {\cal L}$.
Assuming that the friction coefficients of a molten asperity and a non-melting one are given as
\begin{equation}
\label{weakening}
\mu=\left\{
\begin{array}{@{\,}ll}
f_1, \ \ (T < T_w) \\
f_2. \ \ (T > T_w),
\end{array}
\right.
\end{equation}
the average friction coefficient reads
\begin{eqnarray}
\mu &=& f_1 \frac{{\cal L}_{\rm max}}{{\cal L}} + f_2 \left(1-\frac{{\cal L}_{\rm max}}{{\cal L}} \right) \\
\label{fh_mu}
&=& f_2 + (f_1-f_2) \frac{V_w}{V}.
\end{eqnarray}
The friction coefficient decreases as $V^{-1}$ at high slip velocity $V\ge V_w$.
Taking $\alpha=1 {\rm mm^2/s}$, $\rho c_P =4$ ${\rm MJ /m}^3K$, $D=5 \mu{\rm m}$, 
$T_w - T_0=700$K, and $\sigma_Y = 0.02$ to $0.1$ G (shear modulus) = $0.6$ to $3$ GPa, 
the characteristic velocity $V_w$ is $0.5$ to $14$ m/s.
This does not contradict rock experiments on melting-induced weakening.
Also, comparison of Eq. (\ref{fh_mu}) with experiments is not inconsistent, although 
$f_1$ and $f_2$ are fitting parameters.

Note that the above discussion does not depend on the apparent normal pressure, as the pressure 
on asperity is approximately the yield stress (of uniaxial compression) irrespective of the apparent normal pressure.
Thus, flash melting could occur in principle even when the apparent pressure is very low 
as long as the sliding velocity is larger than $V_w$ given by Eq. (\ref{Vw}).
However, in an experiment conducted at relatively low pressures, the threshold velocity 
is an order of magnitude smaller than the prediction of Eq. (\ref{Vw}) \cite{Kuwano2011}.
This may be because other relevant mechanisms are responsible for dynamic weakening observed in experiments, but the answer is yet to be given.

It is also important to notice that in the above discussion the assumption of one dimensional heat conduction 
is essential; i.e., the frictional heat is not transferred in the horizontal directions but only to the normal direction.
This assumption implies that the thermal diffusion length $\sqrt{\alpha t_w}$ must be smaller than 
the height of a protrusion that constitutes an asperity.
Assuming that the height of a protrusion is proportional to a horizontal dimension $L_i$, this condition leads to 
\begin{equation}
\label{fh_condition2}
{\cal L}_i \ge \frac{D_{\rm th} \rho c_P (T_w-T_0)}{\sigma_Y V}.
\end{equation}
Because it is estimated in general that $\rho c_P (T_w-T_0) > \sigma_Y$, Eq. (\ref{fh_condition2}) immediately follows from Eq. (\ref{fh_condition}).
Thus, the assumption of one-dimensional heat conduction may be sound.

If the asperities are sufficiently small so that the thermal diffusion length exceeds the height of protrusions, the assumption of one-dimensional heat conduction is violated.
A good example is friction of nanopowders, in which a typical size of the true contact area is on the order of nanometers \cite{Han2011}.
Interestingly, one can still observe dynamic weakening similar to those caused by flash melting, but they did not attribute this behavior to flash melting, because the duration of contact between nano-grains was too short to cause the significant temperature increase. The physical mechanism of such weakening is still not clear.
(Silica-gel lubrication may be ruled out as the material they used is silica-free.)

\subsubsection{Frictional melting and thermal pressurization}
\noindent There is yet another class of weakening phenomena called frictional melting; 
the melt is squeezed out of asperities to fill the aperture between the two surfaces.
Such a situation can occur if the surfaces are rubbed for sufficiently long time.
If this process occurs, the melt layer supports the apparent normal pressure to reduce the effective pressure at asperities and ultimately hinders solid-solid contact. 
This leads to the disappearance of the asperities; i.e., no solid-solid contacts between 
the surfaces but a thin layer of melt under shear.
There are some analyses of such systems assuming the Arrhenius-type viscosity 
\cite{Fialko2005,Nielsen2008}.
In doing so, one can predict the shear traction is proportional to $P^{1/4}$, 
where $P$ is the normal pressure. The quantitative validation of such theories is yet to be done.

It may be noteworthy here that the viscosity of such a liquid film involves rather different problem; nanofluidics.
The melt may be regarded as nanofluid, the viscosity of which may be very different from the ordinary ones.
The shear flow of very thin layers of melt (under very high pressure) may be unstable due to the partial crystallization \cite{Thompson1992}
Till this date, the effect of nanofluidics on frictional melting is not taken into account and left open to physicists.

Meanwhile, the evidence of frictional melting of a fault is not very often found in core samples or in outcrops.
As faults generally contain fluid, frictional heat increases the fluid temperature as well.
As a result, the fluid pressure increases and the effective pressure on solid-solid contact decreases. Therefore, the frictional heat production generally decreases in the presence of fluid.
In the simplest cases where the fault zone is impermeable, the effective friction (and the produced 
frictional heat)  may vanish as the fluid pressure can be as large as the rock pressure \cite{Sibson1973}.
This is referred to as thermal pressurization, and a large number of work has been devoted to such dynamic interaction between frictional heat and the fluid pressure.
More detailed formulations incorporate the effect of fluid diffusion with nonzero permeability of host rocks\cite{Lachenbruch1980,Mase1987}.
In these analyses, the extent of weakening is enhanced if a fault zone has smaller compressibility and permeability.
Although this behavior is rather trivial in a qualitative viewpoint, some nontrivial behaviors are found in a model where the permeability is assumed to be a dynamic quantity coupled with the total displacement \cite{Suzuki2010}.
However, it is generally difficult to judge the validity of a model from observations and thus we do not discuss this problem further.

\subsubsection{Other mechanochemical effects}
\noindent In some systems, anomalous weakening of friction ($\mu\sim0.2$) can be observed 
at sliding velocities much lower than the critical velocity for flash heating (Eq. (\ref{Vw})).
Typically, one can observe weakening at sliding velocity of the order of mm/s.
Thus, there might be mechanisms for drastic weakening other than frictional melting.

Such experiments are typically conducted with complex materials like fault gouge 
taken from a natural fault so that there may be many different mechanisms of 
weakening depending on the specific compositions of rock species.
Among them, the mechanism that might bear some robustness is the lubrication by 
silica-gel production \cite{Goldsby2002,DiToro2004}.
In several experiments on silica-rich rock such as granite, SEM observation of the surfaces 
reveals a silica-gel layer that experienced shear flow.
The generation of silica-gel may be attributed to chemical reactions between silica and 
water in the environment.
This silica-gel is intervened between the surfaces to result in the lubrication of fault.
Although the details of the chemical reactions is not very clear, 
the mechanics of weakening may be essentially the same as that of flash heating and melting,
because in the both cases the cause of weakening is some soft materials (or liquids) that 
are produced by shear and intervened at asperities.
However, in the case of silica-gel formation, the thixotropic nature of silica-gel may 
result in peculiar behaviors of friction, as observed in experiment by Di Toro et al (2004).

In addition, we wish to add several other mechanisms that lead to anomalous weakening.
Han et al. (2007) found friction coefficient as low as $0.06$ in marble under relatively 
high pressure ($1.1$ to $13.4$ MPa) and high sliding velocity ($1.3$ m/s).
Despite the utilization of several techniques for microstructural observation, they could not
observe any evidence for melting such as glass or amorphous texture but only a layer 
of nanoparticles produced by thermal decomposition of calcite due to frictional heating.
Mizoguchi et al. (2006) also found friction coefficient as low as $0.2$ in fault gouge taken from 
a natural fault, where they also could not find any evidences for melting.
To this date, the mechanism of such frictional weakening at higher sliding velocity is not clear.

It might be important to notice that these samples inevitably include a large amount of 
sub-micron grains that are worn out by high-speed friction \cite{Han2007, Hayashi2010}.
They may play an important role in weakening at high sliding velocities.
The grain size distribution of fault gouge is typically well fitted by a power law with 
exponent $-2.6$ to $-3.0$ \cite{Chester1998} so that smaller grains cannot be 
neglected in terms of volume fraction. The exponent appears to be common to laboratory 
\cite{Marone1989} or numerical experiments of wear \cite{Abe2009}.
Rheology of such fractal grains has not been investigated in a systematic manner, 
notwithstanding a pioneering computational work \cite{Morgan1999}.
The influence of grains to friction shall be discussed in detail in the next subsection.
\subsubsection{Effect of the third body: granular friction}
\noindent Previously, we considered the situation where two surfaces were in contact only at asperities.
This is generally not the case if the asperities are worn out to be free particles that are intervened 
between the two surfaces.
In this case, a system can be regarded as granular matter that is sheared by the two surfaces.
The core of a natural fault always consists of powdered rock \cite{Chester1998}, which is produced by the fault motion of the past.
Thus, friction on fault is closely related to the rheology of granular rock.

As is briefly mentioned in the previous subsection, earthquake physics involves a wide range of 
sliding velocities (or shear rate) ranging from tectonic time scale (e.g., nm/s) to coseismic scale (m/s).
It is thus plausible that the rheological properties of granular matter is qualitatively different depending on the range of sliding velocities.
Here we define two regimes for granular friction: quasititatic and dynamic regimes.
In the quasistatic regime, the frictional properties of granular matter is described by the RSF.
However, some important properties will be remarked that are not observed for bare surfaces.
In the dynamic regime, one may expect dynamic strengthening as observed in numerical simulations \cite{GDR2004, DaCruz2005}.
However, at the same time one may also expect weakening due to various mechanochemical reactions \cite{Mizoguchi2006, Hayashi2010}.
The rheological properties that are experimentally observed are determined by the competition of these two ingredients.
Here we review the essential rheological properties of granular matter in these two regimes.

In experiments on quasistatic deformation, friction of granular matter seems to obey the RSF law. However, some important properties that are different from those of bare surfaces should be remarked.
\begin{enumerate}
\item Velocity dependence of steady-state friction appears to be affected by the layer thickness.
In particular, the value of $a'-b'$ in Eq. (\ref{steadystate}) is an increasing function of 
the layer thickness.
\item The value of $a'-b'$ appears to have negative dependence on the total displacement
applied to a system. This is true both for granular matter and bare surfaces.
\item Transient behaviors can be described by either Dieterich's or Ruina's laws, 
as in the case of bare surfaces. The characteristic length in an evolution law is 
proportional to the layer thickness.
\end{enumerate}
These experimental observations are well summarized and discussed in detail by Marone (1998).
We thus shall not repeat them here and just remark the essential points described above.

As to the first point, there is no plausible explanation to this date.
It appears that the second point could be merged into the first point if the effective layer 
thickness (i.e., the width of shear band) decreases as the displacement increases.
However, we wish to remark that it is also true in the case of bare surfaces, 
where the effective layer thickness is not a simple decreasing function of the displacement.
Thus, the second point cannot be explained in terms of the thickness.
The third point indicates that the shear strain is a more appropriate variable than the displacement of the boundary for the description of the time evolution of friction coefficient.
This may be reasonable as the duration of contact between grains is inversely proportional to the shear rate.
However, the derivation of evolution laws (either Dieterich's or Ruina's) from the grain dynamics is not known to this date.
To construct a theory that can explain these three properties based on the nature of 
granular matter is still a challenge to statistical physicists.

Then we discuss the dynamic regime.
Rheology of granular matter in the dynamic regime is extensively investigated in statistical physics \cite{GDR2004, DaCruz2005}.
As to the steady-state friction coefficient, the shear-rate dependence is one of the main interests also in statistical physics.
There are many ingredients that potentially affect the friction coefficient of granular matter: 
the grain shape, degree of inelasticity (coefficient of restitution), friction coefficient 
between grains, the stiffness, pore-fluid, etc.
Shape dependence is very important to granular friction, but theoretical understanding of this effect is still very poor.
Thus, for simplicity, we neglect the shape effect and involve only spherical grains.
Furthermore, we limit ourselves to the effects of shear rate, stiffness, mass and diameter of grains, coefficient of restitution, and intergrain friction.
This means that we neglect time- or slip-dependent deformation of the grain contacts, such as wear \cite{Marone1989} or frictional healing \cite{Bocquet1998}.
The effects of pore-fluid is also neglected; i.e., we discuss only the dry granular matter here.
With such idealization, one can make a general statement on a constitutive law by dimensional analysis.
The friction coefficient of granular matter is formally written as
\begin{equation}
\label{dimension1}
\mu = \mu(P, m, d, \dot{\gamma}, Y, \mu_e, e),
\end{equation}
where $P$ is the normal pressure, $m$ is the mass, $d$ is the diameter, $\dot{\gamma}$ is shear rate, $Y$ is the Young's modulus of grains, $\mu_e$ is the intergrain friction coefficient, and $e$ is the coefficient of restitution.
(It should be noted that one assumes a single characteristic diameter $d$ in Eq. (\ref{dimension1}).)
From the viewpoint of dimensional analysis, the arguments on the left hand side of Eq. (\ref{dimension1}) must be nondimensional numbers.
\begin{equation}
\mu = \mu(I, \kappa, \mu_e, e),
\end{equation}
where $I= \dot{\gamma}\sqrt{m/Pd}$ and $\kappa = Y/P$.
Thus, the friction coefficient of granular matter depends in principle on these four nondimensional parameters.
Many numerical simulations reveal that $\mu$ is rather insensitive to $\kappa$ and $\epsilon$, 
and the shear-rate dependence is mainly described by $I$.
This nondimensional number $I$ is referred to as the inertial number.
Importantly, the dependence on $I$ is positive in numerical simulations \cite{GDR2004, DaCruz2005, Hatano2007}; namely, the shear-rate dependence is positive.
It is important to remark that the negative shear-rate dependence, which is ubiquitously 
observed in experiments, cannot be reproduced in numerical simulation.
This is rather reasonable because the origin of the negative velocity dependence is 
the time dependent increase of true contact, whereas in simulation the parameters are time-independent.

Experiments in the context of earthquake physics are conducted at relatively high pressures at which the frictional heat affects the physical state of granular matter.
In some experiments \cite{Mizoguchi2006, Hayashi2010}, remarkable weakening ($\mu\sim0.1$) is observed.
Because such anomalous behaviors may involve shear-banding as well as various chemical reactions such as thermal decomposition or silica-gel formation, the frictional properties should depend on the detailed composition of rock species contained in granular matter.
These weakening behaviors must be further investigated by extensive experiments.

So far we discussed the steady-state friction coefficient, but the description of transient states are also important in understanding the frictional instability (and earthquake dynamics).
An evolution law for quasistatic regime is indeed essentially the same as that for bare surfaces; namely, aging law or slip law \cite{Marone1998}.
An evolution law in the dynamic regime is well described by the linear relaxation equation even for relatively large velocity change \cite{Hatano2009}.
\begin{equation}
\label{granular_evolution}
\dot{\mu} = - \tau ^{-1} \left[\mu(t) -\mu_{\rm ss}\right],
\end{equation}
where $\mu_{\rm ss}$ is the steady-state friction coefficient that depends on the sliding velocity 
and $\tau$ is the relaxation time.
It is found that $\tau$ is the relaxation time of the velocity profile inside granular matter
and is scaled with $\sqrt{m/Pd}$.
Thus, importantly, the inertial number, which describes steady-state friction may be 
written using $\tau$ as $I\simeq\tau\dot{\gamma}$; i.e., the shear rate multiplied by the velocity relaxation time. 
It may be noteworthy that the inertial number is an example of Deborah number, 
which is in general the internal relaxation time normalized by the experimental time scale.

Note the difference from the conventional evolution law in the framework of the RSF law; 
Eqs. (\ref{dieterich}) and (\ref{ruina}).
It is essential that Eq. (\ref{granular_evolution}) does not contain any length scale but only the time scale.
This means that the relaxation process of high-speed granular friction takes time rather than the slip distance.
However, we wish to stress that the validity of Eq. (\ref{granular_evolution}) is found only in simulation on dry granular matter and is not verified in physical experiment.

\subsection{Microscopic theories of friction}
\noindent Many attempts were made in explaining friction from an atomistic point of view.
Of course, such effort are meaningful only when the surfaces are smooth and the atomistic properties determine friction. 
  This approach has gained importance in
recent years, because of advancement of technology in this field. Due to Atomic Force Microscopy (AFM) etc., sliding surfaces can
now be probed upto atomic scales. Also, present day computers allow large scale molecular dynamics simulation that helps in
understanding the atomic origin of friction. In this approach, the atomic origin of friction forces are investigated (see also \cite{bkc50,bkc48,bkc51}). In this purpose, 
two atomically smooth surfaces are taken and by writing down the equations of motion, friction
forces are calculated. Effect of inhomogeneity, impurity, lubrication and disorder in terms of vacancies of atoms are also 
considered. 

One of the foremost attempts to model friction from atomic origin was of Tomlinson's \cite{bkc19}. In this model, only one atomic layer of the
surfaces in contact are considered. In particular, the lower surface in considered to be rigid and provide a periodic 
(sinusoidal) potential for the upper body. The contact layer of the upper body is modelled by mutually disconnected beads 
(atoms) which are attached elastically to the bulk above. This model is, of course, oversimplified. The main drawback is
that no interaction between the atoms of the upper body is considered. 
\subsubsection{Frenkel Kontorova model}
\noindent Frenkel-Kontorova \cite{bkc20} model overcomes some of these difficulties. In this model the surface of the sliding object is modelled
by a chain of beads (atoms) connected harmonically by springs. The base is again represented by a sinusoidal potential. The Hamiltonian
of the system can, therefore, be written as
\begin{equation}
H=\sum\limits_{i=1}^N[\frac{1}{2}K(x_{i+1}-x_i-a)^2+V(x_i)],
\end{equation} 
where, $x_i$ is the position of the $i$-th atom, $a$ is the equilibrium spacing of the chain and $V(x)=-V_0cos(\frac{2\pi x}{b})$. 
Clearly, there are two competing lengths in this model, viz. the equilibrium spacing of the upper chain ($a$) and 
the period of the substrate potential ($b$). While the first term tries to keep the atoms in their original positions,
the second term tries to bring them in the local minima of the substrate potential. Simultaneous satisfaction of these two
forces is possible when the ratio $a/b$ is commensurate. The chain is then always
pinned to the substrate in the sense that a finite force is always required to initiate sliding. Below 
that force, average velocity vanishes at large time. However, interesting phenomena takes place when the ratio $a/b$ is
incommensurate. In that case, upto a finite value of the amplitude of the substrate potential, the chain remains ``free''. 
In that condition, for arbitrarily small external force, sliding is initiated. The hull function \cite{bkc21} remains analytic. Beyond
the critical value of the amplitude, the hull function is no longer analytic and a finite external force is now required to
initiate sliding. This transition is called the breaking of analyticity transition or the Aubry transition \cite{bkc21} (for extensive details see \citet{bkc49}). 
\subsubsection{Two-chain model}
\noindent The Frenkel-Kontorova model has been generalised in many ways viz., extension in higher dimensions, effect of impurity,
the Frenkel-Kontorova-Tomlinson model and so on (see \citet{bkc48} and references therein). But one major shortcoming of the Frenkel-Kontorova model is that the 
substrate or the surface atoms of the lower substance are considered to be rigidly fixed in their equilibrium position. 
But for the same reason why the upper surface atoms should relax, the lower surface atoms should relax too. In the two chain
model of friction \cite{bkc22} this question is addressed. 
In this model, a harmonically connected chain of atoms is being pulled over another.
The atoms have only one degree of freedom in the direction parallel to the external force. The equations of motion of
the two chains are
\begin{eqnarray}
m_a\gamma_a(\dot{x_i}-\langle\dot{x_i}\rangle)&=& K_a(x_{i+1}+x_{i-1}-2x_i)\nonumber \\
&&+\sum\limits_{j\in b}^{N_b}F_I(x_i-y_j)+F_{ex},
\end{eqnarray}
\begin{eqnarray}
m_b\gamma_b(\dot{y_i}-\langle \dot{y_i}\rangle)&=& K_b(y_{i+1}+y_{i-1}-2y_i)\nonumber \\&+&\sum\limits_{j\in a}^{N_a}F_I(y_i-x_j)
-K_s(y_i-ic_b)
\end{eqnarray}
where, $x_i$ and $y_i$ denotes the equilibrium positions of the upper and lower chain respectively, $m$'s represent the
atomic masses, $\gamma$'s represent the dissipation constant, $K$'s the strength of inter-atomic force and $N$'s the
number of atoms in each chain, $c$'s the lattice spacing, while suffix $a$ denotes upper chain and suffix $b$ denotes the lower chain.
$F_{ex}$ is the external force and $F_I$ is the inter-chain force between the atoms, which is derived from the following potential
\begin{equation}
U_I=-\frac{K_I}{2}\exp(-4(\frac{x}{c_b})),
\end{equation} 
where $K_I$ is the interaction strength. 

It is argued that the frictional force is of the form
\begin{equation}
-\sum\limits_i\sum\limits_j\langle F_I(x_i-y_j)\rangle_t=N_a\langle F_{ex}\rangle_t.
\end{equation}
It is then shown by numerical analysis that the velocity dependence of the kinetic frictional force becomes weaker as the
static friction increases (tuned by different $K$'s). The velocity dependence essentially vanishes when static frictional 
force is increased, giving one of the Amonton-Coulomb laws.

\begin{figure}
\centering \includegraphics[width=6.0cm]{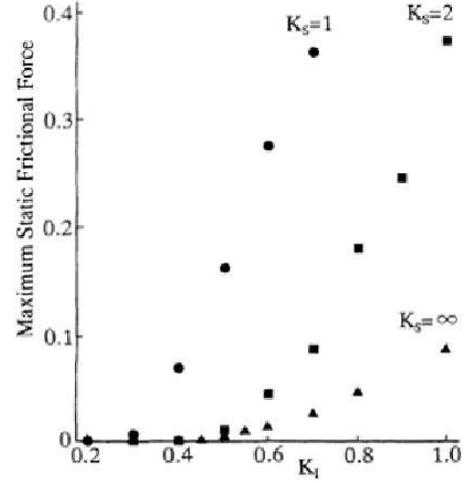}
   \caption{The variations of the maximum static friction with the amplitude ($K_I$) of the inter-chain potential for different values
of the lower-chain stiffness ($K_s$). The limit $K_s\to\infty$ corresponds to Frenkel-Kontorova model. But it is clearly seen that
even for finite $K_s$ (i.e., when the lower chain can relax) there is a finite value of inter-chain potential amplitude upto which
the static friction is practically zero and it increases afterwards, signifying Aubry transition \cite{bkc22}. From \cite{bkc22}.}
\label{twochain-bkc}
\end{figure}

In this case, the lower chain atoms, 
which forms the substrate potential, is no longer rigidly placed. Still, the breaking of analyticity transition is observed.
Fig. \ref{twochain-bkc} shows the variation of the maximum static frictional force with interaction potential strength. For different
values of the rigidity with which the lower chain is bound ($K_s$), different curves are obtained. This indicates a pinned state even for
finite rigidity of the lower chain.  
\subsubsection{Effect of fractal disorder}
\noindent Effects of disorder and impurity have been studied in the microscopic models of friction. Also there have been efforts to
incorporate the effect of self-affine roughness in friction. In Ref.~\cite{bkc23}, the effect of disorder on
static friction is considered. A two-chain version of the Tomlinson model is considered. The self-affine roughness is
introduced by removing atoms and keeping the remaining ones arranged in the form of a Cantor set. The Cantor set, as
is discussed before, is a simple prototype of fractals. Instead of considering the regular Cantor set, here a random version of it is used.
A line segment [0,1] is taken. In each generation, it is divided into $s$ equal segments and $s-r$ of those are randomly removed.
In this way, a self similar disorder is introduced, which is present only in the statistical sense, rather than strict geometric 
arrangement. 
\begin{figure}
\centering \includegraphics[width=6.0cm]{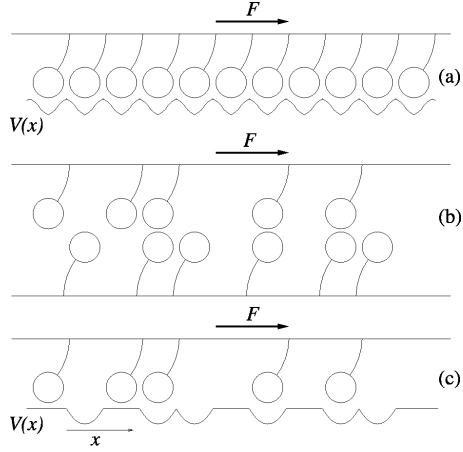}
   \caption{Schematic representation of the two chain version of the Tomlinson model with (a) no disorder, (b) Cantor set 
disorder, (c)the effective substrate potential \cite{bkc23}.}
\label{model}
\end{figure}

\begin{figure}
\centering \includegraphics[width=6.0cm]{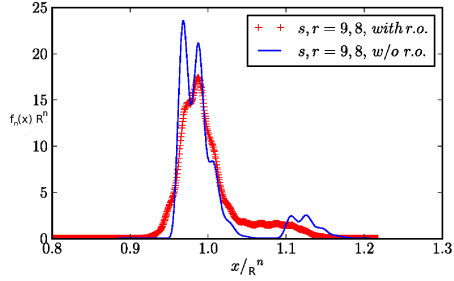}
   \caption{The overlap distribution for $s=9, r=8$ is shown. The dotted curve shows the average distribution with random
off-set and the continuous curve is that without random off set. The distribution is qualitatively different from the Gaussian 
distribution expected for random disorder \cite{bkc23}.}
\label{overlap}
\end{figure}

This kind of roughness is introduced in both the chains. Then the inter-chain interaction is taken to be very short range type.
Only when there is one atom exactly over the other (see Fig.~\ref{model}) there is an attractive interaction. In this way, the maximum static
friction force can be calculated by estimating the overlap of these two chains. It turns out that the static friction force has 
a distribution, which is qualitatively different from what is expected if a random disorder or no disorder is present. The scaled
(independent of generation) distribution of overlap or static friction looks like \cite{bkc23}
\begin{equation}
f^{s,r}(x/R)/R=\sum\limits_{j=1}^r\tilde{c}^{s,r}(j)(f^{s,r}\dots \mbox{$j-1$ terms}\dots f^{s,r})(x),
\end{equation}  
where $R=r^2/s$ and $\tilde{c}^{s,r}(x)=\frac{{}^rC_x{}^{s-r}C_{r-x}}{{}^sC_r}$. For a particular ($s,r$) combination (9,8), the distribution 
function is shown in Fig. \ref{overlap}. It clearly shows that the distribution function is qualitatively different from the Gaussian distribution
expected if the disorder were random.
%
%
%
%
%
%
%
%
%
%
\section{Earthquake models and statistics I: Burridge-Knopoff and Continuum models}
\noindent  In the previous section, we reviewed the basic physics of friction and fracture, which constitutes a ``microscopic'' basis for our study of ``macroscopic'' properties of an earthquake as a large-scale frictional instability. Some emphasis was put on the RSF law now regarded as the standard constitutive law in seismology. In this and following sections, we wish to review the present status of our research on various types of statistical physical models of earthquakes introduced to represent their ``macroscopic'' properties.

\subsection{Statistical properties of the Burridge-Knopoff model}

\subsubsection{The model}

\noindent One of the standard models widely employed in statistical physical study of earthquakes might be the Burridge-Knopoff (BK) model (Rundle, 2003; Ben-Zion, 2008). The model was first introduced in Burridge and Knopoff (1967). Then, Carlson, Langer and collaborators performed a pioneering study of the statistical properties of the model  (Carlson and Langer, 1989a; Carlson and Langer, 1989b; Carlson et al., 1991; Carlson, 1991a; Carlson, 1991b; Carlson, Langer and Shaw, 1994), paying particular attention to the magnitude distribution of earthquake events and its dependence on the friction parameter. 

 In the BK model, an earthquake fault is simulated by an assembly of blocks, each of which is connected via the elastic springs to the neighboring blocks and to the moving plate. Of course, the space discretization in the form of blocks is an approximation to the continuum crust, which could in principle give rise to an artificial effect not realized in the continuum. Indeed, such a criticism against the BK model employing a certain type of friction law, {\it e.g.\/}, the purely velocity-weakening friction law to be defined below in 2[A], was made in the past (Rice, 1993), which we shall return to later. 
%
\begin{figure}[ht]
\begin{center}
\includegraphics[scale=0.45]{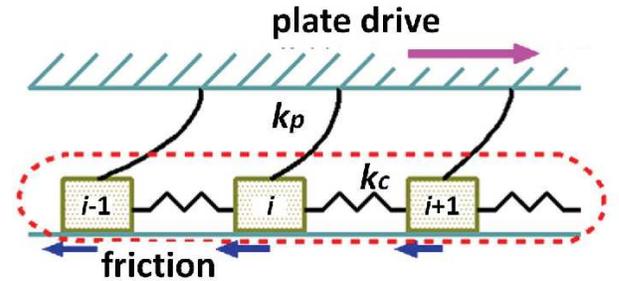}
\end{center}
\caption{
The Burridge-Knopoff (BK) model in one dimension. 
}
\label{BKmodel}
\end{figure}
%

In the BK model, all blocks are assumed to be  subjected to friction force, the source of nonlinearity in the model, which eventually realizes an earthquake-like frictional instability. As mentioned in section II, the standard friction law in modern seismology might be the RSF law. In order to facilitate its computational efficiency, even simpler friction law has also been used in simulation studies made in the past.

 We first introduce the BK model in one dimension (1D). Extension to two dimensions (2D) is straightforward. The 1D BK model consists of a 1D array of $N$ identical blocks, which are mutually connected with the two neighboring blocks via the elastic springs of the elastic constant $k_c$, and are also connected to the moving plate via the springs of the elastic constant $k_p$, and are driven with a constant rate: See Fig.~\ref{BKmodel}. All blocks are subjected to the friction force $\Phi$, which is the only source of nonlinearity in the model. The equation of motion for the $i$-th block can be written as
\begin{equation}
m \ddot U_i=k_p (\nu ' t'-U_i) + k_c (U_{i+1}-2U_i+U_{i-1})-\Phi_i,
\label{eq-1DBK}
\end{equation}
where $t'$ is the time, $U_i$ is the displacement of the $i$-th block, $\nu '$ is the loading rate representing the speed of the moving plate, and $\Phi_i$ is the friction force at the $i$-th block. 

In order to make the equation dimensionless, the time $t'$ is measured in units of the characteristic frequency $\omega =\sqrt{k_p/m}$ and the displacement $U_i$ in units of the length $L^*=\Phi_0/k_p$, $\Phi_0$ being a reference value of the friction force. Then, the equation of motion can be written  in the dimensionless form as
\begin{equation}
\ddot u_i=\nu t-u_i+l^2(u_{i+1}-2u_i+u_{i-1})-\phi_i,
\label{eq-1DBK-normalized}
\end{equation}
where $t=t'\omega $ is the dimensionless time, $u_i\equiv U_i/L^*$ is the dimensionless displacement of the $i$-th block, $l \equiv \sqrt{k_c/k_p}$ is the dimensionless stiffness parameter, $\nu =\nu '/(L^*\omega)$ is the dimensionless loading rate, and $\phi_i\equiv \Phi_i/\Phi_0$ is the dimensionless friction force at the $i$-th block.

The corresponding equation of motion of the 2D BK model is given in the dimensionless form by
\begin{equation}
\begin{array}{ll}
\ddot u_{i,j}=\nu t-u_{i,j}+l^2(u_{i+1,j}+u_{i,j+1} \ \ \ \ \ \\
\ \ \ \ \  +u_{i-1,j}+u_{i,j-1}-4u_{i,j})-\phi_{i,j} ,
\end{array}
\end{equation}
where $u_{i,j}\equiv U_{i,j}/L^*$ is the dimensionless displacement of the block ($i,j$). It is assumed here that the displacement of each block occurs only along the direction of the plate drive. The motion perpendicular to the plate motion is neglected. 

 Often (but not always), the  motion in the direction opposite to the plate drive is also inhibited by imposing an infinitely large friction for $\dot u_i<0$ (or $\dot u_{i,j}<0$) in either case of 1D or 2D. It is also often assumed both in 1D and 2D that the loading rate $\nu$ is infinitesimally small, and put $\nu=0$ during an earthquake event, a very good approximation for real faults (Carlson et al., 1991). Taking this limit ensures that the interval time during successive earthquake events can be measured in units of $\nu^{-1}$ irrespective of particular values of $\nu$. Taking the $\nu \rightarrow 0$ limit also ensures that, during an ongoing event, no other event takes place at a distant place independently of this ongoing event. 
\subsubsection{The friction law}

\noindent The friction force $\Phi$ causing a frictional instability is a crucially important element of the model. Here, we refer to the following two forms for $\Phi$; [A] a velocity weakening friction force (Carlson and Langer, 1989a), and [B] a rate-and-state dependent friction (RSF) law (Dieterich, 1979; Ruina, 1983; Marone, 1998; Scholz, 1998; Scholz, 2002).

\medskip\noindent
[A] In this velocity-weakening friction force, one simply assumes that the friction force $\phi=\phi(\dot u_i)$ is a unique function of the block velocity $\dot u_i$. In order for the model to exhibit a frictional instability corresponding to earthquakes, one needs to assume a velocity-weakening force, {\it i.e.\/}, $\phi (\dot u_i)$ needs to be a decreasing function of $\dot u_i$. The detailed form of $\phi(\dot u_i)$ would be irrelevant. The form originally introduced by Carlson and Langer has widely been used in many subsequent works, that is (Carlson et al, 1991), 
\begin{equation}
\phi(\dot u_i) = \left\{ 
             \begin{array}{ll} 
             (-\infty, 1],  & \ \ \ \ {\rm for}\ \  \dot u_i\leq 0, \\ 
              \frac{1-\delta}{1+2\alpha \dot u_i/(1-\delta )}, &
             \ \ \ \ {\rm for}\ \  \dot u_i>0, 
             \end{array}
\right.
\end{equation}
where the maximum value corresponding to the static friction has been normalized to unity. This normalization condition $\phi(\dot u_i=0)=1$ has been utilized to set the length unit $L^*$. The friction force is characterized by the two parameters, $\delta$ and $\alpha$. The former, $\delta$, introduced in (Carlson et al.,1991) as a technical device facilitating the numerics of simulations, represents an instantaneous drop of the friction force at the onset of the slip, while the latter, $\alpha$, represents the rate of the friction force getting weaker on increasing the sliding velocity. As emphasized by Rice (Rice, 1993), this purely velocity-weakening friction law applied to the discrete BK model did not yield a sensible continuum limit. To achieve the sensible continuum limit, one then needs to introduce an appropriate short-length cutoff by introducing, {\it e.g.\/}, the viscosity term as was done in (Mayers and Langer, 1993): See also the discussion below in subsecton 6.

 We note that, in several simulations on the BK model, the slip-weakening friction force (Ida, 1972; Shaw, 1995; Myers et al., 1996), where the friction force is assumed to be a unique function of the slip distance $\phi(u_i)$,  was utilized instead  of the velocity-weakening friction force. Statistical properties of the corresponding BK model, however, seem not so different from those of the velocity-weakening friction force.  

 Real constitutive relations is of course more complex, neither purely velocity-weakening nor slip-weakening. As discussed in section II, the RSF friction law was introduced to account for such experimental features, which we now refer to.

\medskip\noindent
[B] From Eq. (\ref{rsf}), friction force in the BK model is given  by
\begin{eqnarray}
\phi_i=\{c'
 +a'\log(\frac{v'_i}{v'_*})
 +b'\log\frac{v'_* \theta'_i}{{\cal L}}\}{\cal N},
\end{eqnarray}
where ${\cal N}$ is an effective normal load. See section II.C for the other quantities and parameters.
Among the several evolution laws, we use the aging (slowness) law (Eq.(\ref{dieterich})).
\begin{eqnarray}
\frac{d\theta'_i}{dt'}=1-\frac{v'_i\theta'_i}{{\cal L}}.
\label{slowness}
\end{eqnarray}
Under the evolution law above, the state variable $\theta'_i$ grows linearly with time at a complete halt $v'_i=0$ reaching a very large value just before the seismic rupture, while it decays very rapidly during the seismic rupture.

 The equation of motion can be made dimensionless by taking the length unit to be the characteristic slip distance ${\cal L}$ and the time unit to be the rise time of an earthquake $\omega^{-1}=(m/k_p)^{1/2}$. Then, one has,  
\begin{eqnarray}
\frac{d^2u_i}{dt^2} &=& (\nu t-u_i)+
l^2(u_{i+1}-2u_i+u_{i-1}) \nonumber \\ 
&-& (c+a\log(v_i/v^*)+b\log (v^*\theta_i))
\label{eq-eqmotion1} \\
\frac{d\theta_i}{dt} &=& 1-v_i\theta_i
\label{eq-eqmotion2}
\end{eqnarray}
where the dimensionless variables are defined by $t=\omega t'$, $u_i=u'_i/{\cal L}$,  $v_i=v'_i/({\cal L}\omega )$, $v^*=v'_*/({\cal L}\omega )$, $\theta_i=\omega \theta'_i$, $\nu=\nu'/({\cal L}\omega )$, $a=a'{\cal N}/(k_p{\cal L})$, $b=b'{\cal N}/(k_p{\cal L})$, $c=c'{\cal N}/(k_p{\cal L})$, while $l \equiv (k_c/k_p)^{1/2}$ is the dimensionless stiffness parameter defined above. In some numerical simulations, a slightly different form is used for the $a$-term, where the factor inside the $a$-term, $v/v^*$, is replaced by $1+(v/v^*)$, {\it i.e.\/},
\begin{eqnarray}
\frac{d^2u_i}{dt^2} &=& (\nu t-u_i)+
l^2(u_{i+1}-2u_i+u_{i-1}) \nonumber \\ 
&-&  (c+a\log(1+\frac{v_i}{v^*})+b\log \theta_i),
\label{eq-eqmotion1'}
\end{eqnarray}
where the constant factor $c$ in Eq.(\ref{eq-eqmotion1'}) is shifted by $b\log v^*$ from $c$ in Eq.(\ref{eq-eqmotion2}).

This replacement enables one to describe the system at a complete halt, whereas, without this replacement, the system cannot stop because of the logarithmic anomaly occurring at $v=0$. Similar replacement is sometimes made also for the $b$-term, {\it i.e.\/}, $\theta$ to $1+\theta$.

 The values of various parameters of the model describing natural faults were estimated  (Ohmura and Kawamura, 2007). Typically, $\omega ^{-1}$ corresponds to a rise time of an earthquake event and is estimated to be a few seconds from observations. Though the characteristic slip distance ${\cal L}$ remains to be largely ambiguous, an estimate  of order a few mm or cm was given  by Tse and Rice (Tse and Rice, 1986) and by Scholz (Scholz, 2002). The loading rate associated with the plate motion is typically a few cm/year, and the dimensionless loading rate $\nu=\nu '/({\cal L}\omega)$ is of order $\nu \simeq 10^{-8}$. The dimensionless quantity $k_p{\cal L}/{\cal N}$ was roughly estimated to be of order $10^{-4}$. The dimensionless parameter $c$ should be of order $10^3 \sim 10^4$, and the $a$ and $b$ parameters are one or two orders of magnitude smaller than $c$.
\subsubsection{The 1D BK model with short-range interaction}

\noindent The simplest version of the BK model might be the 1D model with only nearest-neighbor inter-block interaction. Since this model was reviewed in an earlier RMP review article by Carlson, Langer and Shaw, 1994, we keep the discussion here to be minimum, focusing mainly on recent results obtained after the above review article.

 Earlier studies on the 1D BK model have revealed that, while smaller events persistently obeyed the GR law, {\it i.e.\/}, staying critical or near-critical, larger events exhibited a significant deviation from the GR law, being off-critical or ``characteristic'' (Carlson and Langer, 1989a; Carlson and Langer, 1989b; Carlson et al., 1991; Carlson, 1991a; Carlson, 1991b; Schmittbuhl, Vilotte and Roux, 1996).

\begin{figure}[ht]
\begin{center}
\includegraphics[scale=0.65]{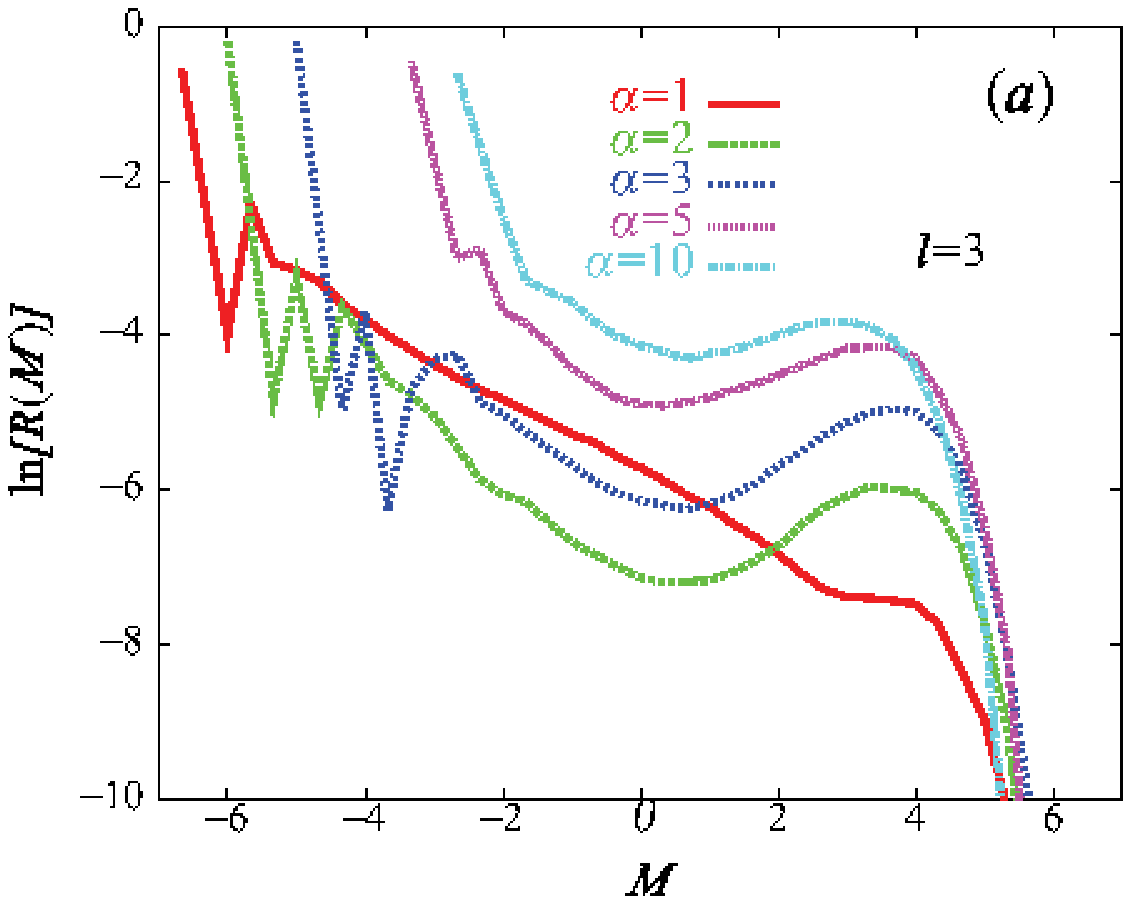}
\includegraphics[scale=0.65]{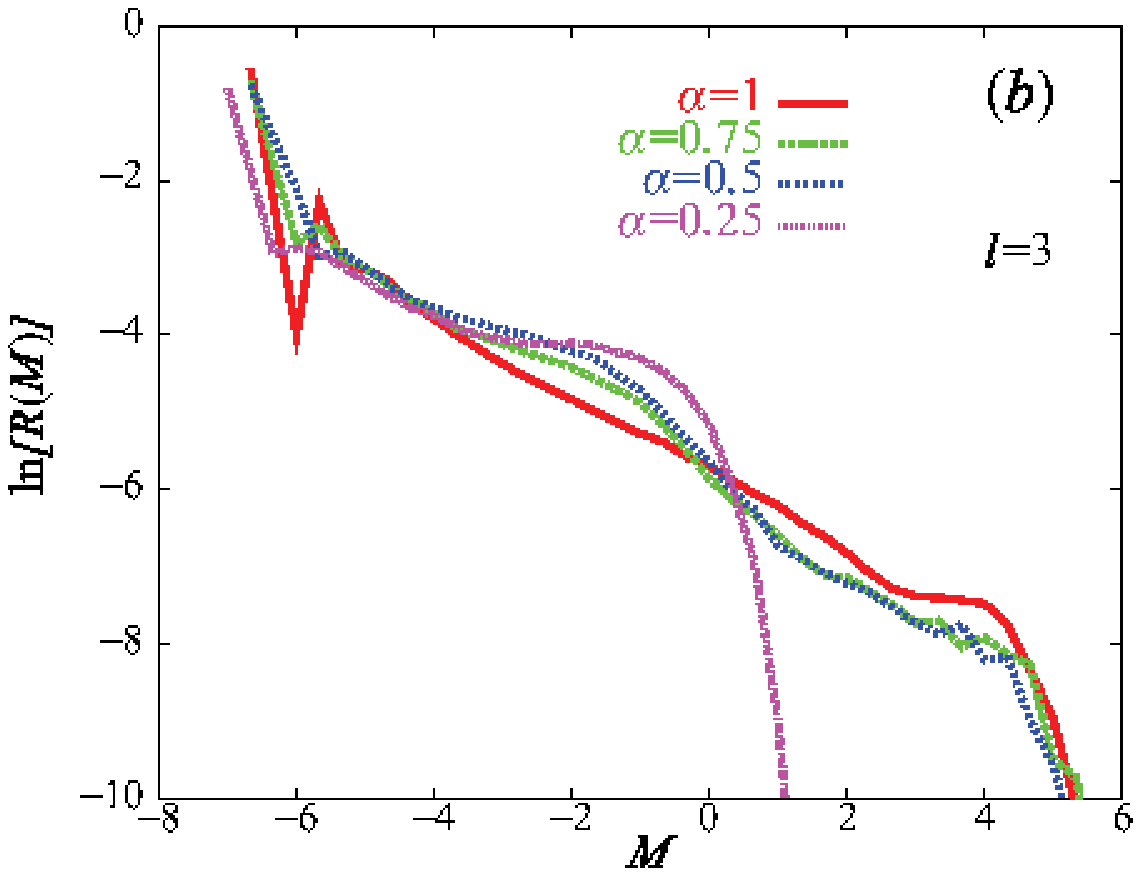}
\end{center}
\caption{
The magnitude distribution of earthquake events of the 1D BK model with nearest-neighbor interaction for various values of the friction parameter $\alpha$; (a) for larger $\alpha =1,2,3,5$ and 10, and (b) for smaller $\alpha =0.25, 0.5, 0.75$ and 1. The parameters $l$ and $\delta$ are fixed to be $l=3$ and $\delta=0.01$. The system size is $N=800$. Taken from (Mori and Kawamura, 2006).
}
\label{magnitude-1DBK}
\end{figure}

 In Fig.~\ref{magnitude-1DBK}, we show the recent data of the magnitude distribution (Mori and Kawamura, 2005; 2006). The magnitude of an event, $M$, is defined by
\begin{equation}
M = \ln \left( \sum_i \Delta u_i \right).
\end{equation}
where the sum is taken over all blocks involved in the event.

%
%

As can be seen from Fig.~\ref{magnitude-1DBK}, the form of the calculated magnitude distribution $R(M)$ depends on the value of the velocity-weakening parameter $\alpha$. The data  for  $\alpha =1$ lie on a straight line fairly well, apparently satisfying the GR law, which may be called ``near-critical'' behavior. The values of the exponent $B$ describing the power-law behavior is estimated to be $B\simeq 0.50$ corresponding to the $b$-value, $b=\frac{3}{2}B\simeq 0.75$. By contrast, the data for larger $\alpha$ deviate from the GR law at larger magnitudes, exhibiting a pronounced peak structure, while the power-law feature still remains for smaller magnitudes: See Fig.~\ref{magnitude-1DBK}(a). These features of the magnitude distribution were observed in many simulations in common (Carlson and Langer,1989a; Carlson and Langer, 1989b; Carlson et al, 1991). It means that, while smaller events exhibit self-similar critical properties,  larger events tend to exhibit off-critical or characteristic properties, which may be called ``supercritical''. Te data for smaller $\alpha <1$ exhibit still considerably different behaviors from those for $\alpha >1$. Large events are rapidly suppressed, which may be called ``subcritical'' behavior. For $\alpha =0.25$, in particular, all events consist almost exclusively of small events only: See Fig.~\ref{magnitude-1DBK}(b). Here the words ``critical'', ``supercritical'' and ``subcritical'' have been defined on the basis of the shape of the magnitude-frequency relationship.
 
\begin{figure}[ht]
\begin{center}
\includegraphics[scale=0.35]{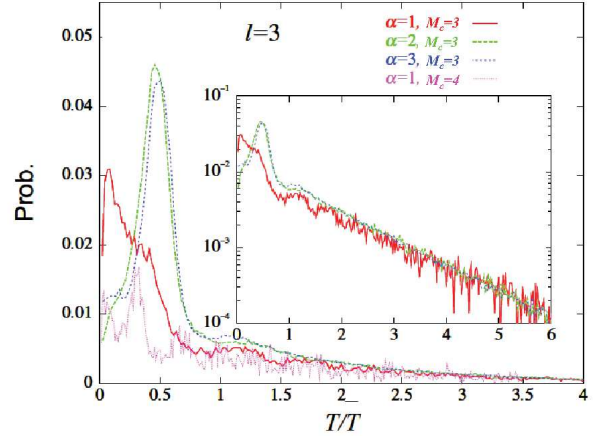}\end{center}
\caption{
The local recurrence-time distribution of the 1D BK model with nearest-neighbor interaction for various values of the frictional parameter $\alpha$. Large events of $M > M_c=3$ or 4 are considered. The parameters are $l$ and $\delta$ are $l=3$ and $\delta=0.01$. The recurrence time $T$ is normalized by its mean $\bar T$. The total number of blocks is $N=800$. The insets represent the semi-logarithmic plots including the tail part of the distribution.  Taken from (Mori and Kawamura, 2005).
\label{recurrence-1DBK}
}
\end{figure}

  As an example of properties other than the magnitude distribution, we show in Fig.~\ref{recurrence-1DBK} the recurrence-time distribution (Mori and Kawamura, 2005; 2006). The recurrence time $T$ is defined here locally for large earthquakes with $M \geq M _c=3$ or $M_c=4$, {\it i.e.\/}, the subsequent large event is counted when a large event occurs with its epicenter in the region within 30 blocks from the epicenter of the previous large event. As can be seen from the figure, the tail of the distribution is exponential at longer $T$ irrespective of the value of $\alpha$. Such an exponential tail of the distribution has also been reported for real seismicity (Corral, 2004). By contrast, the distribution at shorter $T$ is non-exponential and largely differs between for $\alpha =1$ and for $\alpha >1$. For $\alpha>1$, the distribution has an eminent  peak corresponding to a characteristic recurrence time, which suggests the near-periodic recurrence of large events. Such a near-periodic recurrence of large events was reported for several real faults (Nishenko and Buland, 1987; Scholz, 2002). For $\alpha =1$, by contrast, the peak located close to the mean $\bar T$ is hardly discernible. Instead, the distribution has a pronounced peak at a shorter time, just after the previous large event. In other words, large events for $\alpha=1$ tend to occur as ``twins''. A large event for the case of $\alpha=1$ often occurs as a ``unilateral earthquake'' where the rupture propagates only in one direction, hardly propagating in the opposite direction.  

\begin{figure}[ht]
\begin{center}
\includegraphics[scale=0.65]{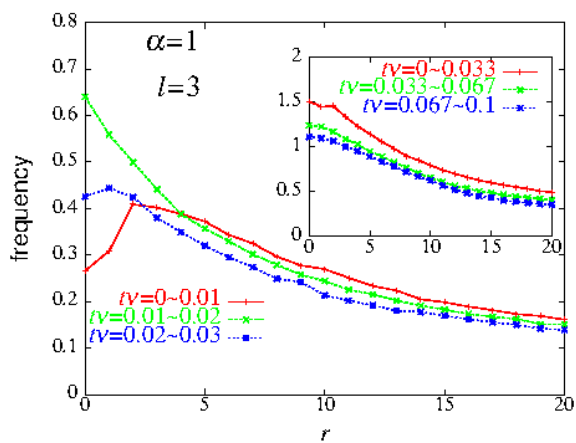}
\end{center}
\caption{The event frequency preceding the large event of $M >M_c=3$ versus the distance from the epicenter of the upcoming mainshock of the 1D BK model with nearest-neighbor interaction.  The parameters $\alpha$, $l$ and $\delta$ are $\alpha=1$, $l=3$ and $\delta=0.01$. The data are shown for several time periods before the mainshock. The insets represent similar plots with longer time intervals. The system size is $N=800$. Taken from (Mori and Kawamura, 2006).
}
\label{mogidoughnut-1DBK}
\end{figure}

 Possible precursory phenomenon exhibited by the model is of much interest, since it might open a way to an earthquake forecast.  In fact, certain precursory features were observed in the 1D BK model. Shaw, Carlson and Langer examined the spatio-temporal patterns of seismic events preceding large events, observing that the seismic activity accelerates as the large event approaches (Shaw, Carlson and Langer, 1992). Mori and Kawamura observed that the  frequency of smaller events was gradually enhanced preceding the mainshock, whereas, just before the mainshock, it is suppressed in a close vicinity of the epicenter of the upcoming event (Mori and Kawamura, 2005; 2006), a phenomenon closely resembling the ``Mogi doughnut'' (Mogi, 1969; 1979; Scholz, 2002).  Fig.~\ref{mogidoughnut-1DBK} represents the space-time correlation function between the large events and the preceding events of arbitrary size (dominated in number by smaller events): It represents the conditional probability that, provided that a large event of $M >M_c =3$ occurs at a time $t_0$ and at a spatial point $r_0$, an event of arbitrary size occurs at a time $t_0-t$ and at a spatial point $r_0\pm r$. As can be seen from the inset of Fig.~\ref{mogidoughnut-1DBK}, seismic activity is gradually accerelated toward the mainshock either spatially or temporally. As can be seen from the main panel, however, the seismic activity is supressed just before the mainshock in a close vicinity of the epicenter of the mainshock: See the dip developing around $r=0$ for $t\leq 0.01$.

 It turned out that the size of the quiescence region was always of only a few blocks, independent of the size of the upcoming mainshock (Mori and Kawamura, 2006). This may suggest that the quiescence is closely related to the discrete nature of the BK model: See subsection III.A.6 below. As such,  the size of the quiescence region cannot be used in predicting the size of the upcoming mainshock. Instead, certain correlation was observed between the size of the upcoming mainshock and the size of the seismically active ``ring'' region surrounding the quiescence region (Pepke, Carlson and Shaw, 1994; Mori and Kawamura, 2006). Such a correlation was also reported in real seismic catalog (Kossobokov and Carlson, 1995).

%
%

 An aftershock sequence obeying the Omori law, although a common observation in real seismicity, is not observed in the BK model, at least in its simplest version (Carlson and Langer, 1989a, 1989b; Mori and Kawamura, 2006). Interestingly, Pelletier reported that the inclusion of the viscosity effect in the form of ``dashpot'' in the 2D BK model, together with the introduction of inhomogeneity of friction parameters, could realize an aftershock sequence obeying the Omori law (Pelletier, 2000). The frictional force employed by Pelletier was a very simple one, {\it i.e.\/}, a constant dynamical vs. static friction coefficient. Further analysis will be desirable to establish the occurrence of the aftershock sequence obeying the Omori law in the BK model.

 We note in passing that the 1D BK model has also been extended in several ways, {\it e.g.\/}, taking account of the effect of the viscosity (Myers and Langer, 1993; Shaw, 1994; De and Ananthakrisna, 2004; Mori and Kawamura, 2008b), modifying the form of the friction force (Myers and Langer, 1993; Shaw, 1995; Cartwright, 1997; De and Ananthakrisna, 2004), and driving the system only at one end of the system (Vieira, 1992; 1996a). The effect of the long-range interactions introduced between blocks was also analyzed, which we will review in  subsection III.A.4. 
\subsubsection{The 2D BK model with short-range interaction}

\noindent  Real earthquake faults are 2D rather than  1D. Hence, it is clearly desirable to study the 2D version of the BK model in order to further clarify the statistical properties of earthquakes. The 2D BK model taken up here is to be understood as representing a 2D fault plane itself, where the direction orthogonal to the fault plane is not considered explicitly in the model (Carlson, 1991b). The other possible version is the one where the second direction of the model is taken to be orthogonal to the fault plane (Myers et al, 1996). 

\begin{figure}[ht]
\begin{center}
\includegraphics[scale=0.5]{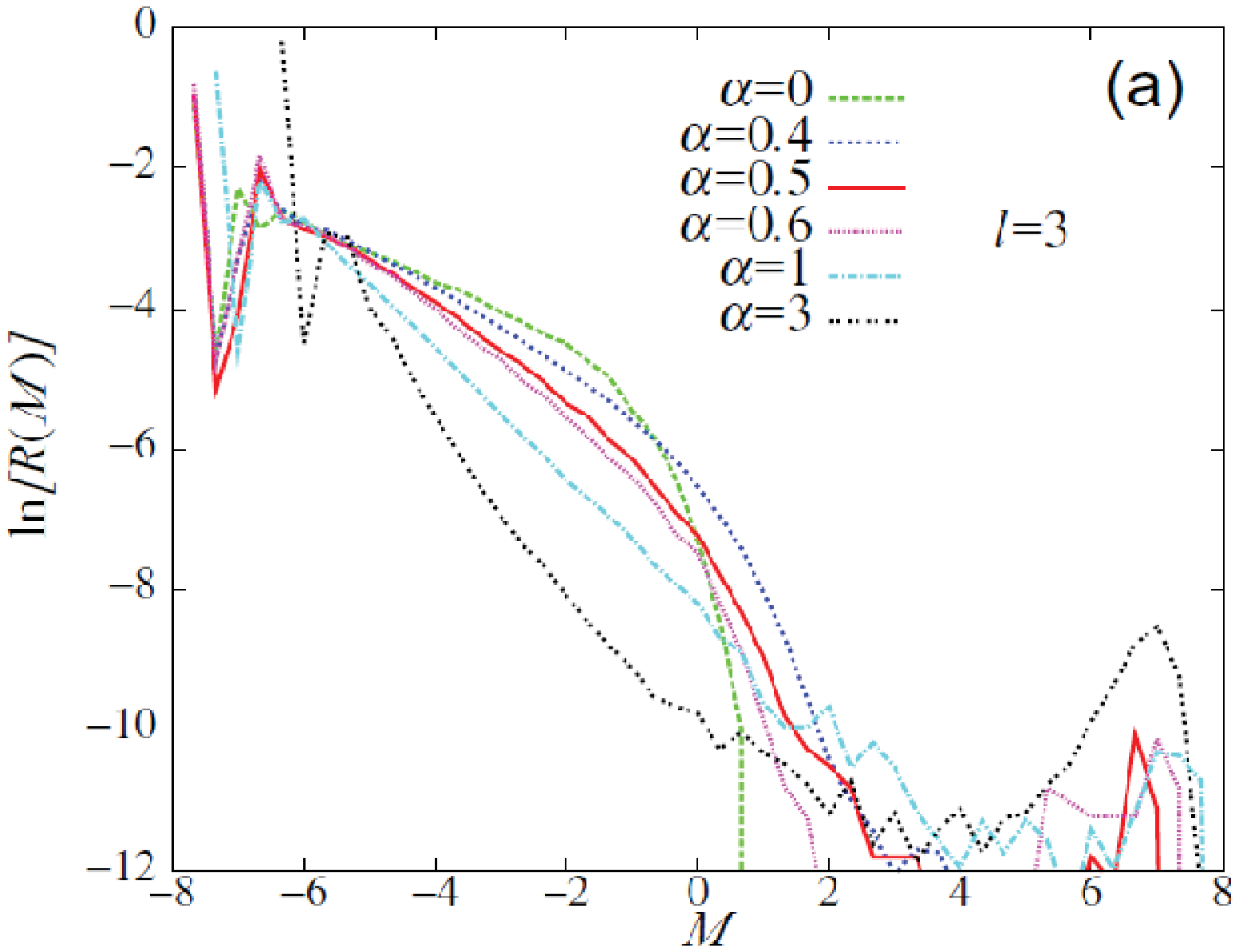}
\includegraphics[scale=0.5]{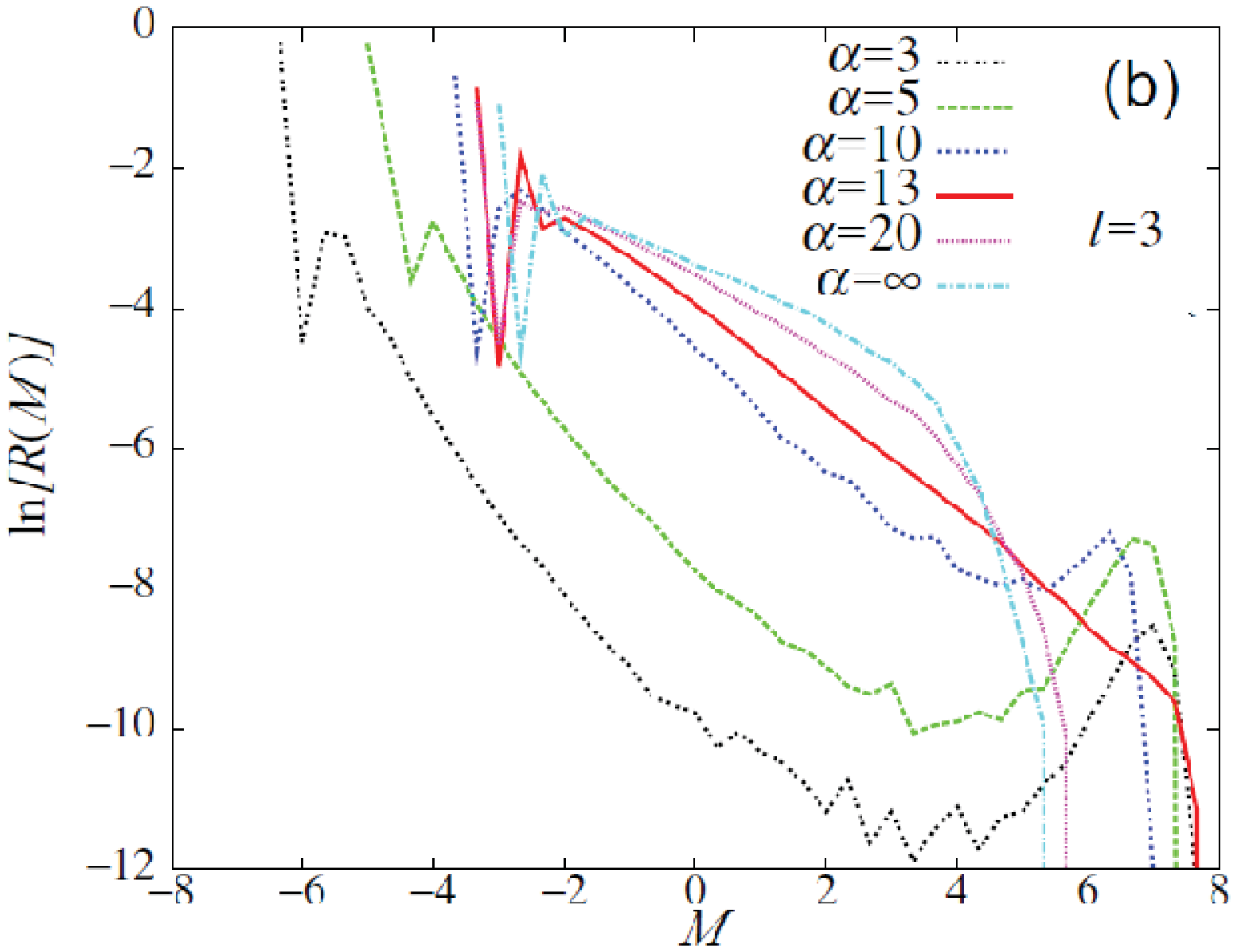}
\end{center}
\caption{
The magnitude distribution $R(M)$ of the 2D BK model with nearest-neighbor interaction for various values of the friction parameter $\alpha$. The other parameters are $l=3$ and $\delta=0.01$. Fig.(a) represents $R(M)$ for  smaller values of the friction parameter $0\leq \alpha \leq 3$, while Fig.(b) represents $R(M)$ for larger values of the friction parameter $3\leq \alpha \leq \infty$. The system size is $60\times 60$.  Taken from (Mori and Kawamura, 2008a).
}
\label{magnitude-2DBK}
\end{figure}

 Extensive numerical studies have revealed that statistical properties of the 2D BK model are more or less similar to those of the 1D BK model reviewed in the previous subsection, at least qualitatively.  The magnitude distribution $R(M)$ of the 2D BK model was studied by several groups (Carlson and Langer, 1989a; Carlson and Langer, 1989b; Carlson et al, 1991; Kumagai, et al, 1999; Mori and Kawamura, 2007).  In Fig.~\ref{magnitude-2DBK}, we show typical behaviors of the magnitude distribution of the 2D BK model with varying the frictional parameter $\alpha$ (Mori and Kawamura, 2007).  For smaller $\alpha \lsim 0.5$, $R(M)$ bends down rapidly at larger magnitudes, exhibiting a ``subcritical'' behavior. Only small events of $M \lsim 2$ occur in this case.  At $\alpha \gsim 0.5$, large earthquakes of their magnitudes $M \simeq 8$  suddenly appear, while earthquakes of intermediate magnitudes, say, $2\lsim M \lsim 6$, remain rather scarce.  Such a sudden appearance of large earthquakes at $\alpha =\alpha_{c1}\simeq 0.5$ coexisting with smaller ones has a feature of a discontinuous or ``first-order'' transition.

 In this context, it might be interesting to point out that Vasconcelos observed that a single block system exhibited a ``first-order transition'' at $\alpha =0.5$ from a stick-slip to a creep (Vasconcelos, 1996), whereas this discontinuous transition becomes apparently continuous in many-block system (Vieira et al, 1993; Clancy and Corcoran, 2005). A ``first-order'' transition observed at $\alpha=\alpha_{c1}\simeq 0.5$ in the 2D model may have some relevance to the first-order transition of a single-block system observed by Vasconcelos, although events observed at  $\alpha<\alpha_{c1}$ in the present 2D model are not really creeps, but rather are stick-slip events of small sizes.

  With increasing $\alpha$ further, earthquakes of intermediate magnitudes gradually increase their frequency.  Fig.~\ref{magnitude-2DBK}(b) exhibits $R(M)$ for larger $\alpha$. In the range of $1\lsim \alpha \lsim 10$,  $R(M)$ exhibits a pronounced peak structure at a larger magnitude, deviating from the GR law, while it exhibits a near straight-line behavior  corresponding to the GR law at smaller magnitudes (``supercritical'' behavior).  As $\alpha$ increases further, the peak  at a larger magnitude becomes less pronounced.  At $\alpha =\alpha_{c2}\simeq 13$, $R(M)$ exhibits a near straight-line behavior for a rather wide magnitude range, though $R(M)$ falls off rapidly at still larger magnitudes $M \gsim 7$, indicating that the ``near-critical'' behavior observed for $\alpha=\alpha_{c2}\simeq 13$ cannot be regarded as a truly asymptotic one, since this rapid fall-off of $R(M)$ at very large magnitudes is  a bulk property, not a finite-size effect. 

%
%

\begin{figure}[ht]
\begin{center}
\includegraphics[scale=0.65]{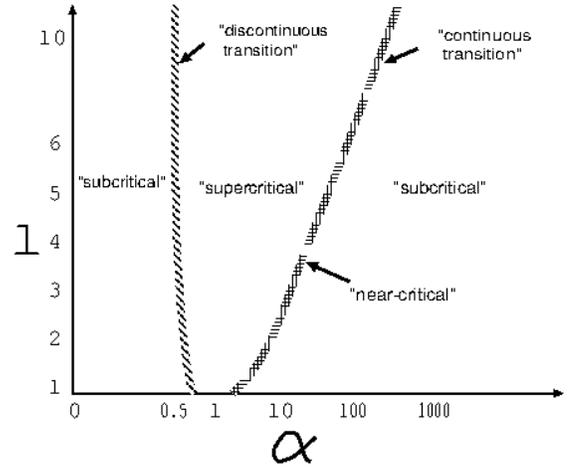}
\end{center}
\caption{
Phase diagram of the 2D BK model with nearest-neighbor interaction in the friction parameter $\alpha$ versus the elastic-parameter $l$ plane. The parameter $\delta$ is  $\delta=0.01$. Taken from (Mori and Kawamura, 2008a).
}
\label{phasediagram-2DBK}
\end{figure}

A ``phase diagram'' of the model in the elasticity parameter $l$ versus the friction parameter $\alpha$, as reported by Mori and Kawamura, 2007 is shown in Fig.~\ref{phasediagram-2DBK}. The region or the ``phase'', called ``supercritical'', ``near-critical'' and ``subcritical'' are observed. The straight-line behavior of $R(M)$, {\it i.e.\/}, the GR law is realized only in the restricted region in the phase diagram along the phase boundary between the supercritical and subcritical regimes. Even along the phase boundary, the GR relation is characterized by a finite cutoff magnitude above which larger earthquakes cease to occur. Hence, the GR relation, as observed in a ubiquitous manner in real faults, is not realized in this model. Since each phase boundary has a finite slope in the $\alpha-l$ plane, one can also induce the ``subcritical''-``supercritical'' transition with varying the $l$-value for a fixed $\alpha$ (Espanol, 1994; Vieira, 1996b).

%
%

 As for other quantities, the recurrence-time distribution of the 2D model exhibits a behavior similar to that of the 1D model (Mori and Kawamura, 2007). As in case of 1D, an aftershock sequence obeying the Omori law is not observed even in the 2D model, at least in its simplest version. The 2D model also exhibits precursory phenomena similar to the ones observed in the 1D model (Mori and Kawamura, 2007). Acceleration of seismic activity prior to mainshock is observed in the supercritical regime, while it is not realized in the subcritical regimes. As in case of 1D, mainshocks are accompanied by the ``Mogi doughnut''-like quiescence in both supercritical and subcritical regimes.

\begin{figure}[ht]
\begin{center}
\includegraphics[scale=0.5]{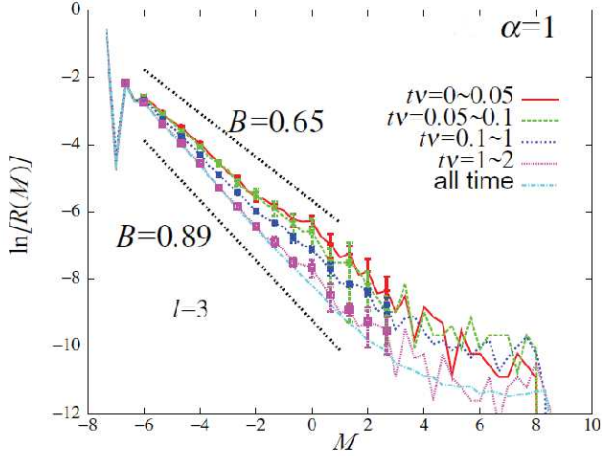}
\end{center}
\caption{
The local magnitude distribution preceding the mainshock of $M >M _c=5$ of the 2D BK model with nearest-neighbor interaction. The parameters are $\alpha=1$, $l=3$ and $\delta=0.01$. The data are shown for several time periods before the mainshock.  The system size is $60\times 60$. Taken from (Mori and Kawamura, 2008a).}
\label{magnitude-2DBK-before}
\end{figure}

 As an other signature of the precursory phenomena, we show in Fig.~\ref{magnitude-2DBK-before} the ``time-resolved''  local magnitude distribution calculated for time periods before the large event in the supercritical regime of $\alpha=1$ and $l=3$ (Mori and Kawamura, 2007). Only events with their epicenters lying within 5 blocks from the upcoming mainshock of its magnitude $M\geq M_c=5$. As can be seen from the figure,  an apparent $B$-value describing the smaller magnitude region gets {\it smaller\/} as the mainshock is approached, {\it i.e.\/}, it changes from $B\simeq 0.89$ of the long-time value to  $B\simeq 0.65$ in the time range $t\nu \leq 0.1$ before the mainshock.  In real seismicity, an appreciable decrease of the $B$-value has been reported preceding large earthquakes (Suyehiro, Asada and Ohtake, 1964; Jaume and Sykes, 1999; Kawamura, 2006). Obviously, a possible change in the magnitude distribution preceding the mainshock possesses a potential importance in earthquake fo
 recast.

\subsubsection{The BK model with long-range interaction}

 \noindent So far, we assumed that the interaction between blocks worked only between  nearest-neighboring blocks. This may correspond to the situation where a thin isolated plate is subject to friction force and is driven by shear force (Clancy and Corcoran, 2006). However, a real fault is not necessarily a thin isolated plate, and the elastic body extends in a direction away from the fault plane. Indeed, the BK model extended in the direction orthogonal to the fault plane was also studied (Myers, Shaw and Langer, 1996).

 Considering the effect of such an extended elastic body adjacent to the fault plane under certain conditions amounts to considering the effective inter-block interaction to be {\it long-ranged\/}. Thus, taking account of the effect of long-range interaction might make the model more realistic.  Rundle {\it et al\/} studied the properties of the 2D cellular automaton version of the BK model with the long-range  interaction decaying as $1/r^3$ (Rundle, et al, 1995). Xia {\it et al\/} studied the 1D BK model with a variable range interaction where a block is  connected to its $R$ neighbors with a rescaled spring constant  proportional to $1/R$ (Xia et al, 2005; Xia et al, 2007). The type of the long-range model considered by Xia {\it et al\/} may be regarded as a mean-field type, since the model reduces to the mean-field infinite-range model in the $R\rightarrow \infty$ limit. 

 One can also derive the relevant long-range interaction based on an elastic theory (Mori and Kawamura, 2008a). Suppose that the 3D elastic body in which the 2D BK model lies is isotropic, homogeneous and infinite, and  a fault surface is a plane lying in this elastic body and slips along one direction only. Then, a static approximation for an elastic equation of motion for the elastic body would give rise to a spring constant between blocks decaying with their distance $r$ as $1/r^3$. This static assumption is justified when the velocity of the seismic-wave propagation is high enough compared with the velocity of the seismic-rupture propagation.

 Properties of the 2D BK model with the long-range power-law interaction derived from an elastic theory, {\it i.e.\/}, the one decaying as $1/r^3$, was investigated (Mori and Kawamura, 2008a).  The interaction between the two blocks at sites ($i,j$) and ($i^{\prime},j^{\prime}$) is given in the dimensionless form by
\begin{equation}
\left(l^2_x\frac{|i^{\prime}-i|^2}{r^5}+l^2_z\frac{|j^{\prime}-j|^2}
{r^5}\right)(u_{i^{\prime},j^{\prime}}-u_{i,j}), 
\end{equation}
which falls off with distance $r$ as $1/r^3$. Then,
the dimensionless equation of motion of the 2D long-range can be written as
\begin{equation}
\begin{array}{ll}
\ddot{u}_{i,j}=\nu t-u_{i,j}
\ \ \ \ \\ \ \ \ \
+ \sum_{(i^{\prime},j^{\prime}) \ne (i,j)}
\left(l^2_x\frac{|i^{\prime}-i|^2}{r^5}+l^2_z\frac{|j^{\prime}-j|^2}
{r^5}\right)(u_{i^{\prime},j^{\prime}}-u_{i,j}) 
\\ \ \ \ \ 
- \phi _{i,j}. 
\end{array}
\label{eq-1DBK-normalized2}
\end{equation}
If one restricts the range of interaction to nearest neighbors and takes the spatially anisotropic spring constant to be isotropic, $l_x=l_z=l$, one recovers the isotropic nearest-neighbor model described by Eq.~\ref{eq-1DBK-normalized}. The ``isotropy'' assumption $l_x=l_z$ is equivalent to putting the Lame's constant to vanish. In fact, in the short-range model, such a spatial anisotropy of the 2D BK model turned out to hardly affect the statistical properties of the model in the sense that the properties of the anisotropic model was quite close to the corresponding isotropic model characterized by the {\it mean\/} spring constant $l=(l_x+l_z)/2$ (Mori and Kawamura, 2008a).

 One might also consider the 1D BK model with the long-range interaction (Mori and Kawamura, 2008a). One possible way to construct the 1D model might be to impose the condition on the corresponding 2D model  that the systems is completely rigid along the $z$-direction corresponding to the depth direction, {\it i.e.\/}, $u(x,z,t)=u(x,t)$. This yields an effective inter-block interaction decaying with distance $r$ as $1/r^2$,
\begin{equation}
l^2\frac{1}{|i-i^{\prime}|^2}(u_{i^{\prime}}-u_{i}), 
\end{equation}
with the dimensionless equation of motion
\begin{equation}
\begin{array}{ll}
\ddot{u}_i=\nu t-u_i+
l^2 \sum_{i^{\prime} \ne i}
\frac{u_{i^{\prime}}-u_{i}}{|i-i^{\prime}|^2}
-\phi_i. 
\end{array}
\end{equation}

\begin{figure}[ht]
\begin{center}
\includegraphics[scale=0.5]{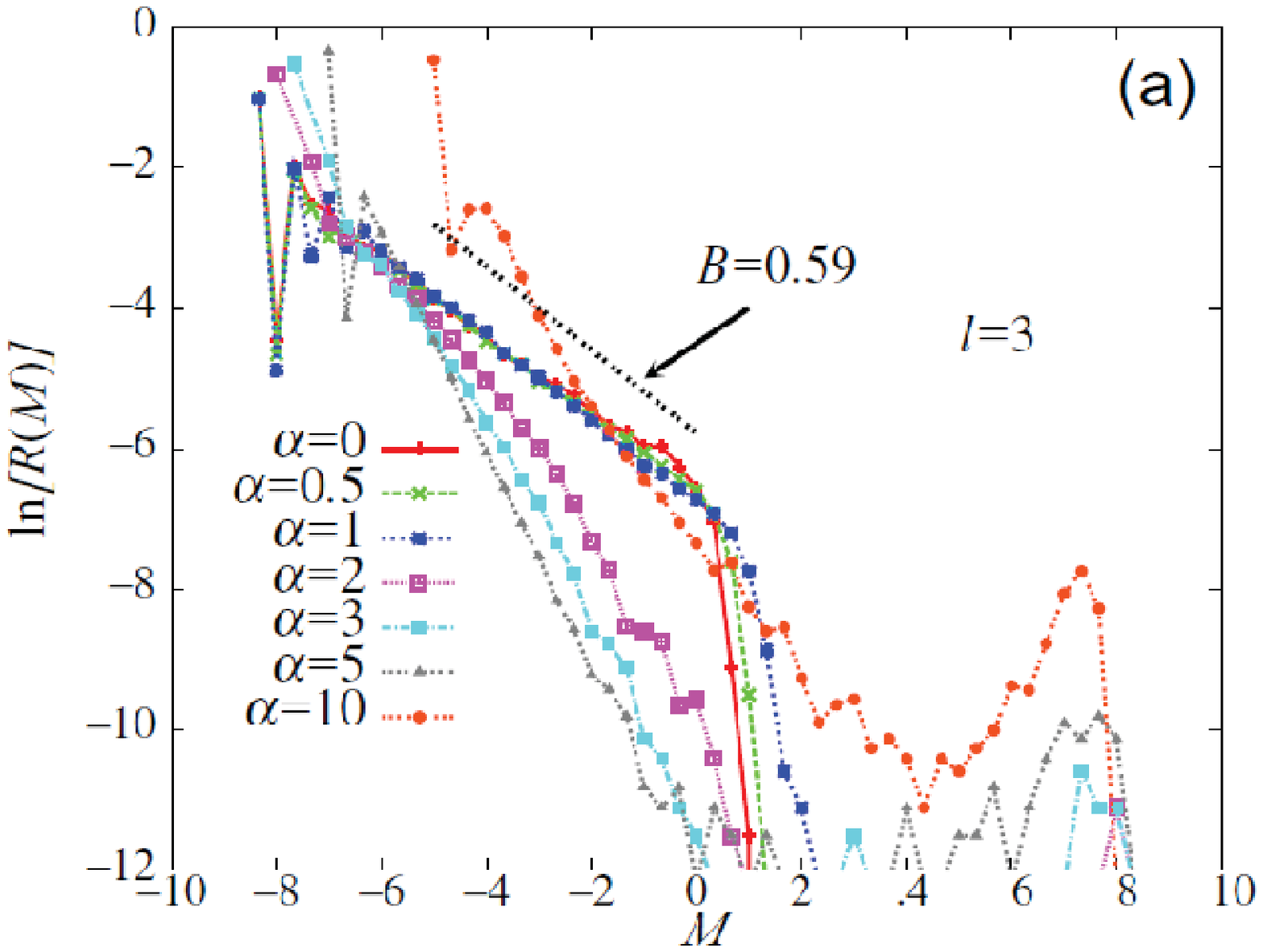}
\includegraphics[scale=0.5]{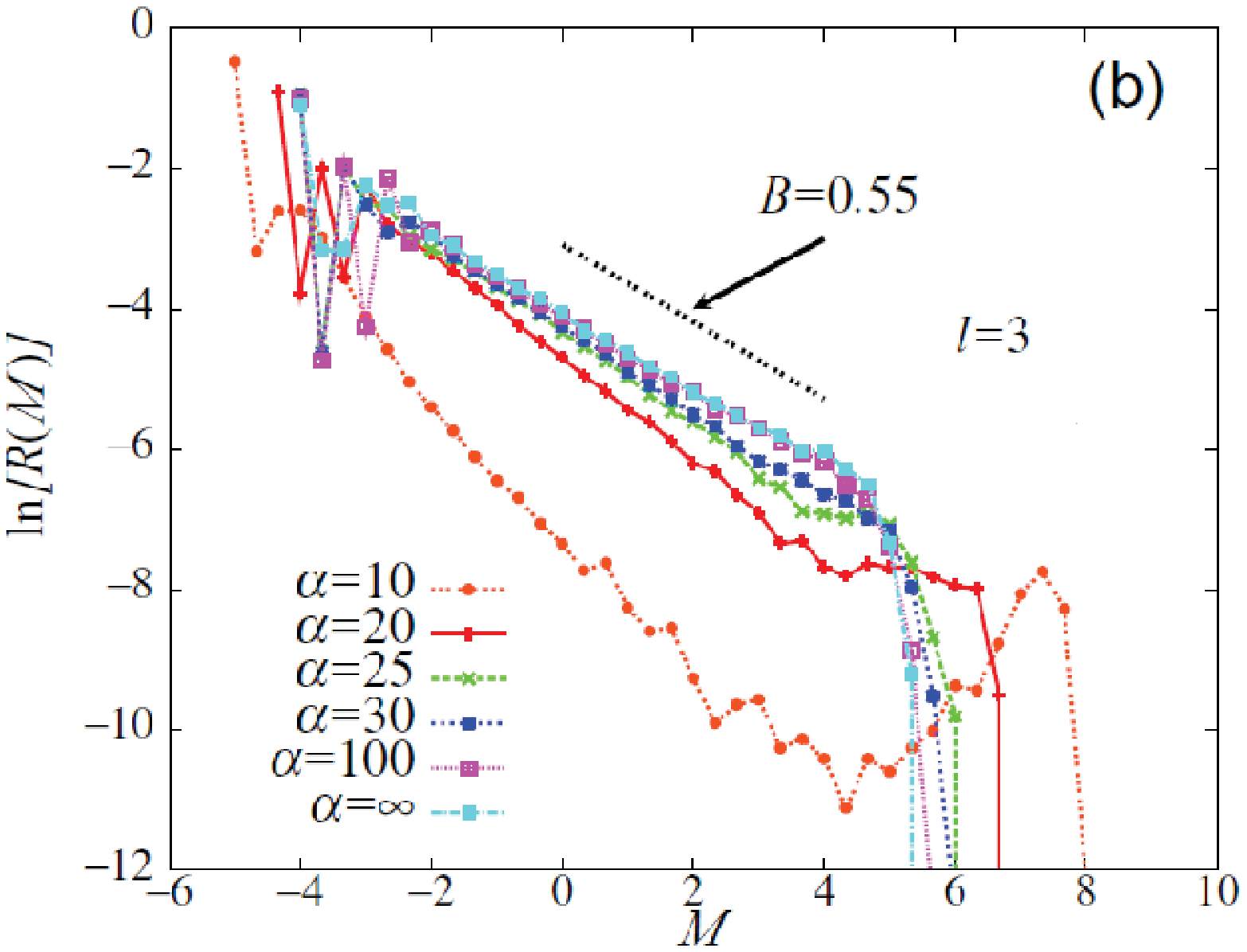}
\end{center}
\caption{
The magnitude distribution $R(M)$ of the 2D BK model with long-range interaction for various values of the friction parameter $\alpha$. The other  parameters are $l=3$ and $\delta=0.01$. Fig.(a) represents $R(M)$ for  smaller values of the frictional parameter $0\leq \alpha \leq 10$, while Fig.(b) represents $R(M)$ for larger values of the frictional parameter $10\leq \alpha \leq \infty$. The system size is $60\times 60$. Taken from (Mori and Kawamura, 2008a).
}
\label{magnitude-BKLR}
\end{figure}

 In  Figs.~\ref{magnitude-BKLR}(a) and (b), we show the magnitude distribution $R(M)$ of the long-range 2D BK model for smaller and larger values of $\alpha$, {\it i.e.\/}, (a) $0\leq \alpha \leq 10$ and (b) $10\leq \alpha \leq \infty$ (Mori and Kawamura, 2008a). Similarly to the short-range case, three distinct regimes are observed depending on the $\alpha$-value. The intermediate-$\alpha$ region corresponds to the supercritical regime where $R(M)$ exhibits a pronounced peak at a larger magnitude, showing a characteristic behavior. Major difference from the short-range case is that the subcritical behavior realized  in the short-range model in the smaller- and larger-$\alpha$ region is now replaced by the near-critical behavior in the long-range model. Namely, for smaller $\alpha< \alpha_{c1}\sim 2$ and for larger $\alpha > \alpha_{c2}\sim 25$,  $R(M)$  exhibits  a  near straight-line behavior over a rather wide magnitude range, and drops off sharply at larger magnitudes. The associated $B$-value  is estimated to be $B\simeq 0.59$ ($\alpha<\alpha_{c1}$) and $B\simeq 0.55$ ($\alpha>\alpha_{c2}$), which is rather insensitive to the $\alpha$-value.  This straight-line behavior of $R(M)$ cannot be regarded as a truly critical one, since $R(M)$ drops off sharply at very large magnitudes. As in the short-range case, the  change from the supercritical to the near-critical behaviors at  $\alpha=\alpha_{c2}\simeq 25$ is continuous, while it is discontinuous at  $\alpha=\alpha_{c1}\simeq 2$.

 Such a near-critical behavior realized over a wide parameter range is in sharp contrast to the behavior of the corresponding short-range model where $R(M)$ at smaller and larger $\alpha$ exhibits only a down-bending subcritical behavior, while a straight-line near-critical behavior is realized only by fine-tuning the $\alpha$-value to a special value $\alpha\simeq \alpha _{c2}$. The robustness of the near-critical behavior of $R(M)$ observed in the 2D long-range model might have an important relevance to real seismicity, since the GR law is ubiquitously observed for different types of faults. Note also that the associated $B$-value observed here turns out to be close to the one observed in real seismicity (Mori and Kawamura, 2008a).  

\begin{figure}[ht]
\begin{center}
\includegraphics[scale=0.58]{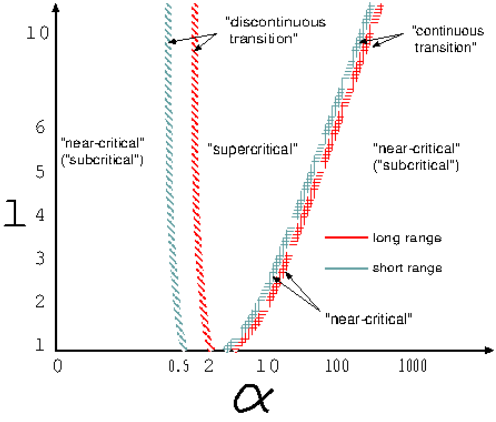}
\end{center}
\caption{
The phase diagram of the 2D BK models with long-range interaction in the friction parameter $\alpha$ versus elastic-parameter $l$ plane, which is compared with the one of the 2D BK model with short-range interaction.  The parameter $\delta$ is set $\delta=0.01$. Taken from (Mori and Kawamura, 2008a).
}
\label{phasediagram-BKLR}
\end{figure}

 In  Fig.~\ref{phasediagram-BKLR}, the behavior of $R(M)$ is summarized in the form of a ``phase diagram'' in the friction parameter $\alpha$ versus the  elastic-parameter $l$ plane (Mori and Kawamura, 2008a). As can be seen from the figure, the phase  diagram of the long-range model consists of three distinct regimes,  two of which are near-critical regimes and one is a supercritical regime. The  ``phase boundary'' between the smaller-$\alpha$  near-critical regime  and the supercritical regime represents a ``discontinuous transition'', while the one between the larger-$\alpha$  near-critical regime and the supercritical regime represents a  ``continuous  transition''. For comparison, the corresponding phase boundary of the short-range model is also shown. The near-critical phases in the long-range model are replaced by the subcritical phases in the short-range model.

\begin{figure}[ht]
\begin{center}
\includegraphics[scale=0.35]{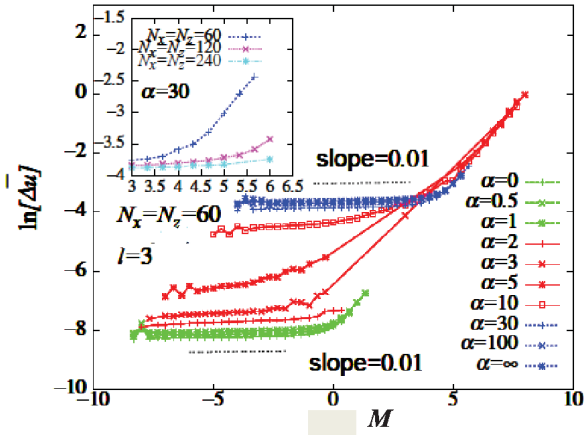}
\end{center}
\caption{
  The magnitude dependence of the mean displacement $\Delta \bar u$ at each seismic event of the 2D BK model with long-range interaction. In the main panel, the friction parameter $\alpha$ is varied with fixing the system size $60\times 60$, while in the inset the system-size $N$ is varied for the case of $\alpha=30$. The parameters $l$ and  $\delta$ are fixed to $l=3$ and $\delta=0.01$.  Taken from (Mori and Kawamura, 2008a).
}
\label{displacement-BKLR}
\end{figure}

 It might be interesting to notice that the system at different ``phases'' of  Fig.~\ref{phasediagram-BKLR} really show different properties. For example, we show in  Fig.~\ref{displacement-BKLR} the magnitude dependence of the mean displacement $\Delta \bar u$ at a seismic event (Mori and Kawamura, 2008a). As can be seen from the figure, the data in the two near-critical regimes (the data in blue and in green) are grouped into two distinct branches, while the data in the supercritical regime (the data in red) exhibit a significantly different behavior. Interestingly, the mean displacement in the near-critical regimes hardly depends on the event magnitude.

 It was observed that the mean stress drop at a seismic event also hardly depends on the event magnitude in the near-critical regimes of the 2D long-range BK  model (Mori and Kawamura, 2008a).
A similar independence was also reported in the mean-field-type 1D long-range BK model (Xia {\it et al\/}, 2005; 2008) and in the 1D long-range BK model (Mori and Kawamura, 2008a).

\subsubsection{Continuum limit of the BK model}

\noindent  Although the BK model has widely been used as a useful tool to investigate statistical properties of earthquakes, the block discretization inherent to the model construction is a crude approximation of the originally continuum earthquake fault. It {\it introduces the short-length cutoff scale into the problem\/}. Therefore, in order to check the validity of the model, it is crucially important to examine the continuum limit of the BK model carefully. Indeed, Rice criticized that the discrete BK model with the velocity-weakening friction law was ``intrinsically discrete'', lacking  in a well-defined continuum limit (Rice, 1993). Rice argued that the spatiotemporal complexity observed in the discrete BK model was due to the inherent discreteness of the model, which should disappear in continuum. Indeed, he  applied the RSF law, which possessed an intrinsic length scale corresponding to the characteristic slip distance, and showed that the system tended to exhibit a quasi-periodic behavior, if the grid spacing $d'$ was taken smaller than the characteristic length scale, while if the grid spacing $d'$ was taken longer than it, the system exhibited an apparently complex or critical behavior. This problem of the continuum limit of the BK model was also addressed by Myers and Langer (Myers and Langer, 1993) within the velocity-weakening friction law, who introduced the Kelvin viscosity term to produce a small length scale which allowed a well-defined continuum limit. Myers and Langer, and subsequently Shaw (Shaw, 1994), observed that the added viscosity term smoothed the rupture dynamics, apparently giving rise to the continuum limit accompanied by the spatiotemporal complexity. More recently, the continuum limit of the 1D BK model with and without the viscosity was examined by Mori and Kawamura within the velocity-weakening friction law (Mori and Kawamura, 2008b).

 Thus, two different ways of taking the continuum limit of the BK model were tried so far, each introducing the short length scale via (A) the viscosity term, or (B) the RSF law. In this subsection, we examine the former (A), while the latter (B) will be discussed in the next subsection. 

 As mentioned, the naive continuum limit of the discrete BK model with a velocity-weakening friction force without viscosity has a problem in that the pulse of slip tends to become increasingly narrow in width in the limit,  {\it i.e.}, the dynamics becomes sensitive to the grid spacing $d^{\prime}\rightarrow 0$. One way to circumvent this problem is to introduce the viscosity term $\eta^{\prime} \partial^3 U_i/(\partial {x^{\prime}}^2 \partial t^{\prime})$ into Eq.\ref{eq-1DBK} to produce a small length scale, where $\eta^{\prime}$ is the viscosity coefficient. Myers and Langer showed that, owing to the added viscosity term, the system became independent of the grid spacing $d^{\prime}$ as long as a new small length scale $\epsilon^{\prime}$, defined by  
\begin{equation}
\epsilon^{\prime} =\pi
\sqrt{\frac{\eta^{\prime}}{\alpha \omega}},
\end{equation}
is sufficiently larger than the grid spacing $d^{\prime}$ (Myers and Langer, 1993).  With $\xi^{\prime}$ being the wave velocity in the continuum limit, this small length scale $\epsilon^{\prime}$ can also be given in the dimensionless form as 
\begin{equation}
\epsilon \equiv \epsilon^{\prime}/(\xi^{\prime}/\omega)
=\pi \sqrt{\frac{\eta}{\alpha}},
\label{eq-smalllength}
\end{equation}
where $\eta \equiv \eta^{\prime}/(\xi^{\prime 2}/\omega)$ is the dimensionless viscosity coefficient. The dimensionless distance $r$ between the block $i$ and $i^{\prime}$ is measured by
\begin{equation}
r=d|i-i^{\prime}|,
\end{equation}
where $d\equiv d^{\prime}/(\xi^{\prime}/\omega)$ is the dimensionless grid spacing. The continuum limit corresponds to taking the limit $d\rightarrow 0$ with fixing $L=Nd$ and $r$, which means $N\rightarrow \infty$ and $l\rightarrow \infty$. Thus, taking the continuum limit in the BK model corresponds to making the model to be infinitely rigid $l\rightarrow \infty$. Numerically, various observables were calculated with successively smaller $d$ to examine its asymptotic $d\rightarrow 0$ limit. 

 Shaw showed, by adding the viscosity term to the 1D BK model, that the magnitude distribution became independent of the grid spacing $d^{\prime}$ for sufficiently small $d^{\prime}$ (Shaw, 1994). Mori and Kawamura studied the 1D BK model with successively smaller grid spacings $d^{\prime}$ to examine how various statistical properties of the model changed and approached the continuum limit for both cases of nonzero ($\eta>0$) and zero ($\eta=0$) viscosity (Mori and Kawamura, 2008b). It was then observed that, in the former viscous case, the results converged to the continuum limit when the condition $d < \epsilon $ was met, whereas, in the latter non-viscous case, such a convergence was obscure.

\begin{figure}[ht]
\begin{center}
\includegraphics[scale=0.32]{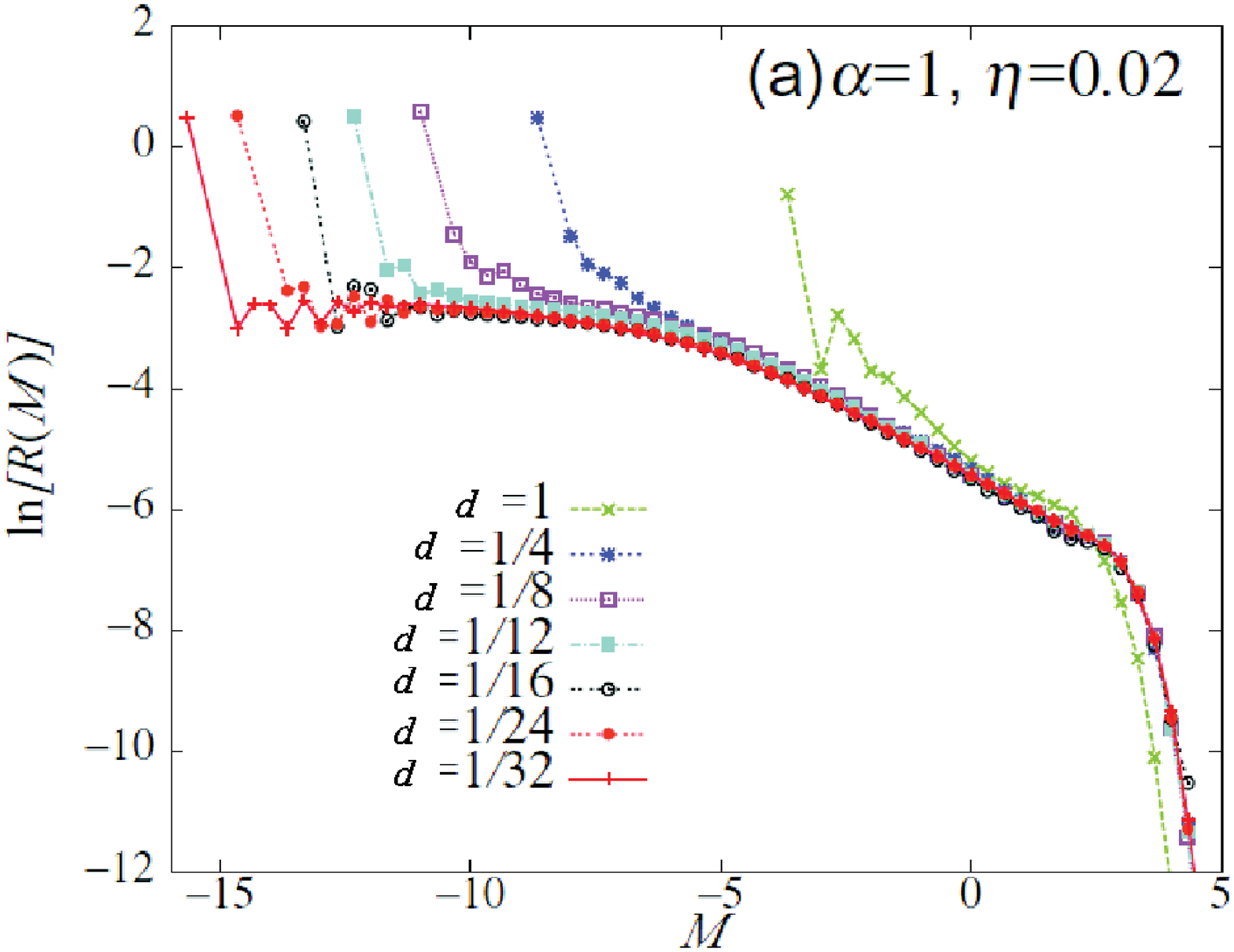}
\includegraphics[scale=0.32]{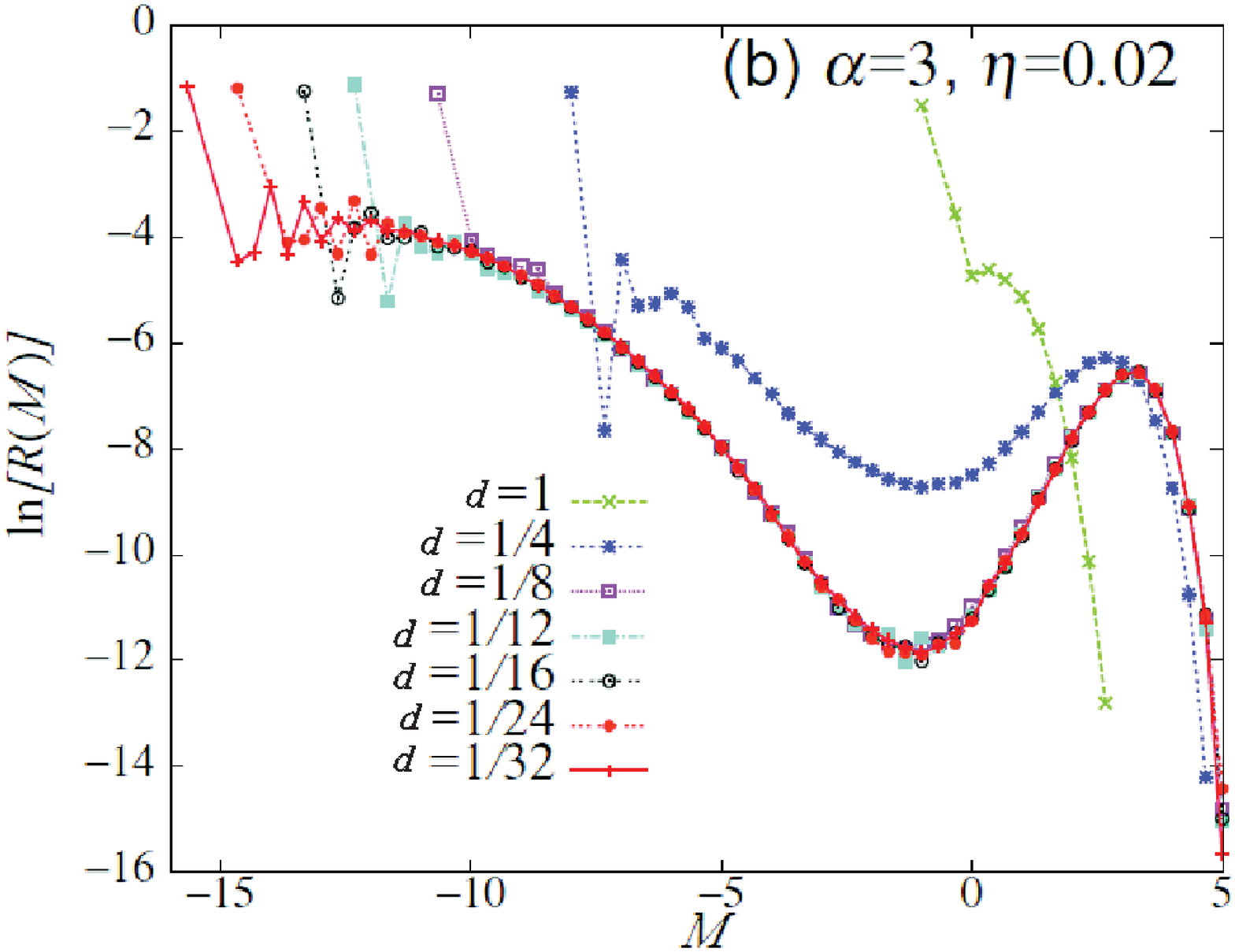}
\end{center}
\caption{
The magnitude distribution $R(M)$ of earthquake events of the 1D viscous BK model ($\eta=0.02$) with $\delta=0.01$. The dimensionless grid spacing  $d$ is varied in the range $1 \geq d \geq 1/32$. Figs.(a) and (b) represent  the cases of $\alpha=1$ and 3, respectively.  The system size is $L=dN=200$. Taken from (Mori and Kawamura, 2008b).
}
\label{magnitude-continuumBK}
\end{figure}

 As an example, we show in  Fig.~\ref{magnitude-continuumBK} the way of convergence of the magnitude distribution function $R(M)$  for $\alpha=1$ (a) and for $\alpha=3$ (b), in the viscous case ($\eta=0.02$). For both cases of $\alpha=1$ and 3, the continuum limit seems to be well reached, {\it i.e.\/}, $R(M)$ seems to converge to an asymptotic form~for smaller $d$, except that the minimum magnitude continuously gets lower as the grid spacing $d$ gets smaller. A similar result was reported by Shaw, 1994.  From  Fig.~\ref{magnitude-continuumBK}(a), one also sees that a nonzero viscosity tends to weaken the GR character of the magnitude distribution somewhat. Such a deviation from the GR law at smaller magnitudes is probably originated from the fact that the viscosity tends to make the relative displacement of neighboring blocks being smoother, enhancing the correlated motion of neighboring blocks, which makes the frequency of smaller events of one or a few blocks considerably 
 reduced (Mori and Kawamura, 2008b).

 The small-length cutoff scale $\epsilon$ as given by Eq.~\ref{eq-smalllength} is estimated here to be $\epsilon \simeq 0.44$ and 0.26 for $\alpha=1$ and 3, respectively. As can be seen from Figs.~\ref{magnitude-continuumBK}(a) and (b), $R(M)$ converges to an asymptotic form for the $\alpha$-values smaller than $d \simeq 1/4$ and 1/8 for $\alpha=1$ and 3, respectively, which is consistent with the expected condition of the continuum limit $d < \epsilon$. 

\begin{figure}[ht]
\begin{center}
\includegraphics[scale=0.32]{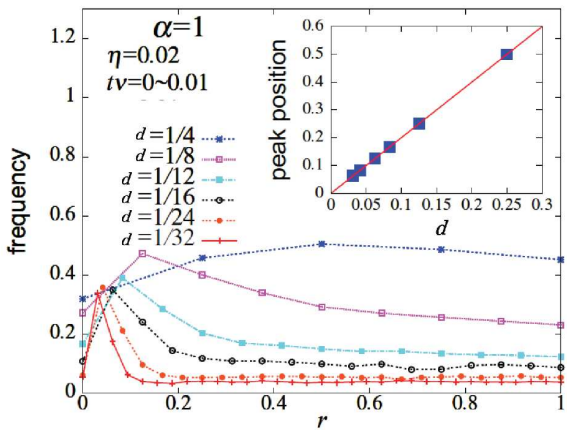}
\end{center}
\caption{
 The event frequency in the  time period $t\nu=0\sim0.01$ immediately before the mainshock of $M >M_c=2$  of the 1D viscous BK model ($\eta=0.02$) with $\alpha=1$ plotted versus $r$, the distance $r$ from the epicenter of the upcoming mainshock. The dimensionless grid spacing $d$ is  varied in the range $1/4 \geq d \geq 1/32$. The parameter $\delta$ is  fixed to $\delta=0.01$. The system size is $L=dN=200$. The insets represent the peak position of the event frequency, corresponding to the range of the doughnut-like quiescence, as a function of the dimensionless grid spacing $d$. The doughnut-like quiescence vanishes in the continuum limit $d\rightarrow 0$. Taken from (Mori and Kawamura, 2008b).
}
\label{mogidoughnut-continuumBK}
\end{figure}

 As mentioned in subsection III.A.3, the BK model generally gives rise to a seismic quiescence phenomenon prior to mainshock, {\it i.e.\/}, the Mogi-doughnut.  Then, a natural question is whether the doughnut-like quiescence observed in the discrete BK model  survives the continuum limit, or it is a phenomenon intrinsically originated from the short cutoff length scale of the model. This question was addressed in (Mori and Kawamura, 2008b).   Fig.~\ref{mogidoughnut-continuumBK} exhibits the time-dependent spatial correlation functions before the mainshock in the case of the viscous model of $\alpha=1$. As the grid spacing $d$ gets smaller,  the spatial range of the  quiescence gets narrower, tending to vanish for small enough $d$: See the inset of Fig.~\ref{mogidoughnut-continuumBK}. This observation strongly suggests that the doughnut-like quiescence might vanish altogether in the continuum limit $d \to 0$.  Thus, the doughnut-like quiescence observed in the discrete BK model  is   likely to be a phenomenon closely related to the short-length cutoff scale of the model. This seems fully consistent with the observation that the one-block events are responsible  for the observed doughnut-like quiescence (Mori and Kawamura, 2006; 2008a). 

%
%

 The observation might have some implications to real seismicity. While the real crust is  obviously a continuum, it is often not  so uniform, possibly with a short-length cutoff. In any case, in real earthquakes  the Mogi-doughnut is occasionally reported to occur (Mogi, 1969; 1979; Scholz, 2002), although establishing its statistical significance is sometimes not easy. Then, our present result may suggest that, if the real crust possesses a cutoff length scale due to the inhomogeneity of the crust, the  ``Mogi-doughnut'' quiescence might occur at such a length scale. In other words, spatial inhomogeneity might be an essential ingredient for the Mogi-doughnut to occur in real seismicity (Mori and Kawamura, 2008b).

\subsubsection{The BK model with RSF law}

\noindent So far, we have mostly assumed a simple velocity-weakening friction law where the friction force is a single-valued function of the velocity. As detailed in section II and in subsection III.A.2, the RSF law is now regarded in seismology as the standard consititutive law.

 Tse and Rice employed this RSF constitutive relation in their numerical simulations of earthquakes (Tse and Rice, 1986). These authors studied the stick-slip motion of the two-dimensional strike-slip fault within an elastic continuum theory, assuming that the fault motion is rigid along strike. It was then observed that large events repeated periodically. Since then, similar RSF constitutive laws have widely been used in numerical simulations (Stuart, 1988; Horowitz and Ruina, 1989; Rice, 1993; Ben-Zion and Rice, 1997; Kato and Hirasawa, 1999; Kato, 2004; Bizzarri and Cocco, 2006). Somewhat different type of slip- and state-dependent constitutive law was also used (Cochard and Madariaga, 1996).

 Cao and Aki performed a numerical simulation of earthquakes by combining the 1D BK model with the RSF law in which various constitutive parameters were set nonuniform over blocks (Cao and Aki, 1986). Ohmura and Kawamura extended an earlier calculation by Cao and Aki to study the statistical properties of the 1D BK model combined with the RSF constitutive law with uniform constitutive parameters (Ohmura and Kawamura, 2007). Clancy and Corcoran also performed a numerical simulation of the 1D BK model based on a modified version of the RSF law (Clancy and Corcoran, 2009). 

 Rice and collaborators argued that the slip complexity of the BK model might be caused by its intrinsic discreteness (Rice, 1993; Ben-Zion and Rice, 1997). In this context, it is important to clarify the statistical properties of the model where the discrete BK structure is combined with the RSF law, to compare its statistical properties with those of the standard BK model with the velocity-weakening or slip-weakening friction law reviewed in the previous subsections.

 Recent study by Morimoto and Kawamura has revealed that the model exhibits largely different behaviors depending on whether the frictional instability is either ``strong'' or ``weak'' (Morimoto and Kawamura, 2011). The condition of strong or weak frictional instability is given by $b>2l^2+1$ or $b<2l^2+1$, respectively, for the 1D BK model. In the case of a weaker frictional instability, the model exhibits a {\it precursory process\/} where a slow nucleation process occurs prior to mainshock.  In the next subsection, we discuss such a precursory process realized in the BK model in more detail. Interestingly, presence or absence of such a nucleation process also affects statistical properties of the model. From a simulation point of view, the case of a weaker friction instability is much harder to deal with, since slow and long-standing nucleation process prior to mainshock generally requires a lot of CPU time.
 
\begin{figure}[ht]
\begin{center}
\includegraphics[scale=0.7]{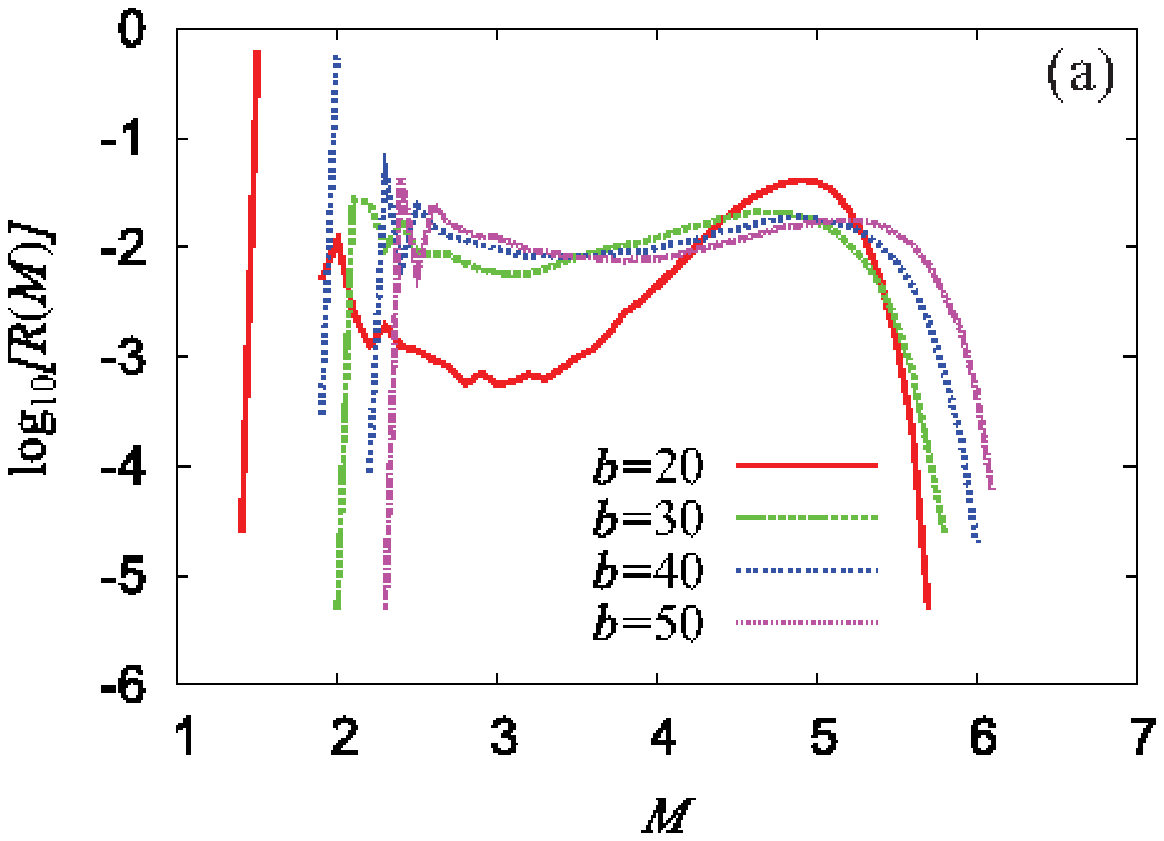}
\includegraphics[scale=1.1]{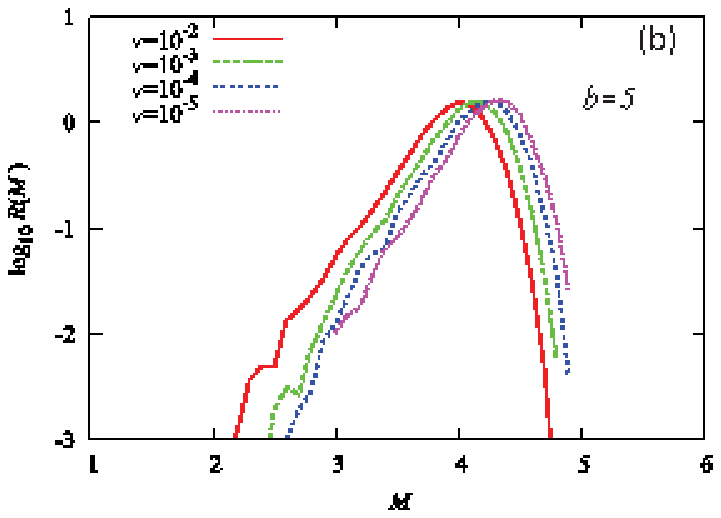}
\end{center}
\caption{
(Color online) The magnitude distribution of the 1D BK model with the RSF law, for the case of (a) a strong frictional instability $b>b_c$, and of (b) a weak frictional instability $b<b_c$, with $b_c=2l^2+1$. The parameter values are $a=0$, $c=1000$, $\nu=10^{-8}$, $v^*=1$ and $l =3$ in (a), and $a=1$, $b=5$, $c=1000$  $v^*=1$ and $l =5$ in (b). The borderline $b$-value is $b_c=19$ in (a), and $b_c=51$ in (b). The system size is $N=800$ in (a), and $N=1200$ in (b). (a) Taken from (Ohmura and Kawamura, 2007). (b) Taken from (Morimoto and Kawamura, 2011).
}
\label{magnitude-1DBKRSF}
\end{figure}

 Statistical properties of the 1D BK model with the RSF law Eq.\ref{eq-eqmotion1} (or Eq.\ref{eq-eqmotion1'}) and Eq.\ref{eq-eqmotion2} was investigated by Ohmura and Kawamura for the case of a strong frictional instability (Ohmura and Kawamura, 2007), and by Yamamoto and Kawamura for the case of a weak frictional instability (Yamamoto and Kawamura, 2011). Typical behaviors of the magnitude distribution are respectively shown in Figs.~\ref{magnitude-1DBKRSF}(a) and (b). As can be seen from the figure, when the frictional instability is strong, almost flat distribution spanning from small to large magnitudes is realized, while, as the critical value is approached, a peak at a larger magnitude becomes more pronounced giving rise to an enhanced characteristic behavior. In the weak frictional instability regime, the distribution has no weight at smaller magnitudes, with a pronounced peak  only at a large magnitude. It means that only large earthquakes of more or less similar magnitude occ
 ur in the regime of a weak frictional instability.

\begin{figure}[ht]
\begin{center}
\includegraphics[scale=0.45]{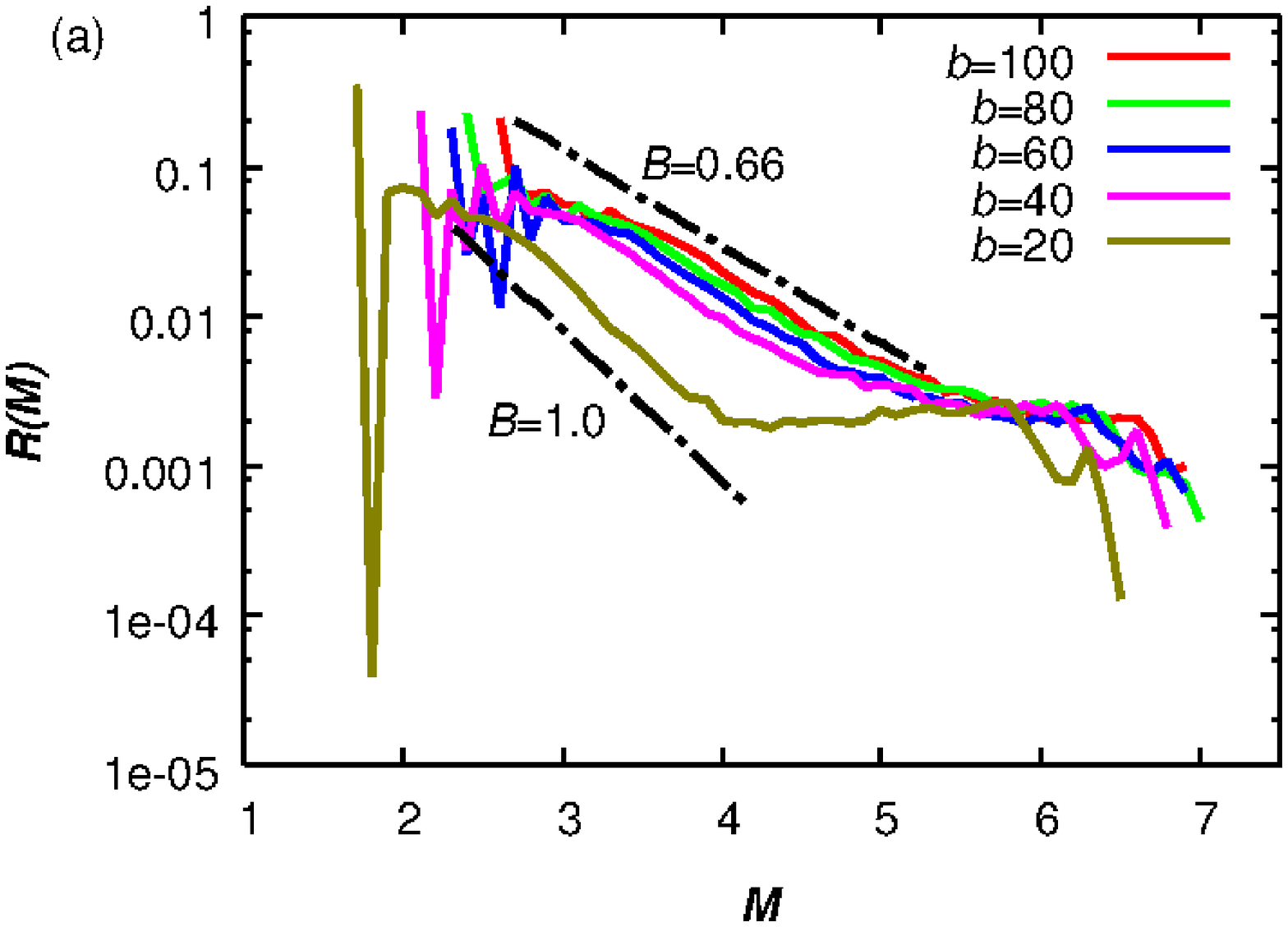}
\includegraphics[scale=0.45]{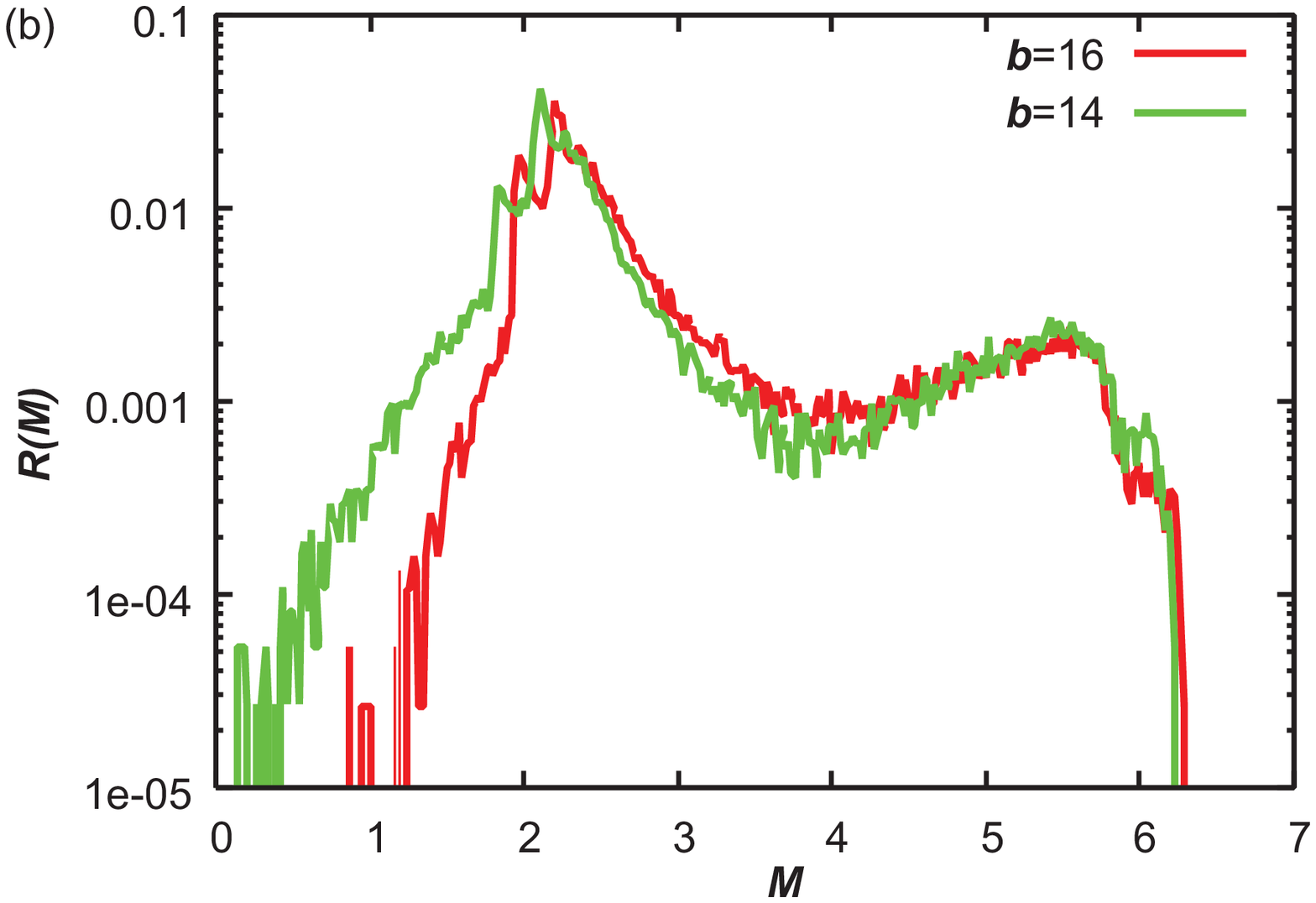}
\end{center}
\caption{
(Color online) The magnitude distribution of the 2D BK model with the RSF law, for the case of (a) a strong frictional instability $b>b_c$, and of (b) a weak frictional instability $b<b_c$, with $b_c=4l^2+1$. The parameter values are $a=1$, $c=1000$, $\nu=10^{-8}$, $v^*=1$ and $l =2$ in (a), and $a=1$, $c=1000$, $\nu=10^{-8}$, $v^*=1$ and $l =2$ in (b). The borderline value is $b_c=17$ in both (a) and (b). The system size is $N=60\times 60$ in (a), and $N=30\times 30$ in (b).  Taken from (Kakui and Kawamura, 2011).
}
\label{magnitude-2DBKRSF}
\end{figure}

 Statistical properties of the corresponding 2D model were investigated by Kakui and Kawamura for both cases of weak and strong frictional instabilities (Kakui and Kawamura, 2011). In the 2D BK model, the condition of strong or weak frictional instability is given by $b>4l^2+1$ or $b<4l^2+1$, respectively. Typical behaviors of the magnitude distribution are shown in Figs.~\ref{magnitude-2DBKRSF}(a) and (b) for the cases of strong and weak instabilities, respectively. As can be seen from the figure, when the frictional instability is strong, a behavior more or less close to the GR law, characterized by the exponent close to $B\sim 2/3$, is realized, although there is a weak shoulder-like structure superimposed at larger magnitudes. The observation of a near-critical behavior close to the GR law would be of much interest in conjunction with real seismicity. As the critical value is approached, on the other hand, a peak at a larger magnitude is further developed, giving rise to  an enhanced characteristic behavior. In the weak frictional instability regime, the distribution has double peaks exhibiting more characteristic behavior: See Fig.~\ref{magnitude-2DBKRSF}(b).

\subsubsection{Nucleation process of the BK model}

\begin{figure}[ht]
\begin{center}
\includegraphics[scale=1.1]{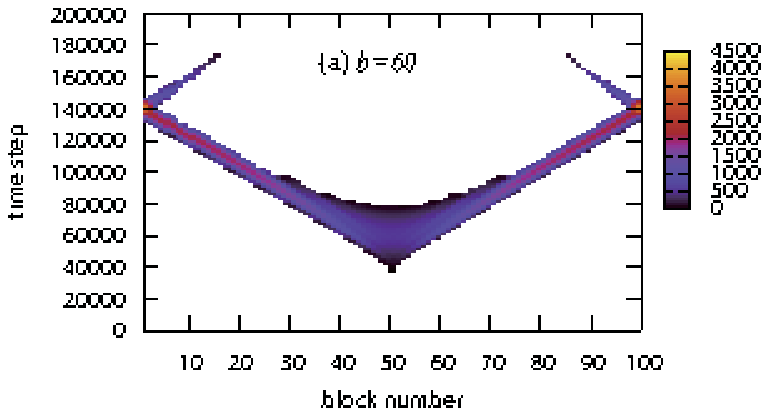}
\includegraphics[scale=0.95]{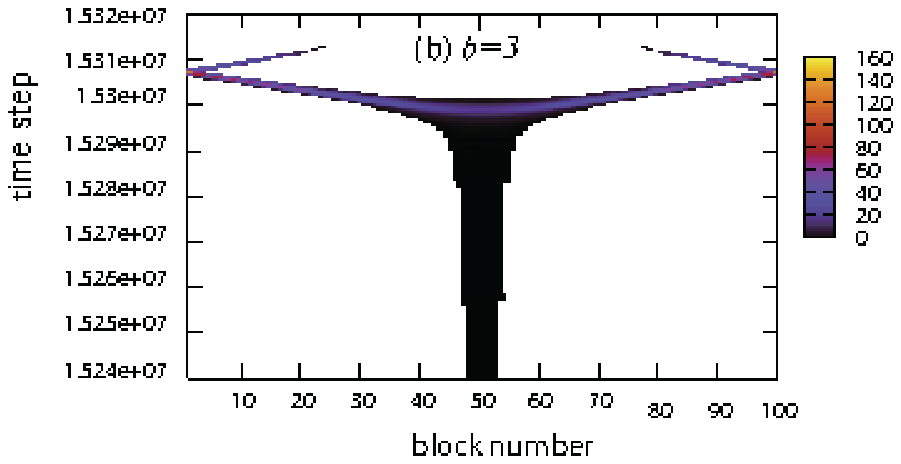}
\end{center}
\caption{
(Color online) The typical rupture process realized in the 1D BK model with the RSF law for (a) a strong and (b) a weak frictional instability, each corresponding to (a) $b>b_c$ and (b) $b<b_c$ with $b_c=2l^2+1$. The color represents the rupture velocity. The parameter values are $a=1$, $c=1000$, $\nu=10^{-2}$, $v^*=1$ and $l =5$ for both (a) and (b) corresponding to $b_c=51$, whereas $b=60$ in (a) and $b=3$ in (b). Taken from (Morimoto and Kawamura, 2011).
}
\label{nucleation}
\end{figure}

\noindent In this subsection, we touch upon the nucleation process as a precursory phenomenon prior to mainshock as realized in the BK model obeying the RSF law. It was observed that the nucleation process is realized even in the BK model with the RSF law for both cases of 1D and 2D, if the model lies in the regime of a weak frictional instability (Morimoto and Kawamura, 2011; Kakui and Kawamura, 2011). Namely, prior to seismic rupture, the system exhibits a slow rupture process localized to a compact ``seed'' area with its rupture velocity orders of magnitude slower than the seismic wave velocity. The system spends a very long time in this nucleation process, and then at some stage, exhibits a rapid acceleration process accompanied by a rapid growth of the rupture velocity and a rapid expansion of the rupture zone, finally getting into a final seismic rupture or a mainshock (Dieterich, 2009). Such a nucleation process has also been observed and extensively studied in the continuum model: See, {\it e.g.\/}, (Ampuero and Rubin, 2008). 
 We illustrate in Fig.~\ref{nucleation} typical example of seismic events realized in the 1D BK model with the RSF law for each case of a weak frictional instability (b), and of a strong frictional instability (a). As can be seen from the figure, a slow nucleation process with a long duration time is observed only in (b), while such a nucleation process is absent in (a).  

 As mentioned, the condition for the appearance of such a nucleation process is given by $b<b_c=2l^2 +1$ in 1D, and by $b<b_c=4l^2 +1$ in 2D (for a square array of blocks).  Indeed, Morimoto and Kawamura found that the critical nucleation size at which the slow nucleation process ends getting into the acceleration stage is given by $X_c=\pi/[\arccos(1-\frac{b-1}{2l^2})]-1$ in units of block size (Morimoto and Kawamura, 2011). Indeed, this length $X_c$ corresponds in its physical meaning to the length $h^*$ of Rice (Rice, 1993), although its detailed functional form, {\it e.g.\/}, the dependence on $b$, is somewhat different from the standard one. The condition of this critical nucleation size being greater than the block size $X_c > 1$ yields the condition of the weak frictional instability $b<b_c$. In other words, when $b>b_c$, the nucleation process cannot be realized in the BK model due to its intrinsic discreteness. Indeed, this is exactly the situation as discussed by Rice (Rice, 1993). 

 The above observation means that, if one takes the continuum limit of the BK model with the RSF law, the system should necessarily lie in the limit of a weak frictional instability, since the continuum limit means $l\rightarrow \infty$. Hence, at least {\it as long as one considers a uniform fault obeying the RSF law without any discretization short-length scale, earthquakes should exhibit characteristic properties rather than critical properties\/}. This fully corroborates an earlier criticism by Rice against the SOC view of earthquakes based on the BK model (Rice, 1993). Indeed, in seismology the concept of earthquake cycle has been used in long-term probabilistic earthquake forecasts (Scholz, 2002; Nishenko, 1987; Working Group on California Earthquake Probabilities, 1995). Of course, a big issue to understand is what is then the true origin of the GR law widely observed in real seismicity.
\subsection{Continuum models}
\noindent As discussed in III.A.6, \textcite{Rice1993} criticized inherently discrete models, where simulated earthquake sequences depend on computation grid size. He confirmed in numerical simulations that complex earthquake sequences disappear when the grid size is sufficiently smaller than the critical size of slip nucleation zone for almost spatially uniform frictional properties. Moreover, he argued that geometrical and/or material disorder is the origin of complexity of earthquakes. The models with sufficiently small grid sizes may be called continuum models, which generate simulation results independent of the grid size, in contrast to inherently discrete models. Note that if a model does not have a finite critical size for nucleating unstable slip, such as a model with constant static and dynamic friction, it is always inherently discrete. In this subsection, we discuss continuum models of earthquakes, especially models using the rate- and state-dependent friction (RSF) law. In the RSF law, the critical size of slip nucleation can be defined as a function of frictional constitutive parameters, and the computation grid sizes are sufficiently smaller than the critical size in the studies mentioned below. We use elastic continuum models below, in contrast to spring-block models in the previous section. "Continuum model" is thus used to express two senses.

The RSF law has commonly been used in models for understanding earthquake phenomena \cite{Scholz2002,Dieterich2007}.These models were sometimes constructed for reproducing and understanding particular earthquakes, earthquake cycles, or sliding processes observed by seismometers, strainmeters, Global Positioning System (GPS), etc. We will see deterministic aspects of earthquake phenomena, in addition to statistical characteristics of earthquakes. Note that comprehensive reviews were presented by \textcite{Rundle2003,Turcotte_etal2007,Ben-Zion2008} for models of statistical properties of earthquakes using friction laws other than the RSF law.

\subsubsection{Earthquake cycles, asperities, and aseismic sliding}
\begin{figure}
\centering \includegraphics[width=6.0cm]{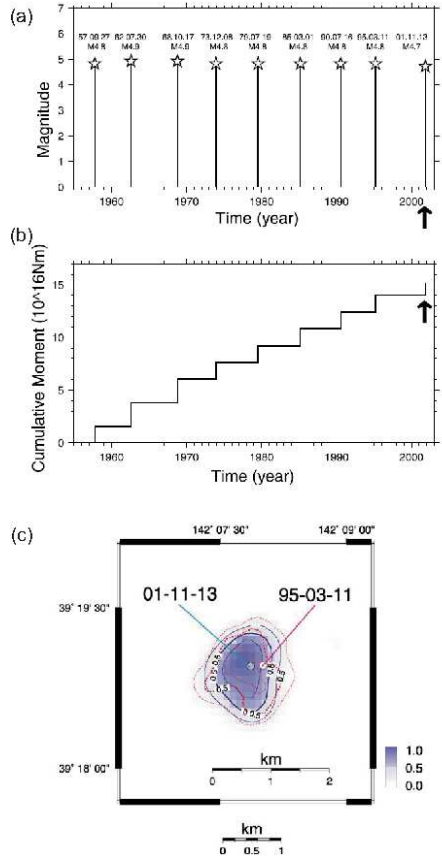}
   \caption{\label{kamaishi} (a) Recurrence of Kamaishi earthquakes of nearly the same magnitudes and recurrence intervals. (b) Cumulative seismic moment of Kamaishi earthquakes. (c) Coseismic slip distribution of the 1995 and 2001 Kamaishi earthquakes estimated from seismic waveforms. Red broken contours and blue contours denote seismic slip of the 1995 and 2001 earthquakes, respectively \cite{Okada_etal2003}.}
\end{figure}

\noindent Before introducing earthquake models, we briefly review observational facts about earthquakes and fault slip behavior. Earthquakes repeatedly occur at the same fault segment. At the Parkfield segment along the San Andreas fault, California, magnitude of about 6 interplate earthquakes have occurred at recurrence intervals of 23 $\pm$ 9 years  since 1857 \cite{SykesMenke2006}. Great earthquakes of magnitude 8 class repeatedly occurred along the Nankai trough, where the Philippine Sea plate subducts beneath southwestern Japan, every one hundred years \cite{SykesMenke2006}. Quasi-periodic earthquake recurrence has been used for long-term forecasts of earthquakes (Working Group on California Earthquake Probabilities, 1995; Matthews et al., 2002).
One of the most remarkable examples of regularity of earthquakes was found off Kamaishi, where the Pacific plate subducts beneath northern Honshu, Japan. Magnitude of 4.8 $\pm$ 0.1 earthquakes have repeatedly occurred at recurrence intervals of 5.5 $\pm$ 0.7 years at the same region since 1957. \textcite{Okada_etal2003} estimated coseismic slip distributions of recent Kamaishi earthquakes from seismic waveform data and found that they overlap with each other (Fig. \ref{kamaishi}). Although many smaller earthquakes occur around the source area of the Kamaishi earthquakes, no comparable or larger earthquakes occur there. This observation suggests that aseismic sliding surrounds the source area of the Kamaishi earthquakes, where stick-slip motion occurs, and steady loading by the surrounding aseismic sliding to the source area leads to the quasi-periodic recurrence of almost the same magnitude earthquakes. The variance in the recurrence interval was suggested to come from temporal variation of aseismic sliding rate surrounding the earthquake source \cite{Uchida_etal2005}. Significant afterslip of the 2011 great Tohoku-oki earthquake (M=9.0) rapidly loaded the source area of the Kamaishi earthquake, generating earthquakes at much shorter recurrence intervals. Recurrences of small earthquakes at the same source areas in mainly creeping (aseissmic sliding) regions were found in many places and these earthquakes are called small repeating earthquakes \cite{NadeauJohnson1998,Igarashi_etal2003}. Although small earthquakes occur, most strain is released by aseismic sliding on these fault planes. The seismic coupling coefficient is defined by the long-term average of the ratio of seismic slip amount to total (seismic and aseismic) slip expected from relative plate motion. The seismic coupling coefficient is variable, dependent on localities. It is close to unity at some segments along Chile and Aleutians, indicating little aseismic sliding and nearly complete locking during interseismic periods, and is nearly equal to zero at Marianas, indicating no or few large interplate earthquakes \cite{Pacheco_etal1993}. These facts show that aseismic sliding is common phenomenon and it plays an important part in strain release at plate boundaries and that frictional properties differ from place to place.

\begin{figure}
\centering \includegraphics[width=6.0cm]{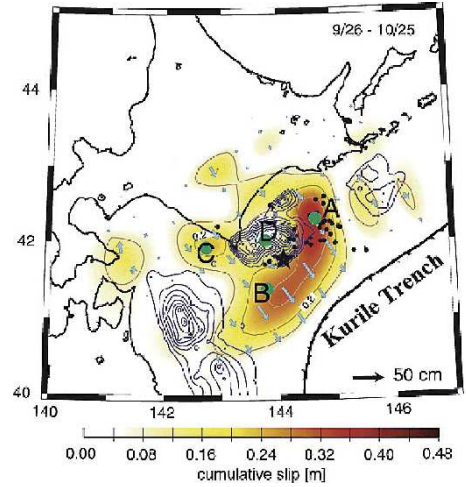}
   \caption{\label{tokachi} Spatial distribution of cumulative slip for 30 days of afterslip of the 2003 Tokachi-oki earthquake (M = 8.0), off Hokkaido, northern Japan, estimated from GPS data (color contours) by \textcite{Miyazaki_etal2004}. Black contours with 0.5m interval show seismic slip in the 1973 Nemuro-oki (right), 1968 Tokachi-oki (left), and 2003 Tokachi-oki (center) earthquakes \cite{YamanakaKikuchi2004}. The black star and small circles denote the epicenter and aftershocks of the 2003 earthquake.}
\end{figure}

\begin{figure}
\centering \includegraphics[width=7.0cm]{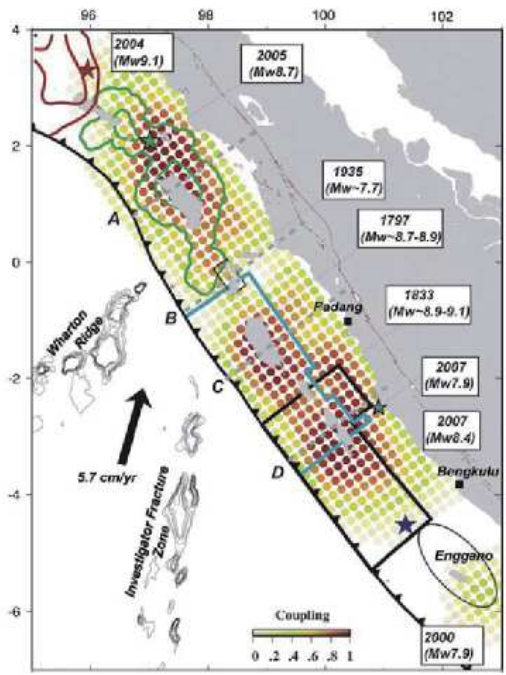}
   \caption{\label{sumatra} Spatial distribution of interplate coupling estimated from geodetic data (colored circles) along the Sunda trench, where the Australian plate subducts beneath the Sumatra island. Red and orange circles indicate that the plate interface is nearly locked and strain is accumulated during an interseismic period, and white and yellow circles indicate that continuous aseismic sliding occurs and strain is not accumulated. Red and green contours with 5m interval show seismic slip in the 2004 Sumatra-Andaman (M = 9.1) and the 2005 Nias-Simeulue (M = 8.7) earthquakes. Blue and black lines show the approximate source areas of the 1797 and 1833 great earthquakes \cite{Chlieh_etal2008}. }
\end{figure}

A patch where stick-slip motion occurs, that is, a fault region where earthquakes repeatedly occur, is often called an asperity, which comes from the rock mechanics term for a contact spot between sliding surfaces as used in II.C. Note that an asperity of an earthquake occupies a considerable part of the earthquake fault area and its size is orders of magnitude larger than seismic slip amount. In contrast, an asperity of a sliding surface is much smaller and its size may be comparable to slip amount. The asperity model has been developed for explaining spatial heterogeneity in seismic slip on faults and complex source processes of earthquakes \cite{KanamoriMcNally1982,Lay_etal1982,Thatcher1990}. When the asperity model was developed around 1980, sliding behavior surrounding asperities was not clarified from observations because aseismic sliding cannot be detected by seismometers. To detect aseismic sliding, geodetic observations such as GPS are required. Since dense GPS networks were established in 1990s \cite{SegallDavis1997}, many aseismic sliding phenomena have been reported such as afterslip (postseismic sliding) and slow (silent) earthquakes. The source areas of afterslip are usually located near coseismic slip areas (asperities), and the afterslip area and the asperity do not overlap as shown in Fig. \ref{tokachi} \cite{Yagi_etal2003,Miyazaki_etal2004, Johnson_etal2006}, which also support spatial heterogeneity of frictional properties. The locations of asperities of large earthquakes were confirmed to be locked during interseismic periods from geodetic observations \cite{Chlieh_etal2008,Hashimoto_etal2009,Perfettini_etal2010}. For instance, Figure \ref{sumatra} clearly shows that seismic slip areas of large interplate earthquakes off the Sumatra island coincide with the locked areas during interseismic periods. For the 2011 great Tohoku-oki earthquake (M = 9.0), a significant peak of seismic slip larger than 30 m was estimated from inversions of seismic waveform and tsunami data \cite{Koketsu_etal2011}. This also suggests nonuniform frictional property on the plate interface. 

Spatial distribution of asperities on plate boundaries has been estimated from source areas of past large interplate earthquakes, and earthquakes repeatedly occurred on the same asperities \cite{YamanakaKikuchi2004}. This suggest that the locations of asperities are unchanged at least a few earthquake cycles. Apparently complex earthquake cycle, where earthquake rupture areas are variable, may be understood by a change in combination of simultaneously ruptured asperities. For example, two adjacent asperities are simultaneously ruptured, resulting in a large earthquake in some cases, and one of them is ruptured to generate a smaller event in other cases. Note that some researchers object against persistent asperities on the basis of seismic waveform analyses \cite{ParkMori2007}.

\subsubsection{Models for nonuniform fault slip using the RSF law}
\noindent The asperity model indicates that spatial heterogeneity of material property is important, and it is compatible with the RSF law discussed in II.C. Regions of velocity-weakening frictional property ($a-b < 0$) correspond to asperities, where stick-slip occurs, and aseismic sliding occurs at regions of velocity-strengthening frictional property ($a-b > 0$). Afterslip occurs in velocity-strengthening areas, and it slowly relaxes stress increases generated by nearby earthquakes. Using a single-degree-of-freedom spring-block model,  \textcite{Marone_etal1991} obtained theoretical slip time function $u(t)$ of afterslip, which occurs on a fault with velocity-strengthening friction ($a-b > 0$), as follows: 
\begin{equation}
\label{uniformslip}
	u(t) = \frac {(a-b)\sigma_n}{k} \ln \biggl[\frac{kV_{cs}}{(a-b)\sigma_n}t + 1\biggr] +V_0 t,
\end{equation}	
where $\sigma_n$ is normal stress on the fault plane, $k$ is spring stiffness, $V_{cs}$ is coseismic slip velocity, $V_0$ is preseismic slip rate, time $t$ is measured from the earthquake occurrence time. Quantitative comparison between afterslip observations and models indicate that the RSF law well explains afterslip \cite{PerfettiniAvouac2004,Freed2007}. 

In case the stiffness is larger than the critical stiffness defined by Eq. (\ref{kc}) for a velocity-weakening fault, it is called conditionally stable (Scholz, 1988). Although aseismic sliding usually occurs under quasi-static loading for conditionally stable case, rapid stress increase may generate seismic slip \cite{Gu_etal_1984}. This fact indicates that sliding behavior at a fault is not determined only by the fault properties but by a loading condition, suggestive of variable sliding behavior of a fault. Note that the effective stiffness of a fault may be related to fault size as will be shown in the next subsection.

\begin{figure*}
\centering \includegraphics[width=16.0cm]{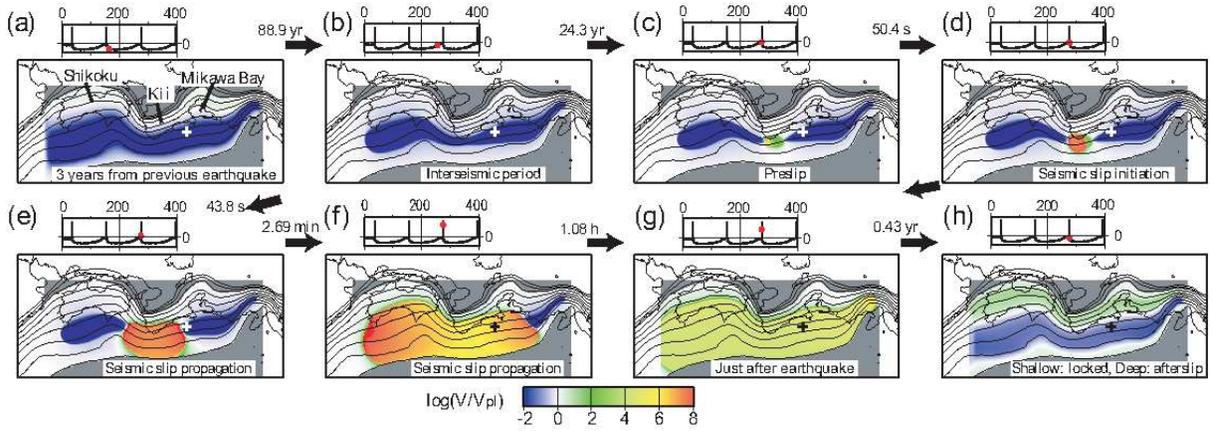}
   \caption{\label{nankai} Snapshots of simulated slip rate $V$ on the model plate interface normalized by the relative plate velocity $V_{pl}$ in a model for recurrence of great earthquakes along the Nankai trough, central Japan. Red, white, and blue show seismic slip rates, stable sliding with sliding velocity nearly equal to the plate velocity, and nearly locked, respectively. Modified from \textcite{Hori_etal2004}. }
\end{figure*}

Since the RSF law takes into consideration time-dependent healing process, it can be used in simulations of earthquake cycles. \textcite{TseRice1986} first published an earthquake cycle model for a strike-slip fault in an elastic continuum using the RSF law to successfully explain stick-slip behavior at a shallower part of a fault, continuous stable sliding at a deeper part, and afterslip at intermediate depths. In the simulation, quasi-dynamic equilibrium between frictional stress and elastic stress generated by fault slip and relative plate motion is numerically solved. Their assumption on depth dependence of $a-b$ is consistent with laboratory data, which indicate $a-b$ changes from negative to positive at about 300$^\circ$C \cite{Blanpied_etal1995}. Similar models have been presented for earthquake cycles at particular regions to compare the simulations with observed earthquake recurrence and/or crustal deformation. Figure \ref{nankai} shows an example simulation result of spatiotemporal evolution of slip velocity on a model plate interface, where great interplate earthquakes repeatedly occur at a shallower part and stable sliding on a deeper part \cite{Hori_etal2004}.  

 If a single asperity exists on a fault plane without any interactions with other asperities, regular stick-slip at a constant recurrence interval is expected to occur. Note that when the asperity size is close to the critical nucleation zone size, irregular stick-slip cycle is observed  even for a single asperity model \cite{LiuRice2007}. When some asperities exist within short distances, they interact with each other, resulting in complex earthquake sequences including single asperity ruptures and multiple asperity ruptures. Numerical simulations of complex earthquake sequences due to interactions between some asperities have been carried out by \textcite{KatoHirasawa1999}, \textcite{Kato2004}, \textcite{LapustaLiu2009}, and \textcite{Kaneko_etal2010}. In these studies, friction obeying the RSF law was assumed and different values of friction parameters ($a', b', {\cal L}$) are assigned for model asperities with velocity-weakening friction, to reproduce compound earthquakes, where some asperities are ruptured simultaneously or with some time delays, which resembles some observations. \textcite{Kato2008}, for instance, reproduced a complex earthquake cycle similar to that observed at the Sanriku-oki region, northeastern Japan, where simulated earthquakes included the 1968 Tokachi-oki earthquake (M=8.2), the 1994 Sanriku-oki earthquake (M=7.7) and its largest aftershock (M=6.9) and afterslip. These studies suggest that spatial distribution of asperities or friction parameters controls regularity and complexity of earthquake recurrence. This further suggests that numerical forecasts of earthquakes may be possible if we can obtain detailed map of friction parameters on a fault. Friction parameters have actually been estimated through comparison of observed data and simulations at California \cite{Johnson_etal2006} and Japan \cite{Miyazaki_etal2004,Fukuda_etal2009} from afterslip data.  

Preseismic sliding, which is aseismic sliding during a slip nucleation process, is expected before earthquake occurrence from the RSF law. 
It is almost ubiquitously observed in laboratory experiments, where the amount of preseismic sliding is of the order of micrometers \cite{OhnakaShen1999}. Using a spring-block 
system implemented with the RSF law, one can show that the preseismic sliding amount 
is approximately given by ${\cal L}$ \cite{Popov2010}.
 Some model studies with the RSF law discussed crustal deformation expected from preseismic sliding for particular earthquakes \cite{StuartTullis1995,Kuroki_etal2002}. However, it is difficult to predict precise amplitudes of crustal deformation, because friction parameters that influence preseismic sliding are not well constrained from presently available data. There are some reports of observations of preseismic sliding, though insignificant or questionable observations are included \cite{Wyss1997}. For example, the close and dense geodetic observation of the Parkfield segment of the 
San Andreas fault could not detect any precursory slip prior to the 2004 earthquake, 
although it should be remarked that an observation of the tremor may suggest 
the accelerated creep on the fault $\sim 16$ km beneath the eventual earthquake 
hypocenter \cite{Shelly2009}. \textcite{KanamoriCipar1974} detected precursory signals in long-period strain seismogram before the occurrence of the 1960 great Chilian earthquake (M=9.5). Since no earthquake that could explain the observed strain signals was detected, they inferred that the signals were caused by preseismic sliding on a deeper extension of the mainshock fault plane. \textcite{LindeSacks2002} examined crustal deformation data before the occurrence of the 1944 Tonankai (M=8.0) and 1946 Nankai (M=8.1) earthquakes, southwestern Japan to construct a model for preseismic sliding of these earthquakes. Their model indicates that preseismic sliding took place at a deeper extension of the main shock fault plane. However, in models with the common RSF law, accelerating preseismic sliding just before earthquake occurrence takes place within the source area of seismic slip because spontaneous accelerating slip can be nucleated only in velocity-weakening region, being inconsistent with these models of preseismic sliding. 
\textcite{Kato2003b} proposed a model for earthquake cycles at a subduction zone to explain large preseismic sliding at the deeper extension of the seismogenic plate interface.
He assumed velocity-weakening friction ($d \mu_{\rm ss}/d \ln V < 0$) at low velocities and velocity-strengthening friction ($d \mu_{\rm ss}/d \ln V > 0$) at high velocities, where $\mu_{\rm ss}$ is a steady-state friction coefficient given by Eq. (\ref{steadystate}).
 Preseismic sliding relaxes regional stresses, which may decrease seismic activity, while it increases stresses around the edges of the slipped region, which tends to increase seismic activity \cite{Kato_etal1997}. This may explain precursory seismic quiescence observed for some large earthquakes \cite{Kanamori1981,Wyss_etal1981}. Preseismic sliding perturbs regional stress field, resulting in an increase or decrease of seismicity. Taking into consideration this effect,\textcite{Ogata2005} systematical searched seismicity changes in Japan to find possible crustal stress changes due to preseismic sliding.

\subsubsection{Slow earthquakes}

\noindent Slow earthquakes are episodic fault slip events that generate little or no seismic waves because their source durations are longer than the periods of observable seismic waves. Slip events without seismic wave radiation are often called silent earthquakes or slow slip events. Slow earthquakes have been studied by using records of very-long-period seismographs \cite{KanamoriStewart1979}, creepmeters that directly detect fault creep at the ground surface \cite{King_etal1973}, and strainmeters \cite{Linde_etal1996}. Afterslip and preseismic sliding mentioned earlier may be included in slow earthquakes. 

Recent development of dense geodetic observation networks including GPS and borehole tiltmeters accelerates studies of slow earthquakes \cite{SchwartzRokosky2007}. \textcite{Hirose_etal1999} found that episodic aseismic slip with duration of about 300 days took place in 1997 on the plate boundary in the Hyuganada region, southwestern Japan, from GPS data. The estimated slip and source area indicated that it released seismic moment corresponding to magnitude of 6.6. Later, almost the same size aseismic slip events occurred at the same area in 2003 and 2010. In the Tokai region, central Japan, another large slow earthquake from 2000 to 2005 released seismic moment nearly equal to that of an M=7.0 earthquake \cite{Ozawa_etal2002,Miyazaki_etal2006}. The source area of this slow earthquake was estimated at the deeper extension of the locked plate boundary, where a magnitude 8 class interplate earthquake is expected to occur. At almost the same area, smaller slow earthquakes, corresponding to moment magnitude of about 6.0, with shorter durations of a few days were found to occur repeatedly \cite{HiroseObara2006}. These slow earthquakes are often called short-term slow slip events (SSEs) to be discriminated from long-term SSEs of durations of several months or longer. Furthermore, \textcite{HiroseObara2006} found low-frequency tremors, which radiate seismic waves with long durations, from high-sensitivity borehole seismometer array data. These events are clearly distinguished from long durations of wave trains and lack of high-frequency components of seismic waves. Short-term SSEs and low-frequency tremors occur simultaneously almost at the same locations. Synchronized occurrence of short-term SSEs and low-frequency tremors were observed in other regions such as the Cascadia subduction zone, North America \cite{RogersDragert2003} and Shikoku, southwestern Japan \cite{Obara_etal2004}. 
 
\begin{figure}
\centering \includegraphics[width=6.0cm]{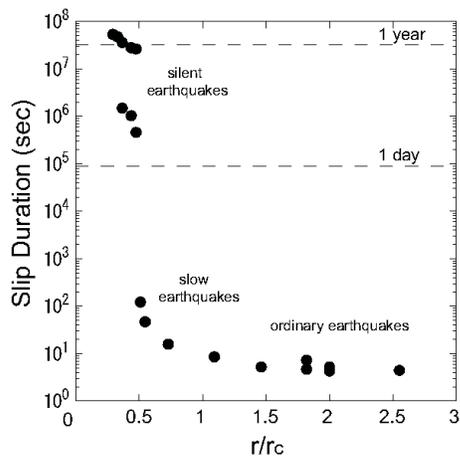}
   \caption{\label{sloweq}The duration of slip events versus $r/r_c$ in numerical simulations using the RSF law, where a circular asperity of radius $r$ with velocity-weakening friction is embedded in a fault of velocity-strengthening frictional property. $r_c$ denotes the critical fault radius for unstable slip defined in the text \cite{Kato2003a}.}
\end{figure}

These findings of slow earthquakes and low-frequency tremors force us to reconsider simple view of earthquakes as brittle fracture. Many mechanical models for slow earthquakes has been proposed. Since both seismic and aseismic slip can be easily modeled with the RSF law, it is natural to consider that slow earthquakes can be modeled with the RSF law. In fact, sustaining aseismic oscillation, similar to recurrence of slow earthquakes, occurs in a single-degree-of-freedom spring-block model if the spring stiffness $k$ is equal to the critical stiffness $k_{\rm crt}$ given by Eq. (\ref{kc}) \cite{Ruina1983}. Using a more realistic elastic continuum model, \textcite{Kato2004} showed that slow earthquakes occur when the size of velocity-weakening region is close to the critical size of slip nucleation zone. An effective stiffness $k_{\rm eff}$ of a fault may be defined by
\begin{equation}
	k_{\rm eff} = \Delta \tau/\Delta u,
\end{equation}							
where $\Delta \tau$ is shear-stress change on the fault due to slip $\Delta u$ \cite{Dieterich1986}. For a circular fault of radius $r$ with a constant stress drop in an infinite uniform elastic medium with Poission ratio = 0.25, $k_{\rm eff}$ is given by
\begin{equation}
	k_{\rm eff} = \frac{7 \pi}{24} \frac{G}{r},
\end{equation}		 	 							
where $G$ is rigidity. Recalling that unstable slip occurs for $k < k_{\rm crt}$ for a spring-block model, unstable slip is expected to occur for $k_{\rm eff} < k_{\rm crt}$ on a fault in an elastic medium. This leads to the condition for occurrence of unstable slip is that the fault radius $r$ is larger than the crtical fault size $r_c$ given by
\begin{equation}
	r_c = \frac{7 \pi}{24} \frac{G}{(b'-a')\sigma_n} {\cal L},
\end{equation}							
where $\sigma_n$ is the normal stress.
Note that the critical nucleation zone size $r_c$ obtained by considering the stability around steady-state sliding may not be realistic in natural conditions during earthquake cycles. Other forms of critical nucleation zone sizes were obtained by considering more realistic conditions \cite{Dieterich1992,RubinAmpuero2005}. It is confirmed in numerical simulations that usual earthquakes with short slip duration occurs for a circular fault with $r > r_c$, continuous stable sliding for $r \ll r_c$, and slow earthquakes for $r \sim r_c$, where slip duration increases with a decrease in $r/r_c$ as shown in Fig. \ref{sloweq} (Kato, 2003b, 2004). The same idea was adopted by \textcite{LiuRice2007} in their model for slow earthquakes at a subduction zone, where they showed that high pore fluid pressure in the fault zone is required to explain observed recurrence intervals and slip amounts of slow earthquakes. Although these models are simple and plausible, slow earthquakes may occur under
  limited conditions of $r \sim r_c$. This seems to be inconsistent with the observations that slow earthquakes are common phenomena at some regions. Using a two-degree-of-freedom spring-block model, \textcite{YoshidaKato2003} showed that slow earthquakes may occur for wider conditions by considering interaction between unstable block where usual earthquakes repeatedly occur and a conditionally stable block where slow earthquakes occur. \textcite{ShibazakiIio2003} and \textcite{ShibazakiShimamoto2007} introduced a cut-off velocity to the state evolution effect in Eq.(\ref{rsf}) to obtain frictional property of velocity weakening ($d \mu_{ss}/d \ln V < 0$) at low velocities and of velocity strengthening ($d \mu_{ss}/d \ln V > 0$) at high velocities, which is similar to the model by \textcite{Kato2003b} for deep preseismic sliding. Similar complex frictional behavior that $d \mu_{ss}/d \ln V$ depends on velocity was actually observed in the laboratory for halite \cite{Shimamoto1986} and for serpentine \cite{Moore_etal1997}. In this case, slip is accelerated at low velocities with $d \mu_{ss}/d \ln V < 0$ and is decelerated at high velocities with $d \mu_{ss}/d \ln V > 0$, leading to slow earthquakes. Repeating slow earthquakes at transition depths from shallow locked zone to deeper stable sliding zone were simulated in \textcite{ShibazakiIio2003} and \textcite{ShibazakiShimamoto2007}. This kind of model was further extended to simulate short- and long-term SSEs and their interaction with shallower large interplate earthquakes \cite{Matsuzawa_etal2010}. Weakness of these models is that experimental data for velocity dependence of $d \mu_{ss}/d \ln V$ are insufficient and frictional properties at depths where slow earthquakes occur are unknown. \textcite{Rubin2008} reviewed models for slow earthquakes based on the RSF law and pointed out that the existing models seem to be difficult to explain common occurrence of slow earthquakes at subduction zone
 s. He suggested that the variation of pore fluid pressure due to inelastic dilation of fault zone and fluid diffusion is required for generating slow earthquakes.  

\subsubsection{Origin of complexities of earthquakes and aftershock decay law}
\noindent \textcite{Rice1993} claimed that complex earthquake sequences simulated in inherently discrete models may be artifact and geometrical and/or material heterogeneity is required to explain observed complexity of earthquakes. Continuum models with relatively homogeneous frictional properties produce simple patterns of earthquakes such as periodic recurrence of large earthquakes that break the entire seismogenic zone. Using a continuum model with the RSF law, \textcite{Ben-ZionRice1995} introduced heterogeneity in effective normal stress on the fault and successfully produces moderately complex earthquake sequences. They pointed out that abrupt change in the effective normal stress is necessary to produce complex earthquakes. \textcite{Hillers_etal2007} introduced spatial heterogeneity in characteristic slip distance ${\cal L}$ in the model of a vertical strike-slip fault to produce complex earthquake sequences that include a wide range of earthquake magnitude. The obtained relation between earthquake magnitude and frequency mimics the Gutenberg-Richter (GR) law and the statistical properties of simulated earthquakes depend on the degree of heterogeneity in ${\cal L}$. They also argued temporal clustering of simulated earthquakes and tendency of nucleation sites of smaller ${\cal L}$. \textcite{HillersMiller2007} introduced spatial variation of pore pressure to generate complex earthquake sequences.   

An important fact about the relation between magnitude and frequency of earthquakes obtained in observations is that the GR law may not always be valid for each individual fault. For some faults and plate boundaries, the number of small earthquakes is too few than that expected from the GR law and the frequency of large earthquakes that rupture the entire fault, indicating violation of the GR law \cite{Stirling_etal1996,IshibeShimazaki2009}. This behavior of fewer small earthquakes than that expected from the frequency of large earthquakes is referred to the characteristic earthquake model. Highly coupled plate interface in the Tokai region, central Japan, is nearly quiescent, while many small earthquakes occur in the overriding plate and subducting oceanic plate \cite{Matsumura1997}. This suggests that except for great earthquakes few small earthquakes occur at the plate interface in the Tokai region. Considering large earthquakes along the San Andreas fault, California, and
  smaller earthquakes at secondary faults around the San Andreas, \textcite{Turcotte1997} argued that the observed GR law comes from a fractal distribution of faults and characteristic earthquakes at each fault.

Another important empirical law that demonstrates complexities of earthquakes is the modified Omori (Omori-Utsu) law for decay in aftershock occurrence rate \cite{Utsu_etal1995}. Aftershock rate $n$ at time $t$ from the occurrence of the mainshock is well approximated by
\begin{equation}
	n(t) = \frac{K}{(t+T_{\rm MOL})^p},
\end{equation}	
where, $K$, $T_{\rm MOL}$, and $p$ are constants. Constant $p$ takes $\sim 1$ for many cases.
In the case of $T_{\rm MOL}=0$, this relation is simply referred to as the Omori law.
Aftershocks have been thought to be manifestation of relaxation of stress generated by the mainshock.
To explain delay times of aftershocks, subcritical cracking due to stress corrosion \cite{YamashitaKnopoff1987} and the variation of effective normal stress due to diffusion of pore fluid, whose pressure is perturbed by the mainshock \cite{BoslNur2002}, were invoked.
\textcite{Dieterich1994b} considered responses of many fault patches, where friction is assumed to obey the RSF law, to instantaneous stress change due to the mainshock. He furtehr assumed that a constant seismicty rate is achieved under a constant loading rate without any stress perturbation. This model successfully explains the power law decay of seismicity rate with $p = 1$, being consistent with observations, and has been applied to analyses of aftershocks of some large earthquakes \cite{Toda_etal1998}.
Another important model for aftershocks using the RSF law is related to afterslip. Afterslip perturbs stresses around its source area, causing aftershocks. Differentiating the slip function Eq.(\ref{uniformslip}) of afterslip with respect to time, we have a slip rate approximately proportional to $(t+c)^{-1}$, which may related to a stress rate and therefore seismicity rate \cite{PerfettiniAvouac2004}. This expected seismicity rate coincides with the Omori-Utsu formula with $p=1$. Moreover, afterslip propagates outward from a mainshock slip area, leading to expansion of aftershock area \cite{Kato2007}. Aftershock expansion pattern obtained from a numerical model with the RSF law is consistent with observed expansions of aftershock ares \cite{TajimaKanamori1985,PengZhao2009}.

\subsubsection{Earthquake dynamics: critical slip distance}
\begin{figure}
\centering \includegraphics[width=6.0cm]{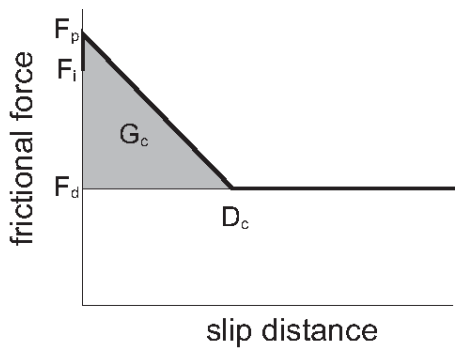}
   \caption{\label{slipweakening} Schematic diagram of the relation between frictional force and slip distance during slip-weakening process, where $F_i$, $F_p$, $F_d$, and $D_c$ denote the initial force, the peak frictional force, the dynamic frictional force, and the critical slip distance, respectively. The shaded area indicates the fracture energy $G_c$.}
\end{figure}

\noindent Here we consider the dynamics of unstable motion. The unstable slip of a spring-block 
system given by Eq. (\ref{onebody}) is accompanied by the drop of frictional force.
If one plots the frictional force as a function of the slip distance (Fig. \ref{slipweakening}), one can define the distance $D_c$ over which the frictional force drops.
This behavior of decreasing frictional force with increasing slip is referred to the slip-weakening model \cite{Ida1972,Andrews1976}, and the slip distance $D_c$ is called the critical slip distance  in seismology.
$D_c$ is on the same order of (or at most several tens of) the characteristic length ${\cal L}$ in an evolution law \cite{BizzarriCocco2003}.
This is so irrespective of the number of degrees of freedom: discrete or continuum.

Importantly, one can estimate $D_c$ of earthquakes by analyzing 
seismic wave. Such analyses show that $D_c$ is on the order of 
several tens of centimeters or a meter \cite{Ide1997}. Note that fracture energy $G_c$, which is equal to twice the surface energy density $\Gamma$, can rather stably be estimated from seismic waveform data, though accurate estimate of $D_c$ is difficult because of  poor resolution of rupture process modeling from seismic waves \cite{GuatteriSpudish2000}.
The characteristic slip distance ${\cal L}$ estimated for afterslip of a large interplate earthquake by GPS data 
is on the order of mm (Fukuda et al. 2009). 
This makes a quite contrast to laboratory experiments, where ${\cal L}$ is typically estimated 
as several micrometers.
Because ${\cal L}$ is a typical longitudinal dimension of true contact patch, application of the RSF law to a natural fault implies that a natural fault also consists of true contact patches, a linear dimension of which is several tens of centimeters.
Although the aperture of a fault is not empty but filled with fluid and gouge, 
a fault generally has the non-planer structure (e.g., jogs) that can interlock to resist 
the displacement. Such jogs may effectively act as the true contact area.
However, it is not obvious at all if the RSF law still holds for such true contact area of macroscopic scale.

At least, we believe that the RSF law should not be used except for very low speed friction. Namely, the RSF law no longer
 holds at seismic slip rate due to physical processes caused by 
frictional heat: flash heating, melting, mechanochemical effects, etc.
In such cases, the critical slip distance $D_c$ is proportional to $\epsilon_c/P$, 
where $\epsilon_c$ is the critical energy per unit area for a weakening process (e.g. melting) to occur and $P$ is the normal stress.
(As the frictional force is proportional to $P$, the produced heat is proportional to $P$ and to the slip distance $D$.
Thus, the weakening process may occur if $PD$ is on the order of $\epsilon_c$.)
Namely, the critical slip distance is inversely proportional to the normal pressure.
This implies that the critical slip distance must be smaller for deeper faults.
However, unfortunately, such depth dependence has not been observed so that 
the mechanism that determines the critical slip distance is different.

Another important process that affect the critical slip distance is the off-fault fracture 
accompanied by the crack propagation on fault.
Andrews (2005) analyzes a model for the slip propagation on a fault supplemented with 
the Coulomb yield condition for off-fault material.
He finds that the effective critical slip distance depends on the distance from 
the crack initiation point. This is because the plastic zone is wider for larger crack.
Thus, the critical slip distance is essentially scale dependent, 
which is consistent with the observation facts.
\section{Earthquake models and statistics II: SOC and other models}

\subsection{Statistical properties of the OFC model}

\subsubsection{The model}

\noindent  In the previous section, we reviewed the properties of statistical physical models of earthquakes such as the spring-block BK model and the continuum model. In the present section, we deal with further simplified statistical physical models of earthquakes (Turcotte 1997; Hergarten, 2002; Turcotte, 2009). Many of them were coupled map lattice models originally introduced as the SOC models (Bak, Tang and Wiesenfeld, 1987; Bak and Tang, 1989, Ito and Matsuzaki, 1990; Nakanishi, 1990; Brown, Scholz and Rundle, 1991; Olami, Feder and Christensen, 1992; Hainzl, Z\"oller and Kurths, 1999; 2000; Hergarten, and Neugebauer, 2000; Helmstetter, Hergarten and Sornette, 2004).

 The one introduced by Olami, Feder and Christensen (OFC) as a further simplification of the BK model, now called the OFC model, is particularly popular (Olami, Feder and Christensen, 1992). It is a two-dimensional coupled map lattice model where the rupture propagates from lattice site to its nearest-neighboring sites  in a non-conservative manner, often causing multi-site ``avalanches''. Extensive numerical studies have also been devoted to this model, mainly in the field of statistical physics, which we wish to review in the present section (Christensen and Olami, 1992; Grassberger, 1994; Middleton and Tang, 1995; Bottani and Delamotte, 1997; de Carvalho and Prado, 2000; Pinho and Prado, 2000; Lise and Paczuski, 2001; Miller and Boulter, 2002; Hergarten and Neugebauer, 2002; Boulter and Miller, 2003; Helmstetter, Hergarten and Sonette, 2004; Peixoto and Prado, 2006; Wissel and Drossel, 2006; Ramos, 2006; Kotani, Yoshino and Kawamura, 2008; Kawamura et al, 2010, Jagla, 2010).

 In the OFC model,  ``stress'' variable $f_i$ ($f_i\geq 0$) is assigned to each site on a square lattice with $L\times L$ sites. Initially,  a random value in the interval [0,1] is assigned to each $f_i$, while $f_i$ is increased with a constant rate uniformly over the lattice until, at a certain site $i$, the $f_i$ value reaches a threshold, $f_c=1$. Then, the site $i$ ``topples'', and a fraction of stress $\alpha f_i$ ($0<\alpha<0.25$) is transmitted to its four nearest neighbors, while $f_i$ itself is reset to zero.  If the stress of some of the neighboring sites $j$ exceeds the threshold, {\it i.e.\/}, $f_j\geq f_c=1$, the site $j$ also topples, distributing a fraction of stress $\alpha f_j$ to its four nearest neighbors. Such a sequence of topplings continues until the stress of all sites on the lattice becomes smaller than the threshold $f_c$. A sequence of toppling events, which is assumed to occur instantaneously, corresponds to one seismic event or an avalanche. After an avalanche, the system goes into an interseismic period where uniform loading of $f$ is resumed, until some of the sites reaches the threshold and the next avalanche starts. 

 The transmission parameter $\alpha$ measures the extent of non-conservation of the model. (This $\alpha$ should not be confused with $\alpha$ describing the velocity-weakening friction force employed in the study of the BK model of subsection III.A. We are using $\alpha$ as a conservation parameter of the OFC model throughout this subsection IV.A). The system is conservative for $\alpha =0.25$, and is non-conservative for $\alpha <0.25$. A unit of of time is taken to be the time required to load $f$ from zero to unity.

 In the OFC model,  boundary conditions play a crucial role. For example,  SOC state is realized under open or free boundary conditions, but is not realized under periodic boundary conditions. Thus, most of the studies made in the past employed open (or free) boundary conditions.

\subsubsection{Properties of the homogeneous model}

\noindent  Earlier studies concentrated mostly on the event size distribution of the model (Olami, Feder and Christensen, 1992; Christensen and Olami, 1994; Grassberger, 1994; de Carvalho and Prado, 2000; Lise and Paczuski, 2001; Miller and Boulter, 2002; Boulter and Miller, 2003; Drossel, 2006). The avalanche size $s$ is defined by the total number of ``topples'' in a given avalanche, which could be larger than the number of toppled sites because multi-toppling is possible in a given avalanche. (In fact, it is observed that multi-toppling rarely occurs in the model except in the conservation limit or in the regime very close to it.)  It turned out that the size distribution of the model exhibited a power-law-like behavior close to the GR law. Yet, there still remains controversy concerning whether the model is strictly critical (Lise and Paczuski, 2001) or only approximately so (de Carvalho and Prado, 2000; Miller and Boulter, 2002; Boulter and Miller, 2003; Drossel, 2006). 

\begin{figure}[ht]
\begin{center}
\includegraphics[scale=0.7]{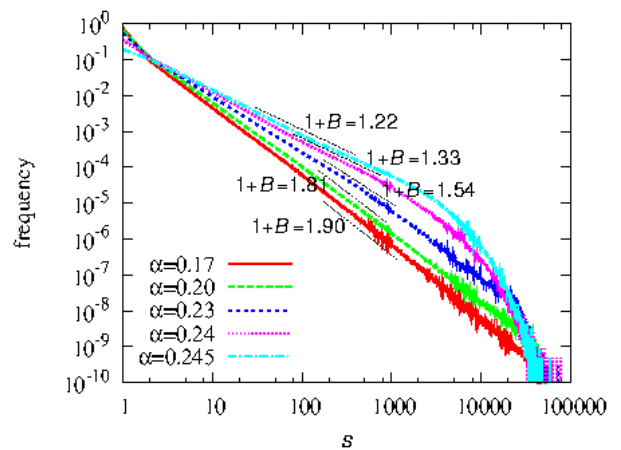}
\end{center}
\caption{
(Color online) The size distribution of the OFC model under open boundary conditions for various values of the transmission parameter $\alpha $. The slope of the data gives the value of $1+B$, which is shown in the figure.  Taken from (Kawamura et al, 2010).
}
\label{magnitude-OFC}
\end{figure}

 In Fig.\ref{magnitude-OFC}, we show the size distribution of the model under open boundary conditions for several values of the transmission parameter $\alpha$ (Kawamura et al, 2010). As can be seen from the figure, a near straight-line behavior corresponding a power-law is observed.  The slope representing the $B$-value is not universal varying from $\simeq 0.90$ to $\simeq 0.22$ as $\alpha$ is varied from 0.17 to 0.245.  The power-law feature is weakened as one approaches the conservation limit.

\begin{figure}[ht]
\begin{center}
\includegraphics[scale=0.8]{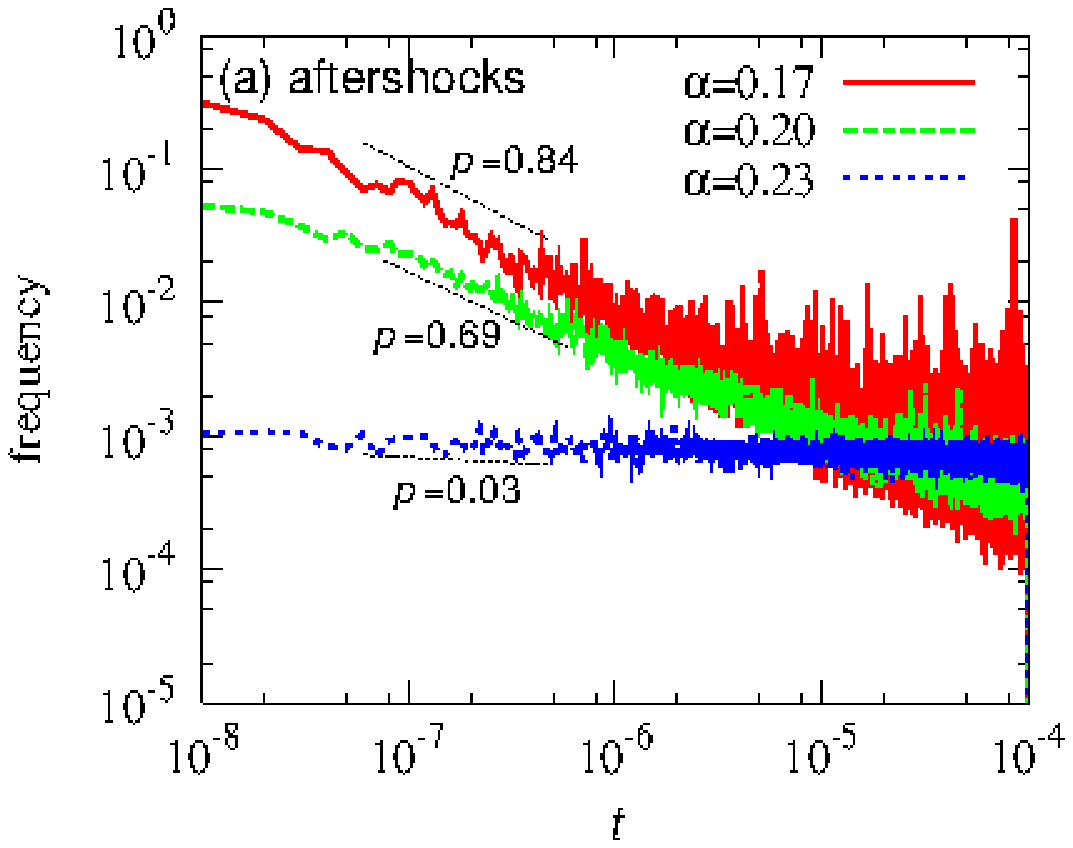}
\includegraphics[scale=0.8]{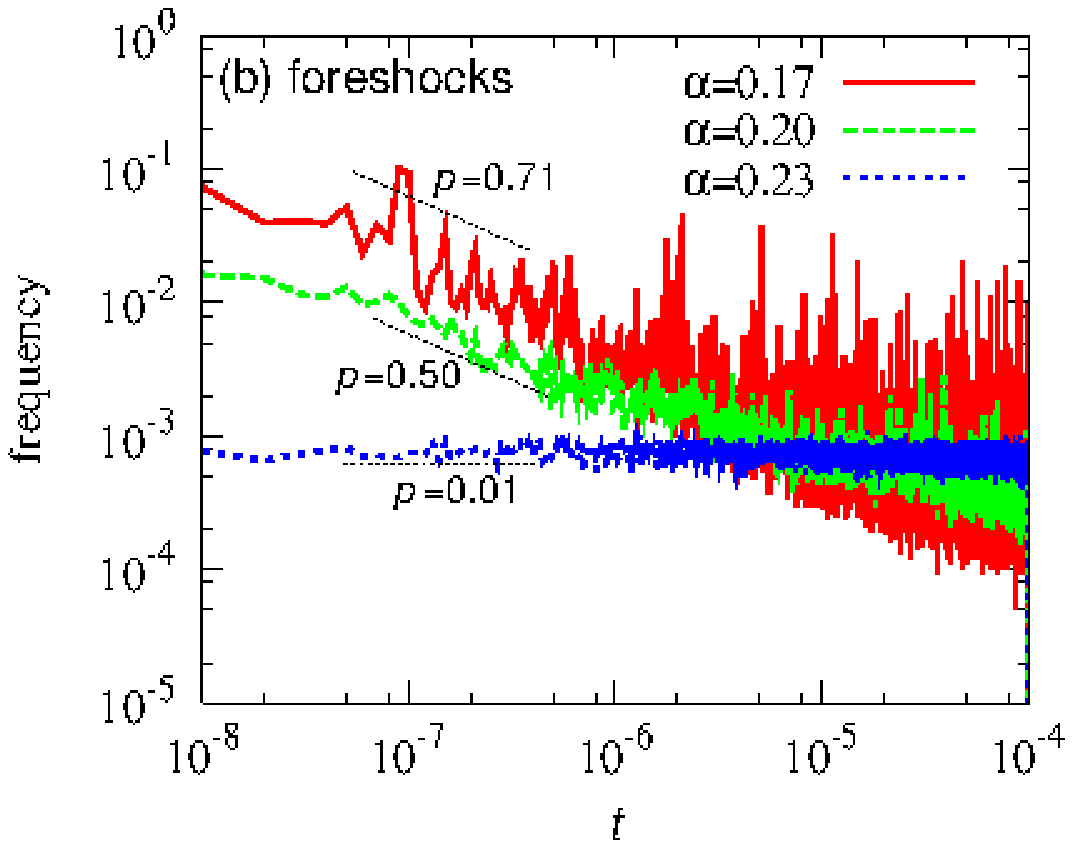}
\end{center}
\caption{
(Color online) The time dependence of the frequency of aftershocks (a), and of foreshocks (b), of the OFC model under open boundary conditions on a log-log plot for several values of the transmission parameter $\alpha$. Mainshocks are the events of their size greater than $s\geq s_c=100$. The time $t$ is measured with the occurrence of a mainshock as an origin. The range parameter is $r_c=10$. Taken from (Kawamura et al, 2010).
}
\label{Omori-OFC}
\end{figure}

 Hergarten {\it et al\/} observed that the OFC model also exhibited another well-known power-law feature of seismicity, {\it i.e.\/}, the Omori law (or the inverse Omori law) describing the time evolution of the frequency of aftershocks (foreshocks) (Hergarten and Neugebauer, 2002; Helmstetter, Hergarten and Sonette, 2004). We show in Fig.\ref{Omori-OFC}(a) on a log-log plot the frequency of aftershocks as a function of the time elapsed after the mainshock $t$ (Kawamura et al, 2010). The slope representing the Omori exponent $p$ is again not universal depending on the parameter $\alpha$ as $p=0.84$, 0.69 and 0.03 for $\alpha=0.17$,0.20 and 0.23, respectively. Since the $p$-value is known to come around unity in real seismicity, the $p$ value of the OFC model is not necessarily close to real observation.  Similar results are obtained also for foreshocks: See Fig.\ref{Omori-OFC}(b). Aftershocks and foreshocks are defined here as events of arbitrary  sizes which occur in the vicinity of mainshock with its epicenter lying with the distance $r\leq r_c$ (the range parameter $r_c$ it taken to be $r_c=10$ in Fig.\ref{Omori-OFC}). As one approaches the conservation limit $\alpha=0.25$, both aftershocks and foreshocks tend to 
 go away. 

 As mentioned, the properties of the model depends on applied boundary conditions. Middleton and Tang observed that the model under open boundary conditions went into a special transient state where events of size 1 (single-site events) repeated periodically with period $1-4\alpha$ (Middleton and Tang, 1995). These single-site events occur in turn in a spatially random manner, but after time $1-4\alpha$, the same site topples repeatedly. Although such a periodic state consisting of single-site events is a steady state  under periodic boundary conditions, it is not a steady state under open boundary conditions because of the boundary. Indeed, clusters are formed near the boundary, within which the stress values are more or less uniform, and gradually invades the interior destroying the periodic state. Eventually, such clusters span the entire lattice, giving rise to an SOC-like steady state. Middleton and Tang pointed out that such clusters might be formed via synchronization  between the interior site and the boundary site, the latter having a slower effective loading rate due to the boundary. Large-scale synchronization occurring in the steady state of the OFC model was further investigated by Bottani and Delamotte (Bottani and Delamotte, 1997). 

\begin{figure}[ht]
\begin{center}
\includegraphics[scale=0.7]{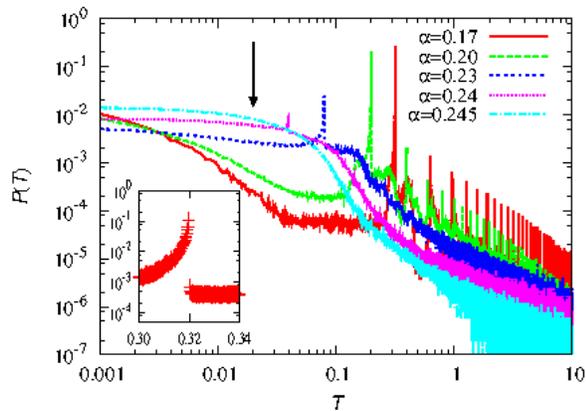}
\end{center}
\caption{
(Color online) Log-log plots of the local recurrence-time distributions of large avalanches of their size $s\geq s_c=100$  for a fixed range parameter $r_c=10$, with varying the transmission parameter $\alpha$. The arrow in the figure represents the expected peak position  for $\alpha=0.245$ corresponding to the period $T^*=1-4\alpha =0.02$. The inset is a magnified view of the main peak for the case of $\alpha=0.17$. Taken from (Kawamura et al, 2010).
}
\label{recurrence-OFC}
\end{figure}

 In contrast to the aforementioned critical properties of the model, recent studies also unraveled the {\it characteristic\/} features of the OFC model (Ramos, 2006; Kotani, Yoshino and Kawamura, 2008; Kawamura et al, 2010). By investigating the time series of events, Ramos found the nearly periodic recurrence of large events (Ramos, 2006). Kotani {\it et al\/} studied the spatiotemporal correlations of the model and identified in the OFC model a phenomenon resembling the ``asperity'' (Kotani, Yoshino and Kawamura, 2008; Kawamura et al, 2010). These authors computed the local recurrence-time distribution, $P(T)$, of the model. The computed $P(T)$, shown in Fig.~\ref{recurrence-OFC}, exhibited a sharp $\delta$-function-like peak at $T=T^*=1-4\alpha$, indicating that many (though not all) events of the OFC model were repeated with a fixed  time-interval $T=T^*$.  While the peak at $T=T^*$ is sharp, it is not infinitely sharp with a finite intrinsic width: See the inset. The  peak position turned out to be independent of the range parameter $r_c$, the size threshold $s_c$, and the lattice size (as long as it was not too small). As $\alpha$ is increased toward $\alpha=0.25$, the $\delta$-function peak is gradually suppressed with keeping its position strictly at $T=1-4\alpha$. The $\delta$-function peak of $P(T)$ goes away toward the conservation limit $\alpha=0.25$: See Fig.~\ref{recurrence-OFC}.

 In the longer time regime $T>T^*$, $P(T)$ exhibits behaviors close to power laws (Kotani, Yoshino and Kawamura, 2008; Kawamura et al, 2010).  Furthermore, the periodic events contributing to a sharp peak of $P(T)$ (``peak events'') possess a power-law-like size distribution very much similar to those observed for other aperiodic events (Kotani, Yoshino and Kawamura, 2008; Kawamura et al, 2010). Hence, in earthquake recurrence of the model,  the characteristic or periodic feature, {\it i.e.\/}, a sharp peak in $P(T)$ at $T=T^*$, and the critical feature, {\it i.e.\/}, power-law-like behaviors of $P(T)$ at $T>T^*$ and power-law-like size distribution, coexist.

\subsubsection{Asperity-like phenomena}

\noindent  In fact, it has turned out that the $\delta$-function peak of $P(T)$ is borne by ``asperity-like'' events, {\it i.e.\/}, the events which rupture repeatedly with almost the same period $1-4\alpha$ and with a common rupture zone and a common epicenter. In seismology, the concept of asperity is now quite popular. A typical example might be the one observed along the subduction zone in northeastern Japan, particularly repeating earthquakes off Kamaishi (Matsuzawa, Igarashi and Hasegawa, 2002; Okada, Matsuzawa and Hasegawa, 2003).

\begin{figure}[ht]
\begin{center}
\includegraphics[scale=0.75]{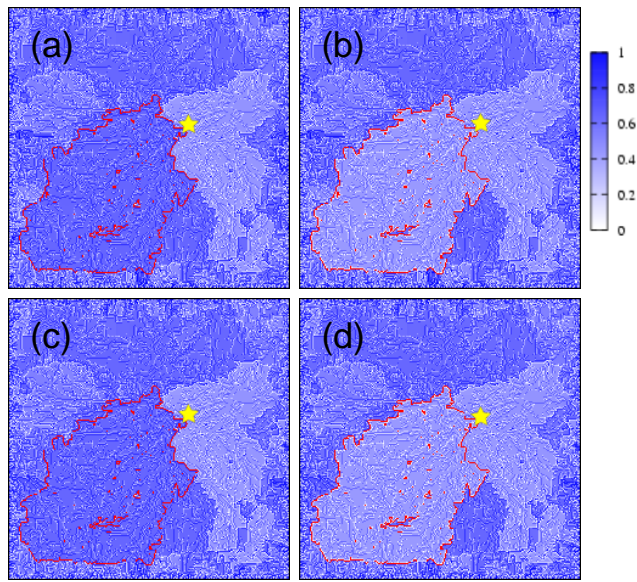}
\end{center}
\caption{
(Color online) Snapshots of the stress-variable distribution of the OFC model under open boundary conditions for the case of $\alpha=0.2$; (a) immediately before a large event at time $t=t_0$, (b) immediately after this event, (c) immediately before the following event which occurs at time $t=t_0+T^* (T^*=0.2)$, and (d) immediately after this second event. Two events are of size $s=15891$ and $s=15910$ on a $L=256$ lattice. The region surrounded by red bold lines represents the rupture zone, while the star symbol represents an epicenter site which is located at the tip of the rupture zone. Taken from (Kawamura et al, 2010).
}
\label{asperity-OFC}
\end{figure}

 In Fig.~\ref{asperity-OFC}, we show typical examples of such asperity-like events as observed in the OFC model (Kawamura et al, 2010).  In the upper panel, we show for the case of $\alpha=0.2$ typical snapshots of the stress-variable distribution immediately before and after a large event which occurs at time $t=t_0$. Discontinuous drop of the stress associated with a rupture of a synchronized cluster is discernible. Then, at time $t=t_0+T^*$, the same cluster (except for a minor difference) ruptures again. In the lower panel, we show snapshots of the stress-variable distribution immediately before and after this subsequent avalanche occurring at $t=t_0+T^*$. In this particular example,  a rhythmic rupture of essentially the same cluster has repeated more than ten times.

 The asperity-like events go away in the conservation limit $\alpha \rightarrow 1/4$ (Kawamura et al, 2010). It is also observed that an epicenter site tends to lie at the tip or at the corner of the rupture zone rather than in its interior (Kawamura et al, 2010). The asperity-like events observed in the OFC model closely resemble those familiar in seismology (Scholz, 2002), in the sense that almost the same spatial region ruptures repeatedly with a common epicenter site and  with a common period. 

 In fact, not all large events of the OFC model occur in the form of asperity.  Many clusters forming large events are left out of the rhythmic recurrence, and rupture more critically with widely-distributed recurrence time, thereby bearing the observed power-law-like part of $P(T)$. 

\begin{figure}[ht]
\begin{center}
\includegraphics[scale=0.7]{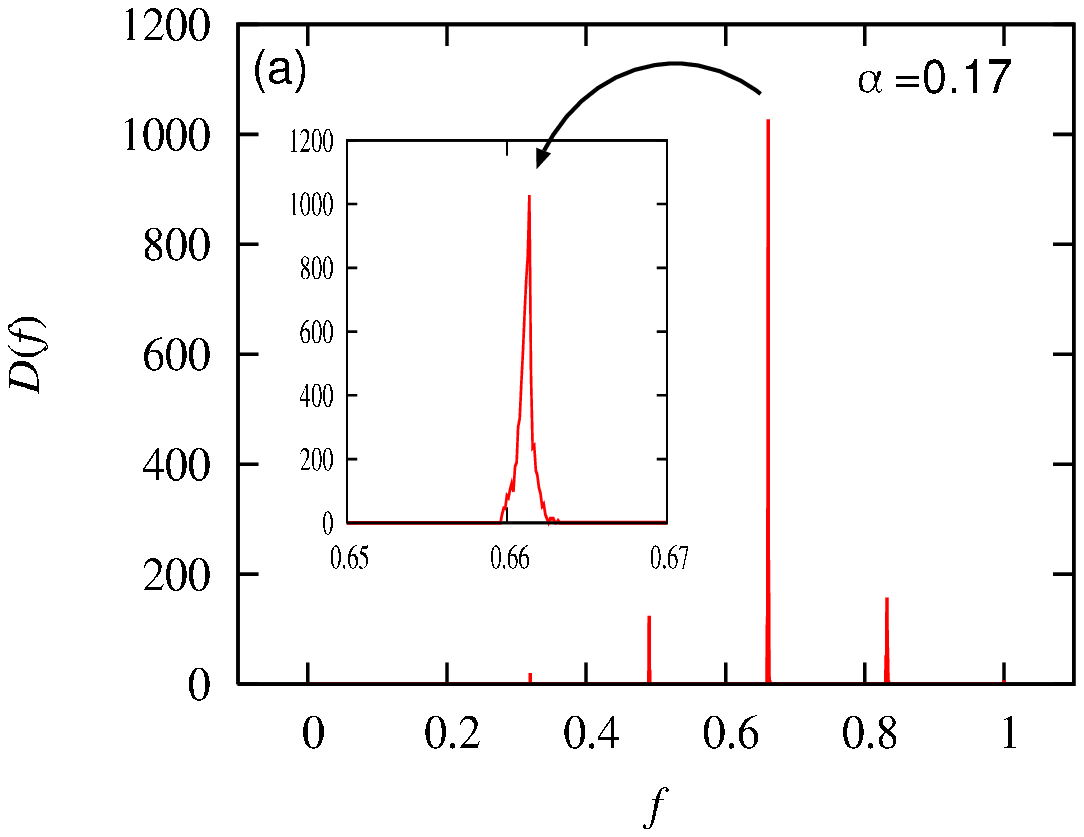}
\includegraphics[scale=0.7]{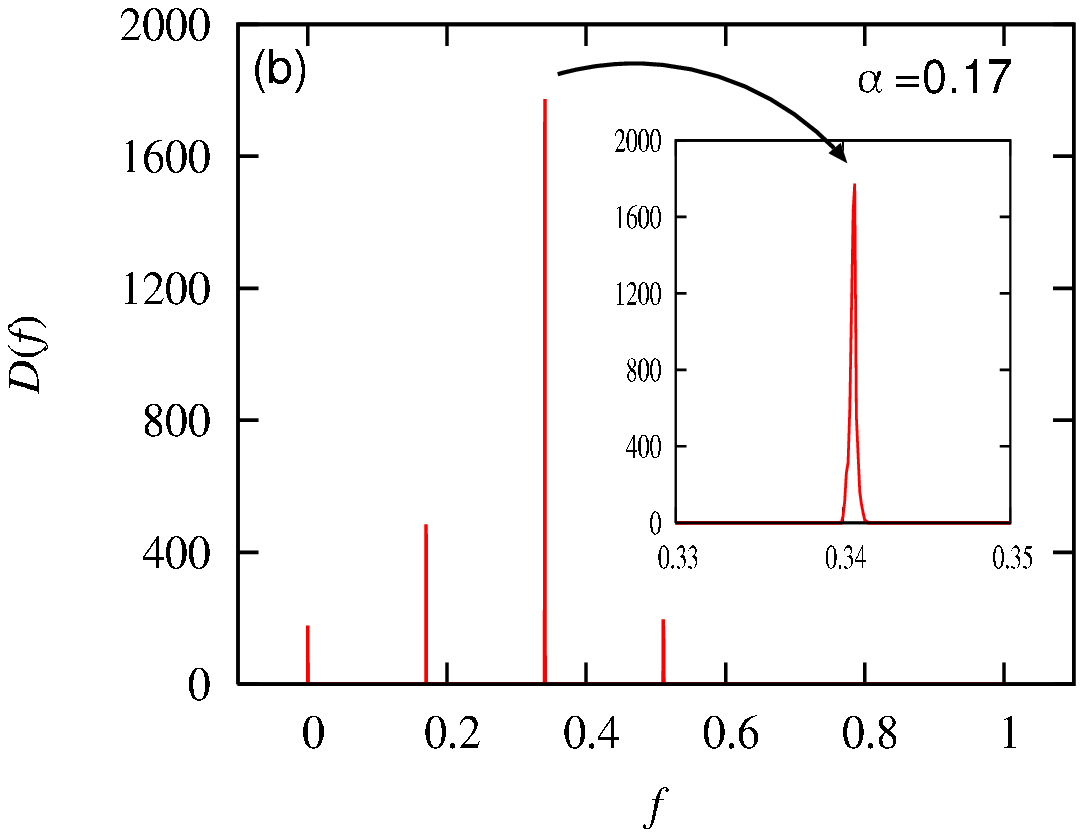}
\end{center}
\caption{
(Color online) The stress-variable distribution $D(f)$ of each site contained in the rupture zone of the asperity event of the OFC model under open boundary conditions, just before (a) and after (b) the asperity event. An asperity event is defined here as an event of its size greater than $s\geq s_c=100$ belonging to the main peak of the local recurrence-time distribution function. The transmission parameter is $\alpha=0.17$. The inset is a magnified view of the main peak. Taken from (Kawamura et al, 2010).
}
\label{stress-OFC}
\end{figure}

\begin{figure}[ht]
\begin{center}
\includegraphics[scale=0.7]{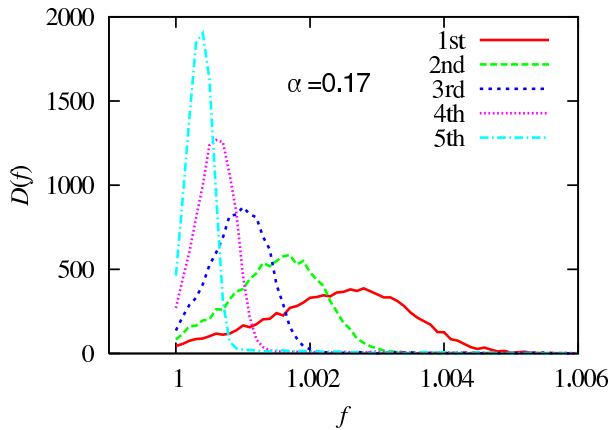}
\end{center}
\caption{
(Color online) The  time sequence of the stress-variable distribution $D(f)$ at the time of toppling of each site contained in the rupture zone of the asperity events. An asperity event is defined here as an event of its size greater than $s\geq s_c=100$ belonging to the main peak of the local recurrence-time distribution function. The transmission parameter is $\alpha=0.17$. As the events repeat, the stress-variable distribution at the time of toppling gets more and more concentrated on the borderline value $f_c=1$. Taken from (Kawamura et al, 2010).
}
\label{stress-evolution-OFC}
\end{figure}

 A key ingredient in the asperity formation is a self-organization of the highly concentrated stress state (Kawamura et al, 2010). The stress-variable distribution in the asperity region tends to be ``discretized'' to certain values. In Fig.~\ref{stress-OFC}, we show for the case of $\alpha=0.17$ the stress-variable distribution $D(f)$ of the asperity sites immediately (a) before and (b) after an avalanche, averaged over asperity events. As can be seen from the figure, $D(f)$ now consists of several ``spikes'' located at appropriate multiples of the transmission parameter $\alpha$, {\it i.e.\/},  at $1-n\alpha$ before the rupture, and at $f=n\alpha$ after the rupture, with $n$ being an integer.  Furthermore, as the asperity events repeat, the tendency of the stress-variable concentration is more and more enhanced. In Fig.~\ref{stress-evolution-OFC}, we show the time sequence of the stress-variable distribution at the time of toppling for the asperity events. As the asperity events repeat, the stress-variable distribution tends to be narrower, being more concentrated on the threshold value $f_c=1$ (Kawamura et al, 2010; Hergarten and Krenn, 2011).

 In fact, one can prove that the stress-variable distribution at the time of toppling tends to be more concentrated on the threshold value $f_c=1$ as the asperity events repeat (Kawamura et al, 2010). Namely, once each site starts to topple more or less at similar stress values close to the threshold value $f_c=1$, this tendency is more and more evolved as the asperity events repeat. {\it The stress-variable concentration tends to be self-organized\/}. Such a stress-variable concentration immediately explains why the interval time of the asperity events is equal to $1-4\alpha$, and why the same site becomes an epicenter in the asperity sequence (Kawamura et al, 2010). For example, the reason why the interval time is $1-4\alpha$  when all sites topple at the stress value close to the threshold  $f_c= 1$ in the asperity events  can easily be seen just by remembering the conservation law of the stress, {\it i.e.\/}, the stress-variable dissipated at the time of toppling, which is $1-4\alpha$ per site if the toppling occurs exactly at $f=1$, should match the stress loaded during the interval time $T$. See (Kawamura et al, 2010), for further details. More recently, Hergarten and Krenn made further analysis of this stress-concentration phenomenon, demonstrating that the mean stress excess representing the extent of the stress concentration approaches zero exponentially with a certain decay time which is dependent on the number of ``internal'' sites (the sites contained in the rupture zone) connected to an epicenter site (Hergarten and Krenn, 2011). Then, the epicenter site with the smallest number of internal nearest-neighbor sites, {\it i.e.\/}, the one lying at the tip of the rupture zone, has the longest decay time and turns out to be most stable. This observation gives an explanation of the finding of (Kawamura et al, 2010) that the majority of epicenter sites of the asprity-like events are located at the tip of the rupture zone.


 Although the origin of the asperity is usually ascribed  in seismology to possible inhomogeneity of the material property of the crust or of the external conditions of that particular region, we stress that, in the present OFC model, there is no built-in inhomogeneity in the model parameters nor in the external conditions. {\it The ``asperity'' in the OFC model is self-generated from the spatially uniform evolution-rule and model parameters.\/}

 As mentioned, the asperity in the OFC model is not a permanent one: In long terms, its position and shape change. After all, the model is uniform. Nevertheless, recovery of spatial uniformity often takes a long time, and the asperity exists stably over many earthquake recurrences. Although one has to be careful in immediately applying the present result for the OFC model to real earthquakes, it might be instructive to recognize that the observation of asperity-like earthquake recurrence does not immediately mean that the asperity region possesses different material properties nor different external conditions from other regions.

 Thus, critical and characteristic features coexist in the OFC model in an intriguing manner. Although the critical features were emphasized in earlier works, the model certainly involves the eminent characteristic features in it as well. Thus, the OFC model, though an extremely simplified model, may capture some of the essential ingredients necessary to understand apparent coexistence of critical and characteristic properties in real earthquakes.

\subsubsection{Effects of inhomogeneity}

\noindent  It should be noticed that the original OFC model is a spatially homogeneous model, where homogeneity of an earthquake fault is implicitly assumed. Needles to say, real earthquake fault is more or less spatially inhomogeneous, which might play an important role in real seismicity. Then, a natural next step is to extend the original homogeneous OFC model to the inhomogeneous one where the evolution rule and/or the model parameters are taken to be random from site to site. 

 Spatial inhomogeneity could be either static or dynamical. As a cause of possible temporal variation of spatial inhomogeneity, one may consider the two distinct processes, {\it i.e.\/}, the fast dynamical process during an earthquake rupture changing the fault state via, {\it e.g.\/},  wear, frictional heating, melting, {\it etc\/} and  many slower processes taking place during a long interseismic period until the next earthquake, {\it e.g.\/}, water migration, plastic deformation, chemical reactions, {\it etc\/} (Scholz, 2002). Thus, in introducing the spatial inhomogeneity into the OFC model, there might be two extreme ways: In one, one may assume that the randomness is quenched in time, namely, spatial inhomogeneity is fixed over many earthquake recurrences. In the other extreme, spatial inhomogeneity is assumed to vary with time in an uncorrelated way over earthquake recurrences.

 Several studies have been made on the inhomogeneous OFC model for both types of inhomogeneities. For the first type of inhomogeneity, {\it i.e.\/}, the quenched or static randomness, Janosi and Kertesz introduced spatial inhomogeneity into the stress threshold and found that the inhomogeneity destroyed the SOC feature of the model (Janosi and Kertesz, 1993). Torvund and Froyland studied the effect of spatial inhomogeneity in the stress threshold, and  observed that the inhomogeneity induced a periodic repetition of system-size avalanches (Torvund and Froyland, 1995). Ceva introduced defects associated with the transmission parameter $\alpha$, and observed that the SOC feature was robust against small number of defects (Ceva, 1995). Mousseau and Bach et al introduced inhomogeneity into the transmission parameter at each site. These authors observed that the bulk sites fully synchronized  in the form of a system-wide avalanche over a wide parameter range of the model (Mousseau
 , 1996; Bach, Wissel and Dressel, 2008). 

 For the second type of inhomogeneity, {\it i.e.\/}, the dynamical randomness, Ramos considered the randomness associated with the stress threshold, and observed that the nearly periodic recurrence of large events persisted (Ramos, 2006). More recently, Jagla studied the same stress-threshold inhomogeneity, to find that the GR law was weakened by randomness (Jagla, 2010). Very interesting observation by Jagla is that, once the slow structural relaxation process is added to the inhomogeneous OFC model, both the GR law and the Omori law are realized with the exponents which are stable against the choice of the model parameter values and are close to the observed values. Yamamoto {\it et al\/} studied the dynamically inhomogeneous model with a variety of implementations of the form of inhomogeneities to find the general tendency that critical features found in the original homogeneous OFC model, {\it e.g.\/},  the Gutenberg-Richter law and the Omori law, were weakened or suppressed in the presence of inhomogeneity, whereas the characteristic features of the original homogeneous OFC model, {\it e.g.\/},  the near-periodic recurrence of large events and the asperity-like phenomena, tended to persist (Yamamoto, Yoshino and Kawamura, 2010). 

 Thus, the properties of the dynamically inhomogeneous models are quite different from those of the static or quenched inhomogeneous models. In the latter case, introduced inhomogeneity often gives rise to a full synchronization and a periodic repetition of system-size events. Such a system-wide synchronization is never realized in the dynamically homogeneous models. Presumably, temporal variation of the spatial inhomogeneity may eventually average out the inhomogeneity over many earthquake recurrences, giving rise to the behavior similar to that of the homogeneous model.

\subsection{Fiber bundle models}
\begin{figure}
\resizebox{7.0cm}{!}{\includegraphics{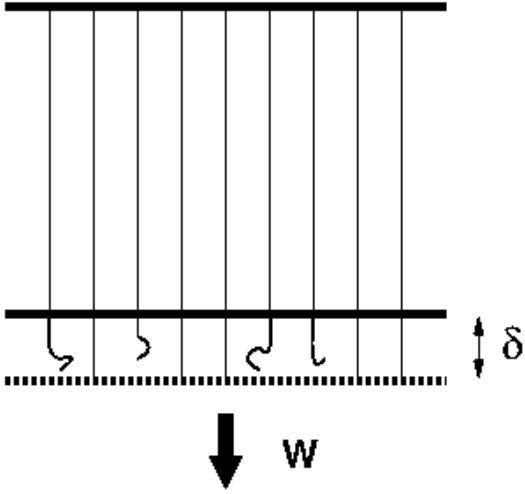}}
\caption{
The fiber bundle consists initially of $N$ fibers attached in parallel
to a fixed and rigid plate at the top and a downwardly movable platform 
from which a load $W$ is suspended at the bottom. In the 
equal load sharing
model considered here, the platform is absolutely rigid and the load $W$ is
consequently shared equally by all the intact fibers.}
\label{bkc-fbm-schematic}
\end{figure}
\noindent The fiber bundle model, initiated by \citet{bkc46} in the context of testing strength of cotton yarns, represents various aspects 
of fracture processes of disordered systems, through its self-organised dynamics (for detailed review see \citet{bkc34}).
The fiber bundle (see Fig.~\ref{bkc-fbm-schematic}) consists of $N$ fibers 
or Hook springs, each having
identical spring constant $\kappa$. The bundle supports a load 
$W=N\sigma$ and the breaking threshold $\left( \sigma _{th}\right) _{i}$
of the fibers are assumed to be different for different fiber ($i$).
For the equal load sharing model we consider here, the lower
platform is absolutely rigid, and therefore no local deformation and hence 
no stress concentration occurs anywhere around the failed fibers. This
ensures equal load sharing, i.e., the intact fibers share
the applied load $W$ equally and the load per fiber increases as
more and more fibers fail. The strength of 
each of the fiber $\left( \sigma_{th}\right)_{i}$ in the bundle is 
given by the stress value it can bear, and beyond which
it fails. The strength of the fibers are taken from a randomly distributed
normalised density $\rho (\sigma _{th})$ within the interval
$0$ and $1$ such that 
\[
\int _{0}^{1}\rho (\sigma _{th})d\sigma _{th}=1.
\] 
The equal load sharing assumption neglects `local' fluctuations
in stress (and its redistribution) and renders the model as a mean-field
one. 

The breaking dynamics starts when an initial stress \( \sigma  \)
(load per fiber) is applied on the bundle. The fibers having strength
less than \( \sigma  \) fail instantly. Due to this rupture, total
number of intact fibers decreases and rest of the (intact) fibers
have to bear the applied load on the bundle. Hence effective stress
on the fibers increases and this compels some more fibers to break.
These two sequential operations, namely the stress redistribution and further
breaking of fibers continue till an equilibrium is reached, where
either the surviving fibers are strong enough to bear the applied
load on the bundle or all fibers fail.

\begin{figure}
\centering\resizebox*{7cm}{!}{\includegraphics{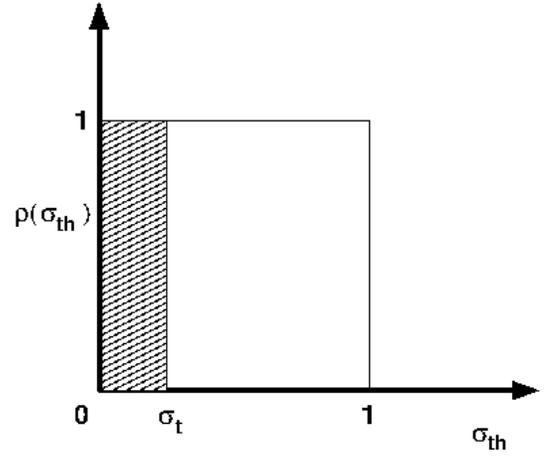}}
\caption{
The simple model considered here assumes uniform density $\rho (\sigma _{th})$
of the fiber strength distribution up to a cutoff strength (normalized to 
unity). At any load per fiber level $\sigma_t$ at time $t$, the fraction 
$\sigma_t$ fails and $1-\sigma_t$ survives.
}
\label{bkc-uniform}
\end{figure}

This breaking dynamics can be represented by recursion
relations in discrete time steps. 
For this, let us consider a very simple model of fiber bundles where
the fibers (having the same spring constant $\kappa$) have a white
or uniform strength distribution  $\rho(\sigma_{th})$ upto a cutoff strength
normalized to unity, as shown in Fig. \ref{bkc-uniform}:
$\rho (\sigma_{th}) = 1$ for $0 \le \sigma_{th} \le 1$ and 
$\rho(\sigma_{th})=0$ for $\sigma_{th} > 1$.
Let us also define $U_t(\sigma)$ to be the fraction of fibers in the bundle
that survive after (discrete) time step $t$, counted from the time $t=0$
when the load is put (time step indicates the number of stress 
re-distributions). As such, $U_t(\sigma=0)=1$ for all $t$ and $U_t(\sigma)=1$
for $t=0$ for any $\sigma$; 
$U_t(\sigma)=U^*(\sigma) \ne 0$ for $t \to \infty$ and 
$\sigma < \sigma_f$, the critical or failure strength of the bundle, and 
$U_t(\sigma)=0$ for $t \to \infty$ if $\sigma > \sigma_f$.

Therefore $U_{t}(\sigma)$ follows a simple recursion relation 
(see Fig. \ref{bkc-uniform})
\[
U_{t+1}= 1-\sigma_t;\ \ \sigma_t = \frac{W}{U_t N}
\]
\begin{equation}
\label{bkc-recU_t}
{\rm or,} \ \ U_{t+1}=1-\frac{\sigma }{U_{t}}.
\end{equation}
At the equilibrium state (\( U_{t+1}=U_{t}=U^{*} \)), the above relation
takes a quadratic form of \( U^{*} \) : 
\[
U^{*^{2}}-U^{*}+\sigma =0.
\]

\noindent The solution is

\[
U^{*}(\sigma )=\frac{1}{2}\pm (\sigma_{f}-\sigma )^{1/2};\sigma_{f}=\frac{1}{4}.
\]

\noindent Here \( \sigma_{f} \) is the critical value of initial
applied stress beyond which the bundle fails completely. The solution
with (\( + \)) sign is the stable one, whereas the one with (\( -) \)
sign gives unstable solution \cite{bkc8, bkc9, bkc10}. The quantity
\( U^{*}(\sigma ) \) must be real valued as it has a physical meaning:
it is the fraction of the original bundle that remains intact under
a fixed applied stress \( \sigma  \) when the applied stress lies
in the range \( 0\leq \sigma \leq \sigma_{f} \). Clearly, \( U^{*}(0)=1 \).
Therefore the stable solution can be written as 
\begin{equation}
\label{bkc-Ustarsigma_c}
U^{*}(\sigma )=U^{*}(\sigma_{f})+(\sigma_{f}-\sigma )^{1/2};
\ U^*(\sigma_f) = \frac{1}{2} \ {\rm and}\ \sigma_{f}=\frac{1}{4}.
\end{equation}
For \( \sigma >\sigma_{f} \) we can not get a real-valued fixed
point as the dynamics never stops until \( U_{t}=0 \) when the bundle
breaks completely.

\vskip.2in
\noindent \textbf{(a) At \(\sigma <\sigma_{f} \)}
\vskip.2in
\noindent It may be noted that the quantity \( U^{*}(\sigma )-U^{*}(\sigma_{f}) \)
behaves like an order parameter that determines a transition from
a state of partial failure (\( \sigma \leq \sigma_{f} \)) to a state
of total failure (\( \sigma >\sigma_{f} \)):
\begin{equation}
\label{bkc-Ustar}
O\equiv U^{*}(\sigma )-U^{*}(\sigma_{f})=(\sigma_{f}-\sigma )^{\beta };\beta =\frac{1}{2}.
\end{equation}

To study the dynamics away from criticality ($\sigma \rightarrow \sigma_{f}$
from below), we replace the recursion relation (\ref{bkc-recU_t}) by a differential
equation 
\[
-\frac{dU}{dt}=\frac{U^{2}-U+\sigma }{U}.
\]

\noindent Close to the fixed point we write \( U_{t}(\sigma )=U^{*}(\sigma ) \)
+ \( \epsilon  \) (where \( \epsilon \rightarrow 0 \)). This, following
Eq.~(\ref{bkc-Ustar}), gives 
\begin{equation}
\label{bkc-epsilon}
\epsilon =U_{t}(\sigma )-U^{*}(\sigma )\approx \exp (-t/\tau ),
\end{equation}

\noindent where \( \tau =\frac{1}{2}\left[ \frac{1}{2}(\sigma_{f}-\sigma )^{-1/2}+1\right]  \).
Near the critical point we can write \begin{equation}
\label{bkc-dec19}
\tau \propto (\sigma_{f}-\sigma )^{-\alpha };\alpha =\frac{1}{2}.
\end{equation}
 Therefore the relaxation time diverges following a power-law as \( \sigma \rightarrow \sigma_{f} \)
from below.

One can also consider the breakdown susceptibility \( \chi  \), defined
as the change of \( U^{*}(\sigma ) \) due to an infinitesimal increment
of the applied stress \( \sigma  \)  \begin{equation}
\label{bkc-sawq}
\chi =\left| \frac{dU^{*}(\sigma )}{d\sigma }\right| =\frac{1}{2}(\sigma_{f}-\sigma )^{-\gamma };\gamma =\frac{1}{2}
\end{equation}

\noindent from Eq.~\ref{bkc-Ustar}. Hence the susceptibility diverges as
the applied stress \( \sigma  \) approaches the critical value \( \sigma_{f}=\frac{1}{4} \).
Such a divergence in \( \chi  \) had already been observed in the
numerical studies.

\vskip.2in
\noindent \textbf{(b) At} \textbf{\large \(\sigma =\sigma_{f} \)}{\large\par}
\vskip.2in
\noindent At the critical point (\( \sigma =\sigma_{f} \)), we observe
a different dynamic critical behavior in the relaxation of the failure process.
From the recursion relation (\ref{bkc-recU_t}), it can be shown
that decay of the fraction \( U_{t}(\sigma_{f}) \) of unbroken fibers
that remain intact at time \( t \) follows a simple power-law decay:
\begin{equation}
\label{bkc-qqq}
U_{t}=\frac{1}{2}(1+\frac{1}{t+1}),
\end{equation}

\noindent starting from \( U_{0}=1 \). For large \( t \) (\( t\rightarrow \infty  \)),
this reduces to \( U_{t}-1/2\propto t^{-\delta } \); \( \delta =1 \);
a strict power law which is a robust characterization of the critical
state (see, however, \citet{bkc62}).

\subsubsection{Universality class of the model}

\noindent The universality class of the model has been checked  taking
two other types of fiber strength distributions: (I) linearly increasing
density distribution and (II) linearly decreasing density distribution
within the ($\sigma_{th}$) limit $0$ and $1$. One can show that while 
$\sigma_{f}$ changes with different strength distributions 
($\sigma_f= \sqrt{4/27}$ for case (I) and $\sigma_f=4/27$ for case II), 
the critical behavior remains unchanged: $\alpha =1/2=\beta =\gamma$, 
$\delta =1$ for all these equal load sharing models \cite{bkc34}.

\subsubsection{Precursors of global failure in the model}

\noindent In any such failure case, it is important to know {\it when} the failure will take place. In this model,
there exist several precursors. The growth of susceptibility $\chi$ with $\sigma$, following Eq.~\ref{bkc-sawq}
indeed suggests one such possibility: $\chi^{-1/2}$ decreases linearly with $\sigma$ to $0$ at $\sigma=\sigma_f$ from below. \citet{bkc47} studied
the rate $R(t)$ ($\equiv -\frac{dU_t}{dt}$) of failure of fibers following the dynamics like in Eq.~\ref{bkc-recU_t} for $\sigma>\sigma_f$ and
found that the rate becomes minimum at a time $t_0$, which is half of the failure time $t_f$ of the bundle. This relation is shown to be 
independent of the breaking strength distribution of the fibers. A similar relation was also found \cite{bkc52} for the rate of energy released
in a bundle. This is, of course, easier to measure using accoustic emmisions. 
\begin{figure} 
\centering\resizebox*{6cm}{!}{\includegraphics{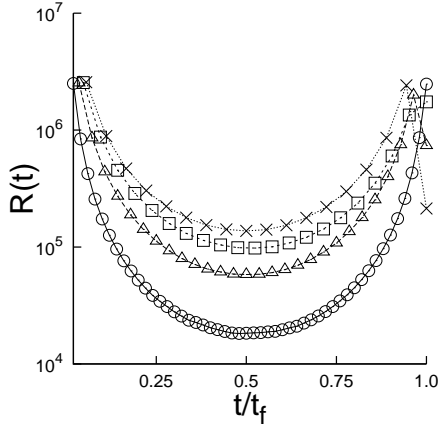}}
\caption{
The breaking rate $R(t)$  vs. the rescaled step variable $t_f /t$  for the
uniform threshold distribution for a bundle of $N = 10^7$ fibers.
Different symbols are  for different excess stress levels
$\sigma-\sigma_f$ : $0.001$ (circles), $0.003$ (triangles), $0.005$ (squares) and
$0.007$ (crosses). From \cite{bkc47}.
}
\label{fbm-failure}
\end{figure}

\subsubsection{Strength of the 
local load sharing fiber bundles}

\noindent So far, we studied models with fibers sharing the external load equally.
This type of model shows (both analytically and
numerically) existence of a critical strength (non zero $\sigma_{f}$)
of the macroscopic bundle  beyond which it collapses. 
The other extreme model, i.e., the local load sharing
model has been proved to be difficult to tackle analytically.

It is clear, however, that the extreme statistics
 comes into play for such
local load sharing models, for which the strength $\sigma_f \to 0$ as the
bundle size ($N$) approaches infinity. Essentially, for any finite load
($\sigma$), depending on the fiber strength distribution, the size of the
defect cluster can be estimated using 
Lifshitz argument (see section \ref{extrm})
as $\ln N$, giving the failure strength $\sigma_f \sim 1/(\ln N)^a$,
where the exponent $a$ assumes a value appropriate for the model 
(see e.g., \citet{bkc14}).
If a fraction $f$ of the load of the failed fiber goes for global
redistribution and the rest (fraction $1-f$) goes to the fibers
neighboring to the failed one, then we see  that there is a crossover from extreme to
self-averaging statistics at a finite value of $f$ (see e.g., \citet{bkc34}).

\subsubsection{Burst distribution: crossover behavior}
\noindent In fiber bundle models, when the load is slowly increased until a new failure occurs, a burst can be  defined 
as the number ($\Delta$) of  fiber failures following that failure. 
The distribution of such bursts ($D(\Delta)$) shows power-law behavior. It was shown for a generic 
case (independent of threshold distribution) that the form of this distribution (for continuous loading) is
\begin{equation}
D(\Delta)/N=C\Delta^{-\zeta}
\end{equation} 
in the limit $N\to \infty$.  

\begin{figure}
\centering \includegraphics[width=6.0cm]{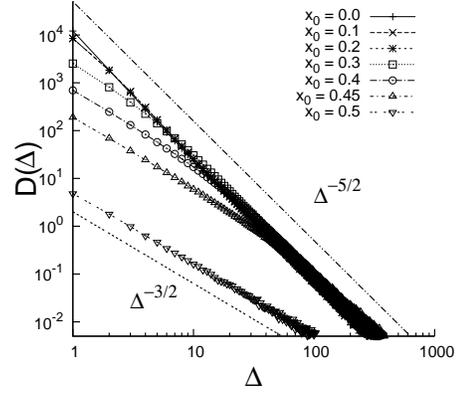}
   \caption{The burst size distribution for different values of $x_0$ in the equal load sharing model with uniform threshold distribution.
The number of fibers is $N=50000$ \cite{bkc36}.}
\label{cross1}
\end{figure}

The burst exponent
$\zeta$ has a value $\frac{5}{2}$ for average over all $\sigma(=0 \quad\mbox{to} \quad \sigma_f)$ and it is universal \cite{bkc37}.
However, the burst exponent value depends, for e.g., on the details of loading process and also from which point 
of the loading the burst statistics are recorded.  If the burst distribution is recorded 
only near the critical point ($\sigma \lesssim \sigma_f$), the burst exponent ($\zeta$) value becomes $3/2$ \cite{bkc35}. 
 For equal load sharing model model with uniform strength distribution, the burst distribution is shown (Fig.~\ref{cross1})
for recording that starts from different points of effective loading which is denoted by $x_0$, where $x_t=\sigma/U_t$ is 
the elongation or  the effective loading (for linear elastic behavior) at any point $t$). 
The crossover behavior is clearly seen. In these studies, the load increase rate is extremely slow and the increase is 
assumed to stop once a fiber fails.  The consequent avalanches are studied at that load. Once the avalanche stops, the
load is increased again. This process is realistic in the case of earthquakes where stress accumulation takes place over years. However,
if the increase in load is fixed ($d\sigma$), then the above exponent value of $\zeta$ becomes $3$: $\Delta \sim \frac{d(1-U^*)}{d\sigma}$,
giving $\Delta^{-2}=\sigma-\sigma_f$ (from Eq.~\ref{bkc-Ustarsigma_c}) and since $D(\Delta)d\Delta\sim d\sigma$, $D(\Delta)\sim\frac{d\sigma}{d\Delta}\sim \Delta^{-\zeta}$,
$\zeta=3$ \cite{bkc9}.

\begin{figure}
\centering \includegraphics[width=7.0cm]{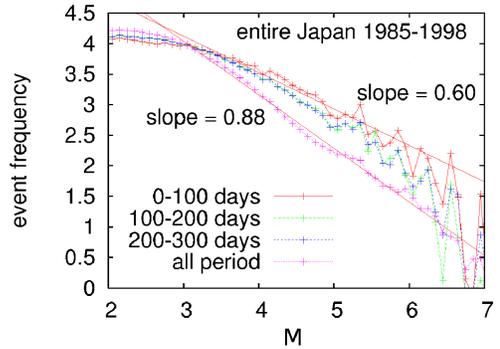}
   \caption{Crossover signature in the local magnitude distribution of earthquakes in Japan. During the 100 days before mainshock
the exponent is 0.60; much smaller than the average value 0.88. \cite{bkc38}}
\label{cross2}
\end{figure}

In fact, the earthquake frequency statistics may indeed show the crossover 
behavior mentioned above: If  event frequency is denoted by $D(M)$, then it is known that $D(M)\sim M^{-\zeta}$,
where $M$ denotes the magnitude (may be assumed to be related to avalanche size $\Delta$ in the models) 
and $\zeta$ value is found  \cite{bkc38} to be more ($\zeta\approx 0.9$) for statistics over 
a smaller time period (before the mainshock),  
compared to the long time average value ($\zeta\approx 0.6$); see Fig.~\ref{cross2}.
\subsection{Two fractal overlap model}
\noindent The common geometrical property observed in seismic faults is its fractal nature. It is now well known that, like other
 fractured surfaces, fault surfaces also posses self-affine roughness (see e.g., \citet{bkc61} and references therein). Therefore, it is worth investigating if earthquake
 phenomena can be modelled as an outcome of relative movement of two self-affine surfaces over each other. \citet{bkc1}, in a simplistic model,  
studied the overlap statistics of two Cantor sets in order to understand the 
underlying physics of such phenomena. 

Cantor set is a prototype example of fractal. In order to construct a triadic Cantor set, in the first step the middle third
 of a base interval [0,1] is removed. In the successive steps, the middle thirds of the remaining intervals ([0,1/3] and [2/3,1] and so
on) are removed. After $n$ such steps, the remaining set is called a Cantor set of generation $n$. When this process
 is continued ad infinitum i.e., in the limit $n\to\infty$, it becomes a true fractal.

\begin{figure}
\centering \includegraphics[width=9.0cm]{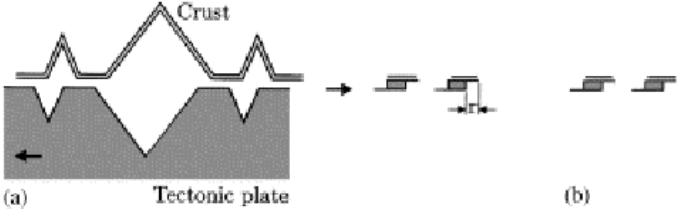}
   \caption{(a) Schematic representation of the rough earth's surface and the tectonic plate. (b) The one-dimensional projection
of the surfaces form overlapping Cantor sets.}
\label{tfo0-bkc}
\end{figure}

In this model, the solid-solid contact surfaces of both the earth's crust and the tectonic plate are considered
as average self-affine surfaces (see Fig.~\ref{tfo0-bkc}). The strain energy grown between the two surfaces due to a stick period is 
taken to be proportional to the overlap between them. During a slip event, this energy is released. Considering that such
 slips occur at intervals proportional to the length corresponding to that area, a power-law for the frequency distribution of the
 energy release is obtained. This compares well with the GR law (see e.g. \citet{bkc30}).

\subsubsection{Renormalisation group study: continuum limit}

\noindent Let the sequence of generators $G_n$ define the Cantor set at the $n$-th generation within the interval [0,1]: 
$ G_0=[0,1]$, $G_1 \equiv RG_0=[0,a] \cup [b,1]$, ... ,$G_{n+1}=RG_n,... $ . The mass density of the set $G_n$ is represented by 
$D_n(r)$ i.e., $D_n(r)=1$ if $r$ is in any of the occupied intervals of $G_n$ and 0 elsewhere. The overlap magnitude between 
the sets at any generation $n$ is then given by the convolution form $s_n(r)=\int dr^{\prime}D_n(r^{\prime})D_n(r-r^{\prime})$ 
(for symmetric fractals). 
\begin{figure}
\centering \includegraphics[width=9.0cm]{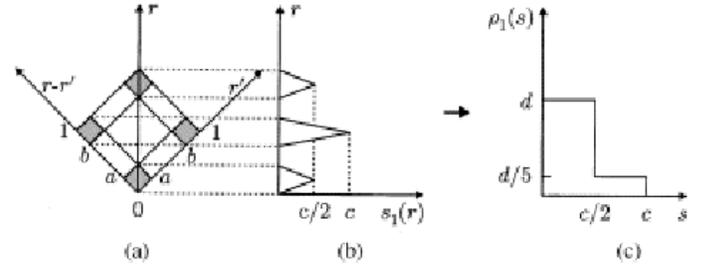}
   \caption{(a) Two Cantor sets along the axes $r$ and $r-r^{\prime}$. (b) The overlap $s_1(r)$ along the diagonal. (c) The 
   corresponding density $\rho_1(s)$.}
\label{tfo1-bkc}
\end{figure}
\begin{figure}
\centering \includegraphics[width=9.0cm]{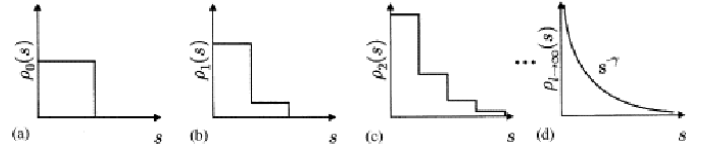}
   \caption{The overlap densities (probabilities) $\rho(s)$ at various generations; (a) zeroth, (b) first, (c) second, and (d)
infinite generation.}
\label{tfo2-bkc}
\end{figure}
One can express the overlap integral $s_1$ in the first generation by the projection of the shaded region along
 the vertical diagonals in Fig.~\ref{tfo1-bkc}(a). That gives the form shown in Fig.~\ref{tfo1-bkc}(b). 
For $a=b \le 1/3$, the non-vanishing $s_1(r)$
 regions do not overlap and are symmetric on both sides with slope of the middle curve being exactly double those on the sides. 
One can then easily check that the distribution $\rho_1(s)$ of overlap $s$ at this generation is given by Fig.~\ref{tfo1-bkc}, with both $c$ and $d$ 
greater than unity, maintaining the normalisation condition with $cd=5/3$. The successive generations of the density $\rho_n(s)$
may therefore be represented by Fig.~\ref{tfo2-bkc}, where

\begin{equation}
\rho_{n+1}(s)=\tilde{R}\rho_n(s)\equiv \frac{d}{5}\rho_n(\frac{s}{c})+\frac{4d}{5}\rho_n(\frac{2s}{c}).
\end{equation}  
   
In the infinite generation limit of the renormalisation group (RG) equation, if $\rho^*(s)$ denotes the fixed point distribution
such that $\rho^*(s)=\tilde{R}\rho^*(s)$, then assuming $\rho^*(s)\sim s^{-\gamma}\tilde{\rho}(s)$, one gets 
$(d/5)c^{\gamma}+(4d/5)(c/2)^{\gamma}=1$. Here $\tilde{\rho}(s)$ represents an arbitrary modular function, which also includes
 a logarithmic correction for large $s$. This agrees with the above mentioned normalisation condition $cd=5/3$ for the choice
$\gamma=1$ giving

\begin{equation}
\rho^*(s)\equiv \rho(s)\sim s^{-\gamma}; \qquad \gamma=1
\end{equation} 

The above analysis is for the continuous relative motion of the overlapping fractals. For discrete steps, the contact area distribution
 can be found exactly for two Cantor sets having same dimension ($log 2/log 3$) \cite{bkc2}. The step size is taken as the minimum element
in the generation at which the distribution is found.  

\subsubsection{Discrete limit}

\noindent Let $s_n(t)$ represent the amount of overlap between the two Cantor sets of generation $n$ at time $t$. Initially ($t=0$) the two 
identical Cantor sets are placed on top of each other, generating the maximum overlap ($2^n$ for the $n$-th generation sets). Then
in every time step (discrete) the length of the shift is chosen to be $1/3^n$ for the $n$-th generation, such that a line segment in one 
set either completely overlaps with one such segment on the other set or does not overlap at all, i.e., partial overlap of two segments
of the two sets are not allowed. Periodic boundary conditions are assigned in both the sets. The magnitude of overlap ($s_n(t)$), 
therefore, in this discrete version, is given by the number of overlapping pairs if line segment of the two sets. Because of 
the structure of the Cantor sets, the overlap magnitudes can only have certain discrete values which are in geometric progression:
$s_n=2^{n-k}$, $k=0,\dots,n$. 

Let $Nr(s_n)$ denote the the number of times an overlap $s_n$ has occurred in one period of the time series for the $n$-th  
generation (i.e. $3^n$ time steps). It can be shown that \cite{bkc30}
\begin{equation}
Nr(2^{n-k})={}^nC_k2^k, \qquad k=0,\dots,n
\end{equation}
Now, if $Prob(s_n)$ denotes the probability that after time $t$ there are $s_n$ overlapping segments, then for the general case of 
$s_n=2^{n-k}$, $k=0,\dots,n$ it is given by
\begin{eqnarray}
Prob(2^{n-k})&=&\frac{Nr(2^{n-k})}{\sum\limits_{k=0}^{n}Nr(2^{n-k})} \nonumber \\
&=& \frac{2^k}{3^n}{}^nC_k \nonumber \\
&=& {}^nC_{n-k}\left(\frac{1}{3}\right)^{n-k}\left(\frac{2}{3}\right)^k
\end{eqnarray}

\subsubsection{Gutenberg-Richter law}
\noindent Since the allowed values of the overlap are $s_n=2^{n-k}$, $k=0,\dots,n$, one can write $\log_2s_n=n-k$. Then the above equation becomes
\begin{eqnarray}
Prob(s_n)&=&{}^nC_{\log_2 s_n}\left(\frac{1}{3}\right)^{\log_2 s_n}\left(\frac{2}{3}\right)^{n-\log_2s_n}\nonumber \\
&&\equiv F(\log_2 s_n).
\end{eqnarray}
Near the maxima it may be written as
\begin{equation}
F(M)=\frac{3}{2\sqrt{n\pi}}\exp[-\frac{9}{4}\frac{(M-n/3)^2}{n}],
\end{equation}
where $M=\log_2 s_n$. To obtain the GR law analog from this distribution we have to integrate $F(M)$ from $M$ to $\infty$.
\begin{eqnarray}
F_{cum}(M) &=& \int\limits_{M}^{\infty}F(M^{\prime})dM^{\prime} \nonumber \\
&=& \int\limits_{M}^{\infty}\frac{3}{2\sqrt{n\pi}}\exp(-\frac{9}{4}\frac{(M^{\prime}-n/3)^2}{n})dM^{\prime}.
\end{eqnarray}
Substituting $p=\frac{3}{2\sqrt{n}}(M^{\prime}-n/3)$ we get
\begin{equation}
F_{cum}(M)=\frac{1}{\sqrt{\pi}}\int\limits_{\frac{3}{2\sqrt{n}}(M-n/3)}^{\infty}\exp(-p^2)dp.
\end{equation}
On simplification, it gives
\begin{equation}
\label{bkc-f-cum}
F_{cum}(M)=\frac{1}{3}\sqrt{\frac{n}{\pi}}\exp\left[-\frac{9}{4}\frac{(M-n/3)^2}{n}\right](M-n/3)^{-1}.
\end{equation}
$F_{cum}(M)$ in the above equation suggests that the `average' quakes are of magnitude $n/3$, while 
\begin{equation}
F_{cum}(M)\sim \exp\left[-(9/4)(M-n/3)^2/n\right]
\end{equation}
can be simplified for large $M$. Using $e^{-a^2}=\left(1/\sqrt{2\pi}\right) \int\limits_{-\infty}^{+\infty}e^{-x^2/2+\sqrt{2}ax}dx$ and $\int\limits_{-\infty}^{+\infty}e^{-f(x)}dx\sim e^{-f(x_0)}$
where $x_0$ refers to the extremal point with $\partial f/\partial x |_{x=x_0}=0$, one finds $F_{cum}(M)\sim e^{-(9/4)[M(m_0/n)-2M/3]}\sim e^{-3M/4}$ where $x_0=\left(\frac{3}{\sqrt{2n}}\right)m_0$; $m_0=n$. 
It gives \cite{bkc26}
\begin{equation}
\log F_{cum}(M)=A-\frac{3}{4}M,
\end{equation}
where $A$ is a constant depending on $n$.
This is the Gutenberg-Richter law in the model and  clearly holds for the high magnitude end of the distribution.
Also, one can equate easily the magnitude $M$ with the released energy $E$ by noting that $M=\log_2 s$ here. The overlap $s$ is related to energy $E$ and hence the
relation $M\sim \log E$, giving $F_{cum}\sim E^{-3/4}$.

Similar to outcome of the simple fractal models considered here, a power law behavior for the overlap distribution also occurs 
for two overlapping random Cantor sets, Sierpinsky gasket and Sierpinsky carpet overlapping on their respective replica \cite{bkc3},
 and a fractional Brownian profile overlapping on another \cite{bkc4}. In view of the generality of the power law distribution and the
fractal geometry of the fault surfaces, it is suggested that the GR law owes its origin significantly to the fractal geometry
of the fault surfaces. It may be noted that identifying the aftershocks as these adjusted overlaps, with average size given by Eq.~({\ref{bkc-f-cum}}),
one can define an average magnitude ($n/3$) dependent on the fractal geometry generator fraction ($=1/3$ here) and the genration number ($n$). This agrees with the observed
data quite satisfactorily (see \citet{bkc63}).

\subsubsection{Omori law}
\noindent Let $N^{(M_0)}(t)$ denote the cumulative number of aftershocks (of magnitude $M \ge M_0$, where $M_0$ is some threshold)
after the mainshock. Then the Omori law states that
\begin{equation}
\frac{dN^{(M_0)}(t)}{dt}=\frac{1}{t^p}.
\end{equation}
The value of the exponent $p$ is close to unity for tectonically active region, although a range of $p$ values are also 
observed (for review see \citet{bkc26}).
In practice, a particular value of $p$ is observed when the threshold $M_0$ is given. For this model, when the threshold is fixed
at the minimum (i.e., $M_0=1$), then $p=0$ due to the fact that aftershock occurs at every step in this model. However, interesting 
facts are seen when the threshold is set at the second highest possible value $n-1$ (recall that the second highest overlap was $2^{n-1}$).
Then for $t=2.3^{r_1}$ (where, $r_1=0,\dots,n-1$) there is an aftershock of magnitude $n-1$. 
Therefore, neglecting the prefactor 2, an aftershock of magnitude $n-1$ occurs in geometric progression with common ratio 3. Therefore
we get the general rule $N(3t)=N(t)+1$, leading to
\begin{equation}
N(t)=\log_3(t).
\end{equation}
On integration,  Omori law gives $N(t)=t^{1-p}$, and therefore from this model we get $p=1$, which is the Omori law suggested value for $p$. The model 
therefore gives a range of $p$ values between 0 and 1 which systematically increases within the range of threshold values. 
\section{Discussions and conclusions}
\noindent Earthquakes, due to their devastating consequences, have been a subjected of extensive studies in various diciplines, ranging from seismology 
to physics. Although the effeorts were not always commensurate (see also \citet{bkc59} for a critical
view of the inherent difficulties of the present approach of theretical physics), in the last decade considerable progress have been made in studying 
different aspects
of this vast topic. In this review, the progresses in such studies are discussed from the point of view of statistical physics. 

Principally being a large scale dynamic failure process, it is necessary to formulate the background of friction and fracture in order to understand 
the physics of earthquake. In Sec. II such issues are discussed: After the Griffith's theory for crack nucleation and the fracture stress statistics
of disordered solids, we discuss the RSF law and microscopic models for solid-solid friction.
Also, the effects that could lead to violations of RSF laws are also discussed. 

Several statistical approaches to model earthquake dynamics are discussed. The BK model is discussed in one and two spatial dimensions as well as its
long range version. In Sec. IIIA6, the continuum limit is also discussed, which gives `characteristic' earthquakes. BK model has also been discussed
in terms of RSF law. 
Apart from relatively complex modelling like that of BK models and continuum models, we also discuss simplistic models such as
OFC models, fiber bundle models and purely geometrical models like the Two Fractal Overlap model. While many details are lost in any such
model, they  still captures the complex nature of the dynamics and the different statistical aspects, helping us to gain new insights. 

As one can easily see that inspite of considerable progress in the study of such an important and complex dynamical phenomenon as earthquake, our knowledge is
far short of any satisfactory level. We believe, major collaborative efforts, involving physicist and seismologiests in particular, are urgently necessary
to unfold the dynamics and employ our knowledge of the precursor events to save us from catastrophic disasters in future. 
\begin{acknowledgements}
\noindent We acknowledge collaborations, at various stages, with P. Bhattacharya, P. Bhattacharyya, J. A. Eriksen, A. Hansen, S. Kakui, T. Kotani, T. Mori, 
S. Morimoto, S. Pradhan, P. Ray, R. B. Stinchcombe, T. Yamamoto and H. Yoshino.
\end{acknowledgements}

\appendix 
\section{Glossary}
\noindent  {\bf  aftershocks:} Small earthquakes that follow a large earthquake (main shock).

\noindent  {\bf  afterslip:} Aseismic sliding that follows an earthquake. 

 \noindent {\bf  asperity:} (a) A region where stick-slip motion occurs on a fault or a plate boundary. Strain energy is accumulated at an asperity during a stick stage between earthquakes and it is released by seismic slip at the occurrence of an earthquake.
(b) Junction of protrusions of the two contacting surfaces.

\noindent {\bf  Cantor set:} One starts with the set of real numbers in the interval [$0:1$] and divide the set in a few subsets and remove one of the  subsets in the first step. As the removal scheme is repeated ad infinitum, one is left with a dust of real numbers called the Cantor set. It is a fractal object.

\noindent {\bf  characteristic earthquakes:} Earthquakes that repeatedly rupture approximately the same segment of a fault.  The magnitudes and slip distributions of characteristic earthquakes are similar to one another.

\noindent {\bf  dynamical critical phenomena:}  Critical behaviors, which are associated with the dynamical properties of the system, rather than the equilibrium properties (e.g., thermal transition is Ising model) are called dynamical critical phenomena (e.g., depinning transition of a fracture front, time dependent field induced transitions in Isng model etc.). 

 \noindent {\bf  fiber bundle model: } Originating from texttile engineering, fiber bundle model is often used as a prototype model for fracture dynamics. In its simplest form, it consists of a large number of fibers or Hooke-springs. The bundle hangs from a rigid ceiling and  supports, via a platform at the bottom, a load. Each fiber has got identical spring constants but the breaking stress for each differs. Depending on the breaking stress of the fibers, the fibers fail and successive failure occur due to load redistribution, showing complex failure dynamics.

\noindent {\bf  fractals:} A fractal is a geometrical object having self-similarity in its internal structure.

\noindent  {\bf  fractional Brownian profile :} Fractional Brownian motion (fBm) is a continuous time random walk with zero mean. However, the directions of the subsequent steps of an fBm are correlated (positively or negatively). A fractional Brownian profile is the trajectory of such a walk. It is self-similar.

\noindent  {\bf  Gutenberg-Richter (GR) law:} 
The power law describing the magnitude-frequency relation of earthquakes. The frequency of earthquakes of its energy (seismic moment) $E$ decays with $E$ according to $\propto E^{-(1+B)}=E^{-(1+\frac{2}{3}b)}$ where $B$ and $b=\frac{3}{2}B$ are appropriate exponents.

\noindent  {\bf  Hamiltonian:} It is essentially the total energy of a system. For a closed system, it would be the sum of kinetic and potential energies.

\noindent  {\bf  Omori law:} 
The power law describing the decay of the number (frequency) of aftershocks with the time elapsed after the mainshock. 

\noindent  {\bf  power-law distribution:} (Also called `scale free distribution') A distribution of the generic form $P(x)\sim x^{\alpha}$. Note that there is no length-scale associated with this type of distribution, since a transformation like $x\to x/b$ would keep the functional form unchanged. Observables (e.g.,  magnetisation, susceptibility etc.) show power-law behavior near criticality. Therefore it is often considered as a signature of critical behavior.

\noindent  {\bf  slow earthquakes:} Fault slip events that radiate little or no seismic wave radiations. Rupture propagation velocities and slip velocities of slow earthquakes are much smaller than those of ordinary earthquakes. Slow earthquakes without seismic wave radiations are often called silent earthquakes. 

\noindent  {\bf  quenched randomness:} The randomness in the system which is not in thermal equilibrium with the same reservoir as the system and does not fluctuate are called quenched randomness.

\noindent  {\bf  rate-and-state friction (RSF) law:} An empirical constitutive law describing the dynamic friction coefficient either at steady states or transient states.

\noindent {\bf  self-organized criticality (SOC):} When the dynamics of a system leads it to a state of criticality (where scale invariance in time and space are observed) without any need of external tuning parameter, the system is said to have self-organized to a critical state. This phenomenon, where a critical point is an attractor of the dynamics, is called self-organized criticality.

\noindent {\bf  self-similarity and self-affinity:} Self similarity refers to the property of an object that it is similar (exactly or approximately) to one or more of its own part(s). Self-affinity refers to the properties of those objects which, in order to be self-similar, are to be scaled by different factor in x and y direction (for 2-d object).

\noindent {\bf  Sierpinski gasket and Sierpinski carpet:} Sierpinski carpet is a fractal object, embedded in a 2-d surface. Its construction is as follows: First a square is taken and it is divided into 9 equal squares. Then the square in the middle is removed. then similar operation is performed upon the 8 remaining squares. This  process is continued ad infinitum to obtain what is called a Sierpinski carpet. Sierpinski gasket (also called Sierpinski triangle) is again a fractal object. Its construction is as follows: First a equilateral triangle is taken. Then it is divided into four equilateral triangle of same sizes and the middle one is removed. Then same operation is performed upon the three remaining triangles. When this process is continued ad infinitum, one is left with what is called the Sierpinski gasket.

\noindent {\bf  universality class:} Phase transitions are characterised by a set of critical exponent values. The values of these exponents are independent of the microscopic details of the system and only depend on the symmetry and dimensionality of the order parameter. Therefore, a large class of systems often have same critical exponent values. A Universality class is a group of systems having same critical exponent values.

\nocite{*}

\bibliography{article}

\begin{thebibliography}{300}
\expandafter\ifx\csname natexlab\endcsname\relax\def\natexlab#1{#1}\fi
\expandafter\ifx\csname bibnamefont\endcsname\relax
  \def\bibnamefont#1{#1}\fi
\expandafter\ifx\csname bibfnamefont\endcsname\relax
  \def\bibfnamefont#1{#1}\fi
\expandafter\ifx\csname citenamefont\endcsname\relax
  \def\citenamefont#1{#1}\fi
\expandafter\ifx\csname url\endcsname\relax
  \def\url#1{\texttt{#1}}\fi
\expandafter\ifx\csname urlprefix\endcsname\relax\def\urlprefix{URL }\fi
\providecommand{\bibinfo}[2]{#2}
\providecommand{\eprint}[2][]{\url{#2}}

\bibitem[{\citenamefont{Abe and Mair}(2009)}]{Abe2009}
\bibinfo{author}{\bibnamefont{Abe}, \bibfnamefont{S.}}, and
  \bibinfo{author}{\bibfnamefont{K.}~\bibnamefont{Mair}}, \bibinfo{year}{2009},
  \bibinfo{journal}{Geophys. Res. Lett.}
  \textbf{\bibinfo{volume}{36}}(\bibinfo{number}{23}),
  \bibinfo{pages}{doi:10.1029/2009GL040684}.

\bibitem[{\citenamefont{Amar and Family}(1990)}]{bkc53}
\bibinfo{author}{\bibnamefont{Amar}, \bibfnamefont{J.~G.}}, and
  \bibinfo{author}{\bibfnamefont{F.}~\bibnamefont{Family}},
  \bibinfo{year}{1990}, \bibinfo{journal}{Phys. Rev. A}
  \textbf{\bibinfo{volume}{41}}, \bibinfo{pages}{3399}.

\bibitem[{\citenamefont{Ampuero and Rubin}(2008)}]{Rubin}
\bibinfo{author}{\bibnamefont{Ampuero}, \bibfnamefont{J.-P.}}, and
  \bibinfo{author}{\bibfnamefont{A.}~\bibnamefont{Rubin}},
  \bibinfo{year}{2008}, \bibinfo{journal}{J.Geophys.Res.}
  \textbf{\bibinfo{volume}{113}}, \bibinfo{pages}{doi:10.1029/2007JB005082}.

\bibitem[{\citenamefont{Andrews}(2005)}]{Andrews2005}
\bibinfo{author}{\bibnamefont{Andrews}, \bibfnamefont{D.}},
  \bibinfo{year}{2005}, \bibinfo{journal}{J. Geophys. Res.}
  \textbf{\bibinfo{volume}{110}}(\bibinfo{number}{B1}),
  \bibinfo{pages}{doi:10.1029/2004JB003191}.

\bibitem[{\citenamefont{Andrews}(1976)}]{Andrews1976}
\bibinfo{author}{\bibnamefont{Andrews}, \bibfnamefont{D.~J.}},
  \bibinfo{year}{1976}, \bibinfo{journal}{J. Geophys. Res.}
  \textbf{\bibinfo{volume}{81}}, \bibinfo{pages}{doi:10.1029/JB081i032p05679}.

\bibitem[{\citenamefont{Bach} \emph{et~al.}(2008)\citenamefont{Bach, Wissel,
  and Drossel}}]{Bach}
\bibinfo{author}{\bibnamefont{Bach}, \bibfnamefont{M.}},
  \bibinfo{author}{\bibfnamefont{F.}~\bibnamefont{Wissel}}, and
  \bibinfo{author}{\bibfnamefont{B.}~\bibnamefont{Drossel}},
  \bibinfo{year}{2008}, \bibinfo{journal}{Phys. Rev. E}
  \textbf{\bibinfo{volume}{77}}, \bibinfo{pages}{067101}.

\bibitem[{\citenamefont{Bak and Tang}(1989)}]{bt89}
\bibinfo{author}{\bibnamefont{Bak}, \bibfnamefont{P.}}, and
  \bibinfo{author}{\bibfnamefont{C.}~\bibnamefont{Tang}}, \bibinfo{year}{1989},
  \bibinfo{journal}{J. Geophys. Res.} \textbf{\bibinfo{volume}{94}},
  \bibinfo{pages}{doi:10.1029/JB094iB11p15635}.

\bibitem[{\citenamefont{Bak} \emph{et~al.}(1987)\citenamefont{Bak, Tang, and
  Wiesenfeld}}]{btw87}
\bibinfo{author}{\bibnamefont{Bak}, \bibfnamefont{P.}},
  \bibinfo{author}{\bibfnamefont{C.}~\bibnamefont{Tang}}, and
  \bibinfo{author}{\bibfnamefont{K.}~\bibnamefont{Wiesenfeld}},
  \bibinfo{year}{1987}, \bibinfo{journal}{Phys. Rev. Lett.}
  \textbf{\bibinfo{volume}{59}}, \bibinfo{pages}{381}.

\bibitem[{\citenamefont{Baraba\'{s}i and Stanley}(1995)}]{bkc41}
\bibinfo{author}{\bibnamefont{Baraba\'{s}i}, \bibfnamefont{A.~L.}}, and
  \bibinfo{author}{\bibfnamefont{H.~E.} \bibnamefont{Stanley}},
  \bibinfo{year}{1995}, \emph{\bibinfo{title}{Fractal Concepts in Surface
  Growth}} (\bibinfo{publisher}{Cambrigde University Press N. Y.}).

\bibitem[{\citenamefont{Ben-Zion}(2008)}]{Ben-Zion2008}
\bibinfo{author}{\bibnamefont{Ben-Zion}, \bibfnamefont{Y.}},
  \bibinfo{year}{2008}, \bibinfo{journal}{Rev. Geophys.}
  \textbf{\bibinfo{volume}{46}}(\bibinfo{number}{RG4006}),
  \bibinfo{pages}{doi:10.1029/2008RG000260}.

\bibitem[{\citenamefont{Ben-Zion and Rice}(1995)}]{Ben-ZionRice1995}
\bibinfo{author}{\bibnamefont{Ben-Zion}, \bibfnamefont{Y.}}, and
  \bibinfo{author}{\bibfnamefont{J.~R.} \bibnamefont{Rice}},
  \bibinfo{year}{1995}, \bibinfo{journal}{J. Geophys. Res.}
  \textbf{\bibinfo{volume}{100}}(\bibinfo{number}{B7}),
  \bibinfo{pages}{doi:10.1029/94JB03037}.

\bibitem[{\citenamefont{Ben-Zion and Rice}(1997)}]{Ben-ZionRice}
\bibinfo{author}{\bibnamefont{Ben-Zion}, \bibfnamefont{Y.}}, and
  \bibinfo{author}{\bibfnamefont{J.~R.} \bibnamefont{Rice}},
  \bibinfo{year}{1997}, \bibinfo{journal}{J. Geophys. Res.}
  \textbf{\bibinfo{volume}{102}}, \bibinfo{pages}{doi:10.1029/97JB01341}.

\bibitem[{\citenamefont{Bergman and Stroud}(1992)}]{bkc27}
\bibinfo{author}{\bibnamefont{Bergman}, \bibfnamefont{D.~J.}}, and
  \bibinfo{author}{\bibfnamefont{D.}~\bibnamefont{Stroud}},
  \bibinfo{year}{1992}, in \emph{\bibinfo{booktitle}{Solid State Phys.}},
  edited by \bibinfo{editor}{\bibfnamefont{H.}~\bibnamefont{Ehrenreich}} and
  \bibinfo{editor}{\bibfnamefont{D.}~\bibnamefont{Turnbull}}
  (\bibinfo{publisher}{Academic press, New York}), volume~\bibinfo{volume}{46},
  p. \bibinfo{pages}{147}.

\bibitem[{\citenamefont{Bhattacharya}
  \emph{et~al.}(2011)\citenamefont{Bhattacharya, Chakrabarti, and
  Kamal}}]{bkc63}
\bibinfo{author}{\bibnamefont{Bhattacharya}, \bibfnamefont{P.}},
  \bibinfo{author}{\bibfnamefont{B.~K.} \bibnamefont{Chakrabarti}}, and
  \bibinfo{author}{\bibnamefont{Kamal}}, \bibinfo{year}{2011},
  \bibinfo{journal}{J. Phys.: Conf. Series} \textbf{\bibinfo{volume}{319}},
  \bibinfo{pages}{012004}.

\bibitem[{\citenamefont{Bhattacharya}
  \emph{et~al.}(2009)\citenamefont{Bhattacharya, Chakrabarti, Kamal, and
  Samanta}}]{bkc26}
\bibinfo{author}{\bibnamefont{Bhattacharya}, \bibfnamefont{P.}},
  \bibinfo{author}{\bibfnamefont{B.~K.} \bibnamefont{Chakrabarti}},
  \bibinfo{author}{\bibnamefont{Kamal}}, and
  \bibinfo{author}{\bibfnamefont{D.}~\bibnamefont{Samanta}},
  \bibinfo{year}{2009}, in \emph{\bibinfo{booktitle}{Rev. Nonlin. Dyn. and
  Complexity}}, edited by \bibinfo{editor}{\bibfnamefont{H.~G.}
  \bibnamefont{Schuster}} (\bibinfo{publisher}{Wiley-VCH, Berlin}),
  volume~\bibinfo{volume}{2}, pp. \bibinfo{pages}{107--158}.

\bibitem[{\citenamefont{Bhattacharyya}(2005)}]{bkc2}
\bibinfo{author}{\bibnamefont{Bhattacharyya}, \bibfnamefont{P.}},
  \bibinfo{year}{2005}, \bibinfo{journal}{Physica A}
  \textbf{\bibinfo{volume}{348}}, \bibinfo{pages}{199}.

\bibitem[{\citenamefont{Bhattacharyya and Chakrabarti}(2006)}]{bkc30}
\bibinfo{editor}{\bibnamefont{Bhattacharyya}, \bibfnamefont{P.}}, and
  \bibinfo{editor}{\bibfnamefont{B.~K.} \bibnamefont{Chakrabarti}} (eds.),
  \bibinfo{year}{2006}, \emph{\bibinfo{title}{{ Modelling Critical and
  Catastrophic Phenomena in Geoscience}}} (\bibinfo{publisher}{Springer-Verlag
  Heidelberg}).

\bibitem[{\citenamefont{Bhattacharyya}
  \emph{et~al.}(2003)\citenamefont{Bhattacharyya, Pradhan, and
  Chakrabarti}}]{bkc10}
\bibinfo{author}{\bibnamefont{Bhattacharyya}, \bibfnamefont{P.}},
  \bibinfo{author}{\bibfnamefont{S.}~\bibnamefont{Pradhan}}, and
  \bibinfo{author}{\bibfnamefont{B.~K.} \bibnamefont{Chakrabarti}},
  \bibinfo{year}{2003}, \bibinfo{journal}{Phys. Rev. E}
  \textbf{\bibinfo{volume}{67}}, \bibinfo{pages}{046122}.

\bibitem[{\citenamefont{Bhushan} \emph{et~al.}(1995)\citenamefont{Bhushan,
  Israelachvili, and Landman}}]{bkc50}
\bibinfo{author}{\bibnamefont{Bhushan}, \bibfnamefont{B.}},
  \bibinfo{author}{\bibfnamefont{J.~N.} \bibnamefont{Israelachvili}}, and
  \bibinfo{author}{\bibfnamefont{U.}~\bibnamefont{Landman}},
  \bibinfo{year}{1995}, \bibinfo{journal}{Nature}
  \textbf{\bibinfo{volume}{374}}, \bibinfo{pages}{607}.

\bibitem[{\citenamefont{Biswas and Chakrabarti}(2011)}]{bkc45}
\bibinfo{author}{\bibnamefont{Biswas}, \bibfnamefont{S.}}, and
  \bibinfo{author}{\bibfnamefont{B.~K.} \bibnamefont{Chakrabarti}},
  \bibinfo{year}{2011}, \bibinfo{journal}{arXiv:1108.1707} .

\bibitem[{\citenamefont{Bizzarri and Cocco}(2003)}]{BizzarriCocco2003}
\bibinfo{author}{\bibnamefont{Bizzarri}, \bibfnamefont{A.}}, and
  \bibinfo{author}{\bibfnamefont{M.}~\bibnamefont{Cocco}},
  \bibinfo{year}{2003}, \bibinfo{journal}{J. Geophys. Res.}
  \textbf{\bibinfo{volume}{108}}(\bibinfo{number}{2373}),
  \bibinfo{pages}{doi:10.1029/2002JB002198}.

\bibitem[{\citenamefont{Bizzarri and Cocco}(2006)}]{Bizzarri}
\bibinfo{author}{\bibnamefont{Bizzarri}, \bibfnamefont{A.}}, and
  \bibinfo{author}{\bibfnamefont{M.}~\bibnamefont{Cocco}},
  \bibinfo{year}{2006}, \bibinfo{journal}{J. Geophys. Res.}
  \textbf{\bibinfo{volume}{111}}, \bibinfo{pages}{doi:10.1029/2005JB003862}.

\bibitem[{\citenamefont{Blanpied} \emph{et~al.}(1995)\citenamefont{Blanpied,
  Lockner, and Byerlee}}]{Blanpied_etal1995}
\bibinfo{author}{\bibnamefont{Blanpied}, \bibfnamefont{M.~L.}},
  \bibinfo{author}{\bibfnamefont{D.~A.} \bibnamefont{Lockner}}, and
  \bibinfo{author}{\bibfnamefont{J.~D.} \bibnamefont{Byerlee}},
  \bibinfo{year}{1995}, \bibinfo{journal}{J. Geophys. Res.}
  \textbf{\bibinfo{volume}{100}}(\bibinfo{number}{B7}),
  \bibinfo{pages}{doi:10.1029/95JB00862}.

\bibitem[{\citenamefont{Bocquet} \emph{et~al.}(1998)\citenamefont{Bocquet,
  Charlaix, Ciliberto, and Crassous}}]{Bocquet1998}
\bibinfo{author}{\bibnamefont{Bocquet}, \bibfnamefont{L.}},
  \bibinfo{author}{\bibfnamefont{E.}~\bibnamefont{Charlaix}},
  \bibinfo{author}{\bibfnamefont{S.}~\bibnamefont{Ciliberto}}, and
  \bibinfo{author}{\bibfnamefont{J.}~\bibnamefont{Crassous}},
  \bibinfo{year}{1998}, \bibinfo{journal}{Nature}
  \textbf{\bibinfo{volume}{396}}, \bibinfo{pages}{735}.

\bibitem[{\citenamefont{Bonamy and Bouchaud}(2010)}]{bkc25}
\bibinfo{author}{\bibnamefont{Bonamy}, \bibfnamefont{D.}}, and
  \bibinfo{author}{\bibfnamefont{E.}~\bibnamefont{Bouchaud}},
  \bibinfo{year}{2010}, \bibinfo{journal}{Phys. Rep.}
  \textbf{\bibinfo{volume}{498}}, \bibinfo{pages}{1}.

\bibitem[{\citenamefont{Bosl and Nur}(2002)}]{BoslNur2002}
\bibinfo{author}{\bibnamefont{Bosl}, \bibfnamefont{W.~J.}}, and
  \bibinfo{author}{\bibfnamefont{A.}~\bibnamefont{Nur}}, \bibinfo{year}{2002},
  \bibinfo{journal}{J. Geophys. Res.}
  \textbf{\bibinfo{volume}{107}}(\bibinfo{number}{2366}),
  \bibinfo{pages}{doi:10.1029/2001JB000155}.

\bibitem[{\citenamefont{Bottani and Delamotte}(1997)}]{Bottani}
\bibinfo{author}{\bibnamefont{Bottani}, \bibfnamefont{S.}}, and
  \bibinfo{author}{\bibfnamefont{B.}~\bibnamefont{Delamotte}},
  \bibinfo{year}{1997}, \bibinfo{journal}{Physica D}
  \textbf{\bibinfo{volume}{103}}, \bibinfo{pages}{430}.

\bibitem[{\citenamefont{Boulter and Miller}(2003)}]{Boulter2}
\bibinfo{author}{\bibnamefont{Boulter}, \bibfnamefont{C.~J.}}, and
  \bibinfo{author}{\bibfnamefont{G.}~\bibnamefont{Miller}},
  \bibinfo{year}{2003}, \bibinfo{journal}{Phys. Rev. E}
  \textbf{\bibinfo{volume}{68}}, \bibinfo{pages}{056108}.

\bibitem[{\citenamefont{Bowden and Tabor}(2001)}]{Bowden2001}
\bibinfo{author}{\bibnamefont{Bowden}, \bibfnamefont{F.~P.}}, and
  \bibinfo{author}{\bibfnamefont{D.}~\bibnamefont{Tabor}},
  \bibinfo{year}{2001}, \emph{\bibinfo{title}{{The Friction and Lubrication of
  Solids (Oxford Classic Texts in the Physical Sciences)}}}
  (\bibinfo{publisher}{Oxford University Press N. Y.}).

\bibitem[{\citenamefont{Braun and Kivshar}(2004)}]{bkc49}
\bibinfo{author}{\bibnamefont{Braun}, \bibfnamefont{O.~M.}}, and
  \bibinfo{author}{\bibfnamefont{Y.~S.} \bibnamefont{Kivshar}},
  \bibinfo{year}{2004}, \emph{\bibinfo{title}{The Frenkel-Kontorova Model:
  Concepts, Methods, and Applications}} (\bibinfo{publisher}{Springer-Verlag
  Berlin}).

\bibitem[{\citenamefont{Braun and Naumovets}(2006)}]{bkc48}
\bibinfo{author}{\bibnamefont{Braun}, \bibfnamefont{O.~M.}}, and
  \bibinfo{author}{\bibfnamefont{A.~G.} \bibnamefont{Naumovets}},
  \bibinfo{year}{2006}, \bibinfo{journal}{Surf. Sc. Rep.}
  \textbf{\bibinfo{volume}{60}}, \bibinfo{pages}{79}.

\bibitem[{\citenamefont{Brechet and Estrin}(1994)}]{Brechet1994}
\bibinfo{author}{\bibnamefont{Brechet}, \bibfnamefont{Y.}}, and
  \bibinfo{author}{\bibfnamefont{Y.}~\bibnamefont{Estrin}},
  \bibinfo{year}{1994}, \bibinfo{journal}{Scripta Metallurgica et Materialia}
  \textbf{\bibinfo{volume}{30}}(\bibinfo{number}{11}), \bibinfo{pages}{1449}.

\bibitem[{\citenamefont{Brown} \emph{et~al.}(1991)\citenamefont{Brown, Scholz,
  and Rundle}}]{Brown}
\bibinfo{author}{\bibnamefont{Brown}, \bibfnamefont{S.~R.}},
  \bibinfo{author}{\bibfnamefont{C.}~\bibnamefont{Scholz}}, and
  \bibinfo{author}{\bibfnamefont{J.~B.} \bibnamefont{Rundle}},
  \bibinfo{year}{1991}, \bibinfo{journal}{Geophys. Res. Lett.}
  \textbf{\bibinfo{volume}{18}}, \bibinfo{pages}{doi:10.1029/91GL00210}.

\bibitem[{\citenamefont{Burridge}(2006)}]{bkc57}
\bibinfo{author}{\bibnamefont{Burridge}, \bibfnamefont{R.}},
  \bibinfo{year}{2006}, in \emph{\bibinfo{booktitle}{{Modelling Critical and
  Catastrophic Phenomena in Geoscience}}}, edited by
  \bibinfo{editor}{\bibfnamefont{P.}~\bibnamefont{Bhattacharyya}} and
  \bibinfo{editor}{\bibfnamefont{B.~K.} \bibnamefont{Chakrabarti}}
  (\bibinfo{publisher}{Springer-Verlag Heidelberg}), pp.
  \bibinfo{pages}{113--154}.

\bibitem[{\citenamefont{Burridge and Knopoff}(1967)}]{burr67}
\bibinfo{author}{\bibnamefont{Burridge}, \bibfnamefont{R.}}, and
  \bibinfo{author}{\bibfnamefont{L.}~\bibnamefont{Knopoff}},
  \bibinfo{year}{1967}, \bibinfo{journal}{Bull. Seismol. Soc. Am.}
  \textbf{\bibinfo{volume}{57}}, \bibinfo{pages}{3411}.

\bibitem[{\citenamefont{Caldarelli}
  \emph{et~al.}(1999)\citenamefont{Caldarelli, Castellano, and Petri}}]{bkc66}
\bibinfo{author}{\bibnamefont{Caldarelli}, \bibfnamefont{G.}},
  \bibinfo{author}{\bibfnamefont{C.}~\bibnamefont{Castellano}}, and
  \bibinfo{author}{\bibfnamefont{A.}~\bibnamefont{Petri}},
  \bibinfo{year}{1999}, \bibinfo{journal}{Physica A}
  \textbf{\bibinfo{volume}{270}}, \bibinfo{pages}{15}.

\bibitem[{\citenamefont{Cao and Aki}(1986)}]{CaoAki}
\bibinfo{author}{\bibnamefont{Cao}, \bibfnamefont{T.}}, and
  \bibinfo{author}{\bibfnamefont{K.}~\bibnamefont{Aki}}, \bibinfo{year}{1986},
  \bibinfo{journal}{Pure Appl. Geophys.} \textbf{\bibinfo{volume}{124}},
  \bibinfo{pages}{487}.

\bibitem[{\citenamefont{Carlson}(1991{\natexlab{a}})}]{Car91a}
\bibinfo{author}{\bibnamefont{Carlson}, \bibfnamefont{J.~M.}},
  \bibinfo{year}{1991}{\natexlab{a}}, \bibinfo{journal}{J. Geophys. Res.}
  \textbf{\bibinfo{volume}{96}}, \bibinfo{pages}{doi:10.1029/90JB02474}.

\bibitem[{\citenamefont{Carlson}(1991{\natexlab{b}})}]{Car91b}
\bibinfo{author}{\bibnamefont{Carlson}, \bibfnamefont{J.~M.}},
  \bibinfo{year}{1991}{\natexlab{b}}, \bibinfo{journal}{Phys. Rev. A}
  \textbf{\bibinfo{volume}{44}}, \bibinfo{pages}{6226}.

\bibitem[{\citenamefont{Carlson and Langer}(1989{\natexlab{a}})}]{car89a}
\bibinfo{author}{\bibnamefont{Carlson}, \bibfnamefont{J.~M.}}, and
  \bibinfo{author}{\bibfnamefont{J.~S.} \bibnamefont{Langer}},
  \bibinfo{year}{1989}{\natexlab{a}}, \bibinfo{journal}{Phys. Rev. lett}
  \textbf{\bibinfo{volume}{62}}, \bibinfo{pages}{2632}.

\bibitem[{\citenamefont{Carlson and Langer}(1989{\natexlab{b}})}]{car89b}
\bibinfo{author}{\bibnamefont{Carlson}, \bibfnamefont{J.~M.}}, and
  \bibinfo{author}{\bibfnamefont{J.~S.} \bibnamefont{Langer}},
  \bibinfo{year}{1989}{\natexlab{b}}, \bibinfo{journal}{Phys. Rev. A}
  \textbf{\bibinfo{volume}{40}}, \bibinfo{pages}{6470}.

\bibitem[{\citenamefont{Carlson} \emph{et~al.}(1994)\citenamefont{Carlson,
  Langer, and Shaw}}]{CL-rev}
\bibinfo{author}{\bibnamefont{Carlson}, \bibfnamefont{J.~M.}},
  \bibinfo{author}{\bibfnamefont{J.~S.} \bibnamefont{Langer}}, and
  \bibinfo{author}{\bibfnamefont{B.~E.} \bibnamefont{Shaw}},
  \bibinfo{year}{1994}, \bibinfo{journal}{Rev. Mod. Phys.}
  \textbf{\bibinfo{volume}{66}}, \bibinfo{pages}{657}.

\bibitem[{\citenamefont{Carlson} \emph{et~al.}(1991)\citenamefont{Carlson,
  Langer, Shaw, and Tang}}]{CL91}
\bibinfo{author}{\bibnamefont{Carlson}, \bibfnamefont{J.~M.}},
  \bibinfo{author}{\bibfnamefont{J.~S.} \bibnamefont{Langer}},
  \bibinfo{author}{\bibfnamefont{B.~E.} \bibnamefont{Shaw}}, and
  \bibinfo{author}{\bibfnamefont{C.}~\bibnamefont{Tang}}, \bibinfo{year}{1991},
  \bibinfo{journal}{Phys. Rev. A} \textbf{\bibinfo{volume}{44}},
  \bibinfo{pages}{884}.

\bibitem[{\citenamefont{Cartwright}
  \emph{et~al.}(1997)\citenamefont{Cartwright, Garcia, and Piro}}]{Cartwright}
\bibinfo{author}{\bibnamefont{Cartwright}, \bibfnamefont{J.~H.}},
  \bibinfo{author}{\bibfnamefont{E.~H.} \bibnamefont{Garcia}}, and
  \bibinfo{author}{\bibfnamefont{O.}~\bibnamefont{Piro}}, \bibinfo{year}{1997},
  \bibinfo{journal}{Phys. Rev. Lett.} \textbf{\bibinfo{volume}{79}},
  \bibinfo{pages}{527}.

\bibitem[{\citenamefont{de~Carvalho and Prado}(2000)}]{Prado}
\bibinfo{author}{\bibnamefont{de~Carvalho}, \bibfnamefont{J.~X.}}, and
  \bibinfo{author}{\bibfnamefont{C.~P.~C.} \bibnamefont{Prado}},
  \bibinfo{year}{2000}, \bibinfo{journal}{Phys. Rev. Lett.}
  \textbf{\bibinfo{volume}{84}}, \bibinfo{pages}{4006}.

\bibitem[{\citenamefont{Ceva}(1995)}]{Ceva}
\bibinfo{author}{\bibnamefont{Ceva}, \bibfnamefont{H.}}, \bibinfo{year}{1995},
  \bibinfo{journal}{Phys. Rev. E} \textbf{\bibinfo{volume}{52}},
  \bibinfo{pages}{154}.

\bibitem[{\citenamefont{Chakrabarti and Benguigui}(1997)}]{bkc6}
\bibinfo{author}{\bibnamefont{Chakrabarti}, \bibfnamefont{B.~K.}}, and
  \bibinfo{author}{\bibfnamefont{L.~G.} \bibnamefont{Benguigui}},
  \bibinfo{year}{1997}, \emph{\bibinfo{title}{Statistical physics of breakdown
  and fracture is disorder systems}} (\bibinfo{publisher}{Oxford University
  Press Oxford}).

\bibitem[{\citenamefont{Chakrabarti and Stinchcombe}(1999)}]{bkc1}
\bibinfo{author}{\bibnamefont{Chakrabarti}, \bibfnamefont{B.~K.}}, and
  \bibinfo{author}{\bibfnamefont{R.~B.} \bibnamefont{Stinchcombe}},
  \bibinfo{year}{1999}, \bibinfo{journal}{Physica A}
  \textbf{\bibinfo{volume}{270}}, \bibinfo{pages}{27}.

\bibitem[{\citenamefont{Chauve} \emph{et~al.}(2001)\citenamefont{Chauve,
  Le~Doussal, and Wiese}}]{bkc17}
\bibinfo{author}{\bibnamefont{Chauve}, \bibfnamefont{P.}},
  \bibinfo{author}{\bibfnamefont{P.}~\bibnamefont{Le~Doussal}}, and
  \bibinfo{author}{\bibfnamefont{K.~J.} \bibnamefont{Wiese}},
  \bibinfo{year}{2001}, \bibinfo{journal}{Phys. Rev. Lett.}
  \textbf{\bibinfo{volume}{86}}, \bibinfo{pages}{1785}.

\bibitem[{\citenamefont{Chester and Chester}(1998)}]{Chester1998}
\bibinfo{author}{\bibnamefont{Chester}, \bibfnamefont{F.}}, and
  \bibinfo{author}{\bibfnamefont{J.}~\bibnamefont{Chester}},
  \bibinfo{year}{1998}, \bibinfo{journal}{Tectonophysics}
  \textbf{\bibinfo{volume}{295}}(\bibinfo{number}{1-2}), \bibinfo{pages}{199}.

\bibitem[{\citenamefont{Chlieh} \emph{et~al.}(2008)\citenamefont{Chlieh,
  Avouac, Sieh, Natawidjaja, and Galetzka}}]{Chlieh_etal2008}
\bibinfo{author}{\bibnamefont{Chlieh}, \bibfnamefont{M.}},
  \bibinfo{author}{\bibfnamefont{J.~P.} \bibnamefont{Avouac}},
  \bibinfo{author}{\bibfnamefont{K.}~\bibnamefont{Sieh}},
  \bibinfo{author}{\bibfnamefont{D.~H.} \bibnamefont{Natawidjaja}}, and
  \bibinfo{author}{\bibfnamefont{J.}~\bibnamefont{Galetzka}},
  \bibinfo{year}{2008}, \bibinfo{journal}{J. Geophys. Res.}
  \textbf{\bibinfo{volume}{113}}(\bibinfo{number}{B05305}),
  \bibinfo{pages}{doi:10.1029/2007JB004981}.

\bibitem[{\citenamefont{Clancy and Corcoran}(2005)}]{Clancy05}
\bibinfo{author}{\bibnamefont{Clancy}, \bibfnamefont{I.}}, and
  \bibinfo{author}{\bibfnamefont{D.}~\bibnamefont{Corcoran}},
  \bibinfo{year}{2005}, \bibinfo{journal}{Phys. Rev. E}
  \textbf{\bibinfo{volume}{71}}, \bibinfo{pages}{0461124}.

\bibitem[{\citenamefont{Clancy and Corcoran}(2006)}]{Clancy06}
\bibinfo{author}{\bibnamefont{Clancy}, \bibfnamefont{I.}}, and
  \bibinfo{author}{\bibfnamefont{D.}~\bibnamefont{Corcoran}},
  \bibinfo{year}{2006}, \bibinfo{journal}{Phys. Rev. E}
  \textbf{\bibinfo{volume}{73}}, \bibinfo{pages}{046115}.

\bibitem[{\citenamefont{Clancy and Corcoran}(2009)}]{Clancy09}
\bibinfo{author}{\bibnamefont{Clancy}, \bibfnamefont{I.}}, and
  \bibinfo{author}{\bibfnamefont{D.}~\bibnamefont{Corcoran}},
  \bibinfo{year}{2009}, \bibinfo{journal}{Phys. Rev. E}
  \textbf{\bibinfo{volume}{80}}, \bibinfo{pages}{016113}.

\bibitem[{\citenamefont{Cochard and Madariaga}(1996)}]{Cochard}
\bibinfo{author}{\bibnamefont{Cochard}, \bibfnamefont{A.}}, and
  \bibinfo{author}{\bibfnamefont{R.}~\bibnamefont{Madariaga}},
  \bibinfo{year}{1996}, \bibinfo{journal}{J. Geophys. Res.}
  \textbf{\bibinfo{volume}{101}}, \bibinfo{pages}{doi:10.1029/96JB02095}.

\bibitem[{\citenamefont{Corral}(2004)}]{Corral}
\bibinfo{author}{\bibnamefont{Corral}, \bibfnamefont{A.}},
  \bibinfo{year}{2004}, \bibinfo{journal}{Phys. Rev. Lett.}
  \textbf{\bibinfo{volume}{92}}, \bibinfo{pages}{108501}.

\bibitem[{\citenamefont{Cristensen and Olami}(1994)}]{OFC2}
\bibinfo{author}{\bibnamefont{Cristensen}, \bibfnamefont{K.}}, and
  \bibinfo{author}{\bibfnamefont{Z.}~\bibnamefont{Olami}},
  \bibinfo{year}{1994}, \bibinfo{journal}{Phys. Rev. A}
  \textbf{\bibinfo{volume}{46}}, \bibinfo{pages}{1829}.

\bibitem[{\citenamefont{da~Cruz} \emph{et~al.}(2005)\citenamefont{da~Cruz,
  Emam, Prochnow, Roux, and Chevoir}}]{DaCruz2005}
\bibinfo{author}{\bibnamefont{da~Cruz}, \bibfnamefont{F.}},
  \bibinfo{author}{\bibfnamefont{S.}~\bibnamefont{Emam}},
  \bibinfo{author}{\bibfnamefont{M.}~\bibnamefont{Prochnow}},
  \bibinfo{author}{\bibfnamefont{J.-N.} \bibnamefont{Roux}}, and
  \bibinfo{author}{\bibfnamefont{F.}~\bibnamefont{Chevoir}},
  \bibinfo{year}{2005}, \bibinfo{journal}{Phys. Rev. E}
  \textbf{\bibinfo{volume}{72}}(\bibinfo{number}{2}), \bibinfo{pages}{021309}.

\bibitem[{\citenamefont{Csahok} \emph{et~al.}(1993)\citenamefont{Csahok, Honda,
  Somfai, Vicsek, and Vicsek}}]{bkc54}
\bibinfo{author}{\bibnamefont{Csahok}, \bibfnamefont{Z.}},
  \bibinfo{author}{\bibfnamefont{K.}~\bibnamefont{Honda}},
  \bibinfo{author}{\bibfnamefont{E.}~\bibnamefont{Somfai}},
  \bibinfo{author}{\bibfnamefont{M.}~\bibnamefont{Vicsek}}, and
  \bibinfo{author}{\bibfnamefont{T.}~\bibnamefont{Vicsek}},
  \bibinfo{year}{1993}, \bibinfo{journal}{Physica A}
  \textbf{\bibinfo{volume}{200}}, \bibinfo{pages}{136}.

\bibitem[{\citenamefont{Daniels}(1945)}]{bkc11}
\bibinfo{author}{\bibnamefont{Daniels}, \bibfnamefont{H.~E.}},
  \bibinfo{year}{1945}, \bibinfo{journal}{Proc. R. Soc. London A}
  \textbf{\bibinfo{volume}{183}}, \bibinfo{pages}{405}.

\bibitem[{\citenamefont{Daub and Carlson}(2010)}]{Daub2010}
\bibinfo{author}{\bibnamefont{Daub}, \bibfnamefont{E.}}, and
  \bibinfo{author}{\bibfnamefont{J.}~\bibnamefont{Carlson}},
  \bibinfo{year}{2010}, \bibinfo{journal}{Annu. Rev. Condens. Matter Phys.}
  \textbf{\bibinfo{volume}{1}}, \bibinfo{pages}{397}.

\bibitem[{\citenamefont{De and Ananthakrisna}(2004)}]{De04}
\bibinfo{author}{\bibnamefont{De}, \bibfnamefont{R.}}, and
  \bibinfo{author}{\bibfnamefont{G.}~\bibnamefont{Ananthakrisna}},
  \bibinfo{year}{2004}, \bibinfo{journal}{Europhys. Lett.}
  \textbf{\bibinfo{volume}{66}}, \bibinfo{pages}{715}.

\bibitem[{\citenamefont{De~Rubeis} \emph{et~al.}(1996)\citenamefont{De~Rubeis,
  Hallgass, Loreto, Paladin, Pietronero, and Tosi}}]{bkc4}
\bibinfo{author}{\bibnamefont{De~Rubeis}, \bibfnamefont{V.}},
  \bibinfo{author}{\bibfnamefont{R.}~\bibnamefont{Hallgass}},
  \bibinfo{author}{\bibfnamefont{V.}~\bibnamefont{Loreto}},
  \bibinfo{author}{\bibfnamefont{G.}~\bibnamefont{Paladin}},
  \bibinfo{author}{\bibfnamefont{L.}~\bibnamefont{Pietronero}}, and
  \bibinfo{author}{\bibfnamefont{P.}~\bibnamefont{Tosi}}, \bibinfo{year}{1996},
  \bibinfo{journal}{Phys. Rev. Lett} \textbf{\bibinfo{volume}{76}},
  \bibinfo{pages}{2599}.

\bibitem[{\citenamefont{De~Rubies} \emph{et~al.}(2006)\citenamefont{De~Rubies,
  Loreto, Pietronero, and Tosi}}]{bkc58}
\bibinfo{author}{\bibnamefont{De~Rubies}, \bibfnamefont{V.}},
  \bibinfo{author}{\bibfnamefont{V.}~\bibnamefont{Loreto}},
  \bibinfo{author}{\bibfnamefont{L.}~\bibnamefont{Pietronero}}, and
  \bibinfo{author}{\bibfnamefont{P.}~\bibnamefont{Tosi}}, \bibinfo{year}{2006},
  in \emph{\bibinfo{booktitle}{{Modelling Critical and Catastrophic Phenomena
  in Geoscience}}}, edited by
  \bibinfo{editor}{\bibfnamefont{P.}~\bibnamefont{Bhattacharyya}} and
  \bibinfo{editor}{\bibfnamefont{B.~K.} \bibnamefont{Chakrabarti}}
  (\bibinfo{publisher}{Springer-Verlag Heidelberg}), pp.
  \bibinfo{pages}{259--280}.

\bibitem[{\citenamefont{{Di Toro}} \emph{et~al.}(2004)\citenamefont{{Di Toro},
  Goldsby, and Tullis}}]{DiToro2004}
\bibinfo{author}{\bibnamefont{{Di Toro}}, \bibfnamefont{G.}},
  \bibinfo{author}{\bibfnamefont{D.~L.} \bibnamefont{Goldsby}}, and
  \bibinfo{author}{\bibfnamefont{T.~E.} \bibnamefont{Tullis}},
  \bibinfo{year}{2004}, \bibinfo{journal}{Nature}
  \textbf{\bibinfo{volume}{427}}(\bibinfo{number}{6973}), \bibinfo{pages}{436}.

\bibitem[{\citenamefont{Dieterich}(1972)}]{Dieterich72}
\bibinfo{author}{\bibnamefont{Dieterich}, \bibfnamefont{J.~H.}},
  \bibinfo{year}{1972}, \bibinfo{journal}{J. Geophys. Res.}
  \textbf{\bibinfo{volume}{77}}, \bibinfo{pages}{doi:10.1029/JB077i020p03690}.

\bibitem[{\citenamefont{Dieterich}(1978)}]{Dieterich1978}
\bibinfo{author}{\bibnamefont{Dieterich}, \bibfnamefont{J.~H.}},
  \bibinfo{year}{1978}, \bibinfo{journal}{J. Geophys. Res.}
  \textbf{\bibinfo{volume}{83}}, \bibinfo{pages}{doi:10.1029/JB083iB08p03940}.

\bibitem[{\citenamefont{Dieterich}(1979)}]{Dieterich1979}
\bibinfo{author}{\bibnamefont{Dieterich}, \bibfnamefont{J.~H.}},
  \bibinfo{year}{1979}, \bibinfo{journal}{J. Geophys. Res.}
  \textbf{\bibinfo{volume}{84}}, \bibinfo{pages}{doi:10.1029/JB084iB05p02161}.

\bibitem[{\citenamefont{Dieterich}(1986)}]{Dieterich1986}
\bibinfo{author}{\bibnamefont{Dieterich}, \bibfnamefont{J.~H.}},
  \bibinfo{year}{1986}, in \emph{\bibinfo{booktitle}{Earthquake Source
  Mechanics}}, edited by \bibinfo{editor}{\bibfnamefont{S.}~\bibnamefont{Das}},
  \bibinfo{editor}{\bibfnamefont{J.}~\bibnamefont{Boarwright}}, and
  \bibinfo{editor}{\bibfnamefont{C.~H.} \bibnamefont{Scholz}}
  (\bibinfo{publisher}{American Geophysical Union}), pp.
  \bibinfo{pages}{37--47}.

\bibitem[{\citenamefont{Dieterich}(1992)}]{Dieterich1992}
\bibinfo{author}{\bibnamefont{Dieterich}, \bibfnamefont{J.~H.}},
  \bibinfo{year}{1992}, \bibinfo{journal}{Tectonophysics}
  \textbf{\bibinfo{volume}{211}}, \bibinfo{pages}{115}.

\bibitem[{\citenamefont{Dieterich}(1994)}]{Dieterich1994b}
\bibinfo{author}{\bibnamefont{Dieterich}, \bibfnamefont{J.~H.}},
  \bibinfo{year}{1994}, \bibinfo{journal}{J. Geophys. Res.}
  \textbf{\bibinfo{volume}{99}}, \bibinfo{pages}{doi:10.1029/93JB02581}.

\bibitem[{\citenamefont{Dieterich}(2009)}]{Dieterich2007}
\bibinfo{author}{\bibnamefont{Dieterich}, \bibfnamefont{J.~H.}},
  \bibinfo{year}{2009}, in \emph{\bibinfo{booktitle}{Treatise on Geophysics
  Vol. 4}}, edited by
  \bibinfo{editor}{\bibfnamefont{H.}~\bibnamefont{Kanamori}}
  (\bibinfo{publisher}{Elsevier, Amsterdam}), pp. \bibinfo{pages}{107--129}.

\bibitem[{\citenamefont{Dieterich and Kilgore}(1994)}]{Dieterich1994}
\bibinfo{author}{\bibnamefont{Dieterich}, \bibfnamefont{J.~H.}}, and
  \bibinfo{author}{\bibfnamefont{B.~D.} \bibnamefont{Kilgore}},
  \bibinfo{year}{1994}, \bibinfo{journal}{Pure and Applied Geophysics}
  \textbf{\bibinfo{volume}{143}}(\bibinfo{number}{1-3}), \bibinfo{pages}{283}.

\bibitem[{\citenamefont{Dieterich and Kilgore}(1996)}]{Dieterich1996}
\bibinfo{author}{\bibnamefont{Dieterich}, \bibfnamefont{J.~H.}}, and
  \bibinfo{author}{\bibfnamefont{B.~D.} \bibnamefont{Kilgore}},
  \bibinfo{year}{1996}, \bibinfo{journal}{Tectonophysics}
  \textbf{\bibinfo{volume}{256}}, \bibinfo{pages}{219}.

\bibitem[{\citenamefont{Duemmer and Krauth}(2007)}]{bkc18}
\bibinfo{author}{\bibnamefont{Duemmer}, \bibfnamefont{O.}}, and
  \bibinfo{author}{\bibfnamefont{W.}~\bibnamefont{Krauth}},
  \bibinfo{year}{2007}, \bibinfo{journal}{J. Stat. Mech.}
  \textbf{\bibinfo{volume}{2007}}, \bibinfo{pages}{P01019}.

\bibitem[{\citenamefont{Edwards and Wilkinson}(1982)}]{bkc39}
\bibinfo{author}{\bibnamefont{Edwards}, \bibfnamefont{S.~F.}}, and
  \bibinfo{author}{\bibfnamefont{D.~R.} \bibnamefont{Wilkinson}},
  \bibinfo{year}{1982}, \bibinfo{journal}{Proc. R. Soc. Lond. A}
  \textbf{\bibinfo{volume}{381}}, \bibinfo{pages}{1780}.

\bibitem[{\citenamefont{Eriksen} \emph{et~al.}(2010)\citenamefont{Eriksen,
  Biswas, and Chakrabarti}}]{bkc23}
\bibinfo{author}{\bibnamefont{Eriksen}, \bibfnamefont{J.~A.}},
  \bibinfo{author}{\bibfnamefont{S.}~\bibnamefont{Biswas}}, and
  \bibinfo{author}{\bibfnamefont{B.~K.} \bibnamefont{Chakrabarti}},
  \bibinfo{year}{2010}, \bibinfo{journal}{Phys. Rev. E}
  \textbf{\bibinfo{volume}{82}}, \bibinfo{pages}{041124}.

\bibitem[{\citenamefont{Espa\={n}ol}(1994)}]{Espanol}
\bibinfo{author}{\bibnamefont{Espa\={n}ol}, \bibfnamefont{P.}},
  \bibinfo{year}{1994}, \bibinfo{journal}{Phys. Rev. E}
  \textbf{\bibinfo{volume}{50}}, \bibinfo{pages}{227}.

\bibitem[{\citenamefont{Fialko and Khazan}(2005)}]{Fialko2005}
\bibinfo{author}{\bibnamefont{Fialko}, \bibfnamefont{Y.}}, and
  \bibinfo{author}{\bibfnamefont{Y.}~\bibnamefont{Khazan}},
  \bibinfo{year}{2005}, \bibinfo{journal}{J. Geophys. Res.}
  \textbf{\bibinfo{volume}{110}}(\bibinfo{number}{B12}),
  \bibinfo{pages}{doi:10.1029/2005JB003869}.

\bibitem[{\citenamefont{Freed}(2007)}]{Freed2007}
\bibinfo{author}{\bibnamefont{Freed}, \bibfnamefont{A.~M.}},
  \bibinfo{year}{2007}, \bibinfo{journal}{Geophys. Res. Lett.}
  \textbf{\bibinfo{volume}{34}}(\bibinfo{number}{L06312}),
  \bibinfo{pages}{doi:10.1029/2006GL029155}.

\bibitem[{\citenamefont{Frenkel and Kontorova}(1938)}]{bkc20}
\bibinfo{author}{\bibnamefont{Frenkel}, \bibfnamefont{Y.}}, and
  \bibinfo{author}{\bibfnamefont{T.}~\bibnamefont{Kontorova}},
  \bibinfo{year}{1938}, \bibinfo{journal}{Zh. Eksp. Teor. Fiz.}
  \textbf{\bibinfo{volume}{8}}, \bibinfo{pages}{1340}.

\bibitem[{\citenamefont{Fukuda} \emph{et~al.}(2009)\citenamefont{Fukuda,
  Johnson, Larson, and Miyazaki}}]{Fukuda_etal2009}
\bibinfo{author}{\bibnamefont{Fukuda}, \bibfnamefont{J.}},
  \bibinfo{author}{\bibfnamefont{K.~M.} \bibnamefont{Johnson}},
  \bibinfo{author}{\bibfnamefont{K.~M.} \bibnamefont{Larson}}, and
  \bibinfo{author}{\bibfnamefont{S.}~\bibnamefont{Miyazaki}},
  \bibinfo{year}{2009}, \bibinfo{journal}{J. Geophys. Res.}
  \textbf{\bibinfo{volume}{114}}(\bibinfo{number}{B04412}),
  \bibinfo{pages}{doi:10.1029/2008JB006166}.

\bibitem[{\citenamefont{GDRMidi}(2004)}]{GDR2004}
\bibinfo{author}{\bibnamefont{GDRMidi}}, \bibinfo{year}{2004},
  \bibinfo{journal}{Eur. Phys. J. E}
  \textbf{\bibinfo{volume}{14}}(\bibinfo{number}{4}), \bibinfo{pages}{341}.

\bibitem[{\citenamefont{Goldsby and Tullis}(2002)}]{Goldsby2002}
\bibinfo{author}{\bibnamefont{Goldsby}, \bibfnamefont{D.~L.}}, and
  \bibinfo{author}{\bibfnamefont{T.~E.} \bibnamefont{Tullis}},
  \bibinfo{year}{2002}, \bibinfo{journal}{Geophys. Res. Lett.}
  \textbf{\bibinfo{volume}{29}}(\bibinfo{number}{17}),
  \bibinfo{pages}{doi:10.1029/2002GL015240}.

\bibitem[{\citenamefont{Grassberger}(1994)}]{Grassberger}
\bibinfo{author}{\bibnamefont{Grassberger}, \bibfnamefont{P.}},
  \bibinfo{year}{1994}, \bibinfo{journal}{Phys. Rev. E}
  \textbf{\bibinfo{volume}{49}}, \bibinfo{pages}{2436}.

\bibitem[{\citenamefont{Gu} \emph{et~al.}(1984)\citenamefont{Gu, Rice, Ruina,
  and Tse}}]{Gu_etal_1984}
\bibinfo{author}{\bibnamefont{Gu}, \bibfnamefont{J.-C.}},
  \bibinfo{author}{\bibfnamefont{J.~R.} \bibnamefont{Rice}},
  \bibinfo{author}{\bibfnamefont{A.~L.} \bibnamefont{Ruina}}, and
  \bibinfo{author}{\bibfnamefont{S.~T.} \bibnamefont{Tse}},
  \bibinfo{year}{1984}, \bibinfo{journal}{J. Mech. Phys. Solids}
  \textbf{\bibinfo{volume}{32}}, \bibinfo{pages}{167}.

\bibitem[{\citenamefont{Guatteri and Spudish}(2000)}]{GuatteriSpudish2000}
\bibinfo{author}{\bibnamefont{Guatteri}, \bibfnamefont{M.}}, and
  \bibinfo{author}{\bibfnamefont{P.}~\bibnamefont{Spudish}},
  \bibinfo{year}{2000}, \bibinfo{journal}{Bull. Seismol. Soc. Am.}
  \textbf{\bibinfo{volume}{90}}, \bibinfo{pages}{98}.

\bibitem[{\citenamefont{Hainzl} \emph{et~al.}(1999)\citenamefont{Hainzl,
  Z\"ollar, and Kurths}}]{Hainzl99}
\bibinfo{author}{\bibnamefont{Hainzl}, \bibfnamefont{S.}},
  \bibinfo{author}{\bibfnamefont{G.}~\bibnamefont{Z\"ollar}}, and
  \bibinfo{author}{\bibfnamefont{J.}~\bibnamefont{Kurths}},
  \bibinfo{year}{1999}, \bibinfo{journal}{J. Geophys. Res.}
  \textbf{\bibinfo{volume}{104}}, \bibinfo{pages}{doi:10.1029/1998JB900122}.

\bibitem[{\citenamefont{Hainzl} \emph{et~al.}(2000)\citenamefont{Hainzl,
  Z\"oller, Kurths, and Zschau}}]{Hainzl100}
\bibinfo{author}{\bibnamefont{Hainzl}, \bibfnamefont{S.}},
  \bibinfo{author}{\bibfnamefont{G.}~\bibnamefont{Z\"oller}},
  \bibinfo{author}{\bibfnamefont{J.}~\bibnamefont{Kurths}}, and
  \bibinfo{author}{\bibfnamefont{J.}~\bibnamefont{Zschau}},
  \bibinfo{year}{2000}, \bibinfo{journal}{Geophys. Res. Lett.}
  \textbf{\bibinfo{volume}{27}}, \bibinfo{pages}{doi:10.1029/1999GL011000}.

\bibitem[{\citenamefont{Han} \emph{et~al.}(2011)\citenamefont{Han, Hirose,
  Shimamoto, Lee, and Ando}}]{Han2011}
\bibinfo{author}{\bibnamefont{Han}, \bibfnamefont{R.}},
  \bibinfo{author}{\bibfnamefont{T.}~\bibnamefont{Hirose}},
  \bibinfo{author}{\bibfnamefont{T.}~\bibnamefont{Shimamoto}},
  \bibinfo{author}{\bibfnamefont{Y.}~\bibnamefont{Lee}}, and
  \bibinfo{author}{\bibfnamefont{J.}~\bibnamefont{Ando}}, \bibinfo{year}{2011},
  \bibinfo{journal}{J. Ando. Geology} \textbf{\bibinfo{volume}{39}},
  \bibinfo{pages}{599}.

\bibitem[{\citenamefont{Han} \emph{et~al.}(2007)\citenamefont{Han, Shimamoto,
  Hirose, Ree, and Ando}}]{Han2007}
\bibinfo{author}{\bibnamefont{Han}, \bibfnamefont{R.}},
  \bibinfo{author}{\bibfnamefont{T.}~\bibnamefont{Shimamoto}},
  \bibinfo{author}{\bibfnamefont{T.}~\bibnamefont{Hirose}},
  \bibinfo{author}{\bibfnamefont{J.}~\bibnamefont{Ree}}, and
  \bibinfo{author}{\bibfnamefont{J.}~\bibnamefont{Ando}}, \bibinfo{year}{2007},
  \bibinfo{journal}{Science} \textbf{\bibinfo{volume}{316}},
  \bibinfo{pages}{878}.

\bibitem[{\citenamefont{Hashimoto} \emph{et~al.}(2009)\citenamefont{Hashimoto,
  Noda, Sagiya, and Matsu'ura}}]{Hashimoto_etal2009}
\bibinfo{author}{\bibnamefont{Hashimoto}, \bibfnamefont{C.}},
  \bibinfo{author}{\bibfnamefont{A.}~\bibnamefont{Noda}},
  \bibinfo{author}{\bibfnamefont{T.}~\bibnamefont{Sagiya}}, and
  \bibinfo{author}{\bibfnamefont{M.}~\bibnamefont{Matsu'ura}},
  \bibinfo{year}{2009}, \bibinfo{journal}{Nature Geosci.}
  \textbf{\bibinfo{volume}{2}}, \bibinfo{pages}{141}.

\bibitem[{\citenamefont{Hatano}(2007)}]{Hatano2007}
\bibinfo{author}{\bibnamefont{Hatano}, \bibfnamefont{T.}},
  \bibinfo{year}{2007}, \bibinfo{journal}{Phys. Rev. E}
  \textbf{\bibinfo{volume}{75}}, \bibinfo{pages}{060301(R)}.

\bibitem[{\citenamefont{Hatano}(2009)}]{Hatano2009}
\bibinfo{author}{\bibnamefont{Hatano}, \bibfnamefont{T.}},
  \bibinfo{year}{2009}, \bibinfo{journal}{Geophys. Res. Lett.}
  \textbf{\bibinfo{volume}{36}}(\bibinfo{number}{18}),
  \bibinfo{pages}{doi:10.1029/2009GL039665}.

\bibitem[{\citenamefont{Hayashi and Tsutsumi}(2010)}]{Hayashi2010}
\bibinfo{author}{\bibnamefont{Hayashi}, \bibfnamefont{N.}}, and
  \bibinfo{author}{\bibfnamefont{A.}~\bibnamefont{Tsutsumi}},
  \bibinfo{year}{2010}, \bibinfo{journal}{Geophys. Res. Lett.}
  \textbf{\bibinfo{volume}{37}}(\bibinfo{number}{12}),
  \bibinfo{pages}{doi:10.1029/2010GL042943}.

\bibitem[{\citenamefont{Helmstetter}
  \emph{et~al.}(2004)\citenamefont{Helmstetter, Hergarten, and
  Sornette}}]{Helstetter04}
\bibinfo{author}{\bibnamefont{Helmstetter}, \bibfnamefont{A.}},
  \bibinfo{author}{\bibfnamefont{S.}~\bibnamefont{Hergarten}}, and
  \bibinfo{author}{\bibfnamefont{D.}~\bibnamefont{Sornette}},
  \bibinfo{year}{2004}, \bibinfo{journal}{Phys. Rev. E}
  \textbf{\bibinfo{volume}{70}}, \bibinfo{pages}{046120}.

\bibitem[{\citenamefont{Hemmer and Hansen}(1992)}]{bkc37}
\bibinfo{author}{\bibnamefont{Hemmer}, \bibfnamefont{P.~C.}}, and
  \bibinfo{author}{\bibfnamefont{A.}~\bibnamefont{Hansen}},
  \bibinfo{year}{1992}, \bibinfo{journal}{ASME J. Appl. Mech.}
  \textbf{\bibinfo{volume}{59}}, \bibinfo{pages}{909}.

\bibitem[{\citenamefont{Hergarten}(2002)}]{Hergarten2002}
\bibinfo{author}{\bibnamefont{Hergarten}, \bibfnamefont{S.}},
  \bibinfo{year}{2002}, \emph{\bibinfo{title}{Self-Organized Criticality in
  Earth Systems}} (\bibinfo{publisher}{Springer Berlin}).

\bibitem[{\citenamefont{Hergarten and Krenn}(2011)}]{kawamura-rev1}
\bibinfo{author}{\bibnamefont{Hergarten}, \bibfnamefont{S.}}, and
  \bibinfo{author}{\bibfnamefont{R.}~\bibnamefont{Krenn}},
  \bibinfo{year}{2011}, \bibinfo{journal}{Nonlin. Processes Geophys.}
  \textbf{\bibinfo{volume}{18}}, \bibinfo{pages}{635}.

\bibitem[{\citenamefont{Hergarten and Neugebauer}(2000)}]{Hergarten-OFC}
\bibinfo{author}{\bibnamefont{Hergarten}, \bibfnamefont{S.}}, and
  \bibinfo{author}{\bibfnamefont{H.}~\bibnamefont{Neugebauer}},
  \bibinfo{year}{2000}, \bibinfo{journal}{Phys. Rev. E}
  \textbf{\bibinfo{volume}{61}}, \bibinfo{pages}{2382}.

\bibitem[{\citenamefont{Hergarten and Neugebauer}(2002)}]{Hergarten}
\bibinfo{author}{\bibnamefont{Hergarten}, \bibfnamefont{S.}}, and
  \bibinfo{author}{\bibfnamefont{H.}~\bibnamefont{Neugebauer}},
  \bibinfo{year}{2002}, \bibinfo{journal}{Phys. Rev. Lett.}
  \textbf{\bibinfo{volume}{88}}, \bibinfo{pages}{238501}.

\bibitem[{\citenamefont{Herrmann and Roux}(1990)}]{bkc29}
\bibinfo{editor}{\bibnamefont{Herrmann}, \bibfnamefont{H.~J.}}, and
  \bibinfo{editor}{\bibfnamefont{S.}~\bibnamefont{Roux}} (eds.),
  \bibinfo{year}{1990}, \emph{\bibinfo{title}{Statistical models for the
  fracture of disordered media}} (\bibinfo{publisher}{Elsevier Amsterdam}).

\bibitem[{\citenamefont{Heslot} \emph{et~al.}(1994)\citenamefont{Heslot,
  Baumberger, Perrin, Caroli, and Caroli}}]{Heslot1994}
\bibinfo{author}{\bibnamefont{Heslot}, \bibfnamefont{F.}},
  \bibinfo{author}{\bibfnamefont{T.}~\bibnamefont{Baumberger}},
  \bibinfo{author}{\bibfnamefont{B.}~\bibnamefont{Perrin}},
  \bibinfo{author}{\bibfnamefont{B.}~\bibnamefont{Caroli}}, and
  \bibinfo{author}{\bibfnamefont{C.}~\bibnamefont{Caroli}},
  \bibinfo{year}{1994}, \bibinfo{journal}{Phys. Rev. E}
  \textbf{\bibinfo{volume}{49}}(\bibinfo{number}{6}), \bibinfo{pages}{4973}.

\bibitem[{\citenamefont{Hillers} \emph{et~al.}(2006)\citenamefont{Hillers,
  Ben-Zion, and Mai}}]{Hillers_etal2006}
\bibinfo{author}{\bibnamefont{Hillers}, \bibfnamefont{G.}},
  \bibinfo{author}{\bibfnamefont{Y.}~\bibnamefont{Ben-Zion}}, and
  \bibinfo{author}{\bibfnamefont{P.~M.} \bibnamefont{Mai}},
  \bibinfo{year}{2006}, \bibinfo{journal}{J. Geophys. Res.}
  \textbf{\bibinfo{volume}{111}}(\bibinfo{number}{B01403}),
  \bibinfo{pages}{doi:10.1029/2005JB003859}.

\bibitem[{\citenamefont{Hillers} \emph{et~al.}(2007)\citenamefont{Hillers, Mai,
  Ben-Zion, and Ampuero}}]{Hillers_etal2007}
\bibinfo{author}{\bibnamefont{Hillers}, \bibfnamefont{G.}},
  \bibinfo{author}{\bibfnamefont{P.~M.} \bibnamefont{Mai}},
  \bibinfo{author}{\bibfnamefont{Y.}~\bibnamefont{Ben-Zion}}, and
  \bibinfo{author}{\bibfnamefont{J.-P.} \bibnamefont{Ampuero}},
  \bibinfo{year}{2007}, \bibinfo{journal}{Geophys. J. Int.}
  \textbf{\bibinfo{volume}{169}}, \bibinfo{pages}{515}.

\bibitem[{\citenamefont{Hillers and Miller}(2007)}]{HillersMiller2007}
\bibinfo{author}{\bibnamefont{Hillers}, \bibfnamefont{G.}}, and
  \bibinfo{author}{\bibfnamefont{S.~A.} \bibnamefont{Miller}},
  \bibinfo{year}{2007}, \bibinfo{journal}{Geophys. J. Int.}
  \textbf{\bibinfo{volume}{168}}, \bibinfo{pages}{431}.

\bibitem[{\citenamefont{Hirose} \emph{et~al.}(1999)\citenamefont{Hirose,
  Hirahara, Kimata, Fujii, and Miyazaki}}]{Hirose_etal1999}
\bibinfo{author}{\bibnamefont{Hirose}, \bibfnamefont{H.}},
  \bibinfo{author}{\bibfnamefont{K.}~\bibnamefont{Hirahara}},
  \bibinfo{author}{\bibfnamefont{F.}~\bibnamefont{Kimata}},
  \bibinfo{author}{\bibfnamefont{N.}~\bibnamefont{Fujii}}, and
  \bibinfo{author}{\bibfnamefont{S.}~\bibnamefont{Miyazaki}},
  \bibinfo{year}{1999}, \bibinfo{journal}{Geophys. Res. Lett.}
  \textbf{\bibinfo{volume}{26}}(\bibinfo{number}{21}),
  \bibinfo{pages}{doi:10.1029/1999GL010999}.

\bibitem[{\citenamefont{Hirose and Obara}(2006)}]{HiroseObara2006}
\bibinfo{author}{\bibnamefont{Hirose}, \bibfnamefont{H.}}, and
  \bibinfo{author}{\bibfnamefont{K.}~\bibnamefont{Obara}},
  \bibinfo{year}{2006}, \bibinfo{journal}{Geophys. Res. Lett.}
  \textbf{\bibinfo{volume}{33}}(\bibinfo{number}{L17311}),
  \bibinfo{pages}{doi:10.1029/2006GL026579}.

\bibitem[{\citenamefont{Hirose and Shimamoto}(2005)}]{Hirose2005}
\bibinfo{author}{\bibnamefont{Hirose}, \bibfnamefont{T.}}, and
  \bibinfo{author}{\bibfnamefont{T.}~\bibnamefont{Shimamoto}},
  \bibinfo{year}{2005}, \bibinfo{journal}{J. Geophys. Res.}
  \textbf{\bibinfo{volume}{110}}(\bibinfo{number}{B5}),
  \bibinfo{pages}{doi:10.1029/2004JB003207}.

\bibitem[{\citenamefont{H\"{o}lscher}
  \emph{et~al.}(2008)\citenamefont{H\"{o}lscher, Schirmeisen, and
  Schwarz}}]{bkc51}
\bibinfo{author}{\bibnamefont{H\"{o}lscher}, \bibfnamefont{H.}},
  \bibinfo{author}{\bibfnamefont{A.}~\bibnamefont{Schirmeisen}}, and
  \bibinfo{author}{\bibfnamefont{U.~D.} \bibnamefont{Schwarz}},
  \bibinfo{year}{2008}, \bibinfo{journal}{Phil. Trans. R. Soc. A}
  \textbf{\bibinfo{volume}{366}}, \bibinfo{pages}{1383}.

\bibitem[{\citenamefont{Hori} \emph{et~al.}(2004)\citenamefont{Hori, Kato,
  Hirahara, Baba, and Kaneda}}]{Hori_etal2004}
\bibinfo{author}{\bibnamefont{Hori}, \bibfnamefont{T.}},
  \bibinfo{author}{\bibfnamefont{N.}~\bibnamefont{Kato}},
  \bibinfo{author}{\bibfnamefont{K.}~\bibnamefont{Hirahara}},
  \bibinfo{author}{\bibfnamefont{T.}~\bibnamefont{Baba}}, and
  \bibinfo{author}{\bibfnamefont{Y.}~\bibnamefont{Kaneda}},
  \bibinfo{year}{2004}, \bibinfo{journal}{Earth Planet. Sci. Lett.}
  \textbf{\bibinfo{volume}{228}}, \bibinfo{pages}{215}.

\bibitem[{\citenamefont{Horowitz and Ruina}(1989)}]{Horowitz}
\bibinfo{author}{\bibnamefont{Horowitz}, \bibfnamefont{F.}}, and
  \bibinfo{author}{\bibfnamefont{A.}~\bibnamefont{Ruina}},
  \bibinfo{year}{1989}, \bibinfo{journal}{J. Geophys. Res.}
  \textbf{\bibinfo{volume}{94}}, \bibinfo{pages}{doi:10.1029/JB094iB08p10279}.

\bibitem[{\citenamefont{Hyun} \emph{et~al.}(2004)\citenamefont{Hyun, Pei,
  Molinari, and Robbins}}]{Hyun2004}
\bibinfo{author}{\bibnamefont{Hyun}, \bibfnamefont{S.}},
  \bibinfo{author}{\bibfnamefont{L.}~\bibnamefont{Pei}},
  \bibinfo{author}{\bibfnamefont{J.-F.} \bibnamefont{Molinari}}, and
  \bibinfo{author}{\bibfnamefont{M.}~\bibnamefont{Robbins}},
  \bibinfo{year}{2004}, \bibinfo{journal}{Phys. Rev. E}
  \textbf{\bibinfo{volume}{70}}(\bibinfo{number}{2}), \bibinfo{pages}{026117}.

\bibitem[{\citenamefont{Ida}(1972)}]{Ida1972}
\bibinfo{author}{\bibnamefont{Ida}, \bibfnamefont{Y.}}, \bibinfo{year}{1972},
  \bibinfo{journal}{J. Geophys. Res.} \textbf{\bibinfo{volume}{77}},
  \bibinfo{pages}{doi:10.1029/JB077i020p03796}.

\bibitem[{\citenamefont{Ide and Takeo}(1997)}]{Ide1997}
\bibinfo{author}{\bibnamefont{Ide}, \bibfnamefont{S.}}, and
  \bibinfo{author}{\bibfnamefont{M.}~\bibnamefont{Takeo}},
  \bibinfo{year}{1997}, \bibinfo{journal}{J. Geophys. Res}
  \textbf{\bibinfo{volume}{102}}(\bibinfo{number}{B12}),
  \bibinfo{pages}{doi:10.1029/97JB02675}.

\bibitem[{\citenamefont{Igarashi} \emph{et~al.}(2003)\citenamefont{Igarashi,
  Matsuzawa, and Hasegawa}}]{Igarashi_etal2003}
\bibinfo{author}{\bibnamefont{Igarashi}, \bibfnamefont{T.}},
  \bibinfo{author}{\bibfnamefont{T.}~\bibnamefont{Matsuzawa}}, and
  \bibinfo{author}{\bibfnamefont{A.}~\bibnamefont{Hasegawa}},
  \bibinfo{year}{2003}, \bibinfo{journal}{J. Geophys. Res.}
  \textbf{\bibinfo{volume}{108}}(\bibinfo{number}{2249}),
  \bibinfo{pages}{doi:10.1029/2002JB001920}.

\bibitem[{\citenamefont{Ishibe and Shimazaki}(2009)}]{IshibeShimazaki2009}
\bibinfo{author}{\bibnamefont{Ishibe}, \bibfnamefont{T.}}, and
  \bibinfo{author}{\bibfnamefont{K.}~\bibnamefont{Shimazaki}},
  \bibinfo{year}{2009}, \bibinfo{journal}{Earth Planets Space}
  \textbf{\bibinfo{volume}{61}}, \bibinfo{pages}{1041}.

\bibitem[{\citenamefont{Ito and Matsuzaki}(1990)}]{Ito}
\bibinfo{author}{\bibnamefont{Ito}, \bibfnamefont{K.}}, and
  \bibinfo{author}{\bibfnamefont{M.}~\bibnamefont{Matsuzaki}},
  \bibinfo{year}{1990}, \bibinfo{journal}{J. Geophys. Res.}
  \textbf{\bibinfo{volume}{95}}, \bibinfo{pages}{doi:10.1029/JB095iB05p06853}.

\bibitem[{\citenamefont{Jagla}(2010)}]{Jagla}
\bibinfo{author}{\bibnamefont{Jagla}, \bibfnamefont{E.~A.}},
  \bibinfo{year}{2010}, \bibinfo{journal}{Phys. Rev. E}
  \textbf{\bibinfo{volume}{81}}, \bibinfo{pages}{046117}.

\bibitem[{\citenamefont{J\'{a}nosi and Kert\'{e}sz}(1993)}]{Janosi}
\bibinfo{author}{\bibnamefont{J\'{a}nosi}, \bibfnamefont{I.~M.}}, and
  \bibinfo{author}{\bibfnamefont{J.}~\bibnamefont{Kert\'{e}sz}},
  \bibinfo{year}{1993}, \bibinfo{journal}{Physica A}
  \textbf{\bibinfo{volume}{200}}, \bibinfo{pages}{179}.

\bibitem[{\citenamefont{Jaume and Sykes}(1999)}]{Jaume99}
\bibinfo{author}{\bibnamefont{Jaume}, \bibfnamefont{S.~C.}}, and
  \bibinfo{author}{\bibfnamefont{L.~R.} \bibnamefont{Sykes}},
  \bibinfo{year}{1999}, \bibinfo{journal}{Pure Appl. Geophys.}
  \textbf{\bibinfo{volume}{155}}, \bibinfo{pages}{279}.

\bibitem[{\citenamefont{J.B.} \emph{et~al.}(2000)\citenamefont{J.B., Turcotte,
  and Klein}}]{Rundle2000}
\bibinfo{author}{\bibnamefont{J.B.}, \bibfnamefont{R.}},
  \bibinfo{author}{\bibfnamefont{D.~L.} \bibnamefont{Turcotte}}, and
  \bibinfo{author}{\bibfnamefont{W.}~\bibnamefont{Klein}},
  \bibinfo{year}{2000}, \emph{\bibinfo{title}{GeoComplexity and the Physics of
  Earthquakes, Geophysical Monograph 120}} (\bibinfo{publisher}{American
  Geophysical Union, Washington DC}).

\bibitem[{\citenamefont{Johnson} \emph{et~al.}(2009)\citenamefont{Johnson,
  Bürgmann, and Freymueller}}]{Johnson_etal2009}
\bibinfo{author}{\bibnamefont{Johnson}, \bibfnamefont{K.~M.}},
  \bibinfo{author}{\bibfnamefont{R.}~\bibnamefont{Bürgmann}}, and
  \bibinfo{author}{\bibfnamefont{J.~T.} \bibnamefont{Freymueller}},
  \bibinfo{year}{2009}, \bibinfo{journal}{Geophys. J. Int.}
  \textbf{\bibinfo{volume}{176}}, \bibinfo{pages}{670–682}.

\bibitem[{\citenamefont{Johnson} \emph{et~al.}(2006)\citenamefont{Johnson,
  Bürgmann, and Larson}}]{Johnson_etal2006}
\bibinfo{author}{\bibnamefont{Johnson}, \bibfnamefont{K.~M.}},
  \bibinfo{author}{\bibfnamefont{R.}~\bibnamefont{Bürgmann}}, and
  \bibinfo{author}{\bibfnamefont{K.}~\bibnamefont{Larson}},
  \bibinfo{year}{2006}, \bibinfo{journal}{Bull. Seismol. Soc. Am.}
  \textbf{\bibinfo{volume}{96}}, \bibinfo{pages}{S321}.

\bibitem[{\citenamefont{Kagan}(2006)}]{bkc59}
\bibinfo{author}{\bibnamefont{Kagan}, \bibfnamefont{Y.~Y.}},
  \bibinfo{year}{2006}, in \emph{\bibinfo{booktitle}{{Modelling Critical and
  Catastrophic Phenomena in Geoscience}}}, edited by
  \bibinfo{editor}{\bibfnamefont{P.}~\bibnamefont{Bhattacharyya}} and
  \bibinfo{editor}{\bibfnamefont{B.~K.} \bibnamefont{Chakrabarti}}
  (\bibinfo{publisher}{Springer-Verlag Heidelberg}), pp.
  \bibinfo{pages}{303--361}.

\bibitem[{\citenamefont{Kakui and Kawamura}(2011)}]{KakuiKawamura}
\bibinfo{author}{\bibnamefont{Kakui}, \bibfnamefont{S.}}, and
  \bibinfo{author}{\bibfnamefont{H.}~\bibnamefont{Kawamura}},
  \bibinfo{year}{2011}, \bibinfo{journal}{in preparation} .

\bibitem[{\citenamefont{Kanamori}(1981)}]{Kanamori1981}
\bibinfo{author}{\bibnamefont{Kanamori}, \bibfnamefont{H.}},
  \bibinfo{year}{1981}, in \emph{\bibinfo{booktitle}{Earthquake prediction: an
  international review}}, edited by \bibinfo{editor}{\bibfnamefont{D.~W.}
  \bibnamefont{Simpson}} and \bibinfo{editor}{\bibfnamefont{P.~G.}
  \bibnamefont{Richards}} (\bibinfo{publisher}{American Geophysical Union}),
  pp. \bibinfo{pages}{1--19}.

\bibitem[{\citenamefont{Kanamori}(2009)}]{Kanamori2009}
\bibinfo{editor}{\bibnamefont{Kanamori}, \bibfnamefont{H.}} (ed.),
  \bibinfo{year}{2009}, \emph{\bibinfo{title}{Earthquake Seismology, Treatise
  on Geophysics vol.4}} (\bibinfo{publisher}{Elsevier Amsterdam}).

\bibitem[{\citenamefont{Kanamori and Cipar}(1974)}]{KanamoriCipar1974}
\bibinfo{author}{\bibnamefont{Kanamori}, \bibfnamefont{H.}}, and
  \bibinfo{author}{\bibfnamefont{J.~J.} \bibnamefont{Cipar}},
  \bibinfo{year}{1974}, \bibinfo{journal}{Phys. Earth Planet. Inter.}
  \textbf{\bibinfo{volume}{9}}, \bibinfo{pages}{128}.

\bibitem[{\citenamefont{Kanamori and McNally}(1982)}]{KanamoriMcNally1982}
\bibinfo{author}{\bibnamefont{Kanamori}, \bibfnamefont{H.}}, and
  \bibinfo{author}{\bibfnamefont{K.~C.} \bibnamefont{McNally}},
  \bibinfo{year}{1982}, \bibinfo{journal}{Bull. Seismol. Soc. Am.}
  \textbf{\bibinfo{volume}{72}}, \bibinfo{pages}{1241}.

\bibitem[{\citenamefont{Kanamori and Stewart}(1979)}]{KanamoriStewart1979}
\bibinfo{author}{\bibnamefont{Kanamori}, \bibfnamefont{H.}}, and
  \bibinfo{author}{\bibfnamefont{G.~S.} \bibnamefont{Stewart}},
  \bibinfo{year}{1979}, \bibinfo{journal}{Phys. Earth Planet. Inter.}
  \textbf{\bibinfo{volume}{18}}, \bibinfo{pages}{167}.

\bibitem[{\citenamefont{Kaneko} \emph{et~al.}(2010)\citenamefont{Kaneko,
  Avouac, and Lapusta}}]{Kaneko_etal2010}
\bibinfo{author}{\bibnamefont{Kaneko}, \bibfnamefont{Y.}},
  \bibinfo{author}{\bibfnamefont{J.-P.} \bibnamefont{Avouac}}, and
  \bibinfo{author}{\bibfnamefont{N.}~\bibnamefont{Lapusta}},
  \bibinfo{year}{2010}, \bibinfo{journal}{Nature Geosci.}
  \textbf{\bibinfo{volume}{3}}, \bibinfo{pages}{363}.

\bibitem[{\citenamefont{Kardar} \emph{et~al.}(1986)\citenamefont{Kardar,
  Parisi, and Zhang}}]{bkc40}
\bibinfo{author}{\bibnamefont{Kardar}, \bibfnamefont{M.}},
  \bibinfo{author}{\bibfnamefont{G.}~\bibnamefont{Parisi}}, and
  \bibinfo{author}{\bibfnamefont{Y.~C.} \bibnamefont{Zhang}},
  \bibinfo{year}{1986}, \bibinfo{journal}{Phys. Rev. Lett.}
  \textbf{\bibinfo{volume}{56}}, \bibinfo{pages}{889}.

\bibitem[{\citenamefont{Kato}(2003{\natexlab{a}})}]{Kato2003b}
\bibinfo{author}{\bibnamefont{Kato}, \bibfnamefont{N.}},
  \bibinfo{year}{2003}{\natexlab{a}}, \bibinfo{journal}{Earth Planet. Sci.
  Lett.} \textbf{\bibinfo{volume}{216}}, \bibinfo{pages}{17}.

\bibitem[{\citenamefont{Kato}(2003{\natexlab{b}})}]{Kato2003a}
\bibinfo{author}{\bibnamefont{Kato}, \bibfnamefont{N.}},
  \bibinfo{year}{2003}{\natexlab{b}}, \bibinfo{journal}{Bull. Earthq. Res.
  Inst., Univ. Tokyo} \textbf{\bibinfo{volume}{78}}, \bibinfo{pages}{155}.

\bibitem[{\citenamefont{Kato}(2004)}]{Kato2004}
\bibinfo{author}{\bibnamefont{Kato}, \bibfnamefont{N.}}, \bibinfo{year}{2004},
  \bibinfo{journal}{J. Geophys. Res.} \textbf{\bibinfo{volume}{109}},
  \bibinfo{pages}{doi:10.1029/2004JB003001}.

\bibitem[{\citenamefont{Kato}(2007)}]{Kato2007}
\bibinfo{author}{\bibnamefont{Kato}, \bibfnamefont{N.}}, \bibinfo{year}{2007},
  \bibinfo{journal}{Geophys. J. Int.} \textbf{\bibinfo{volume}{168}},
  \bibinfo{pages}{286}.

\bibitem[{\citenamefont{Kato}(2008)}]{Kato2008}
\bibinfo{author}{\bibnamefont{Kato}, \bibfnamefont{N.}}, \bibinfo{year}{2008},
  \bibinfo{journal}{J. Geophys. Res.}
  \textbf{\bibinfo{volume}{113}}(\bibinfo{number}{B06302}),
  \bibinfo{pages}{doi:10.1029/2007JB005515}.

\bibitem[{\citenamefont{Kato and
  Hirasawa}(1999{\natexlab{a}})}]{KatoHirasawa1999a}
\bibinfo{author}{\bibnamefont{Kato}, \bibfnamefont{N.}}, and
  \bibinfo{author}{\bibfnamefont{T.}~\bibnamefont{Hirasawa}},
  \bibinfo{year}{1999}{\natexlab{a}}, \bibinfo{journal}{Bull. Seismol. Soc.
  Am.} \textbf{\bibinfo{volume}{89}}, \bibinfo{pages}{1401}.

\bibitem[{\citenamefont{Kato and
  Hirasawa}(1999{\natexlab{b}})}]{KatoHirasawa1999}
\bibinfo{author}{\bibnamefont{Kato}, \bibfnamefont{N.}}, and
  \bibinfo{author}{\bibfnamefont{T.}~\bibnamefont{Hirasawa}},
  \bibinfo{year}{1999}{\natexlab{b}}, \bibinfo{journal}{Pure Appl. Geophys.}
  \textbf{\bibinfo{volume}{155}}, \bibinfo{pages}{93}.

\bibitem[{\citenamefont{Kato} \emph{et~al.}(1997)\citenamefont{Kato, Ohatake,
  and Hirasawa}}]{Kato_etal1997}
\bibinfo{author}{\bibnamefont{Kato}, \bibfnamefont{N.}},
  \bibinfo{author}{\bibfnamefont{M.}~\bibnamefont{Ohatake}}, and
  \bibinfo{author}{\bibfnamefont{T.}~\bibnamefont{Hirasawa}},
  \bibinfo{year}{1997}, \bibinfo{journal}{Pure Appl. Geophys.}
  \textbf{\bibinfo{volume}{150}}, \bibinfo{pages}{249}.

\bibitem[{\citenamefont{Kato and Tullis}(2001)}]{Kat}
\bibinfo{author}{\bibnamefont{Kato}, \bibfnamefont{N.}}, and
  \bibinfo{author}{\bibfnamefont{T.~E.} \bibnamefont{Tullis}},
  \bibinfo{year}{2001}, \bibinfo{journal}{Geophys. Res. Lett.}
  \textbf{\bibinfo{volume}{28}}, \bibinfo{pages}{doi:10.1029/2000GL012060}.

\bibitem[{\citenamefont{Kawamura}(2006)}]{bkc38}
\bibinfo{author}{\bibnamefont{Kawamura}, \bibfnamefont{H.}},
  \bibinfo{year}{2006}, in \emph{\bibinfo{booktitle}{{Modelling Critical and
  Catastrophic Phenomena in Geoscience}}}, edited by
  \bibinfo{editor}{\bibfnamefont{P.}~\bibnamefont{Bhattacharyya}} and
  \bibinfo{editor}{\bibfnamefont{B.~K.} \bibnamefont{Chakrabarti}}
  (\bibinfo{publisher}{Springer-Verlag Heidelberg}), pp.
  \bibinfo{pages}{223--257}.

\bibitem[{\citenamefont{Kawamura} \emph{et~al.}(2010)\citenamefont{Kawamura,
  Yamamoto, Kotani, and Yoshino}}]{KawamurOFC}
\bibinfo{author}{\bibnamefont{Kawamura}, \bibfnamefont{H.}},
  \bibinfo{author}{\bibfnamefont{T.}~\bibnamefont{Yamamoto}},
  \bibinfo{author}{\bibfnamefont{T.}~\bibnamefont{Kotani}}, and
  \bibinfo{author}{\bibfnamefont{H.}~\bibnamefont{Yoshino}},
  \bibinfo{year}{2010}, \bibinfo{journal}{Phys. Rev. E}
  \textbf{\bibinfo{volume}{81}}, \bibinfo{pages}{031119}.

\bibitem[{\citenamefont{King} \emph{et~al.}(1973)\citenamefont{King, Nason, and
  Tocher}}]{King_etal1973}
\bibinfo{author}{\bibnamefont{King}, \bibfnamefont{C.-Y.}},
  \bibinfo{author}{\bibfnamefont{R.~D.} \bibnamefont{Nason}}, and
  \bibinfo{author}{\bibfnamefont{D.}~\bibnamefont{Tocher}},
  \bibinfo{year}{1973}, \bibinfo{journal}{Phil. Trans. Roy. Soc. London, Ser.
  A} \textbf{\bibinfo{volume}{274}}, \bibinfo{pages}{355}.

\bibitem[{\citenamefont{Koivisto} \emph{et~al.}(2007)\citenamefont{Koivisto,
  Rosti, and Alava}}]{bkc16}
\bibinfo{author}{\bibnamefont{Koivisto}, \bibfnamefont{J.}},
  \bibinfo{author}{\bibfnamefont{J.}~\bibnamefont{Rosti}}, and
  \bibinfo{author}{\bibfnamefont{M.~J.} \bibnamefont{Alava}},
  \bibinfo{year}{2007}, \bibinfo{journal}{Phys. Rev. Lett.}
  \textbf{\bibinfo{volume}{99}}, \bibinfo{pages}{145504}.

\bibitem[{\citenamefont{Koketsu} \emph{et~al.}(2011)\citenamefont{Koketsu,
  Yokota, Nishimura, Yagi, Miyazaki, Satake, Fujii, Miyake, Sakai, Yamanaka,
  and Okada}}]{Koketsu_etal2011}
\bibinfo{author}{\bibnamefont{Koketsu}, \bibfnamefont{K.}},
  \bibinfo{author}{\bibfnamefont{Y.}~\bibnamefont{Yokota}},
  \bibinfo{author}{\bibfnamefont{N.}~\bibnamefont{Nishimura}},
  \bibinfo{author}{\bibfnamefont{Y.}~\bibnamefont{Yagi}},
  \bibinfo{author}{\bibfnamefont{S.}~\bibnamefont{Miyazaki}},
  \bibinfo{author}{\bibfnamefont{K.}~\bibnamefont{Satake}},
  \bibinfo{author}{\bibfnamefont{Y.}~\bibnamefont{Fujii}},
  \bibinfo{author}{\bibfnamefont{H.}~\bibnamefont{Miyake}},
  \bibinfo{author}{\bibfnamefont{S.}~\bibnamefont{Sakai}},
  \bibinfo{author}{\bibfnamefont{Y.}~\bibnamefont{Yamanaka}}, and
  \bibinfo{author}{\bibfnamefont{T.}~\bibnamefont{Okada}},
  \bibinfo{year}{2011}, \bibinfo{journal}{Earth Planet. Sci. Lett.}
  \textbf{\bibinfo{volume}{310}}, \bibinfo{pages}{480}.

\bibitem[{\citenamefont{Kossobokov and Carlson}(1995)}]{Kos95}
\bibinfo{author}{\bibnamefont{Kossobokov}, \bibfnamefont{V.~G.}}, and
  \bibinfo{author}{\bibfnamefont{J.~M.} \bibnamefont{Carlson}},
  \bibinfo{year}{1995}, \bibinfo{journal}{J. Geophys. Res.}
  \textbf{\bibinfo{volume}{100}}, \bibinfo{pages}{doi:10.1029/94JB02868}.

\bibitem[{\citenamefont{Kotani} \emph{et~al.}(2008)\citenamefont{Kotani,
  Yoshino, and Kawamura}}]{Kotani}
\bibinfo{author}{\bibnamefont{Kotani}, \bibfnamefont{T.}},
  \bibinfo{author}{\bibfnamefont{H.}~\bibnamefont{Yoshino}}, and
  \bibinfo{author}{\bibfnamefont{H.}~\bibnamefont{Kawamura}},
  \bibinfo{year}{2008}, \bibinfo{journal}{Phys. Rev. E}
  \textbf{\bibinfo{volume}{77}}, \bibinfo{pages}{R010102}.

\bibitem[{\citenamefont{Kumagai} \emph{et~al.}(1999)\citenamefont{Kumagai,
  Fukao, Watanabe, and Baba}}]{Kumagai}
\bibinfo{author}{\bibnamefont{Kumagai}, \bibfnamefont{H.}},
  \bibinfo{author}{\bibfnamefont{Y.}~\bibnamefont{Fukao}},
  \bibinfo{author}{\bibfnamefont{S.}~\bibnamefont{Watanabe}}, and
  \bibinfo{author}{\bibfnamefont{Y.}~\bibnamefont{Baba}}, \bibinfo{year}{1999},
  \bibinfo{journal}{Geophys. Res. Lett.} \textbf{\bibinfo{volume}{26}},
  \bibinfo{pages}{doi:10.1029/1999GL005383}.

\bibitem[{\citenamefont{Kuroki} \emph{et~al.}(2002)\citenamefont{Kuroki, Ito,
  and Yoshida}}]{Kuroki_etal2002}
\bibinfo{author}{\bibnamefont{Kuroki}, \bibfnamefont{H.}},
  \bibinfo{author}{\bibfnamefont{H.~M.} \bibnamefont{Ito}}, and
  \bibinfo{author}{\bibfnamefont{A.}~\bibnamefont{Yoshida}},
  \bibinfo{year}{2002}, \bibinfo{journal}{Phys. Earth Planet. Inter.}
  \textbf{\bibinfo{volume}{132}}, \bibinfo{pages}{39}.

\bibitem[{\citenamefont{Kuwano and Hatano}(2011)}]{Kuwano2011}
\bibinfo{author}{\bibnamefont{Kuwano}, \bibfnamefont{O.}}, and
  \bibinfo{author}{\bibfnamefont{T.}~\bibnamefont{Hatano}},
  \bibinfo{year}{2011}, \bibinfo{journal}{Geophys. Res. Lett.}
  \textbf{\bibinfo{volume}{38}}, \bibinfo{pages}{L17305}.

\bibitem[{\citenamefont{Lachenbruch}(1980)}]{Lachenbruch1980}
\bibinfo{author}{\bibnamefont{Lachenbruch}, \bibfnamefont{A.~H.}},
  \bibinfo{year}{1980}, \bibinfo{journal}{J. Geophys. Res.}
  \textbf{\bibinfo{volume}{85}}, \bibinfo{pages}{6097}.

\bibitem[{\citenamefont{Lapusta and Liu}(2009)}]{LapustaLiu2009}
\bibinfo{author}{\bibnamefont{Lapusta}, \bibfnamefont{N.}}, and
  \bibinfo{author}{\bibfnamefont{Y.}~\bibnamefont{Liu}}, \bibinfo{year}{2009},
  \bibinfo{journal}{J. Geophys. Res.}
  \textbf{\bibinfo{volume}{114}}(\bibinfo{number}{B09303}),
  \bibinfo{pages}{doi:10.1029/2008JB005934}.

\bibitem[{\citenamefont{Lawn}(1993)}]{bkc5}
\bibinfo{author}{\bibnamefont{Lawn}, \bibfnamefont{B.}}, \bibinfo{year}{1993},
  \emph{\bibinfo{title}{{ Fracture of brittle solids}}}
  (\bibinfo{publisher}{Cambridge Univ. Press, Cambridge}).

\bibitem[{\citenamefont{Lay} \emph{et~al.}(1982)\citenamefont{Lay, Kanamori,
  and Ruff}}]{Lay_etal1982}
\bibinfo{author}{\bibnamefont{Lay}, \bibfnamefont{T.}},
  \bibinfo{author}{\bibfnamefont{H.}~\bibnamefont{Kanamori}}, and
  \bibinfo{author}{\bibfnamefont{L.}~\bibnamefont{Ruff}}, \bibinfo{year}{1982},
  \bibinfo{journal}{Earthq. Predict. Res.} \textbf{\bibinfo{volume}{1}},
  \bibinfo{pages}{3}.

\bibitem[{\citenamefont{Leschhorn}(1992)}]{bkc42}
\bibinfo{author}{\bibnamefont{Leschhorn}, \bibfnamefont{H.}},
  \bibinfo{year}{1992}, \bibinfo{journal}{J. Phys. A}
  \textbf{\bibinfo{volume}{25}}, \bibinfo{pages}{L255}.

\bibitem[{\citenamefont{Linde} \emph{et~al.}(1996)\citenamefont{Linde, Gladwin,
  Johnston, Gwyther, and Bilham}}]{Linde_etal1996}
\bibinfo{author}{\bibnamefont{Linde}, \bibfnamefont{A.~T.}},
  \bibinfo{author}{\bibfnamefont{M.~T.} \bibnamefont{Gladwin}},
  \bibinfo{author}{\bibfnamefont{M.~J.~S.} \bibnamefont{Johnston}},
  \bibinfo{author}{\bibfnamefont{R.~L.} \bibnamefont{Gwyther}}, and
  \bibinfo{author}{\bibfnamefont{R.~G.} \bibnamefont{Bilham}},
  \bibinfo{year}{1996}, \bibinfo{journal}{Nature}
  \textbf{\bibinfo{volume}{383}}, \bibinfo{pages}{65}.

\bibitem[{\citenamefont{Linde and Sacks}(2002)}]{LindeSacks2002}
\bibinfo{author}{\bibnamefont{Linde}, \bibfnamefont{A.~T.}}, and
  \bibinfo{author}{\bibfnamefont{I.~S.} \bibnamefont{Sacks}},
  \bibinfo{year}{2002}, \bibinfo{journal}{Earth Planet. Sci. Lett.}
  \textbf{\bibinfo{volume}{203}}, \bibinfo{pages}{265}.

\bibitem[{\citenamefont{Lise and Paczuski}(2001)}]{Lise}
\bibinfo{author}{\bibnamefont{Lise}, \bibfnamefont{S.}}, and
  \bibinfo{author}{\bibfnamefont{M.}~\bibnamefont{Paczuski}},
  \bibinfo{year}{2001}, \bibinfo{journal}{Phys. Rev. E}
  \textbf{\bibinfo{volume}{63}}, \bibinfo{pages}{036111}.

\bibitem[{\citenamefont{Liu and Rice}(2007)}]{LiuRice2007}
\bibinfo{author}{\bibnamefont{Liu}, \bibfnamefont{Y.}}, and
  \bibinfo{author}{\bibfnamefont{J.~R.} \bibnamefont{Rice}},
  \bibinfo{year}{2007}, \bibinfo{journal}{J. Geophys. Res.}
  \textbf{\bibinfo{volume}{112}}(\bibinfo{number}{B09404}),
  \bibinfo{pages}{doi:10.1029/2007JB004930}.

\bibitem[{\citenamefont{Lowry} \emph{et~al.}(2001)\citenamefont{Lowry, Larson,
  Kostoglodov, and Bilham}}]{Lowry_etal2001}
\bibinfo{author}{\bibnamefont{Lowry}, \bibfnamefont{A.~R.}},
  \bibinfo{author}{\bibfnamefont{K.~M.} \bibnamefont{Larson}},
  \bibinfo{author}{\bibfnamefont{V.}~\bibnamefont{Kostoglodov}}, and
  \bibinfo{author}{\bibfnamefont{R.}~\bibnamefont{Bilham}},
  \bibinfo{year}{2001}, \bibinfo{journal}{Geophys. Res. Lett.}
  \textbf{\bibinfo{volume}{28}}(\bibinfo{number}{19}),
  \bibinfo{pages}{doi:10.1029/2001GL013238}.

\bibitem[{\citenamefont{Marone}(1998)}]{Marone1998}
\bibinfo{author}{\bibnamefont{Marone}, \bibfnamefont{C.}},
  \bibinfo{year}{1998}, \bibinfo{journal}{Annual Review of Earth and Planetary
  Sciences} \textbf{\bibinfo{volume}{26}}(\bibinfo{number}{1}),
  \bibinfo{pages}{643}.

\bibitem[{\citenamefont{Marone} \emph{et~al.}(1991)\citenamefont{Marone,
  Scholtz, and Bilham}}]{Marone_etal1991}
\bibinfo{author}{\bibnamefont{Marone}, \bibfnamefont{C.}},
  \bibinfo{author}{\bibfnamefont{C.}~\bibnamefont{Scholtz}}, and
  \bibinfo{author}{\bibfnamefont{R.}~\bibnamefont{Bilham}},
  \bibinfo{year}{1991}, \bibinfo{journal}{J. Geophys. Res.}
  \textbf{\bibinfo{volume}{96}}(\bibinfo{number}{B9}),
  \bibinfo{pages}{doi:10.1029/91JB01588}.

\bibitem[{\citenamefont{Marone and Scholz}(1989)}]{Marone1989}
\bibinfo{author}{\bibnamefont{Marone}, \bibfnamefont{C.}}, and
  \bibinfo{author}{\bibfnamefont{C.~H.} \bibnamefont{Scholz}},
  \bibinfo{year}{1989}, \bibinfo{journal}{J. Struct. Geo.}
  \textbf{\bibinfo{volume}{11}}, \bibinfo{pages}{799}.

\bibitem[{\citenamefont{Mase and Smith}(1987)}]{Mase1987}
\bibinfo{author}{\bibnamefont{Mase}, \bibfnamefont{C.}}, and
  \bibinfo{author}{\bibfnamefont{L.}~\bibnamefont{Smith}},
  \bibinfo{year}{1987}, \bibinfo{journal}{J. Geophys. Res.}
  \textbf{\bibinfo{volume}{92}}, \bibinfo{pages}{6249}.

\bibitem[{\citenamefont{Matsukawa and Fukuyama}(1994)}]{bkc22}
\bibinfo{author}{\bibnamefont{Matsukawa}, \bibfnamefont{H.}}, and
  \bibinfo{author}{\bibfnamefont{H.}~\bibnamefont{Fukuyama}},
  \bibinfo{year}{1994}, \bibinfo{journal}{Phys. Rev. B}
  \textbf{\bibinfo{volume}{49}}, \bibinfo{pages}{17286}.

\bibitem[{\citenamefont{Matsumura}(1997)}]{Matsumura1997}
\bibinfo{author}{\bibnamefont{Matsumura}, \bibfnamefont{S.}},
  \bibinfo{year}{1997}, \bibinfo{journal}{Tectonophysics}
  \textbf{\bibinfo{volume}{273}}, \bibinfo{pages}{271}.

\bibitem[{\citenamefont{Matsuzawa} \emph{et~al.}(2010)\citenamefont{Matsuzawa,
  Hirose, Shibazaki, and Obara}}]{Matsuzawa_etal2010}
\bibinfo{author}{\bibnamefont{Matsuzawa}, \bibfnamefont{T.}},
  \bibinfo{author}{\bibfnamefont{H.}~\bibnamefont{Hirose}},
  \bibinfo{author}{\bibfnamefont{B.}~\bibnamefont{Shibazaki}}, and
  \bibinfo{author}{\bibfnamefont{K.}~\bibnamefont{Obara}},
  \bibinfo{year}{2010}, \bibinfo{journal}{J. Geophys. Res.}
  \textbf{\bibinfo{volume}{115}}(\bibinfo{number}{B12301}),
  \bibinfo{pages}{doi:10.1029/2010JB007566}.

\bibitem[{\citenamefont{Matsuzawa} \emph{et~al.}(2002)\citenamefont{Matsuzawa,
  Igarashi, and Hasegawa}}]{Matsuzawa}
\bibinfo{author}{\bibnamefont{Matsuzawa}, \bibfnamefont{T.}},
  \bibinfo{author}{\bibfnamefont{T.}~\bibnamefont{Igarashi}}, and
  \bibinfo{author}{\bibfnamefont{A.}~\bibnamefont{Hasegawa}},
  \bibinfo{year}{2002}, \bibinfo{journal}{Geophys. Res. Lett.}
  \textbf{\bibinfo{volume}{29}}, \bibinfo{pages}{doi:10.1029/2001GL014632}.

\bibitem[{\citenamefont{Matthews} \emph{et~al.}(2002)\citenamefont{Matthews,
  Ellsworth, and Reasenberg}}]{Matthews_etal2002}
\bibinfo{author}{\bibnamefont{Matthews}, \bibfnamefont{M.~V.}},
  \bibinfo{author}{\bibfnamefont{W.~L.} \bibnamefont{Ellsworth}}, and
  \bibinfo{author}{\bibfnamefont{P.~A.} \bibnamefont{Reasenberg}},
  \bibinfo{year}{2002}, \bibinfo{journal}{Bull. Seismol. Soc. Am.}
  \textbf{\bibinfo{volume}{92}}, \bibinfo{pages}{2233}.

\bibitem[{\citenamefont{Middleton and Tang}(1995)}]{Middleton}
\bibinfo{author}{\bibnamefont{Middleton}, \bibfnamefont{A.~A.}}, and
  \bibinfo{author}{\bibfnamefont{C.}~\bibnamefont{Tang}}, \bibinfo{year}{1995},
  \bibinfo{journal}{Phys. Rev. Lett.} \textbf{\bibinfo{volume}{74}},
  \bibinfo{pages}{742}.

\bibitem[{\citenamefont{Miller and Boulter}(2002)}]{Boulter}
\bibinfo{author}{\bibnamefont{Miller}, \bibfnamefont{G.}}, and
  \bibinfo{author}{\bibfnamefont{C.~J.} \bibnamefont{Boulter}},
  \bibinfo{year}{2002}, \bibinfo{journal}{Phys. Rev. E}
  \textbf{\bibinfo{volume}{66}}, \bibinfo{pages}{016123}.

\bibitem[{\citenamefont{Miyazaki} \emph{et~al.}(2004)\citenamefont{Miyazaki,
  Segall, Fukuda, and Kato}}]{Miyazaki_etal2004}
\bibinfo{author}{\bibnamefont{Miyazaki}, \bibfnamefont{S.}},
  \bibinfo{author}{\bibfnamefont{P.}~\bibnamefont{Segall}},
  \bibinfo{author}{\bibfnamefont{J.}~\bibnamefont{Fukuda}}, and
  \bibinfo{author}{\bibfnamefont{T.}~\bibnamefont{Kato}}, \bibinfo{year}{2004},
  \bibinfo{journal}{Geophys. Res. Lett.}
  \textbf{\bibinfo{volume}{31}}(\bibinfo{number}{L06623}),
  \bibinfo{pages}{doi:10.1029/2003GL019410}.

\bibitem[{\citenamefont{Miyazaki} \emph{et~al.}(2006)\citenamefont{Miyazaki,
  Segall, McGuire, Kato, and Hatanaka}}]{Miyazaki_etal2006}
\bibinfo{author}{\bibnamefont{Miyazaki}, \bibfnamefont{S.}},
  \bibinfo{author}{\bibfnamefont{P.}~\bibnamefont{Segall}},
  \bibinfo{author}{\bibfnamefont{J.~J.} \bibnamefont{McGuire}},
  \bibinfo{author}{\bibfnamefont{T.}~\bibnamefont{Kato}}, and
  \bibinfo{author}{\bibfnamefont{Y.}~\bibnamefont{Hatanaka}},
  \bibinfo{year}{2006}, \bibinfo{journal}{J. Geophys. Res.}
  \textbf{\bibinfo{volume}{111}}(\bibinfo{number}{B03409}),
  \bibinfo{pages}{doi:10.1029/2004JB003426}.

\bibitem[{\citenamefont{Mizoguchi} \emph{et~al.}(2006)\citenamefont{Mizoguchi,
  Hirose, Shimamoto, and Fukuyama}}]{Mizoguchi2006}
\bibinfo{author}{\bibnamefont{Mizoguchi}, \bibfnamefont{K.}},
  \bibinfo{author}{\bibfnamefont{T.}~\bibnamefont{Hirose}},
  \bibinfo{author}{\bibfnamefont{T.}~\bibnamefont{Shimamoto}}, and
  \bibinfo{author}{\bibfnamefont{E.}~\bibnamefont{Fukuyama}},
  \bibinfo{year}{2006}, \bibinfo{journal}{Geophys. Res. Lett.}
  \textbf{\bibinfo{volume}{33}}(\bibinfo{number}{16}),
  \bibinfo{pages}{doi:10.1029/2006GL026980}.

\bibitem[{\citenamefont{Mogi}(1969)}]{Mogi69}
\bibinfo{author}{\bibnamefont{Mogi}, \bibfnamefont{K.}}, \bibinfo{year}{1969},
  \bibinfo{journal}{Bull. Earthquake Res. Inst. Univ. Tokyo}
  \textbf{\bibinfo{volume}{47}}, \bibinfo{pages}{395}.

\bibitem[{\citenamefont{Mogi}(1979)}]{Mogi79}
\bibinfo{author}{\bibnamefont{Mogi}, \bibfnamefont{K.}}, \bibinfo{year}{1979},
  \bibinfo{journal}{Pure Appl. Geophys.} \textbf{\bibinfo{volume}{117}},
  \bibinfo{pages}{1172}.

\bibitem[{\citenamefont{Moore} \emph{et~al.}(1997)\citenamefont{Moore, Lockner,
  Shengli, Summers, and Byerlee}}]{Moore_etal1997}
\bibinfo{author}{\bibnamefont{Moore}, \bibfnamefont{D.~E.}},
  \bibinfo{author}{\bibfnamefont{D.~A.} \bibnamefont{Lockner}},
  \bibinfo{author}{\bibfnamefont{M.}~\bibnamefont{Shengli}},
  \bibinfo{author}{\bibfnamefont{R.}~\bibnamefont{Summers}}, and
  \bibinfo{author}{\bibfnamefont{J.~D.} \bibnamefont{Byerlee}},
  \bibinfo{year}{1997}, \bibinfo{journal}{J. Geophys. Res.}
  \textbf{\bibinfo{volume}{102}}(\bibinfo{number}{B7}),
  \bibinfo{pages}{doi:10.1029/97JB00995}.

\bibitem[{\citenamefont{Morgan}(1999)}]{Morgan1999}
\bibinfo{author}{\bibnamefont{Morgan}, \bibfnamefont{J.~K.}},
  \bibinfo{year}{1999}, \bibinfo{journal}{J. Geophys. Res.}
  \textbf{\bibinfo{volume}{104}}(\bibinfo{number}{B2}),
  \bibinfo{pages}{doi:10.1029/1998JB900055}.

\bibitem[{\citenamefont{Mori and Kawamura}(2005)}]{MK05}
\bibinfo{author}{\bibnamefont{Mori}, \bibfnamefont{T.}}, and
  \bibinfo{author}{\bibfnamefont{H.}~\bibnamefont{Kawamura}},
  \bibinfo{year}{2005}, \bibinfo{journal}{Phys. Rev. Lett.}
  \textbf{\bibinfo{volume}{94}}, \bibinfo{pages}{058501}.

\bibitem[{\citenamefont{Mori and Kawamura}(2006)}]{MK06}
\bibinfo{author}{\bibnamefont{Mori}, \bibfnamefont{T.}}, and
  \bibinfo{author}{\bibfnamefont{H.}~\bibnamefont{Kawamura}},
  \bibinfo{year}{2006}, \bibinfo{journal}{J. Geophys. Res.}
  \textbf{\bibinfo{volume}{111}}, \bibinfo{pages}{doi:10.1029/2005JB003942}.

\bibitem[{\citenamefont{Mori and Kawamura}(2008{\natexlab{a}})}]{MK2D07}
\bibinfo{author}{\bibnamefont{Mori}, \bibfnamefont{T.}}, and
  \bibinfo{author}{\bibfnamefont{H.}~\bibnamefont{Kawamura}},
  \bibinfo{year}{2008}{\natexlab{a}}, \bibinfo{journal}{J. Geophys. Res.}
  \textbf{\bibinfo{volume}{113}}, \bibinfo{pages}{doi:10.1029/2007JB005219}.

\bibitem[{\citenamefont{Mori and Kawamura}(2008{\natexlab{b}})}]{MKLR08}
\bibinfo{author}{\bibnamefont{Mori}, \bibfnamefont{T.}}, and
  \bibinfo{author}{\bibfnamefont{H.}~\bibnamefont{Kawamura}},
  \bibinfo{year}{2008}{\natexlab{b}}, \bibinfo{journal}{Phys. Rev. E}
  \textbf{\bibinfo{volume}{77}}, \bibinfo{pages}{051123}.

\bibitem[{\citenamefont{Mori and Kawamura}(2008{\natexlab{c}})}]{MKcontinuum08}
\bibinfo{author}{\bibnamefont{Mori}, \bibfnamefont{T.}}, and
  \bibinfo{author}{\bibfnamefont{H.}~\bibnamefont{Kawamura}},
  \bibinfo{year}{2008}{\natexlab{c}}, \bibinfo{journal}{J. Geophys. Res.}
  \textbf{\bibinfo{volume}{113}}, \bibinfo{pages}{doi:10.1029/2008JB005725}.

\bibitem[{\citenamefont{Morimoto and Kawamura}(2011)}]{MorimotoKawamura}
\bibinfo{author}{\bibnamefont{Morimoto}, \bibfnamefont{S.}}, and
  \bibinfo{author}{\bibfnamefont{H.}~\bibnamefont{Kawamura}},
  \bibinfo{year}{2011}, \bibinfo{journal}{in preparation} .

\bibitem[{\citenamefont{Moser} \emph{et~al.}(1991)\citenamefont{Moser,
  Kert\'{e}sz, and Wolf}}]{bkc55}
\bibinfo{author}{\bibnamefont{Moser}, \bibfnamefont{K.}},
  \bibinfo{author}{\bibfnamefont{J.}~\bibnamefont{Kert\'{e}sz}}, and
  \bibinfo{author}{\bibfnamefont{D.~E.} \bibnamefont{Wolf}},
  \bibinfo{year}{1991}, \bibinfo{journal}{Physica A}
  \textbf{\bibinfo{volume}{178}}, \bibinfo{pages}{215}.

\bibitem[{\citenamefont{Mousseau}(1996)}]{Mousseau}
\bibinfo{author}{\bibnamefont{Mousseau}, \bibfnamefont{N.}},
  \bibinfo{year}{1996}, \bibinfo{journal}{Phys. Rev. Lett.}
  \textbf{\bibinfo{volume}{77}}, \bibinfo{pages}{968}.

\bibitem[{\citenamefont{Myers and Langer}(1993)}]{ML93}
\bibinfo{author}{\bibnamefont{Myers}, \bibfnamefont{C.~R.}}, and
  \bibinfo{author}{\bibfnamefont{J.~S.} \bibnamefont{Langer}},
  \bibinfo{year}{1993}, \bibinfo{journal}{Phys. Rev. E}
  \textbf{\bibinfo{volume}{47}}, \bibinfo{pages}{3048}.

\bibitem[{\citenamefont{Myers} \emph{et~al.}(1996)\citenamefont{Myers, Shaw,
  and Langer}}]{MLS96}
\bibinfo{author}{\bibnamefont{Myers}, \bibfnamefont{C.~R.}},
  \bibinfo{author}{\bibfnamefont{B.~E.} \bibnamefont{Shaw}}, and
  \bibinfo{author}{\bibfnamefont{J.~S.} \bibnamefont{Langer}},
  \bibinfo{year}{1996}, \bibinfo{journal}{Phys. Rev. Lett.}
  \textbf{\bibinfo{volume}{77}}, \bibinfo{pages}{972}.

\bibitem[{\citenamefont{Nadeau and Johnson}(1998)}]{NadeauJohnson1998}
\bibinfo{author}{\bibnamefont{Nadeau}, \bibfnamefont{R.~M.}}, and
  \bibinfo{author}{\bibfnamefont{L.~R.} \bibnamefont{Johnson}},
  \bibinfo{year}{1998}, \bibinfo{journal}{Bull. Seismol. Soc. Am.}
  \textbf{\bibinfo{volume}{88}}, \bibinfo{pages}{790}.

\bibitem[{\citenamefont{Nakanishi}(1990)}]{Nakanishi}
\bibinfo{author}{\bibnamefont{Nakanishi}, \bibfnamefont{H.}},
  \bibinfo{year}{1990}, \bibinfo{journal}{Phys. Rev. A}
  \textbf{\bibinfo{volume}{41}}, \bibinfo{pages}{7086}.

\bibitem[{\citenamefont{Nakatani}(2001)}]{Nakatani2001}
\bibinfo{author}{\bibnamefont{Nakatani}, \bibfnamefont{M.}},
  \bibinfo{year}{2001}, \bibinfo{journal}{J. Geophys. Res.}
  \textbf{\bibinfo{volume}{106}}(\bibinfo{number}{B7}),
  \bibinfo{pages}{doi:10.1029/2000JB900453}.

\bibitem[{\citenamefont{Nielsen} \emph{et~al.}(2008)\citenamefont{Nielsen, {Di
  Toro}, Hirose, and Shimamoto}}]{Nielsen2008}
\bibinfo{author}{\bibnamefont{Nielsen}, \bibfnamefont{S.}},
  \bibinfo{author}{\bibfnamefont{G.}~\bibnamefont{{Di Toro}}},
  \bibinfo{author}{\bibfnamefont{T.}~\bibnamefont{Hirose}}, and
  \bibinfo{author}{\bibfnamefont{T.}~\bibnamefont{Shimamoto}},
  \bibinfo{year}{2008}, \bibinfo{journal}{J. Geophys. Res.}
  \textbf{\bibinfo{volume}{113}}, \bibinfo{pages}{doi:10.1029/2007JB005122}.

\bibitem[{\citenamefont{Nishenko and Buland}(1987)}]{Nishenko}
\bibinfo{author}{\bibnamefont{Nishenko}, \bibfnamefont{S.~P.}}, and
  \bibinfo{author}{\bibfnamefont{R.}~\bibnamefont{Buland}},
  \bibinfo{year}{1987}, \bibinfo{journal}{Bull. Seismol. Soc. Am.}
  \textbf{\bibinfo{volume}{77}}, \bibinfo{pages}{1382}.

\bibitem[{\citenamefont{Obara} \emph{et~al.}(2004)\citenamefont{Obara, Hirose,
  Yamamizu, and Kasahara}}]{Obara_etal2004}
\bibinfo{author}{\bibnamefont{Obara}, \bibfnamefont{K.}},
  \bibinfo{author}{\bibfnamefont{H.}~\bibnamefont{Hirose}},
  \bibinfo{author}{\bibfnamefont{F.}~\bibnamefont{Yamamizu}}, and
  \bibinfo{author}{\bibfnamefont{K.}~\bibnamefont{Kasahara}},
  \bibinfo{year}{2004}, \bibinfo{journal}{Geophys. Res. Lett.}
  \textbf{\bibinfo{volume}{31}}(\bibinfo{number}{L23602}),
  \bibinfo{pages}{doi:10.1029/2004GL020848}.

\bibitem[{\citenamefont{Ogata}(2005)}]{Ogata2005}
\bibinfo{author}{\bibnamefont{Ogata}, \bibfnamefont{Y.}}, \bibinfo{year}{2005},
  \bibinfo{journal}{J. Geophys. Res.}
  \textbf{\bibinfo{volume}{110}}(\bibinfo{number}{B05S06}),
  \bibinfo{pages}{doi:10.1029/2004JB003245}.

\bibitem[{\citenamefont{Ohmura and Kawamura}(2007)}]{OhmuraKawamura}
\bibinfo{author}{\bibnamefont{Ohmura}, \bibfnamefont{A.}}, and
  \bibinfo{author}{\bibfnamefont{H.}~\bibnamefont{Kawamura}},
  \bibinfo{year}{2007}, \bibinfo{journal}{Europhys. Lett.}
  \textbf{\bibinfo{volume}{77}}, \bibinfo{pages}{69001}.

\bibitem[{\citenamefont{Ohnaka and Shen}(1999)}]{OhnakaShen1999}
\bibinfo{author}{\bibnamefont{Ohnaka}, \bibfnamefont{M.}}, and
  \bibinfo{author}{\bibfnamefont{L.}~\bibnamefont{Shen}}, \bibinfo{year}{1999},
  \bibinfo{journal}{J. Geophys. Res.} \textbf{\bibinfo{volume}{104}},
  \bibinfo{pages}{doi:10.1029/1998JB900007}.

\bibitem[{\citenamefont{Okada} \emph{et~al.}(2003)\citenamefont{Okada,
  Matsuzawa, and Hasegawa}}]{Okada_etal2003}
\bibinfo{author}{\bibnamefont{Okada}, \bibfnamefont{T.}},
  \bibinfo{author}{\bibfnamefont{T.}~\bibnamefont{Matsuzawa}}, and
  \bibinfo{author}{\bibfnamefont{A.}~\bibnamefont{Hasegawa}},
  \bibinfo{year}{2003}, \bibinfo{journal}{Earth Planet. Sci. Lett.}
  \textbf{\bibinfo{volume}{213}}, \bibinfo{pages}{361}.

\bibitem[{\citenamefont{Olami} \emph{et~al.}(1992)\citenamefont{Olami, Feder,
  and Christensen}}]{olm92}
\bibinfo{author}{\bibnamefont{Olami}, \bibfnamefont{Z.}},
  \bibinfo{author}{\bibfnamefont{H.}~\bibnamefont{Feder}}, and
  \bibinfo{author}{\bibfnamefont{K.}~\bibnamefont{Christensen}},
  \bibinfo{year}{1992}, \bibinfo{journal}{Phys. Rev. Lett.}
  \textbf{\bibinfo{volume}{68}}, \bibinfo{pages}{1244}.

\bibitem[{\citenamefont{Ozawa} \emph{et~al.}(2002)\citenamefont{Ozawa,
  Murakami, Kaidzu, Tada, Sagiya, Hatanaka, Yarai, and
  Nishimura}}]{Ozawa_etal2002}
\bibinfo{author}{\bibnamefont{Ozawa}, \bibfnamefont{S.}},
  \bibinfo{author}{\bibfnamefont{M.}~\bibnamefont{Murakami}},
  \bibinfo{author}{\bibfnamefont{M.}~\bibnamefont{Kaidzu}},
  \bibinfo{author}{\bibfnamefont{T.}~\bibnamefont{Tada}},
  \bibinfo{author}{\bibfnamefont{T.}~\bibnamefont{Sagiya}},
  \bibinfo{author}{\bibfnamefont{Y.}~\bibnamefont{Hatanaka}},
  \bibinfo{author}{\bibfnamefont{H.}~\bibnamefont{Yarai}}, and
  \bibinfo{author}{\bibfnamefont{T.}~\bibnamefont{Nishimura}},
  \bibinfo{year}{2002}, \bibinfo{journal}{Science}
  \textbf{\bibinfo{volume}{298}}, \bibinfo{pages}{1009}.

\bibitem[{\citenamefont{Pacheco} \emph{et~al.}(1993)\citenamefont{Pacheco,
  Sykes, and Scholz}}]{Pacheco_etal1993}
\bibinfo{author}{\bibnamefont{Pacheco}, \bibfnamefont{J.~F.}},
  \bibinfo{author}{\bibfnamefont{L.~R.} \bibnamefont{Sykes}}, and
  \bibinfo{author}{\bibfnamefont{C.~H.} \bibnamefont{Scholz}},
  \bibinfo{year}{1993}, \bibinfo{journal}{J. Geophys. Res.}
  \textbf{\bibinfo{volume}{98}}(\bibinfo{number}{B8}),
  \bibinfo{pages}{doi:10.1029/93JB00349}.

\bibitem[{\citenamefont{Park and Mori}(2007)}]{ParkMori2007}
\bibinfo{author}{\bibnamefont{Park}, \bibfnamefont{S.-C.}}, and
  \bibinfo{author}{\bibfnamefont{J.}~\bibnamefont{Mori}}, \bibinfo{year}{2007},
  \bibinfo{journal}{J. Geophys. Res.}
  \textbf{\bibinfo{volume}{112}}(\bibinfo{number}{B03302}),
  \bibinfo{pages}{doi:10.1029/2006JB004480}.

\bibitem[{\citenamefont{Peirce}(1926)}]{bkc46}
\bibinfo{author}{\bibnamefont{Peirce}, \bibfnamefont{F.~T.}},
  \bibinfo{year}{1926}, \bibinfo{journal}{J. Text. Ind.}
  \textbf{\bibinfo{volume}{17}}, \bibinfo{pages}{355}.

\bibitem[{\citenamefont{Peixoto and Prado}(2006)}]{Peixoto}
\bibinfo{author}{\bibnamefont{Peixoto}, \bibfnamefont{T.~P.}}, and
  \bibinfo{author}{\bibfnamefont{C.~P.~C.} \bibnamefont{Prado}},
  \bibinfo{year}{2006}, \bibinfo{journal}{Phys. Rev. E}
  \textbf{\bibinfo{volume}{69}}, \bibinfo{pages}{R025101}.

\bibitem[{\citenamefont{Pelletier}(2000)}]{Pelletier}
\bibinfo{author}{\bibnamefont{Pelletier}, \bibfnamefont{J.~D.}},
  \bibinfo{year}{2000}, in \emph{\bibinfo{booktitle}{Geophysical Monograph}},
  volume \bibinfo{volume}{120}, p.~\bibinfo{pages}{27}.

\bibitem[{\citenamefont{Peng and Zhao}(2009)}]{PengZhao2009}
\bibinfo{author}{\bibnamefont{Peng}, \bibfnamefont{Z.~G.}}, and
  \bibinfo{author}{\bibfnamefont{P.}~\bibnamefont{Zhao}}, \bibinfo{year}{2009},
  \bibinfo{journal}{Nature Geosci.} \textbf{\bibinfo{volume}{2}},
  \bibinfo{pages}{877}.

\bibitem[{\citenamefont{Pepke} \emph{et~al.}(1994)\citenamefont{Pepke, Carlson,
  and Shaw}}]{Pepke}
\bibinfo{author}{\bibnamefont{Pepke}, \bibfnamefont{S.~L.}},
  \bibinfo{author}{\bibfnamefont{J.~M.} \bibnamefont{Carlson}}, and
  \bibinfo{author}{\bibfnamefont{B.~E.} \bibnamefont{Shaw}},
  \bibinfo{year}{1994}, \bibinfo{journal}{J. Geophys. Res.}
  \textbf{\bibinfo{volume}{99}}, \bibinfo{pages}{doi:10.1029/93JB03125}.

\bibitem[{\citenamefont{Perfettini and Avouac}(2004)}]{PerfettiniAvouac2004}
\bibinfo{author}{\bibnamefont{Perfettini}, \bibfnamefont{H.}}, and
  \bibinfo{author}{\bibfnamefont{J.-P.} \bibnamefont{Avouac}},
  \bibinfo{year}{2004}, \bibinfo{journal}{J. Geophys. Res.}
  \textbf{\bibinfo{volume}{109}}(\bibinfo{number}{B02304}),
  \bibinfo{pages}{doi:10.1029/2003JB002488}.

\bibitem[{\citenamefont{Perfettini}
  \emph{et~al.}(2010)\citenamefont{Perfettini, Avouac, Tavera, Kositsky,
  Nocquet, Bondoux, Chlieh, Sladen, Audin, Farber, and
  Soler}}]{Perfettini_etal2010}
\bibinfo{author}{\bibnamefont{Perfettini}, \bibfnamefont{H.}},
  \bibinfo{author}{\bibfnamefont{J.-P.} \bibnamefont{Avouac}},
  \bibinfo{author}{\bibfnamefont{H.}~\bibnamefont{Tavera}},
  \bibinfo{author}{\bibfnamefont{A.}~\bibnamefont{Kositsky}},
  \bibinfo{author}{\bibfnamefont{J.-M.} \bibnamefont{Nocquet}},
  \bibinfo{author}{\bibfnamefont{F.}~\bibnamefont{Bondoux}},
  \bibinfo{author}{\bibfnamefont{M.}~\bibnamefont{Chlieh}},
  \bibinfo{author}{\bibfnamefont{A.}~\bibnamefont{Sladen}},
  \bibinfo{author}{\bibfnamefont{L.}~\bibnamefont{Audin}},
  \bibinfo{author}{\bibfnamefont{D.~L.} \bibnamefont{Farber}}, and
  \bibinfo{author}{\bibfnamefont{P.}~\bibnamefont{Soler}},
  \bibinfo{year}{2010}, \bibinfo{journal}{Nature}
  \textbf{\bibinfo{volume}{465}}, \bibinfo{pages}{78}.

\bibitem[{\citenamefont{Petri} \emph{et~al.}(1994)\citenamefont{Petri,
  Vespignani, Alippi, and Constantini}}]{bkc65}
\bibinfo{author}{\bibnamefont{Petri}, \bibfnamefont{A.}},
  \bibinfo{author}{\bibfnamefont{A.}~\bibnamefont{Vespignani}},
  \bibinfo{author}{\bibfnamefont{A.}~\bibnamefont{Alippi}}, and
  \bibinfo{author}{\bibfnamefont{M.}~\bibnamefont{Constantini}},
  \bibinfo{year}{1994}, \bibinfo{journal}{Phys. Rev. Lett.}
  \textbf{\bibinfo{volume}{73}}, \bibinfo{pages}{3423}.

\bibitem[{\citenamefont{Peyrard and Aubry}(1983)}]{bkc21}
\bibinfo{author}{\bibnamefont{Peyrard}, \bibfnamefont{M.}}, and
  \bibinfo{author}{\bibfnamefont{S.}~\bibnamefont{Aubry}},
  \bibinfo{year}{1983}, \bibinfo{journal}{J. Phys. C}
  \textbf{\bibinfo{volume}{16}}, \bibinfo{pages}{1593}.

\bibitem[{\citenamefont{Phoenix}(1978)}]{bkc12}
\bibinfo{author}{\bibnamefont{Phoenix}, \bibfnamefont{S.~L.}},
  \bibinfo{year}{1978}, \bibinfo{journal}{SIAM J. Appl. Math.}
  \textbf{\bibinfo{volume}{34}}, \bibinfo{pages}{227}.

\bibitem[{\citenamefont{Phoenix}(1979)}]{bkc13}
\bibinfo{author}{\bibnamefont{Phoenix}, \bibfnamefont{S.~L.}},
  \bibinfo{year}{1979}, \bibinfo{journal}{Adv. Appl. Prob.}
  \textbf{\bibinfo{volume}{11}}, \bibinfo{pages}{153}.

\bibitem[{\citenamefont{Politi} \emph{et~al.}(2002)\citenamefont{Politi,
  Ciliberto, and Scorretti}}]{bkc33}
\bibinfo{author}{\bibnamefont{Politi}, \bibfnamefont{A.}},
  \bibinfo{author}{\bibfnamefont{S.}~\bibnamefont{Ciliberto}}, and
  \bibinfo{author}{\bibfnamefont{R.}~\bibnamefont{Scorretti}},
  \bibinfo{year}{2002}, \bibinfo{journal}{Phys. Rev. E}
  \textbf{\bibinfo{volume}{66}}, \bibinfo{pages}{026107}.

\bibitem[{\citenamefont{Ponson}(2009)}]{bkc15}
\bibinfo{author}{\bibnamefont{Ponson}, \bibfnamefont{L.}},
  \bibinfo{year}{2009}, \bibinfo{journal}{Phys. Rev. Lett.}
  \textbf{\bibinfo{volume}{103}}, \bibinfo{pages}{055501}.

\bibitem[{\citenamefont{Ponson and Bonamy}(2010)}]{bkc64}
\bibinfo{author}{\bibnamefont{Ponson}, \bibfnamefont{L.}}, and
  \bibinfo{author}{\bibfnamefont{D.}~\bibnamefont{Bonamy}},
  \bibinfo{year}{2010}, \bibinfo{journal}{Int. J. Frac.}
  \textbf{\bibinfo{volume}{162}}, \bibinfo{pages}{21}.

\bibitem[{\citenamefont{Popov} \emph{et~al.}(2010)\citenamefont{Popov, Grzemba,
  Starcevic, and Fabry}}]{Popov2010}
\bibinfo{author}{\bibnamefont{Popov}, \bibfnamefont{V.}},
  \bibinfo{author}{\bibfnamefont{B.}~\bibnamefont{Grzemba}},
  \bibinfo{author}{\bibfnamefont{J.}~\bibnamefont{Starcevic}}, and
  \bibinfo{author}{\bibfnamefont{C.}~\bibnamefont{Fabry}},
  \bibinfo{year}{2010}, \bibinfo{journal}{Physical Mesomechanics}
  \textbf{\bibinfo{volume}{13}}(\bibinfo{number}{5-6}), \bibinfo{pages}{283}.

\bibitem[{\citenamefont{Pradhan} \emph{et~al.}(2002)\citenamefont{Pradhan,
  Bhattacharyya, and Chakrabarti}}]{bkc9}
\bibinfo{author}{\bibnamefont{Pradhan}, \bibfnamefont{S.}},
  \bibinfo{author}{\bibfnamefont{P.}~\bibnamefont{Bhattacharyya}}, and
  \bibinfo{author}{\bibfnamefont{B.~K.} \bibnamefont{Chakrabarti}},
  \bibinfo{year}{2002}, \bibinfo{journal}{Phys. Rev. E}
  \textbf{\bibinfo{volume}{66}}, \bibinfo{pages}{016116}.

\bibitem[{\citenamefont{Pradhan and Chakrabarti}(2001)}]{bkc8}
\bibinfo{author}{\bibnamefont{Pradhan}, \bibfnamefont{S.}}, and
  \bibinfo{author}{\bibfnamefont{B.~K.} \bibnamefont{Chakrabarti}},
  \bibinfo{year}{2001}, \bibinfo{journal}{Phys. Rev. E}
  \textbf{\bibinfo{volume}{65}}, \bibinfo{pages}{016113}.

\bibitem[{\citenamefont{Pradhan and Chakrabarti}(2003{\natexlab{a}})}]{bkc31}
\bibinfo{author}{\bibnamefont{Pradhan}, \bibfnamefont{S.}}, and
  \bibinfo{author}{\bibfnamefont{B.~K.} \bibnamefont{Chakrabarti}},
  \bibinfo{year}{2003}{\natexlab{a}}, \bibinfo{journal}{Phys. Rev. E}
  \textbf{\bibinfo{volume}{67}}, \bibinfo{pages}{046124}.

\bibitem[{\citenamefont{Pradhan and Chakrabarti}(2003{\natexlab{b}})}]{bkc14}
\bibinfo{author}{\bibnamefont{Pradhan}, \bibfnamefont{S.}}, and
  \bibinfo{author}{\bibfnamefont{B.~K.} \bibnamefont{Chakrabarti}},
  \bibinfo{year}{2003}{\natexlab{b}}, \bibinfo{journal}{Int. J. Mod. Phys. B}
  \textbf{\bibinfo{volume}{17}}, \bibinfo{pages}{5565}.

\bibitem[{\citenamefont{Pradhan} \emph{et~al.}(2003)\citenamefont{Pradhan,
  Chakrabarti, Ray, and Dey}}]{bkc3}
\bibinfo{author}{\bibnamefont{Pradhan}, \bibfnamefont{S.}},
  \bibinfo{author}{\bibfnamefont{B.~K.} \bibnamefont{Chakrabarti}},
  \bibinfo{author}{\bibfnamefont{P.}~\bibnamefont{Ray}}, and
  \bibinfo{author}{\bibfnamefont{M.~K.} \bibnamefont{Dey}},
  \bibinfo{year}{2003}, \bibinfo{journal}{Physica Scripta}
  \textbf{\bibinfo{volume}{T106}}, \bibinfo{pages}{77}.

\bibitem[{\citenamefont{Pradhan} \emph{et~al.}(2010)\citenamefont{Pradhan,
  Hansen, and Chakrabarti}}]{bkc34}
\bibinfo{author}{\bibnamefont{Pradhan}, \bibfnamefont{S.}},
  \bibinfo{author}{\bibfnamefont{A.}~\bibnamefont{Hansen}}, and
  \bibinfo{author}{\bibfnamefont{B.~K.} \bibnamefont{Chakrabarti}},
  \bibinfo{year}{2010}, \bibinfo{journal}{Rev. Mod. Phys.}
  \textbf{\bibinfo{volume}{82}}, \bibinfo{pages}{499}.

\bibitem[{\citenamefont{Pradhan} \emph{et~al.}(2005)\citenamefont{Pradhan,
  Hansen, and Hemmer}}]{bkc35}
\bibinfo{author}{\bibnamefont{Pradhan}, \bibfnamefont{S.}},
  \bibinfo{author}{\bibfnamefont{A.}~\bibnamefont{Hansen}}, and
  \bibinfo{author}{\bibfnamefont{P.~C.} \bibnamefont{Hemmer}},
  \bibinfo{year}{2005}, \bibinfo{journal}{Phys. Rev. Lett.}
  \textbf{\bibinfo{volume}{95}}, \bibinfo{pages}{125501}.

\bibitem[{\citenamefont{Pradhan} \emph{et~al.}(2006)\citenamefont{Pradhan,
  Hansen, and Hemmer}}]{bkc36}
\bibinfo{author}{\bibnamefont{Pradhan}, \bibfnamefont{S.}},
  \bibinfo{author}{\bibfnamefont{A.}~\bibnamefont{Hansen}}, and
  \bibinfo{author}{\bibfnamefont{P.~C.} \bibnamefont{Hemmer}},
  \bibinfo{year}{2006}, \bibinfo{journal}{Phys. Rev. E}
  \textbf{\bibinfo{volume}{74}}, \bibinfo{pages}{016122}.

\bibitem[{\citenamefont{Pradhan and Hemmer}(2009)}]{bkc47}
\bibinfo{author}{\bibnamefont{Pradhan}, \bibfnamefont{S.}}, and
  \bibinfo{author}{\bibfnamefont{P.~C.} \bibnamefont{Hemmer}},
  \bibinfo{year}{2009}, \bibinfo{journal}{Phys. Rev. E}
  \textbf{\bibinfo{volume}{79}}, \bibinfo{pages}{041148}.

\bibitem[{\citenamefont{Pradhan and Hemmer}(2011)}]{bkc52}
\bibinfo{author}{\bibnamefont{Pradhan}, \bibfnamefont{S.}}, and
  \bibinfo{author}{\bibfnamefont{P.~C.} \bibnamefont{Hemmer}},
  \bibinfo{year}{2011}, \bibinfo{journal}{Phys. Rev. E}
  \textbf{\bibinfo{volume}{83}}, \bibinfo{pages}{041116}.

\bibitem[{\citenamefont{Rabinowicz}(1965)}]{Rabinowicz1965}
\bibinfo{author}{\bibnamefont{Rabinowicz}, \bibfnamefont{E.}},
  \bibinfo{year}{1965}, \emph{\bibinfo{title}{Friction and wear of materials}}
  (\bibinfo{publisher}{Wiley New York}).

\bibitem[{\citenamefont{Ramos} \emph{et~al.}(2006)\citenamefont{Ramos,
  Altshuler, and Maloy}}]{Ramos}
\bibinfo{author}{\bibnamefont{Ramos}, \bibfnamefont{O.}},
  \bibinfo{author}{\bibfnamefont{E.}~\bibnamefont{Altshuler}}, and
  \bibinfo{author}{\bibfnamefont{K.~J.} \bibnamefont{Maloy}},
  \bibinfo{year}{2006}, \bibinfo{journal}{Phys. Rev. Lett.}
  \textbf{\bibinfo{volume}{96}}, \bibinfo{pages}{098501}.

\bibitem[{\citenamefont{Ray and Chakrabarti}(1985)}]{bkc60}
\bibinfo{author}{\bibnamefont{Ray}, \bibfnamefont{P.}}, and
  \bibinfo{author}{\bibfnamefont{B.~K.} \bibnamefont{Chakrabarti}},
  \bibinfo{year}{1985}, \bibinfo{journal}{Sol. St. Comm.}
  \textbf{\bibinfo{volume}{53}}, \bibinfo{pages}{477}.

\bibitem[{\citenamefont{Rice}(1993)}]{Rice1993}
\bibinfo{author}{\bibnamefont{Rice}, \bibfnamefont{J.~R.}},
  \bibinfo{year}{1993}, \bibinfo{journal}{J. Geophys. Res.}
  \textbf{\bibinfo{volume}{98}}(\bibinfo{number}{B6}),
  \bibinfo{pages}{doi:10.1029/93JB00191}.

\bibitem[{\citenamefont{Rice}(2006)}]{Rice2006}
\bibinfo{author}{\bibnamefont{Rice}, \bibfnamefont{J.~R.}},
  \bibinfo{year}{2006}, \bibinfo{journal}{J. Geophys. Res.}
  \textbf{\bibinfo{volume}{111}}(\bibinfo{number}{B5}),
  \bibinfo{pages}{doi:10.1029/2005JB004006}.

\bibitem[{\citenamefont{Rice} \emph{et~al.}(2001)\citenamefont{Rice, Lapusta,
  and Ranjith}}]{Rice2001}
\bibinfo{author}{\bibnamefont{Rice}, \bibfnamefont{J.~R.}},
  \bibinfo{author}{\bibfnamefont{N.}~\bibnamefont{Lapusta}}, and
  \bibinfo{author}{\bibfnamefont{K.}~\bibnamefont{Ranjith}},
  \bibinfo{year}{2001}, \bibinfo{journal}{J. Mech. Phys. of Solids}
  \textbf{\bibinfo{volume}{49}}, \bibinfo{pages}{1865}.

\bibitem[{\citenamefont{Rogers and Dragert}(2003)}]{RogersDragert2003}
\bibinfo{author}{\bibnamefont{Rogers}, \bibfnamefont{G.}}, and
  \bibinfo{author}{\bibfnamefont{H.}~\bibnamefont{Dragert}},
  \bibinfo{year}{2003}, \bibinfo{journal}{Science}
  \textbf{\bibinfo{volume}{300}}, \bibinfo{pages}{1942}.

\bibitem[{\citenamefont{Rubin}(2008)}]{Rubin2008}
\bibinfo{author}{\bibnamefont{Rubin}, \bibfnamefont{A.~M.}},
  \bibinfo{year}{2008}, \bibinfo{journal}{J. Geophys. Res.}
  \textbf{\bibinfo{volume}{113}}(\bibinfo{number}{B11414}),
  \bibinfo{pages}{doi:10.1029/2008JB005642}.

\bibitem[{\citenamefont{Rubin and Ampuero}(2005)}]{RubinAmpuero2005}
\bibinfo{author}{\bibnamefont{Rubin}, \bibfnamefont{A.~M.}}, and
  \bibinfo{author}{\bibfnamefont{J.-P.} \bibnamefont{Ampuero}},
  \bibinfo{year}{2005}, \bibinfo{journal}{J. Geophys. Res.}
  \textbf{\bibinfo{volume}{110}}(\bibinfo{number}{B11312}),
  \bibinfo{pages}{doi:10.1029/2005JB003686}.

\bibitem[{\citenamefont{Ruina}(1983)}]{Ruina1983}
\bibinfo{author}{\bibnamefont{Ruina}, \bibfnamefont{A.~L.}},
  \bibinfo{year}{1983}, \bibinfo{journal}{J. Geophys. Res.}
  \textbf{\bibinfo{volume}{88}}(\bibinfo{number}{B12}),
  \bibinfo{pages}{doi:10.1029/JB088iB12p10359}.

\bibitem[{\citenamefont{Rundle} \emph{et~al.}(2003)\citenamefont{Rundle,
  Shcherbakov, Klein, and Sammis}}]{Rundle2003}
\bibinfo{author}{\bibnamefont{Rundle}, \bibfnamefont{D.~L., J.
  B.and~Turcotte}},
  \bibinfo{author}{\bibfnamefont{R.}~\bibnamefont{Shcherbakov}},
  \bibinfo{author}{\bibfnamefont{W.}~\bibnamefont{Klein}}, and
  \bibinfo{author}{\bibfnamefont{C.}~\bibnamefont{Sammis}},
  \bibinfo{year}{2003}, \bibinfo{journal}{Rev. Geophys.}
  \textbf{\bibinfo{volume}{41}}, \bibinfo{pages}{1019}.

\bibitem[{\citenamefont{Rundle} \emph{et~al.}(1995)\citenamefont{Rundle, Klein,
  Gross, and Turcotte}}]{Rundlel95}
\bibinfo{author}{\bibnamefont{Rundle}, \bibfnamefont{J.~B.}},
  \bibinfo{author}{\bibfnamefont{W.}~\bibnamefont{Klein}},
  \bibinfo{author}{\bibfnamefont{S.}~\bibnamefont{Gross}}, and
  \bibinfo{author}{\bibfnamefont{D.~L.} \bibnamefont{Turcotte}},
  \bibinfo{year}{1995}, \bibinfo{journal}{Phys. Rev. Lett.}
  \textbf{\bibinfo{volume}{75}}, \bibinfo{pages}{1658}.

\bibitem[{\citenamefont{Sahimi}(2003)}]{bkc28}
\bibinfo{author}{\bibnamefont{Sahimi}, \bibfnamefont{M.}},
  \bibinfo{year}{2003}, \emph{\bibinfo{title}{{Heterogeneous Materials}}}
  (\bibinfo{publisher}{Vol II, Springer N.Y.}).

\bibitem[{\citenamefont{Santucci} \emph{et~al.}(2007)\citenamefont{Santucci,
  M\aa{}l\o{}y, Delaplace, Mathiesen, Hansen, Haavig~Bakke, Schmittbuhl, Vanel,
  and Ray}}]{bkc61}
\bibinfo{author}{\bibnamefont{Santucci}, \bibfnamefont{S.}},
  \bibinfo{author}{\bibfnamefont{K.~J.} \bibnamefont{M\aa{}l\o{}y}},
  \bibinfo{author}{\bibfnamefont{A.}~\bibnamefont{Delaplace}},
  \bibinfo{author}{\bibfnamefont{J.}~\bibnamefont{Mathiesen}},
  \bibinfo{author}{\bibfnamefont{A.}~\bibnamefont{Hansen}},
  \bibinfo{author}{\bibfnamefont{J.~O.} \bibnamefont{Haavig~Bakke}},
  \bibinfo{author}{\bibfnamefont{J.}~\bibnamefont{Schmittbuhl}},
  \bibinfo{author}{\bibfnamefont{L.}~\bibnamefont{Vanel}}, and
  \bibinfo{author}{\bibfnamefont{P.}~\bibnamefont{Ray}}, \bibinfo{year}{2007},
  \bibinfo{journal}{Phys. Rev. E} \textbf{\bibinfo{volume}{75}},
  \bibinfo{pages}{016104}.

\bibitem[{\citenamefont{Sasamoto and Spohn}(2010)}]{bkc56}
\bibinfo{author}{\bibnamefont{Sasamoto}, \bibfnamefont{T.}}, and
  \bibinfo{author}{\bibfnamefont{H.}~\bibnamefont{Spohn}},
  \bibinfo{year}{2010}, \bibinfo{journal}{Phys. Rev. Lett.}
  \textbf{\bibinfo{volume}{104}}, \bibinfo{pages}{230602}.

\bibitem[{\citenamefont{Schmittbuhl}
  \emph{et~al.}(1996)\citenamefont{Schmittbuhl, Vilotte, and
  Roux}}]{Schmittbuhl}
\bibinfo{author}{\bibnamefont{Schmittbuhl}, \bibfnamefont{J.}},
  \bibinfo{author}{\bibfnamefont{J.-P.} \bibnamefont{Vilotte}}, and
  \bibinfo{author}{\bibfnamefont{S.}~\bibnamefont{Roux}}, \bibinfo{year}{1996},
  \bibinfo{journal}{J. Geophys. Res.} \textbf{\bibinfo{volume}{101}},
  \bibinfo{pages}{doi:10.1029/96JB02294}.

\bibitem[{\citenamefont{Scholz}(1998)}]{sch98}
\bibinfo{author}{\bibnamefont{Scholz}, \bibfnamefont{C.}},
  \bibinfo{year}{1998}, \bibinfo{journal}{Nature}
  \textbf{\bibinfo{volume}{391}}, \bibinfo{pages}{37}.

\bibitem[{\citenamefont{Scholz}(2002)}]{Scholz2002}
\bibinfo{author}{\bibnamefont{Scholz}, \bibfnamefont{C.~H.}},
  \bibinfo{year}{2002}, \emph{\bibinfo{title}{The Mechanics of Earthquakes and
  Faulting, 2nd Ed.}} (\bibinfo{publisher}{Cambridge Univ. Press, N. Y.}).

\bibitem[{\citenamefont{Schwartz and Rokosky}(2007)}]{SchwartzRokosky2007}
\bibinfo{author}{\bibnamefont{Schwartz}, \bibfnamefont{S.~Y.}}, and
  \bibinfo{author}{\bibfnamefont{J.~M.} \bibnamefont{Rokosky}},
  \bibinfo{year}{2007}, \bibinfo{journal}{Rev. Geophys.}
  \textbf{\bibinfo{volume}{45}}(\bibinfo{number}{RG3004}),
  \bibinfo{pages}{doi:10.1029/2006RG000208}.

\bibitem[{\citenamefont{Segall and Davis}(1997)}]{SegallDavis1997}
\bibinfo{author}{\bibnamefont{Segall}, \bibfnamefont{P.}}, and
  \bibinfo{author}{\bibfnamefont{J.~L.} \bibnamefont{Davis}},
  \bibinfo{year}{1997}, \bibinfo{journal}{Annu. Rev. Earth Planet. Sci.}
  \textbf{\bibinfo{volume}{25}}, \bibinfo{pages}{301}.

\bibitem[{\citenamefont{Shaw}(1994)}]{Shaw94}
\bibinfo{author}{\bibnamefont{Shaw}, \bibfnamefont{B.~E.}},
  \bibinfo{year}{1994}, \bibinfo{journal}{Geophys. Res. Lett.}
  \textbf{\bibinfo{volume}{21}}, \bibinfo{pages}{doi:10.1029/94GL01685}.

\bibitem[{\citenamefont{Shaw}(1995)}]{Shaw95}
\bibinfo{author}{\bibnamefont{Shaw}, \bibfnamefont{B.~E.}},
  \bibinfo{year}{1995}, \bibinfo{journal}{J. Geophys. Res.}
  \textbf{\bibinfo{volume}{100}}, \bibinfo{pages}{doi:10.1029/95JB01306}.

\bibitem[{\citenamefont{Shaw} \emph{et~al.}(1992)\citenamefont{Shaw, Carlson,
  and Langer}}]{Shaw92}
\bibinfo{author}{\bibnamefont{Shaw}, \bibfnamefont{B.~E.}},
  \bibinfo{author}{\bibfnamefont{J.~M.} \bibnamefont{Carlson}}, and
  \bibinfo{author}{\bibfnamefont{J.~S.} \bibnamefont{Langer}},
  \bibinfo{year}{1992}, \bibinfo{journal}{J. Geophys. Res.}
  \textbf{\bibinfo{volume}{97}}, \bibinfo{pages}{doi:10.1029/91JB01796}.

\bibitem[{\citenamefont{Shelly}(2009)}]{Shelly2009}
\bibinfo{author}{\bibnamefont{Shelly}, \bibfnamefont{D.~R.}},
  \bibinfo{year}{2009}, \bibinfo{journal}{Geophys. Res. Lett.}
  \textbf{\bibinfo{volume}{36}}(\bibinfo{number}{L17318}),
  \bibinfo{pages}{doi:10.1029/2009GL039589}.

\bibitem[{\citenamefont{Shibazaki and Iio}(2003)}]{ShibazakiIio2003}
\bibinfo{author}{\bibnamefont{Shibazaki}, \bibfnamefont{B.}}, and
  \bibinfo{author}{\bibfnamefont{Y.}~\bibnamefont{Iio}}, \bibinfo{year}{2003},
  \bibinfo{journal}{Geophys. Res. Lett.}
  \textbf{\bibinfo{volume}{30}}(\bibinfo{number}{1489}),
  \bibinfo{pages}{doi:10.1029/2003GL017047}.

\bibitem[{\citenamefont{Shibazaki and
  Shimamoto}(2007)}]{ShibazakiShimamoto2007}
\bibinfo{author}{\bibnamefont{Shibazaki}, \bibfnamefont{B.}}, and
  \bibinfo{author}{\bibfnamefont{T.}~\bibnamefont{Shimamoto}},
  \bibinfo{year}{2007}, \bibinfo{journal}{Geophys. J. Int.}
  \textbf{\bibinfo{volume}{171}}, \bibinfo{pages}{191}.

\bibitem[{\citenamefont{Shimamoto}(1986)}]{Shimamoto1986}
\bibinfo{author}{\bibnamefont{Shimamoto}, \bibfnamefont{T.}},
  \bibinfo{year}{1986}, \bibinfo{journal}{Science}
  \textbf{\bibinfo{volume}{231}}, \bibinfo{pages}{711}.

\bibitem[{\citenamefont{Sibson}(1973)}]{Sibson1973}
\bibinfo{author}{\bibnamefont{Sibson}, \bibfnamefont{R.}},
  \bibinfo{year}{1973}, \bibinfo{journal}{Nature}
  \textbf{\bibinfo{volume}{243}}, \bibinfo{pages}{66}.

\bibitem[{\citenamefont{Sornette}(2004)}]{bkc32}
\bibinfo{author}{\bibnamefont{Sornette}, \bibfnamefont{D.}},
  \bibinfo{year}{2004}, \emph{\bibinfo{title}{Critical phenomena in natural
  sciences, 2nd Ed.}} (\bibinfo{publisher}{Springer Heidelberg, Berlin}).

\bibitem[{\citenamefont{Stauffer and Aharony}(1992)}]{bkc7}
\bibinfo{author}{\bibnamefont{Stauffer}, \bibfnamefont{D.}}, and
  \bibinfo{author}{\bibfnamefont{A.}~\bibnamefont{Aharony}},
  \bibinfo{year}{1992}, \emph{\bibinfo{title}{Introduction to percolation
  theory}} (\bibinfo{publisher}{Taylor \& Francis London}).

\bibitem[{\citenamefont{Stirling} \emph{et~al.}(1996)\citenamefont{Stirling,
  Wesnousky, and Shimazaki}}]{Stirling_etal1996}
\bibinfo{author}{\bibnamefont{Stirling}, \bibfnamefont{M.~W.}},
  \bibinfo{author}{\bibfnamefont{S.~G.} \bibnamefont{Wesnousky}}, and
  \bibinfo{author}{\bibfnamefont{K.}~\bibnamefont{Shimazaki}},
  \bibinfo{year}{1996}, \bibinfo{journal}{Geophys. J. Int.}
  \textbf{\bibinfo{volume}{124}}, \bibinfo{pages}{833}.

\bibitem[{\citenamefont{Stuart}(1988)}]{Stuart1988}
\bibinfo{author}{\bibnamefont{Stuart}, \bibfnamefont{W.~D.}},
  \bibinfo{year}{1988}, \bibinfo{journal}{Pure Appl. Geophys}
  \textbf{\bibinfo{volume}{126}}, \bibinfo{pages}{619}.

\bibitem[{\citenamefont{Stuart and Tullis}(1995)}]{StuartTullis1995}
\bibinfo{author}{\bibnamefont{Stuart}, \bibfnamefont{W.~D.}}, and
  \bibinfo{author}{\bibfnamefont{T.~E.} \bibnamefont{Tullis}},
  \bibinfo{year}{1995}, \bibinfo{journal}{J. Geophys. Res.}
  \textbf{\bibinfo{volume}{100}}(\bibinfo{number}{B12}),
  \bibinfo{pages}{doi:10.1029/95JB02517}.

\bibitem[{\citenamefont{Suyehiro} \emph{et~al.}(1964)\citenamefont{Suyehiro,
  Asada, and Ohtake}}]{Suyeetal64}
\bibinfo{author}{\bibnamefont{Suyehiro}, \bibfnamefont{S.}},
  \bibinfo{author}{\bibfnamefont{T.}~\bibnamefont{Asada}}, and
  \bibinfo{author}{\bibfnamefont{M.}~\bibnamefont{Ohtake}},
  \bibinfo{year}{1964}, \bibinfo{journal}{Pap. Meteorol. Geophys.}
  \textbf{\bibinfo{volume}{19}}, \bibinfo{pages}{427}.

\bibitem[{\citenamefont{Suzuki and Yamashita}(2010)}]{Suzuki2010}
\bibinfo{author}{\bibnamefont{Suzuki}, \bibfnamefont{T.}}, and
  \bibinfo{author}{\bibfnamefont{T.}~\bibnamefont{Yamashita}},
  \bibinfo{year}{2010}, \bibinfo{journal}{J. Geophys. Res.}
  \textbf{\bibinfo{volume}{115}}, \bibinfo{pages}{B02303}.

\bibitem[{\citenamefont{Sykes and Menke}(2006)}]{SykesMenke2006}
\bibinfo{author}{\bibnamefont{Sykes}, \bibfnamefont{L.~R.}}, and
  \bibinfo{author}{\bibfnamefont{W.}~\bibnamefont{Menke}},
  \bibinfo{year}{2006}, \bibinfo{journal}{Bull. Seismol. Soc. Am.}
  \textbf{\bibinfo{volume}{96}}, \bibinfo{pages}{1569}.

\bibitem[{\citenamefont{Tajima and Kanamori}(1985)}]{TajimaKanamori1985}
\bibinfo{author}{\bibnamefont{Tajima}, \bibfnamefont{F.}}, and
  \bibinfo{author}{\bibfnamefont{H.}~\bibnamefont{Kanamori}},
  \bibinfo{year}{1985}, \bibinfo{journal}{Phys. Earth Planet. Inter.}
  \textbf{\bibinfo{volume}{40}}, \bibinfo{pages}{77}.

\bibitem[{\citenamefont{Thatcher}(1990)}]{Thatcher1990}
\bibinfo{author}{\bibnamefont{Thatcher}, \bibfnamefont{W.}},
  \bibinfo{year}{1990}, \bibinfo{journal}{J. Geophys. Res.}
  \textbf{\bibinfo{volume}{95}}, \bibinfo{pages}{doi:10.1029/JB095iB03p02609}.

\bibitem[{\citenamefont{Thompson} \emph{et~al.}(1992)\citenamefont{Thompson,
  Grest, and Robbins}}]{Thompson1992}
\bibinfo{author}{\bibnamefont{Thompson}, \bibfnamefont{P.~A.}},
  \bibinfo{author}{\bibfnamefont{G.~S.} \bibnamefont{Grest}}, and
  \bibinfo{author}{\bibfnamefont{M.~O.} \bibnamefont{Robbins}},
  \bibinfo{year}{1992}, \bibinfo{journal}{Phys. Rev. Lett.}
  \textbf{\bibinfo{volume}{68}}, \bibinfo{pages}{3448}.

\bibitem[{\citenamefont{Toda} \emph{et~al.}(1998)\citenamefont{Toda, Stein,
  Reasenberg, Dieterich, and Yoshida}}]{Toda_etal1998}
\bibinfo{author}{\bibnamefont{Toda}, \bibfnamefont{S.}},
  \bibinfo{author}{\bibfnamefont{R.~S.} \bibnamefont{Stein}},
  \bibinfo{author}{\bibfnamefont{P.~A.} \bibnamefont{Reasenberg}},
  \bibinfo{author}{\bibfnamefont{J.~H.} \bibnamefont{Dieterich}}, and
  \bibinfo{author}{\bibfnamefont{A.}~\bibnamefont{Yoshida}},
  \bibinfo{year}{1998}, \bibinfo{journal}{J. Geophys. Res.}
  \textbf{\bibinfo{volume}{103}}(\bibinfo{number}{B10}),
  \bibinfo{pages}{doi:10.1029/98JB00765}.

\bibitem[{\citenamefont{Tomlinson}(1929)}]{bkc19}
\bibinfo{author}{\bibnamefont{Tomlinson}, \bibfnamefont{G.~A.}},
  \bibinfo{year}{1929}, \bibinfo{journal}{Philos. Mag.}
  \textbf{\bibinfo{volume}{7}}, \bibinfo{pages}{905}.

\bibitem[{\citenamefont{Torvund and Froyland}(1995)}]{Torvund}
\bibinfo{author}{\bibnamefont{Torvund}, \bibfnamefont{F.}}, and
  \bibinfo{author}{\bibfnamefont{J.}~\bibnamefont{Froyland}},
  \bibinfo{year}{1995}, \bibinfo{journal}{Physica Scripta}
  \textbf{\bibinfo{volume}{52}}, \bibinfo{pages}{624}.

\bibitem[{\citenamefont{Tse and Rice}(1986)}]{TseRice1986}
\bibinfo{author}{\bibnamefont{Tse}, \bibfnamefont{S.~T.}}, and
  \bibinfo{author}{\bibfnamefont{J.~R.} \bibnamefont{Rice}},
  \bibinfo{year}{1986}, \bibinfo{journal}{J. Geophys. Res.}
  \textbf{\bibinfo{volume}{91}}(\bibinfo{number}{B9}),
  \bibinfo{pages}{doi:10.1029/JB091iB09p09452}.

\bibitem[{\citenamefont{Tsutsumi and Shimamoto}(1997)}]{Tsutsumi1997}
\bibinfo{author}{\bibnamefont{Tsutsumi}, \bibfnamefont{A.}}, and
  \bibinfo{author}{\bibfnamefont{T.}~\bibnamefont{Shimamoto}},
  \bibinfo{year}{1997}, \bibinfo{journal}{Geophys. Res. Lett.}
  \textbf{\bibinfo{volume}{24}}(\bibinfo{number}{6}),
  \bibinfo{pages}{doi:10.1029/97GL00503}.

\bibitem[{\citenamefont{Tullis}(2009)}]{kawamura-rev2}
\bibinfo{author}{\bibnamefont{Tullis}, \bibfnamefont{T.~E.}},
  \bibinfo{year}{2009}, in \emph{\bibinfo{booktitle}{Earthquake Seismology,
  Treatise on Geophysics}}, edited by
  \bibinfo{editor}{\bibfnamefont{H.}~\bibnamefont{Kanamori}}
  (\bibinfo{publisher}{Elsevier, Amsterdam}), volume~\bibinfo{volume}{4}, p.
  \bibinfo{pages}{131}.

\bibitem[{\citenamefont{Turcotte}(1997)}]{Turcotte1997}
\bibinfo{author}{\bibnamefont{Turcotte}, \bibfnamefont{D.~L.}},
  \bibinfo{year}{1997}, \emph{\bibinfo{title}{Fractals and Chaos in Geology and
  Geophysics, 2nd Ed.}} (\bibinfo{publisher}{Cambridge Univ. Press, N. Y.}).

\bibitem[{\citenamefont{Turcotte} \emph{et~al.}(2009)\citenamefont{Turcotte,
  Shcherbakov, and Rundle}}]{Turcotte_etal2007}
\bibinfo{author}{\bibnamefont{Turcotte}, \bibfnamefont{D.~L.}},
  \bibinfo{author}{\bibfnamefont{R.}~\bibnamefont{Shcherbakov}}, and
  \bibinfo{author}{\bibfnamefont{J.~B.} \bibnamefont{Rundle}},
  \bibinfo{year}{2009}, in \emph{\bibinfo{booktitle}{Treatise on Geophysics
  Vol. 4}}, edited by
  \bibinfo{editor}{\bibfnamefont{H.}~\bibnamefont{Kanamori}}
  (\bibinfo{publisher}{Elsevier, Amsterdam}), pp. \bibinfo{pages}{675--700}.

\bibitem[{\citenamefont{Uchida} \emph{et~al.}(2005)\citenamefont{Uchida,
  Matsuzawaa, Hasegawaa, and Igarashi}}]{Uchida_etal2005}
\bibinfo{author}{\bibnamefont{Uchida}, \bibfnamefont{N.}},
  \bibinfo{author}{\bibfnamefont{T.}~\bibnamefont{Matsuzawaa}},
  \bibinfo{author}{\bibfnamefont{A.}~\bibnamefont{Hasegawaa}}, and
  \bibinfo{author}{\bibfnamefont{T.}~\bibnamefont{Igarashi}},
  \bibinfo{year}{2005}, \bibinfo{journal}{Earth Planet. Sci. Lett.}
  \textbf{\bibinfo{volume}{233}}, \bibinfo{pages}{155}.

\bibitem[{\citenamefont{Utsu} \emph{et~al.}(1995)\citenamefont{Utsu, Ogata, and
  Matsu’ura}}]{Utsu_etal1995}
\bibinfo{author}{\bibnamefont{Utsu}, \bibfnamefont{T.}},
  \bibinfo{author}{\bibfnamefont{Y.}~\bibnamefont{Ogata}}, and
  \bibinfo{author}{\bibfnamefont{R.~S.} \bibnamefont{Matsu’ura}},
  \bibinfo{year}{1995}, \bibinfo{journal}{J. Phys. Earth}
  \textbf{\bibinfo{volume}{43}}, \bibinfo{pages}{1}.

\bibitem[{\citenamefont{Vannimenus}(2002)}]{bkc44}
\bibinfo{author}{\bibnamefont{Vannimenus}, \bibfnamefont{J.}},
  \bibinfo{year}{2002}, \bibinfo{journal}{Physica A}
  \textbf{\bibinfo{volume}{314}}, \bibinfo{pages}{264}.

\bibitem[{\citenamefont{Vannimenus and Derrida}(2001)}]{bkc43}
\bibinfo{author}{\bibnamefont{Vannimenus}, \bibfnamefont{J.}}, and
  \bibinfo{author}{\bibfnamefont{B.}~\bibnamefont{Derrida}},
  \bibinfo{year}{2001}, \bibinfo{journal}{J. Stat. Phys.}
  \textbf{\bibinfo{volume}{105}}, \bibinfo{pages}{1}.

\bibitem[{\citenamefont{Vasconcelos}(1996)}]{Vasconcelos}
\bibinfo{author}{\bibnamefont{Vasconcelos}, \bibfnamefont{G.~L.}},
  \bibinfo{year}{1996}, \bibinfo{journal}{Phys. Rev. Lett.}
  \textbf{\bibinfo{volume}{76}}, \bibinfo{pages}{4865}.

\bibitem[{\citenamefont{Vieira}(1992)}]{Vieira92}
\bibinfo{author}{\bibnamefont{Vieira}, \bibfnamefont{M.~S.}},
  \bibinfo{year}{1992}, \bibinfo{journal}{Phys. Rev. A}
  \textbf{\bibinfo{volume}{46}}, \bibinfo{pages}{6288}.

\bibitem[{\citenamefont{Vieira}(1996)}]{Vieira96}
\bibinfo{author}{\bibnamefont{Vieira}, \bibfnamefont{M.~S.}},
  \bibinfo{year}{1996}, \bibinfo{journal}{Phys. Rev. E}
  \textbf{\bibinfo{volume}{54}}, \bibinfo{pages}{5925}.

\bibitem[{\citenamefont{Vieira and Lichtenberg}(1996)}]{VieiraLichtenberg}
\bibinfo{author}{\bibnamefont{Vieira}, \bibfnamefont{M.~S.}}, and
  \bibinfo{author}{\bibfnamefont{A.~J.} \bibnamefont{Lichtenberg}},
  \bibinfo{year}{1996}, \bibinfo{journal}{Phys. Rev. E}
  \textbf{\bibinfo{volume}{53}}, \bibinfo{pages}{1441}.

\bibitem[{\citenamefont{Vieira} \emph{et~al.}(1993)\citenamefont{Vieira,
  Vasconcelos, and Nagel}}]{Vieira93}
\bibinfo{author}{\bibnamefont{Vieira}, \bibfnamefont{M.~S.}},
  \bibinfo{author}{\bibfnamefont{G.~L.} \bibnamefont{Vasconcelos}}, and
  \bibinfo{author}{\bibfnamefont{S.~R.} \bibnamefont{Nagel}},
  \bibinfo{year}{1993}, \bibinfo{journal}{Phys. Rev. E}
  \textbf{\bibinfo{volume}{47}}, \bibinfo{pages}{R2221}.

\bibitem[{\citenamefont{Wallace and Beavan}(2006)}]{WallaceBeavan2006}
\bibinfo{author}{\bibnamefont{Wallace}, \bibfnamefont{L.~M.}}, and
  \bibinfo{author}{\bibfnamefont{J.}~\bibnamefont{Beavan}},
  \bibinfo{year}{2006}, \bibinfo{journal}{Geophys. Res. Lett.}
  \textbf{\bibinfo{volume}{33}}(\bibinfo{number}{L10301}),
  \bibinfo{pages}{doi:10.1029/2006GL025775}.

\bibitem[{\citenamefont{Wissel and Drossel}(2006)}]{Drossel}
\bibinfo{author}{\bibnamefont{Wissel}, \bibfnamefont{F.}}, and
  \bibinfo{author}{\bibfnamefont{B.}~\bibnamefont{Drossel}},
  \bibinfo{year}{2006}, \bibinfo{journal}{Phys. Rev. E}
  \textbf{\bibinfo{volume}{74}}, \bibinfo{pages}{066109}.

\bibitem[{\citenamefont{Working~Group}(1995)}]{Working1995}
\bibinfo{author}{\bibnamefont{Working~Group}, \bibfnamefont{C.~E.~P.}},
  \bibinfo{year}{1995}, \bibinfo{journal}{Bull. Seismol. Soc. Am.}
  \textbf{\bibinfo{volume}{85}}, \bibinfo{pages}{378}.

\bibitem[{\citenamefont{Wyss}(1997)}]{Wyss1997}
\bibinfo{author}{\bibnamefont{Wyss}, \bibfnamefont{M.}}, \bibinfo{year}{1997},
  \bibinfo{journal}{Pure Appl. Geophys} \textbf{\bibinfo{volume}{149}},
  \bibinfo{pages}{3}.

\bibitem[{\citenamefont{Wyss} \emph{et~al.}(1981)\citenamefont{Wyss, Klein, and
  Johnston}}]{Wyss_etal1981}
\bibinfo{author}{\bibnamefont{Wyss}, \bibfnamefont{M.}},
  \bibinfo{author}{\bibfnamefont{F.~W.} \bibnamefont{Klein}}, and
  \bibinfo{author}{\bibfnamefont{A.~C.} \bibnamefont{Johnston}},
  \bibinfo{year}{1981}, \bibinfo{journal}{J. Geophys. Res.}
  \textbf{\bibinfo{volume}{86}}(\bibinfo{number}{B5}),
  \bibinfo{pages}{doi:10.1029/JB086iB05p03881}.

\bibitem[{\citenamefont{Xia} \emph{et~al.}(2005)\citenamefont{Xia, Gould,
  Klein, and Rundle}}]{Xia05}
\bibinfo{author}{\bibnamefont{Xia}, \bibfnamefont{J.}},
  \bibinfo{author}{\bibfnamefont{H.}~\bibnamefont{Gould}},
  \bibinfo{author}{\bibfnamefont{W.}~\bibnamefont{Klein}}, and
  \bibinfo{author}{\bibfnamefont{J.~B.} \bibnamefont{Rundle}},
  \bibinfo{year}{2005}, \bibinfo{journal}{Phys. Rev. Lett.}
  \textbf{\bibinfo{volume}{95}}, \bibinfo{pages}{248501}.

\bibitem[{\citenamefont{Xia} \emph{et~al.}(2008)\citenamefont{Xia, Gould,
  Klein, and Rundle}}]{Xia08}
\bibinfo{author}{\bibnamefont{Xia}, \bibfnamefont{J.}},
  \bibinfo{author}{\bibfnamefont{H.}~\bibnamefont{Gould}},
  \bibinfo{author}{\bibfnamefont{W.}~\bibnamefont{Klein}}, and
  \bibinfo{author}{\bibfnamefont{J.~B.} \bibnamefont{Rundle}},
  \bibinfo{year}{2008}, \bibinfo{journal}{Phys. Rev. E}
  \textbf{\bibinfo{volume}{77}}, \bibinfo{pages}{031132}.

\bibitem[{\citenamefont{Yagi} \emph{et~al.}(2003)\citenamefont{Yagi, Kikuchi,
  and Nishimura}}]{Yagi_etal2003}
\bibinfo{author}{\bibnamefont{Yagi}, \bibfnamefont{Y.}},
  \bibinfo{author}{\bibfnamefont{M.}~\bibnamefont{Kikuchi}}, and
  \bibinfo{author}{\bibfnamefont{T.}~\bibnamefont{Nishimura}},
  \bibinfo{year}{2003}, \bibinfo{journal}{Geophys. Res. Lett.}
  \textbf{\bibinfo{volume}{30}}(\bibinfo{number}{2177}),
  \bibinfo{pages}{doi:10.1029/2003GL018189}.

\bibitem[{\citenamefont{Yamamoto and Kawamura}(2011)}]{YamamotoKawamura}
\bibinfo{author}{\bibnamefont{Yamamoto}, \bibfnamefont{T.}}, and
  \bibinfo{author}{\bibfnamefont{H.}~\bibnamefont{Kawamura}},
  \bibinfo{year}{2011}, \bibinfo{journal}{in preparation} .

\bibitem[{\citenamefont{Yamamoto} \emph{et~al.}(2010)\citenamefont{Yamamoto,
  Yoshino, and Kawamura}}]{Yamamoto}
\bibinfo{author}{\bibnamefont{Yamamoto}, \bibfnamefont{T.}},
  \bibinfo{author}{\bibfnamefont{H.}~\bibnamefont{Yoshino}}, and
  \bibinfo{author}{\bibfnamefont{H.}~\bibnamefont{Kawamura}},
  \bibinfo{year}{2010}, \bibinfo{journal}{Eur. J. Phys. B}
  \textbf{\bibinfo{volume}{77}}, \bibinfo{pages}{559}.

\bibitem[{\citenamefont{Yamanaka and Kikuchi}(2004)}]{YamanakaKikuchi2004}
\bibinfo{author}{\bibnamefont{Yamanaka}, \bibfnamefont{Y.}}, and
  \bibinfo{author}{\bibfnamefont{M.}~\bibnamefont{Kikuchi}},
  \bibinfo{year}{2004}, \bibinfo{journal}{J. Geophys. Res.}
  \textbf{\bibinfo{volume}{109}}(\bibinfo{number}{B07307}),
  \bibinfo{pages}{doi:10.1029/2003JB002683}.

\bibitem[{\citenamefont{Yamashita and Knopoff}(1987)}]{YamashitaKnopoff1987}
\bibinfo{author}{\bibnamefont{Yamashita}, \bibfnamefont{T.}}, and
  \bibinfo{author}{\bibfnamefont{L.}~\bibnamefont{Knopoff}},
  \bibinfo{year}{1987}, \bibinfo{journal}{Geophys. J. R. Astr. Soc.}
  \textbf{\bibinfo{volume}{91}}, \bibinfo{pages}{13}.

\bibitem[{\citenamefont{Yoshida and Kato}(2003)}]{YoshidaKato2003}
\bibinfo{author}{\bibnamefont{Yoshida}, \bibfnamefont{S.}}, and
  \bibinfo{author}{\bibfnamefont{N.}~\bibnamefont{Kato}}, \bibinfo{year}{2003},
  \bibinfo{journal}{Geophys. Res. Lett.}
  \textbf{\bibinfo{volume}{30}}(\bibinfo{number}{1681}),
  \bibinfo{pages}{doi:10.1029/2003GL017439}.

\bibitem[{\citenamefont{Yoshioka}(1997)}]{Yoshioka1997}
\bibinfo{author}{\bibnamefont{Yoshioka}, \bibfnamefont{N.}},
  \bibinfo{year}{1997}, \bibinfo{journal}{Tectonophysics}
  \textbf{\bibinfo{volume}{277}}, \bibinfo{pages}{29}.

\bibitem[{\citenamefont{Zapperi} \emph{et~al.}(1997)\citenamefont{Zapperi, Ray,
  Stanley, and Vespignani}}]{bkc62}
\bibinfo{author}{\bibnamefont{Zapperi}, \bibfnamefont{S.}},
  \bibinfo{author}{\bibfnamefont{P.}~\bibnamefont{Ray}},
  \bibinfo{author}{\bibfnamefont{H.~E.} \bibnamefont{Stanley}}, and
  \bibinfo{author}{\bibfnamefont{A.}~\bibnamefont{Vespignani}},
  \bibinfo{year}{1997}, \bibinfo{journal}{Phys. Rev. Lett.}
  \textbf{\bibinfo{volume}{78}}, \bibinfo{pages}{1408}.

\end{thebibliography}

\end{document}